%% file: main.tex
\documentclass[conference]{IEEEtran}

\usepackage{xspace}
\usepackage{xparse}
\let\labelindent\relax
\usepackage{enumitem}
\usepackage{mathtools}
\usepackage{tikz}
\usetikzlibrary{arrows}
\usetikzlibrary{patterns}
\usepackage{amsthm}
\usepackage{amsmath}

\usepackage{stackrel}
\usepackage{pstricks}
\usepackage{subcaption}

\usepackage{wrapfig}
\usepackage[utf8]{inputenc}
\usepackage{pgfplots}
\usepackage{booktabs}
\usepackage{expl3}
\usepackage{collcell}
\usepackage{xurl}
\usepackage[hidelinks]{hyperref}
\usepackage{breakurl}
\usepackage{graphicx}
\usepackage{multirow}
\usepackage{framed}
\usepackage{txfonts}
\usepackage{colortbl} 

\input{macros}

\begin{document}

\title{Bernoulli Honeywords}

\author{
  \IEEEauthorblockN{Ke Coby Wang}
  \IEEEauthorblockA{Duke University\\
     coby.wang@ieee.org}
  \and
  \IEEEauthorblockN{Michael K.\ Reiter}
  \IEEEauthorblockA{Duke University\\
    michael.reiter@duke.edu}
}

\IEEEoverridecommandlockouts
\makeatletter\def\@IEEEpubidpullup{6.5\baselineskip}\makeatother
\IEEEpubid{\parbox{\columnwidth}{
    Network and Distributed System Security (NDSS) Symposium 2024\\
    26 February - 1 March 2024, San Diego, CA, USA\\
    ISBN 1-891562-93-2\\
    https://dx.doi.org/10.14722/ndss.2024.23295\\
    www.ndss-symposium.org
}
\hspace{\columnsep}\makebox[\columnwidth]{}}

\maketitle

\begin{abstract}
Decoy passwords, or ``honeywords,'' planted in a credential database
can alert a site to its breach if ever submitted in a login attempt.
To be effective, some honeywords must appear at least as likely to be
user-chosen passwords as the real ones, and honeywords must be very
difficult to guess without having breached the database, to prevent
false breach alarms.  These goals have proved elusive, however, for
heuristic honeyword generation algorithms.  In this paper we explore
an alternative strategy in which the defender treats honeyword
selection as a Bernoulli process in which each possible password
(except the user-chosen one) is selected as a honeyword independently
with a fixed probability.  We show how Bernoulli honeywords can be
integrated into two existing system designs for leveraging honeywords:
one based on a honeychecker that stores the secret index of the
user-chosen password in the list of account passwords, and another
that does not leverage secret state at all.  We show that Bernoulli
honeywords enable analytic derivation of false breach-detection
probabilities irrespective of what information the attacker gathers
about the sites' users; that their true and false breach-detection
probabilities demonstrate compelling efficacy; and that they can even
enable performance improvements in modern honeyword system designs.
\end{abstract}

\section{Introduction}
\label{sec:intro}

In the Colonial Pipeline ransomware attack in May 2021, unauthorized
access to the company network was gained via an employee's VPN account
with a (complicated) password that was found in a password database
breached from a different site~\cite{ikeda2021:colonial,
  culafi2021:mandiant}.  After gaining access, the attackers went on
to disable part of the company's network and demanded a \$5 million
ransom to recover it, leading to fuel shortages across the U.S.\ and
emergency declarations in a number of states.

This example was not an isolated incident.  Breached credential
databases are the source of most passwords leveraged in credential
stuffing campaigns~\cite{thomas2017:credential}, which are themselves
the cause of the vast majority of account
takeovers~\cite{shape2018:spill}, owing to the tendency of users to
reuse passwords across sites~\cite{das2014:tangled,
  pearman2017:habitat, wang2018:domino}.  Password managers (PMs) can
mitigate password reuse, but they are not a panacea: A recent
survey~\cite{mayer2022:managers} at a U.S.\ university found that
though respondents using \textit{third-party} PMs were less likely to
reuse passwords, 47\% still did so.  Moreover, third-party PMs were
less prevalent than password-management strategies more prone to
password reuse---e.g., 77\% and 84\% of OS- or browser-built-in PM
users reported reusing passwords---yielding an overall password-reuse
rate of 77\%.  Attackers often use leaked credentials to harvest vast
numbers of accounts from sites, e.g.,~\cite{reuter2016:alibaba}.
Discovering a credential database breach takes an average of between
seven~\cite{ibm2021:breach} and fifteen months~\cite{shape2018:spill},
during which time the attacker can access accounts with little
accountability.

To discover credential database breaches more quickly, Juels and
Rivest~\cite{juels2013:honeywords} proposed that (hashes of) decoy
passwords, or \textit{honeywords}, be included alongside the
user-chosen password for each account in the credential database.  In
this way, the attacker attempting to access an account at a site where
it breached the credential database risks alerting the site to its
breach, since only the attacker (and not the user) has any chance of
knowing the honeywords.  To be effective, however, honeyword
generation faces at least two requirements.  The first is
\textit{flatness}, namely that an attacker cannot reliably
guess which password in the set of passwords stored for the account is
the user-chosen one.  The second, and arguably more critical,
requirement (see \secref{sec:related}) is a quantifiable and low
false-alarm rate; i.e., it must be quantifiably difficult for an
attacker who has \textit{not} breached a site's database to guess
honeywords for an account at that site.  The discovery of a credential
database breach is an urgent, disruptive, and costly event, generally
requiring that all passwords be reset and that a breach investigation
commence.  IBM put the average cost of a breach detection and
escalation at \$1.24 million~\cite[p.~16]{ibm2021:breach}.  Without
quantifying the risk of a false alarm and minimizing it, breach
detections will be disregarded by operators.

Meeting these requirements has proved difficult.  Prevailing honeyword
generation techniques, which are based on heuristic methods for
explicitly generating these honeywords, seem unlikely to succeed.  For
example, to render honeywords seemingly as likely as the user-chosen
passwords to an attacker who gathers information about users (perhaps
from the same database it breached to obtain the passwords), recent
advice is that honeywords should include personal
information~\cite{wang2022:honeywords}.  However, state-of-the-art
heuristics to do so come with significant risk of false breach alarms
(see \secref{sec:related}).  Moreover, against an attacker who knows
\textit{more} personal information about users represented in the
breached database than the defender---e.g., if it identifies the same
users in another breached dataset---there seems to be little hope for
achieving flatness on a per-account basis. Even if the defender knows
the same user information as the attacker, it may be reluctant to
increase the exposure of its users' information by leveraging it to
create honeywords.

Here we explore an alternative strategy that implements honeyword
selection as a Bernoulli process in which \textit{each password in the
  password space} (aside from the user-chosen one) is selected
independently with a fixed probability to be a honeyword for this
account.  Intuitively, provided that these honeywords sufficiently
often intersect the passwords that the attacker finds at least as
likely as the user-chosen password, the breach will be detected with
significant probability, particularly as the number of accounts the
attacker accesses grows.  Of course, explicitly selecting and storing
so many honeywords would be intractable, and so we instead select them
implicitly, by randomly configuring a data structure that stores them.

We adapt two existing systems to use this idea: the design of Juels \&
Rivest~\cite{juels2013:honeywords} and a design called
Amnesia~\cite{wang2021:amnesia}.  A central feature of Bernoulli
honeywords is that, unlike honeyword generation heuristics, they
enable analytic estimation of the false and true breach-detection
rates, which we provide.  Critically, \textit{the false
  breach-detection rate is independent of the attacker's knowledge
  about users' passwords}, which we argue is essential for breach
detection.  Moreover, Bernoulli honeywords lead to cost
\textit{improvements} in a feature of Amnesia, namely the
ability for one site to monitor for the entry of its honeywords at
\textit{other} sites, which is important since stuffing breached
credentials at another site enables the attacker to identify the
user-chosen password if it was reused there.

To summarize, our contributions are as follows.
\begin{itemize}[nosep,leftmargin=1em,labelwidth=*,align=left]
\item We explore the use of \textit{Bernoulli honeywords} that are not
  constructed but rather are sampled independently from all possible
  passwords.  Bernoulli honeywords do not depend on
  honeyword-generation heuristics that, we argue, will continue to
  struggle against attackers who know as much or more information
  about a site's users than the site does.
\item We describe a realization of Bernoulli honeywords in the
  original honeyword system design of Juels \&
  Rivest~\cite{juels2013:honeywords}.  We analyze
  the false- and true-detection probabilities for this construction,
  showing that it can be highly effective as a breach-detection
  mechanism.  In particular, the false-detection probability is
  independent of the attacker's knowledge about the site's users
  (including even their passwords), which we argue is essential.
\item We describe a second realization of Bernoulli honeywords in
  Amnesia~\cite{wang2021:amnesia}, a system design that can detect
  breaches without requiring that any secret state survives the
  breach.  We analytically estimate the false- and true-detection
  probabilities of this design, as well; again, the former is
  independent of the attacker's knowledge about the site's users.  We
  further show that our design accommodates a site monitoring for
  entry of its honeywords at remote sites, at an expense that is lower
  than in Amnesia in several important measures.
\end{itemize}

\section{Background and Related Work}
\label{sec:related}

\myparagraph{Honeyword system designs}
We are aware of only a handful of system designs for detecting
password database breaches using honeywords.  We separate these
proposals into two camps: \textit{asymmetric} and \textit{symmetric}.
Asymmetric designs leverage an information asymmetry between the
defender and the attacker, in the form of a secret datum that the site
stores but assumes will not be captured by the attacker when he
breaches the site.  As we will discuss in \secref{sec:honeychecker},
in the original proposal of Juels \&
Rivest~\cite{juels2013:honeywords} this information is the index of
the user-chosen password within a list of passwords per account,
stored in an unbreachable \textit{honeychecker}.  In
Lethe~\cite{dionysiou2022:lethe}, this secret is the seed to a
pseudorandom number generator that is used to service logins within an
interval of time and then later reused by an unbreachable
\textit{checking server} to detect entry of a honeyword.  A third
example is that of Almeshekah et al.~\cite{almeshekah2015:ersatz},
which leverages a machine-dependent function for password hashing; if
the attacker who breaches the site is unaware of this design, then its
attempts to crack the database offline will yield decoy passwords that
the site can detect using its unbreachable
machine-dependent function.

We know of only one symmetric design to date, which is
Amnesia~\cite{wang2021:amnesia}.  Amnesia permits the attacker to
learn the entire state of a breached site.  In exchange for allowing
this, Amnesia enables the site to detect its breach only as legitimate
users log into accounts that the attacker previously logged into.
Moreover, detection is only probabilistic, though as the number of
accounts the attacker logs into grows, the probability of breach
detection also grows.  We defer a detailed introduction of Amnesia to
\secref{sec:amnesia}.

Bernoulli honeywords can be used in both asymmetric and symmetric
designs, as we will demonstrate by realizing them within both the
original honeychecker design~\cite{juels2013:honeywords}, which is
asymmetric, and Amnesia~\cite{wang2021:amnesia}, which is symmetric.
In both cases, the integration reveals the need for careful additional
analysis, which we provide.

\myparagraph{Honeyword generation heuristics}
A mostly distinct line of research (e.g.,~\cite{erguler2016:flatness,
  chakraborty2016:honeyword, wang2018:honeywords,
  akshima2019:honeywords, wang2022:honeywords}) has developed on
generating honeywords to be flat, so that the attacker cannot easily
select the user-chosen password from the passwords associated with an
account in the breached database.  One of the most difficult aspects
of ensuring flatness is that users tend to incorporate personal
information in their passwords (birth year, favorite team mascot,
etc.).  An attacker who can mine such information about users that the
site does not take into account in generating honeywords will
generally be able to distinguish the user-chosen password from the
decoys by selecting the one that includes personal
information~\cite{wang2018:honeywords}. Recent progress has therefore
advocated that personal information be incorporated into
honeywords~\cite{wang2022:honeywords}.

We contend that trying to match the attacker's
knowledge about users in order to generate flat honeywords for them
might be difficult, at best.  Even a site that knows a considerable
amount of personal information about its users might not wish to risk
further exposure of that information by importing it into the password
(re)setting pipeline or results.  As such, here we explore an approach
different from creating a small number of explicit honeywords via
tuned heuristics.  Instead, our idea here (see \secref{sec:insight})
is to include a fraction of all passwords as honeywords, in the hopes
of there being some that the attacker finds at least as likely to be
user-chosen as the actual ones.

\myparagraph{False alarms in breach detection}
The importance of a quantifiably low false-alarm rate, particularly
for breach detection, was detailed in stark terms by the
Tripwire study~\cite{deblasio2017:tripwire}.  In this study,
researchers worked with an email provider to monitor for logins to
fake email accounts, each used to register a \textit{decoy account}
with the same password at another site.  Any login to an email account
suggested that the site hosting its decoy account had been
breached---assuming the email provider itself had not been
breached---since the only places where that password (or a hash
thereof) existed were the email provider and the site hosting that
decoy account.  Despite DeBlasio, et al.\ disclosing 18 apparent site
breaches (and the Tripwire methodology) to the relevant site
administrators, only one-third responded at all, only one indicated
that it would force a password reset, and none notified their
users~\cite[\secrefstatic{6.3}]{deblasio2017:tripwire}.  DeBlasio et
al.\ concluded, ``a major open question ... is how much (probative,
but not particularly illustrative) evidence produced by an external
monitoring system like Tripwire is needed to convince operators to
act, such as notifying their users and forcing a password
reset''~\cite[\secrefstatic{8}]{deblasio2017:tripwire}.
This is compelling evidence that a quantifiable, tunable
false-alarm rate is the core requirement for breach detection.

Unfortunately, honeyword generation heuristics come with significant
risk of false breach alarms.  For example, the best proposal of Wang
et al.~\cite{wang2022:honeywords}, even after blocklisting $10^5$
common passwords from being selected as either honeywords or
user-chosen passwords, still enables an attacker who has not breached
a site to guess one of only $20$ honeywords for an account with $>6\%$
chance within only 100 online
guesses~\cite[\figrefstatic{4}]{wang2022:honeywords}.  Wang et
al.\ thus speculated that a threshold of three honeywords entered to
raise an alarm might be more appropriate than only one, as a way to
mitigate false-alarm risk. However, modern estimates of online
guessing attacks (see \secref{sec:honeychecker:eval}) suggest that
resilience to $100$ online guesses is $10^4\times$ \textit{too small},
raising doubts as to whether there is any suitable threshold for
honeywords entered that would permit satisfactory quantification of
the resulting false and true breach-detection probabilities for such
heuristic approaches.

Another often overlooked subtlety is that a low false-alarm rate
should be guaranteed even if an attacker learns user-chosen passwords
for some accounts at the target site by other means (e.g., phishing),
or even against legitimate users of the site themselves.  Some
compromised user-chosen passwords and a wholesale breach of the
credential database should not be confused, as the reactions warranted
by each are qualitatively different.  Juels and
Rivest~\cite[\secrefstatic{7.5}]{juels2013:honeywords} propose to
reduce the likelihood of false breach alarms by selecting the
honeywords for an account randomly from a pool of honeywords for that
user-chosen password.  However, we know of no work that has
demonstrated explicit generation of a honeyword pool sufficiently
large to achieve a suitably low false-detection probability in the
face of realistic threats as we do here, particularly against an
attacker knowing user-chosen passwords.  Bernoulli honeywords resolve
this difficulty, ensuring a quantifiable and tunable false
breach-detection probability even in this case.

\section{Bernoulli Honeywords}
\label{sec:insight}

We begin by abstractly describing Bernoulli honeywords and their
properties, in terms of one user at one site.  We defer constructions
of systems using them to later sections.

Let \pwdList{\pwdRank}, $\pwdRank \ge 1$, denote the list of all
allowable passwords ranked in order of non-increasing likelihood for
the user to have chosen at this site \textit{from the perspective of
  the attacker}.  Formally, if \rvPwdRankChosen is a random variable
with value the rank of the password chosen by the user at this site,
then for $\pwdRankChosen < \pwdRankChosenAlt$, $\prob{\rvPwdRankChosen
  = \pwdRankChosen} \ge \prob{\rvPwdRankChosen =
  \pwdRankChosenAlt}$. If the attacker knows little about the user,
then the \pwdList{\cdot} list might be simply a list of passwords
ranked in order of popularity.  However, if the attacker knows
personal information about the user, then this order might reflect
that personal information.  We stress that the defender (the site)
will generally not know the distribution of \rvPwdRankChosen.

At its core, our key idea is simple: Suppose that during password
(re)set, each possible password other than the one the user chose is
selected independently with probability \honeyProb as a honeyword for
the user at this site.  Per account, let \rvHoneywordInd{\pwdRank} be
an indicator random variable such that $\rvHoneywordInd{\pwdRank} = 1$
if and only if either $\rvPwdRankChosen = \pwdRank$ or
\pwdList{\pwdRank} is chosen as a honeyword (with probability
\honeyProb), and $\rvHoneywordInd{\pwdRank} = 0$ otherwise.  We
explore the implications of this idea to two attackers.

\subsection{\underline{R}aising \underline{a}larm \underline{at}tacker
  (\raisingAlarmAttacker)}
\label{sec:insight:raat}

We first consider an attacker that does \textit{not} breach the site
but that wishes to enter a honeyword to induce a (false) alarm at it.
In this threat model, we permit the \raisingAlarmAttacker to know the
user-chosen password and so its rank.  While this permits the
\raisingAlarmAttacker to access the account as the intended user
could---indeed, the \raisingAlarmAttacker might \textit{be} the
intended user---that is not our concern here.  Rather, our concern is
the \raisingAlarmAttacker's ability to input a honeyword despite no
breach having occurred.

Let
\honeyEntryProb{\raisingAlarmAttacker}{\honeyProb}{\nmbrLoginsPerAccount}
denote the probability with which at least one \underline{h}oney\underline{w}ord is
\underline{in}put by a \raisingAlarmAttacker when it attempts
\nmbrLoginsPerAccount distinct logins on one account.  Since any
password other than the user-chosen one that the \raisingAlarmAttacker
enters in a login attempt is a honeyword with probability \honeyProb,
we immediately have
\[
\honeyEntryProb{\raisingAlarmAttacker}{\honeyProb}{\nmbrLoginsPerAccount}
= 1-(1-\honeyProb)^{\nmbrLoginsPerAccount}
\]
Note,
\honeyEntryProb{\raisingAlarmAttacker}{\honeyProb}{\nmbrLoginsPerAccount}
is independent of the distribution of \rvPwdRankChosen.

\subsection{\underline{Br}eaching \underline{at}tacker (\breachingAttacker)}
\label{sec:insight:brat}

The next attacker we consider is one who breaches the credential
database for the site, thereby obtaining the per-account values
\honeywordInd{\pwdRank} taken on by \rvHoneywordInd{\pwdRank} for all
$\pwdRank \ge 1$, and then attempts to access accounts at that site.
For each account,
\[
\cprob{\big}{\rvPwdRankChosen = \pwdRankChosen}{\textstyle{\bigwedge_{\pwdRank}
\rvHoneywordInd{\pwdRank} = \honeywordInd{\pwdRank}}} =
\frac{\prob{\rvPwdRankChosen = \pwdRankChosen \wedge
    \left(\textstyle{\bigwedge_{\pwdRank} \rvHoneywordInd{\pwdRank} = \honeywordInd{\pwdRank}}\right)}}{\prob{\textstyle{\bigwedge_{\pwdRank} \rvHoneywordInd{\pwdRank} = \honeywordInd{\pwdRank}}}}
\]
with the numerator being 
\begin{align*}
  \lefteqn{\prob{\rvPwdRankChosen = \pwdRankChosen \wedge
    \left(\textstyle{\bigwedge_{\pwdRank} \rvHoneywordInd{\pwdRank} = \honeywordInd{\pwdRank}}\right)}} \nonumber\\
  & = \prob{\rvPwdRankChosen = \pwdRankChosen} \times
  \cprob{\big}{\textstyle{\bigwedge_{\pwdRank \neq \pwdRankChosen}
  \rvHoneywordInd{\pwdRank} =
  \honeywordInd{\pwdRank}}}{\rvPwdRankChosen = \pwdRankChosen}
\end{align*}
Now consider distinct ranks \pwdRankChosen and \pwdRankChosenAlt for
which $\honeywordInd{\pwdRankChosen} =
\honeywordInd{\pwdRankChosenAlt} = 1$. Since
$\cprob{\big}{\bigwedge_{\pwdRank \neq \pwdRankChosen}
  \rvHoneywordInd{\pwdRank} =
  \honeywordInd{\pwdRank}}{\rvPwdRankChosen = \pwdRankChosen} =
\cprob{\big}{\bigwedge_{\pwdRank \neq \pwdRankChosenAlt}
  \rvHoneywordInd{\pwdRank} =
  \honeywordInd{\pwdRank}}{\rvPwdRankChosen = \pwdRankChosenAlt}$, we
see
\[
\frac{\cprob{\big}{\rvPwdRankChosen = \pwdRankChosen}{\textstyle{\bigwedge_{\pwdRank} \rvHoneywordInd{\pwdRank} = \honeywordInd{\pwdRank}}}}
     {\cprob{\big}{\rvPwdRankChosen = \pwdRankChosenAlt}{\textstyle{\bigwedge_{\pwdRank} \rvHoneywordInd{\pwdRank} = \honeywordInd{\pwdRank}}}} =
     \frac{\prob{\rvPwdRankChosen = \pwdRankChosen}}{\prob{\rvPwdRankChosen = \pwdRankChosenAlt}}
\]
In other words, observing $\{\honeywordInd{\pwdRank}\}_{\pwdRank\ge
  1}$ helps the \breachingAttacker only in limiting his attention to
those ranks \pwdRank for which $\honeywordInd{\pwdRank} = 1$.  Among
those for which $\honeywordInd{\pwdRank} = 1$, their relative
likelihoods are unchanged by $\{\honeywordInd{\pwdRank}\}_{\pwdRank
  \ge 1}$ from the \breachingAttacker's perspective.

Thus, the best the \breachingAttacker can do is to try to access the
account using the password with lowest rank \pwdRank (i.e., the most
likely password) for which $\honeywordInd{\pwdRank} = 1$.  More
specifically, let
\[
\rvPwdRank{\pwdCount} \defined \min\cset{\Big}{\pwdRank \ge 1}{\textstyle{\sum_{\pwdRankAlt = 1}^{\pwdRank} \rvHoneywordInd{\pwdRankAlt} = \pwdCount}}
\]
be a random variable denoting the minimum rank \pwdRank for which
there are \pwdCount passwords of at most rank \pwdRank that remain
possible as the user-chosen password from the \breachingAttacker's
perspective.  Then, \textit{the best move for the \breachingAttacker
  attempting to access this account without entering a honeyword is to
  login using \pwdList{\pwdRank{1}} where \pwdRank{1} is the value
  taken on by \rvPwdRank{1}}.  When the \breachingAttacker does so,
the probability of it entering the user-chosen password is
\begin{align*}
  \cprob{\Big}{\rvPwdRankChosen \!=\! \pwdRank{1}}{\textstyle{\bigwedge_{\pwdCountIdx=1}^{\pwdCount} \!\rvPwdRank{\pwdCountIdx} \!=\! \pwdRank{\pwdCountIdx}}}
  & = \frac{\prob{\rvPwdRankChosen = \pwdRank{1}}}{\textstyle{\sum_{\pwdCountIdxAlt=1}^{\pwdCount} \prob{\rvPwdRankChosen = \pwdRank{\pwdCountIdxAlt}} + \prob{\rvPwdRankChosen > \pwdRank{\pwdCount}} \honeyProb}}
\end{align*}
which is computed as the ratio of
\begin{align*}
  \prob{\rvPwdRankChosen = \pwdRank{1} \wedge \textstyle{\bigwedge_{\pwdCountIdx=1}^{\pwdCount} \rvPwdRank{\pwdCountIdx} = \pwdRank{\pwdCountIdx}}}
  = \prob{\rvPwdRankChosen = \pwdRank{1}}\honeyProb^{\pwdCount-1}(1\!-\!\honeyProb)^{\pwdRank{\pwdCount}-\pwdCount}
\end{align*}
and
\begin{align}
\prob{\textstyle{\bigwedge_{\pwdCountIdx=1}^{\pwdCount}\! \rvPwdRank{\pwdCountIdx}\!=\!\pwdRank{\pwdCountIdx}}}
& = \begin{array}[t]{@{}l@{}}
  \textstyle{\sum_{\pwdCountIdxAlt=1}^{\pwdCount} \prob{\rvPwdRankChosen = \pwdRank{\pwdCountIdxAlt}} \cprob{\Big}{\textstyle{\bigwedge_{\pwdCountIdx=1}^{\pwdCount} \rvPwdRank{\pwdCountIdx} = \pwdRank{\pwdCountIdx}}}{\rvPwdRankChosen = \pwdRank{\pwdCountIdxAlt}}} \\
  +~\prob{\rvPwdRankChosen > \pwdRank{\pwdCount}} \cprob{\Big}{\textstyle{\bigwedge_{\pwdCountIdx=1}^{\pwdCount} \rvPwdRank{\pwdCountIdx} = \pwdRank{\pwdCountIdx}}}{\rvPwdRankChosen > \pwdRank{\pwdCount}}
\end{array}
\nonumber\\
& = \mathrlap{\begin{array}[t]{@{}l@{}}
  \sum_{\pwdCountIdxAlt=1}^{\pwdCount}\! \prob{\rvPwdRankChosen = \pwdRank{\pwdCountIdxAlt}} \honeyProb^{\pwdCount-1}(1\!-\!\honeyProb)^{\pwdRank{\pwdCount}-\pwdCount} \\
  +~\prob{\rvPwdRankChosen > \pwdRank{\pwdCount}} \honeyProb^{\pwdCount}(1\!-\!\honeyProb)^{\pwdRank{\pwdCount}-\pwdCount}
\end{array}}
\label{eqn:rank-set}
\end{align}
Then, for $\pwdRankSet =
\{\pwdRank{\pwdCountIdx}\}_{\pwdCountIdx=1}^{\pwdCount}$, the
probability of entering a honeyword is
\begin{align}
  \honeyEntryProb{\breachingAttacker}{\honeyProb}{\pwdRankSet}
  & \defined \cprob{\Big}{\rvPwdRankChosen \neq \pwdRank{1}}{\textstyle{\bigwedge_{\pwdCountIdx=1}^{\pwdCount} \rvPwdRank{\pwdCountIdx} = \pwdRank{\pwdCountIdx}}} \nonumber\\
  & = \frac{\sum_{\pwdCountIdxAlt=2}^{\pwdCount} \prob{\rvPwdRankChosen = \pwdRank{\pwdCountIdxAlt}} + \prob{\rvPwdRankChosen > \pwdRank{\pwdCount}} \honeyProb}{\sum_{\pwdCountIdxAlt=1}^{\pwdCount} \prob{\rvPwdRankChosen = \pwdRank{\pwdCountIdxAlt}} + \prob{\rvPwdRankChosen > \pwdRank{\pwdCount}} \honeyProb}
  \label{eqn:hwin-brat}
\end{align}
Below, we refer to an account for which
$\textstyle{\bigwedge_{\pwdCountIdx=1}^{\pwdCount}
  \rvPwdRank{\pwdCountIdx} = \pwdRank{\pwdCountIdx}}$ as an
\pwdRankSet-account for $\pwdRankSet =
\{\pwdRank{\pwdCountIdx}\}_{\pwdCountIdx=1}^{\pwdCount}$.

While in general the defender will not know the distribution of
\rvPwdRankChosen, in the next section we will incorporate known
results from previous research to estimate that distribution when
evaluating the true-detection probabilities for specific system
realizations based on the insights in this section.

\section{Integration with a Honeychecker}
\label{sec:honeychecker}

In this section we present a practical realization of the insights
presented in \secref{sec:insight}, which we obtain by modifying the
original design of Juels \& Rivest~\cite{juels2013:honeywords}.  One
challenge in such adaptations is finding a way to represent the
honeyword status (i.e., honeyword or not) for every possible password
in a compact way.  As we will show, this representation can come with
other consequences to the properties offered by the designs into which
they are integrated.

\subsection{Bloom filters}
\label{sec:honeychecker:Bloom}

An ingredient of our realization below is a Bloom
filter~\cite{bloom1970:space}, which is a data structure for compactly
storing a set of elements.  A Bloom filter supports two operations,
namely element insertion and membership testing.  In brief, a Bloom
filter is defined by a set $\bfHashFnSet =
\{\bfHashFn{\bfHashFnIdx}\}_{\bfHashFnIdx=1}^{\bfNmbrHashFns}$ of
\bfNmbrHashFns uniform hash functions where each
$\bfHashFn{\bfHashFnIdx}: \{0,1\}^{\bfHashFnDomainBits} \rightarrow
\{1,\ldots,\bfHashFnRange\}$, and a set $\bfIndices \subseteq
\{1,\ldots,\bfHashFnRange\}$, initially empty.  To insert an element
\genericElmt into a Bloom filter $\langle \bfHashFnSet,
\bfIndices\rangle$, the set \bfIndices is updated as $\bfIndices \gets
\bfIndices \cup \bfHashFnSet(\genericElmt)$ where
$\bfHashFnSet(\genericElmt) =
\{\bfHashFn{\bfHashFnIdx}(\genericElmt)\}_{\bfHashFnIdx=1}^{\bfNmbrHashFns}$.
A membership test for \genericElmt, denoted $\genericElmt \bfTestIn
\langle \bfHashFnSet, \bfIndices\rangle$, returns true if and only if
$\bfHashFnSet(\genericElmt) \subseteq \bfIndices$.  For simplicity, we
write $\genericElmt \bfIn \langle \bfHashFnSet, \bfIndices\rangle$
when this test returns \boolTrue, and $\genericElmt \bfNotIn \langle
\bfHashFnSet, \bfIndices\rangle$ when it returns \boolFalse.  As such,
\begin{align}
\cprob{\Big}{\genericElmt \bfIn \langle
    \bfHashFnSet,
    \bfIndices\rangle}{\genericElmt \getsr
  \{0,1\}^{\bfHashFnDomainBits}} =
\left(\frac{\setSize{\bfIndices}}{\bfHashFnRange}\right)^{\bfNmbrHashFns}
\label{eqn:honeyProb}
\end{align}
where \setSize{\bfIndices} is the cardinality of \bfIndices and
\getsr denotes uniform sampling.

Note that $\setSize{\bfHashFnSet(\genericElmt)} < \bfNmbrHashFns$ if
$\bfHashFn{\bfHashFnIdx}(\genericElmt) =
\bfHashFn{\bfHashFnIdxAlt}(\genericElmt)$ for some $\bfHashFnIdx \neq
\bfHashFnIdxAlt$.  Below, we will leverage the facts that for
$\genericElmt, \genericElmtAlt \getsr \{0,1\}^{\bfHashFnDomainBits}$,
\begin{align}
  \expv{\setSize{\bfHashFnSet(\genericElmt)}}
  & = \bfHashFnRange - \bfHashFnRange\left(1-\frac{1}{\bfHashFnRange}\right)^{\bfNmbrHashFns} \label{eqn:oneBFInsertion}\\
  \expv{\setSize{\bfHashFnSet(\genericElmt) \cup \bfHashFnSet(\genericElmtAlt)}}
  & = \bfHashFnRange - \bfHashFnRange\left(1-\frac{1}{\bfHashFnRange}\right)^{2\bfNmbrHashFns} \label{eqn:twoBFInsertions}
\end{align}
and, more importantly, the distributions of
\setSize{\bfHashFnSet(\genericElmt)} and
\setSize{\bfHashFnSet(\genericElmt) \cup
  \bfHashFnSet(\genericElmtAlt)} are tightly concentrated around these
expected values
(e.g.,~\cite[\secrefstatic{12.5.3}]{mitzenmacher2005:probability}).
The distribution of \setSize{\bfHashFnSet(\genericElmt) \setminus
  \bfHashFnSet(\genericElmtAlt)}, then, is tightly concentrated around
\begin{align}
\expv{\setSize{\bfHashFnSet(\genericElmt) \setminus
    \bfHashFnSet(\genericElmtAlt)}}
& = \bfHashFnRange\left[\left(1-\frac{1}{\bfHashFnRange}\right)^{\bfNmbrHashFns}
  - \left(1-\frac{1}{\bfHashFnRange}\right)^{2\bfNmbrHashFns}\right]
\label{eqn:bfSetDifference}
\end{align}
by subtracting \eqnref{eqn:oneBFInsertion} from
\eqnref{eqn:twoBFInsertions}.

\subsection{Background on honeycheckers}
\label{sec:honeychecker:background}

Juels and Rivest~\cite{juels2013:honeywords} introduced honeywords in
the context of a design that detected the entry of a honeyword using a
trusted component called a \textit{honeychecker}.  For each account,
the site holds in its credential database a list of (hashes of)
passwords, one user-chosen and the others honeywords.  The index of
the user-chosen password in the list is stored in the honeychecker.
In a login attempt to the account using password \loginPassword, the
attempt fails if \loginPassword (i.e., its hash) is not in the list.
If \loginPassword is in the list, its index in the list is sent to the
honeychecker.  If this index matches the index stored for this account
in the honeychecker, then the login succeeds; otherwise, a breach
alarm is raised.

Because the index of the user-chosen password is what enables
detection of a login by a \breachingAttacker, it is necessary that the
\breachingAttacker not learn the contents of the honeychecker despite
breaching the site.  In other words, the honeychecker must not be
breachable, even if the site is.  

\subsection{Adapting a system using a honeychecker}
\label{sec:honeychecker:design}

To adapt a system leveraging a honeychecker to leverage Bernoulli
honeywords, the site will store a Bloom filter per account to hold
elements in $\{0,1\}^{\bfHashFnDomainBits}$.  The elements stored in
the Bloom filter will be outputs of a password hashing function
$\pwdHashFn: \{0,1\}^\ast \rightarrow \{0,1\}^{\bfHashFnDomainBits}$
modeled as a random oracle~\cite{bellare1993:oracles}.  This hash
function can be salted, as is standard in good password
management~\cite{florencio2014:guide}.  Specifically, for an account
with user-chosen password \userPassword, the Bloom filter $\langle
\bfHashFnSet, \bfIndices\rangle$ is selected first by choosing the
uniform hash functions \bfHashFnSet randomly, and then by choosing
\bfIndices randomly subject to (i) $\pwdHashFn(\userPassword) \bfIn
\langle \bfHashFnSet, \bfIndices\rangle$, and (ii)
$\setSize{\bfIndices} =
(\honeyProb)^{1/\bfNmbrHashFns}\bfHashFnRange$, so that
\eqnref{eqn:honeyProb} equals \honeyProb.  This Bloom filter is stored
for the account at the (potentially breachable) server.

The (unbreachable) honeychecker stores the indices
$\bfHashFnSet(\pwdHashFn(\userPassword))$ for the account.  A login
attempt with password \loginPassword is evaluated as follows:
\[
\loginResult{\loginPassword} =
\left\{\begin{array}{ll}
\loginFailure & \mbox{if $\pwdHashFn(\loginPassword) \bfNotIn \langle
  \bfHashFnSet,
  \bfIndices\rangle$} \\
\loginSuccess & \mbox{if $\bfHashFnSet(\pwdHashFn(\loginPassword)) = \bfHashFnSet(\pwdHashFn(\userPassword))$} \\
\loginAlarm & \mbox{otherwise}
\end{array}\right.
\]
Note that only the honeychecker can decide between \loginSuccess or
\loginAlarm, since only the honeychecker holds
$\bfHashFnSet(\pwdHashFn(\userPassword))$.  That is, the server first
checks for \loginFailure and, if its condition does not hold, the
server invokes the honeychecker with
$\bfHashFnSet(\pwdHashFn(\loginPassword))$ to decide between
\loginSuccess or \loginAlarm.

\subsection{Security against a \raisingAlarmAttacker}
\label{sec:honeychecker:raat}

Since a \raisingAlarmAttacker does not breach the site, the
\raisingAlarmAttacker learns nothing about the Bloom filter $\langle
\bfHashFnSet, \bfIndices\rangle$ for an account except by submitting
passwords in login attempts to the account.  Note that the Bloom
filter is configured to hold each password with probability \honeyProb
(except for the user-chosen one, which it holds with probability
$1.0$).  If the \raisingAlarmAttacker has \nmbrLoginsPerAccount login
attempts per account to enter a honeyword, and performs these attempts
on \nmbrAccountsAttempted accounts, then the false-detection
probability is
\begin{align}
\falseAlarmProb(\nmbrLoginsPerAccount,\nmbrAccountsAttempted) \le
1-\left(1-\honeyEntryProb{\raisingAlarmAttacker}{\honeyProb}{\nmbrLoginsPerAccount}\right)^{\nmbrAccountsAttempted}
\label{eqn:falseAlarmProb}
\end{align}

\subsection{Security against a \breachingAttacker}
\label{sec:honeychecker:brat}

Since a \breachingAttacker breaches the site, it knows the Bloom
filter $\langle \bfHashFnSet, \bfIndices\rangle$ for each account.  As
proved in \secref{sec:insight:brat}, the best the \breachingAttacker
can do to login to an account is to attempt the most likely password
\loginPassword for which $\pwdHashFn(\loginPassword) \bfIn \langle
\bfHashFnSet, \bfIndices\rangle$; i.e., the password $\loginPassword =
\pwdList{\pwdRank}$ for the lowest \pwdRank for which
$\pwdHashFn(\pwdList{\pwdRank})$ \bfIn $\langle \bfHashFnSet,
\bfIndices\rangle$.  If \loginPassword is a honeyword (which happens
with probability
\honeyEntryProb{\breachingAttacker}{\honeyProb}{\pwdRank}), then the
probability of the breach going undetected is the probability that
$\bfHashFnSet(\pwdHashFn(\loginPassword)) =
\bfHashFnSet(\pwdHashFn(\userPassword))$ for the user-chosen password
\userPassword.  Note that
\begin{align*}
& \cprob{\Big}{\bfHashFnSet(\pwdHashFn(\loginPassword)) =
  \bfHashFnSet(\pwdHashFn(\userPassword))}{\pwdHashFn(\loginPassword) \neq \pwdHashFn(\userPassword)} \\
& \le \cprob{\Big}{\bfHashFnSet(\pwdHashFn(\loginPassword)) \subseteq
  \bfHashFnSet(\pwdHashFn(\userPassword))}{\pwdHashFn(\loginPassword) \neq \pwdHashFn(\userPassword)}
 \approx \left(\frac{\expv{\setSize{\bfHashFnSet(\pwdHashFn(\userPassword))}}}{\setSize{\bfIndices}}\right)^{\bfNmbrHashFns}
\end{align*}
Disregarding the possibility that $\pwdHashFn(\loginPassword) =
\pwdHashFn(\userPassword)$ even though $\loginPassword \neq
\userPassword$ (which happens with probability
$2^{-\bfHashFnDomainBits}$), we thus estimate the true-detection
probability for a \breachingAttacker attempting to login to an
\pwdRankSet-account to be:
\begin{align}
  \trueAlarmProb{\pwdRankSet}
  & \approx \honeyEntryProb{\breachingAttacker}{\honeyProb}{\pwdRankSet} ~\times \nonumber\\
  & \hspace{1.5em} \cprob{\Big}{\bfHashFnSet(\pwdHashFn(\loginPassword)) \neq
  \bfHashFnSet(\pwdHashFn(\userPassword))}{\pwdHashFn(\loginPassword) \neq \pwdHashFn(\userPassword)} \nonumber\\
  & \gtrapprox
  \honeyEntryProb{\breachingAttacker}{\honeyProb}{\pwdRankSet} \times
  \left(1-\left(\frac{\expv{\setSize{\bfHashFnSet(\pwdHashFn(\userPassword))}}}{\setSize{\bfIndices}}\right)^{\bfNmbrHashFns}\right)
\label{eqn:honeychecker:trueAlarmProbOne}
\end{align}
with the expected value instantiated as in
\eqnref{eqn:oneBFInsertion}.  Let \pwdRankSet{\accountId} be the rank
set such that account \accountId is an
\pwdRankSet{\accountId}-account.  The true-detection probability for a
\breachingAttacker who attacks accounts $\accountsSubset \subseteq
\accountsSet$, where \accountsSet is the set of accounts at the
breached site, is
\begin{align}
  \trueAlarmProb(\accountsSubset)
  = 1- \prod_{\accountId \in \accountsSubset}
  (1 - \trueAlarmProb{\pwdRankSet{\accountId}})
\label{eqn:honeychecker:trueAlarmProb}
\end{align}
and the \textit{minimum} true-detection probability for
\breachingAttacker who attacks \nmbrAccountsAttempted accounts is
\begin{align}
  \trueAlarmProb(\nmbrAccountsAttempted) = \min_{\accountsSubset \subseteq \accountsSet: \setSize{\accountsSubset} = \nmbrAccountsAttempted} \trueAlarmProb(\accountsSubset)
\label{eqn:honeychecker:minTrueAlarmProb}
\end{align}

\subsection{Security evaluation}
\label{sec:honeychecker:eval}

Having provided closed-form estimates for the false- and
true-detection probabilities in
\secsref{sec:honeychecker:raat}{sec:honeychecker:brat} for our design
in \secref{sec:honeychecker:design}, we now illustrate the efficacy of
this design using empirical data.  To do so, there are two more types
of information we need.

\subsubsection{Bounding $\falseAlarmProb(\nmbrLoginsPerAccount,\nmbrAccountsAttempted)$}
\label{sec:honeychecker:eval:fdp}

The first information we need is an estimate of how
$\falseAlarmProb(\nmbrLoginsPerAccount,\nmbrAccountsAttempted)$ should
be bounded in practice, as this will permit us to select other
parameters of our system to best meet that bound.  To find such an
estimate, we turned to Flor\^{e}ncio et
al.~\cite{florencio2014:guide}, who categorize online guessing attacks
into ``depth-first'' ones that submit many login attempts to few
accounts over time and ``breadth-first'' ones that submit login
attempts to many accounts over time (but necessarily fewer per
account).  Assuming a guessing campaign over a four-month period, they
estimate that an account targeted in a depth-first attack should
withstand $\nmbrLoginsPerAccount = 10^6$ guesses, while an account
included in a breadth-first attack should withstand
$\nmbrLoginsPerAccount = 10^4$
guesses~\cite[\tblrefstatic{5}]{florencio2014:guide}.  While a
\raisingAlarmAttacker in our context is not trying to guess the
user-chosen password for an account (which it might already know), it
is instead trying to guess a honeyword; nevertheless, we take the
characterization of online guessing campaigns by Flor\^{e}ncio et
al.\ as suitable for \raisingAlarmAttackers, as well.  Unfortunately,
Flor\^{e}ncio et al.\ did not attempt to precisely characterize with
what probability an account subjected to \nmbrLoginsPerAccount login
attempts should remain uncompromised, nor did they specify the number
\nmbrAccountsAttempted of attacked accounts that distinguish
breadth-first from depth-first attacks.  As such, noticing that
$\falseAlarmProb(10^6, 10) \approx
\falseAlarmProb(10^4, 1000)$, it is convenient to require
\begin{equation}
  \falseAlarmProb(10^6, 10) \le \fdpBound \mbox{~~and~~}
  \falseAlarmProb(10^4, 1000) \le \fdpBound
\label{eqn:fdpBounds}
\end{equation}
for a fixed $\fdpBound \le 10^{-1}$ as consistent with the analysis of
Flor\^{e}ncio et al.  We stress that this requirement is not per login
attempt or per account, but \textit{per online-guessing campaign}.  If
the campaigns envisioned by these authors (each four months long)
were mounted consecutively, $\fdpBound \le 10^{-1}$ would imply less
than one false detection \textit{every 3 years} ($= \mbox{four months
  per campaign} \times 9 \mbox{ campaigns}$) in expectation.

\subsubsection{Estimating the distribution of \rvPwdRankChosen}
\label{sec:honeychecker:eval:rank}
The second type of information we need is the distribution of
\rvPwdRankChosen, i.e., the random variable that records the rank of
the user-chosen password for an account in the list \pwdList{\cdot} of
passwords ranked in order of the user's likelihood of choosing them,
from a \breachingAttacker's perspective.  We need this distribution to
calculate \eqnref{eqn:hwin-brat}.  We use three sources for
distributions of \rvPwdRankChosen below.

Wang et al.~\cite{wang2016:targeted} studied algorithms to guess user
passwords from personal information about users.  For a breached
Chinese train-ticketing dataset of $129{,}303$ passwords with
accompanying personal information (name, username, national
identification number, phone number, birth date, and email address),
they reported the fraction \wangCracked of accounts they cracked in
half the dataset ($\nmbrAccounts = 64{,}651$ accounts) after training
with the other half, as a function the ($\log_{10}$ of the) number
\passwordGuesses of guesses by the attacker up to $\passwordGuesses
\le 1000$~\cite[\figrefstatic{8}]{wang2016:targeted}.  Due to the
unavailability of Wang et al.'s source data or algorithm
implementations, we extracted the source data underlying this plot
using WebPlotDigitizer~\cite{rohatgi2021:webplotdigitizer} for the
following algorithms listed there: TarGuess-I (\TGI), which utilizes
all of the personal data; TarGuess-I$''$ (\TGIPP), which uses only
name and birth date; TarGuess-I$'''$ (\TGIPPP), which uses only name;
and \PCFG, which leverages no personal information.  We fit
lines\footnote{Password frequencies are sometimes modeled using a Zipf
distribution (e.g.,~\cite{bonneau2012:guessing,
  malone2012:investigating, wang2017:zipf}).  However, we know of no
studies on the effects of attacker knowledge (as we consider here) on
this modeling.  Moreover, our fitting CDFs achieve better $R^2$ and
RMSE measures on both datasets than our attempts using a Zipf
distribution or power regression did.} to each dataset, obtaining
estimates for \prob{\rvPwdRankChosen{\TGI} \le \pwdRank},
\prob{\rvPwdRankChosen{\TGIPP} \le \pwdRank},
\prob{\rvPwdRankChosen{\TGIPPP} \le \pwdRank}, and
\prob{\rvPwdRankChosen{\PCFG} \le \pwdRank}
(\subfigsref{fig:fitting:data}{fig:fitting:lines}).

\input{figures/fig_fitting.tex}

As discussed in \secref{sec:intro}, a motivation for Bernoulli
honeywords is detecting a \breachingAttacker who knows more about
users than the defending site does.  To show this benefit with these
distributions, we allow the \breachingAttacker to attack using \TGI
(the most user information), but the defending site to blocklist
passwords guessable in $10^6$ guesses using \TGIPP, \TGIPPP, or \PCFG
(i.e., less user information), preventing the user from setting such a
password.  We select a per-user blocklist of size $10^6$ in accordance
with blocklist size recommendations (e.g.,~\cite{tan2020:blocklist})
and since passwords guessable in $10^6$ guesses are typically
categorized as \textit{weak} by password strength meters
(e.g.,~\cite{golla2018:psm, xu2021:chunk}).  To estimate the effects
of this blocklisting, we formulate \prob{\rvPwdRankChosen \le
  \pwdRank} using two line segments, one for $\pwdRank \le 10^6$ in
which \prob{\rvPwdRankChosen \le \pwdRank} is suppressed by the
blocklist, and one for $\pwdRank > 10^6$ in which the ground lost when
$\pwdRank \le 10^6$ is recovered.  That is, for blocklisting algorithm
$\blistAlg \in \{\TGIPP, \TGIPPP, \PCFG\}$, we evaluate
$\trueAlarmProb(\nmbrAccountsAttempted)$ using
\begin{align}
  \hspace{-0.75em}
  \prob{\rvPwdRankChosen \le \pwdRank} \!=\!
  \left\{\begin{array}{@{\hspace{0.15em}}l@{\hspace{0.75em}}l@{}}
  \prob{\rvPwdRankChosen{\TGI} \le \pwdRank} - \prob{\rvPwdRankChosen{\blistAlg} \le \pwdRank} & \mbox{if $1\!\le\!\pwdRank\!\le\!10^6$} \\[6pt]
  \begin{array}{@{}r@{}}
    \left(\frac{1-\prob{\rvPwdRankChosen \le 10^6}}{\log_{10}(\pwdRankMax)-6}\right)(\log_{10}(\pwdRank)-6) \\
    +~\prob{\rvPwdRankChosen \le 10^6}
  \end{array}
  & \mbox{if $10^6\!<\!\pwdRank\!\le\!\pwdRankMax$}
  \end{array}\right.
  \label{eqn:wangCracked}
\end{align}
where \pwdRankMax is the minimum \pwdRank satisfying
$\prob{\rvPwdRankChosen{\TGI} \le \pwdRank} = 1$ (see
\figref{fig:fitting:blocklist}).  We concede that this estimate is
likely very rough, but we know of no better way to estimate the
effects of blocklisting on subsequent password choices, i.e., once
one's initial choice has been declined and the user has been
told to avoid including personal information in her password.

The second source from which we estimate \prob{\rvPwdRankChosen \le
  \pwdRank} is Mazurek et al.~\cite{mazurek2013:guessability}, who
studied $>25{,}000$ passwords in use at Carnegie Mellon University
(CMU).  They analyzed these passwords' guessing resistance up to $3.8
\times 10^{14}$ guesses by an ``extensive knowledge'' attacker trained
on a subset of the passwords in use.  This attacker was thus partially
trained on passwords that presumably reflected an affiliation at CMU,
which constitutes a type of personal information.  Fitting a line to
points extracted by WebPlotDigitizer from the CDF for the guessing
success of this attacker against $\nmbrAccounts = 5{,}459$
accounts~\cite[\figrefstatic{7}]{mazurek2013:guessability}, we
obtained an estimate for \prob{\rvPwdRankChosen \le \pwdRank},
also shown in \subfigsref{fig:fitting:data}{fig:fitting:lines}.  

We get a third estimate for $\prob{\rvPwdRankChosen \le \pwdRank}$
from Xu et al.~\cite{xu2021:chunk}, who propose password guessing
models based on ``chunk-level'' password characteristics and show that
their models' guessing accuracies outperform counterparts working at
character-level granularity.  For example, their ``CKL-PCFG'' model
guessed on average 51.2\% more passwords than state-of-the-art PCFG
models, including the PCFG model used by Wang et
al.~\cite{wang2016:targeted}.  Their evaluation included a ``Neopets''
dataset of $\nmbrAccounts = 67{,}672{,}205$ account-password pairs
breached from a virtual pets website.  Since they trained their
password-guessing algorithm using some of the passwords in the Neopets
dataset, it implicitly incorporates some private information (e.g.,
interest in virtual pets) about the site's users.  Again, we estimate
\prob{\rvPwdRankChosen \le \pwdRank} by fitting a line to data points
extracted by WebPlotDigitizer from their subfigure labeled
``CKL-PCFG''~\cite[\figrefstatic{5}]{xu2021:chunk}
(\subfigsref{fig:fitting:data}{fig:fitting:lines}).  We did not
explore blocklisting in tests using the CMU or CKL-PCFG estimates,
though since these passwords were much stronger than the passwords
studied by Wang et al.~\cite{wang2016:targeted}, we will see that they
nevertheless yield far higher estimates for
$\trueAlarmProb(\nmbrAccountsAttempted)$.

\subsubsection{Results}
\label{sec:honeychecker:eval:results}

Note that $\trueAlarmProb(\nmbrAccountsAttempted)$ is itself a random
variable, since it depends on the rank sets $\pwdRankSet{\accountId}$
selected per account \accountId.  Computing statistics of the
distribution of $\trueAlarmProb(\nmbrAccountsAttempted)$ is costly,
however, since it involves summing over values for the set
$\pwdRankSet =
\{\pwdRank{\pwdCountIdx}\}_{\pwdCountIdx=1}^{\pwdCount}$ and
accumulating their probabilities per \eqnref{eqn:rank-set}, with each
\pwdRank{\pwdCountIdx} ranging beyond $10^{30}$ for some of our
datasets in \secref{sec:honeychecker:eval:rank}.  For this reason,
here we simulate results by sampling $\pwdRankSet{\accountId}$ for
each account \accountId, for $\pwdCount = 1000$.  We do this for each
of the datasets described in
\secref{sec:honeychecker:eval:rank}---i.e., sampling
$\pwdRankSet{\accountId}$ for each of $\nmbrAccounts = 64{,}651$,
$\nmbrAccounts = 5{,}459$, and $\nmbrAccounts = 67{,}672{,}205$
accounts in the three datasets---and then simulate the
\breachingAttacker attacking these accounts in increasing order of
\eqnref{eqn:hwin-brat} (and so
\eqnref{eqn:honeychecker:trueAlarmProbOne}).  We repeat this
sample-then-attack experiment for $10{,}000$ trials and report the
fraction in which detection occurred within the first
\nmbrAccountsAttempted accounts attempted as
$\trueAlarmProb(\nmbrAccountsAttempted)$.

We show the results in \figref{fig:roc_hc_cmp}. These curves were
plotted by setting the Bloom-filter dimensions to $\bfNmbrHashFns =
20$ and $\bfHashFnRange = 128$ and then setting \honeyProb so to
ensure \eqnref{eqn:fdpBounds}.  We chose \bfNmbrHashFns and
\bfHashFnRange to get adequate granularity of datapoints for plotting
these curves.  In practice, smaller values \bfHashFnRange and
\bfNmbrHashFns, e.g., $\bfHashFnRange = 64$ or/and $\bfNmbrHashFns =
10$, could ensure roughly the same accuracy.

\input{figures/fig_roc_hc_cmp.tex}

\figref{fig:roc_hc_cmp:tdp} shows true-detection probabilities using
the distributions shown in \figref{fig:fitting}, as a function of the
fraction $\nmbrAccountsAttempted / \nmbrAccounts$ of accounts accessed
by the \breachingAttacker, when $\fdpBound = 10^{-1}$.  This figure
shows that using \eqref{eqn:wangCracked-TGI} for
\prob{\rvPwdRankChosen \le \pwdRank}, the true-detection probability
reaches $0.5$ when the \breachingAttacker accesses $\approx 20.4\%$ of
the accounts, and it reaches $1.0$ when that fraction reaches $\approx
26.1\%$.  We reiterate that since the \breachingAttacker accesses
accounts in increasing order of \eqnref{eqn:hwin-brat}, this curve
represents the best the \breachingAttacker can do to evade detection
subject to the number of accounts he accesses.  The total number of
accessed accounts to make detection likely, however, is somewhat
large, since these passwords are quite weak and so easily predictable
by the \breachingAttacker.  For this reason, blocklisting helps
considerably: e.g., when blocklisting (i.e., using
\eqnref{eqn:wangCracked}) with \TGIPPP, the true-detection probability
passes $0.5$ at only $\approx 13.6\%$, and with \TGIPP, the
probability surpasses $0.5$ at $\approx 6.4\%$.  The results using
\eqnref{eqn:mazurekCracked} or \eqnref{eqn:xuCracked} for
\prob{\rvPwdRankChosen \le \pwdRank} are stronger still: after the
\breachingAttacker accesses only one account, the true-detection
probability is already $\approx 1.0$.  That is, due to the strength of
these datasets, even the accounts with the smallest
\eqnref{eqn:hwin-brat} have rank-sets with the lowest-ranked password
being a honeyword with near certainty.

\input{figures/fig_pwdcount_hc.tex}

\figref{fig:roc_hc_cmp:fdp} shows the impact of constraining the
false-positive probability even more stringently than $\fdpBound =
10^{-1}$.  This graph shows that driving \fdpBound lower decreases the
true-detection probability in some cases.  More specifically, when
$\fdpBound = 10^{-5}$ and for the weakest password distribution we
consider \eqref{eqn:wangCracked-TGI}, a \breachingAttacker can access
more than $65\%$ of the accounts before the breach is detected with
probability $\ge 0.5$.  That said, such a stringent \fdpBound implies
less than one false detection per several millennia, in expectation
(see \secref{sec:honeychecker:eval:fdp}).  The strongest datasets
(\eqnref{eqn:mazurekCracked}, \eqnref{eqn:xuCracked}) withstand even
such a stringent \fdpBound with no impact on true-detection
probability.

Our true-detection results in \figref{fig:roc_hc_cmp} were
computed using $\pwdCount = 1000$, i.e., assuming the
\breachingAttacker had determined the 1000 lowest-ranked passwords
present in the Bloom filter for each account.  A natural question is
whether setting $\pwdCount = 1000$ in our analyses is sufficiently
conservative.  \figref{fig:pwdcount:hc} confirms that it clearly is.
Specifically, this figure demonstrates that true-detection
probabilities decrease noticeably when increasing \pwdCount from
$\pwdCount = 1$ to $\pwdCount = 2$, but increasing it further has
essentially no effect on true detections.  Intuitively, the absence of
an effect for $\pwdCount > 2$ indicates that while the difference
between the probabilities of the two lowest-ranked passwords in an
account's Bloom filter---i.e., how ``isolated'' the first-ranked
password is---provides guidance for which account to attack first,
additional passwords provide the \breachingAttacker little additional
information.

\section{Integration with Amnesia}
\label{sec:amnesia}

The Amnesia design~\cite{wang2021:amnesia} improves on that of Juels
\& Rivest~\cite{juels2013:honeywords} in two ways.  First, it
eliminates the assumption that honeychecker state remains secret past
the breach of a target site; indeed, the target site in Amnesia has no
honeychecker at all.  Second, it enables a target to request that
another site monitor for the entry of its honeywords, without
disclosing them (or the user-chosen password) to the monitoring site
and without exposing login attempts at the monitoring site to the
target site unless the login attempt actually involves one of the
target's honeywords.  This remote monitoring is important since a
\breachingAttacker can distinguish a user-chosen password from
honeywords by stuffing them at other sites; since users often reuse a
password across sites~\cite{das2014:tangled, pearman2017:habitat,
  wang2018:domino, mayer2022:managers}, the user-chosen password at
the target emerges as the one that works elsewhere.

In this section we describe an adaptation of the Amnesia framework
using the insights of \secref{sec:insight}.  As we will see, this
adaptation does come with some consequences in terms of security
against \breachingAttackers, which we will detail.  This section will
also leverage Bloom filters, as presented in
\secref{sec:honeychecker:Bloom}.

\subsection{Detecting a breach locally}
\label{sec:amnesia:local}

\subsubsection{Background on local detection in Amnesia}
\label{sec:amnesia:local:background}

In Amnesia, the site forgoes a honeychecker and indeed does not have
any ability to distinguish the user-chosen password from honeywords
for an account.  So, to detect a breach using honeywords, the site
needs some other way to determine that the account has been accessed
using a honeyword.  Amnesia does so by detecting probabilistically if
an account has been successfully accessed by two distinct
passwords---one of which must be a honeyword.  Note, moreover, that
Amnesia must do so without storing any state that would indicate to a
\breachingAttacker what password was previously used to access the
account, since that would reveal the user-chosen password to the
\breachingAttacker, enabling it to avoid using a honeyword.

To achieve this, Amnesia attaches one-bit \textit{marks} to the
(hashes of) passwords for an account, so that the password with which
the account was last accessed is marked (i.e., its mark is set to $1$)
and each other password is marked with a certain probability.  The
user-chosen password is the only one that the intended user should
ever use, and so her accesses leave this one marked all the time.  If
a \breachingAttacker accesses the account using a honeyword, however,
then the user-chosen password becomes \textit{unmarked} with some
probability.  In that case, the legitimate user's next login triggers
a breach alert, due to using an unmarked password.

Several subtleties in the security of Amnesia against a
\breachingAttacker were explored in its original
analysis~\cite{wang2021:amnesia}.  The first is that a
\breachingAttacker that continues to watch the persistent storage of
the breached site as its users log in over time---essentially
breaching the site repeatedly across some number \nmbrLoginsSeen of
legitimate logins per account---can narrow in on the user-chosen
password as one of those that remain marked across those
\nmbrLoginsSeen logins.  To do so, however, the \breachingAttacker
must remain in the system and exfiltrate this data over time, which
presumably leaves the \breachingAttacker at greater risk of exposure.
Amnesia therefore assumes that \nmbrLoginsSeen can be reasonably
bounded (or that the \breachingAttacker will be noticed by other means
if not).

The second subtlety in the Amnesia analysis is that once the
\breachingAttacker decides to access an account using one of the
passwords \loginPassword that remained marked through the
\nmbrLoginsSeen logins, it can do so many times---these logins are
indexed $\nmbrLoginsSeen+1, \ldots, \nmbrLoginsTotal$ below---in an
attempt to ensure that the \textit{next} most-likely password
\userPassword, from the \breachingAttacker's perspective, remains
marked after login \nmbrLoginsTotal.  If the \breachingAttacker
succeeds and if \userPassword is in fact the user-chosen password,
then the user entering it will not trigger an alarm.  Amnesia
therefore additionally requires accounts to be monitored for an
unusually high frequency of \textit{successful} logins, e.g.,
triggering a second-factor or backup authentication challenge if that
frequency becomes abnormally large.

Characterizing the effects of \nmbrLoginsSeen and \nmbrLoginsTotal on
\breachingAttacker detection is challenging.  In the Amnesia
paper~\cite{wang2021:amnesia}, the authors resorted to probabilistic
model-checking to analyze these effects and, indeed, did not quantify
a true-detection rate for their design.  In contrast, our design below
supports the first closed-form (albeit approximate) solution for the
impact of \nmbrLoginsSeen and \nmbrLoginsTotal on its true-detection
probability.  We use this solution to show the influence of these
parameters on the design.  

\subsubsection{Adapting local detection in Amnesia}
\label{sec:amnesia:local:design}

In adapting Amnesia to leverage the insights of \secref{sec:insight},
we adapt a Bloom filter to accommodate marks.  Specifically, we
subsequently denote a \textit{marked Bloom filter} as a triple
$\langle \bfHashFnSet, \bfIndices, \bfMarkedIndices\rangle$ where
$\bfHashFnSet$ and \bfIndices are as before and where
$\bfMarkedIndices \subseteq \bfIndices$ includes the indices in
\bfIndices that are marked.  Upon a login attempt with password
\loginPassword, the result is determined as follows:
\[
\loginResult{\loginPassword} =
\left\{\begin{array}{ll}
\loginFailure & \mbox{if $\pwdHashFn(\loginPassword) \bfNotIn \langle
  \bfHashFnSet,
  \bfIndices\rangle$} \\
\loginSuccess & \mbox{if $\pwdHashFn(\loginPassword) \bfIn \langle
  \bfHashFnSet,
  \bfMarkedIndices\rangle$} \\
\loginAlarm & \mbox{otherwise}
\end{array}\right.
\]
If $\loginResult{\loginPassword} = \loginSuccess$, a remarking occurs
with probability \remarkProb. In a remarking, the set \bfMarkedIndices
is reset to include $\bfHashFnSet(\pwdHashFn(\loginPassword))$ (i.e.,
$\bfHashFnSet(\pwdHashFn(\loginPassword))$ $\subseteq$ \bfMarkedIndices
with probability $1.0$) and each element of $\bfIndices \setminus
\bfHashFnSet(\pwdHashFn(\loginPassword))$ independently with
probability \markProb.

\subsubsection{Security against a \raisingAlarmAttacker}
\label{sec:amnesia:local:raat}

As in \secref{sec:honeychecker}, the false-detection probability is
simple to characterize for this design.  By definition, a
\raisingAlarmAttacker does \textit{not} breach the system and so
cannot observe $\langle \bfHashFnSet, \bfIndices, \bfMarkedIndices
\rangle$ directly for an account.  So, when attempting to enter any
honeyword in \nmbrLoginsPerAccount logins per account, for
\nmbrAccountsAttempted accounts, the false-detection probability
$\falseAlarmProb(\nmbrLoginsPerAccount,\nmbrAccountsAttempted)$ can
again be bounded as in \eqnref{eqn:falseAlarmProb}.

\subsubsection{Security against a \breachingAttacker}
\label{sec:amnesia:local:brat}

The threat model envisioned by Amnesia permits the \breachingAttacker,
in our case, to capture snapshots of the marked Bloom filter $\langle
\bfHashFnSet, \bfIndices, \bfMarkedIndices\rangle$ after multiple
logins by the legitimate user.  If the \breachingAttacker breaches the
site, capturing the current marked Bloom filter $\langle \bfHashFnSet,
\bfIndices, \bfMarkedIndices{0}\rangle$, and then continues to monitor
the site while the user successfully logs in \nmbrLoginsSeen times,
then the \breachingAttacker observes consecutive marked Bloom filters
$\{\langle \bfHashFnSet, \bfIndices,
\bfMarkedIndices{\loginIdx}\rangle\}_{\loginIdx=0}^{\nmbrLoginsSeen}$,
and a password remains viable only if it is contained in all of them.
That is, in the terminology of \secref{sec:insight},
$\honeywordInd{\pwdRank} = 1$ iff
$\bigwedge_{\loginIdx=0}^{\nmbrLoginsSeen}
\left(\pwdHashFn(\pwdList{\pwdRank}) \bfIn \langle \bfHashFnSet,
\bfMarkedIndices{\loginIdx}\rangle\right)$.
Since for $\genericElmt
\getsr \{0,1\}^{\bfHashFnDomainBits}$,
\begin{align*}
\cprob{\Big}
      {\genericElmt \bfIn \langle
        \bfHashFnSet,
        \bfMarkedIndices{0}\rangle}
      {\genericElmt \bfIn \langle
        \bfHashFnSet,
        \bfIndices\rangle}
      & \approx (\markProb)^{\expv{\setSize{\bfHashFnSet(\genericElmt)}}}
\end{align*}
and for any $\loginIdx \ge 0$,
\begin{align*}
& \cprob{\Big}
      {\genericElmt \bfIn \langle
          \bfHashFnSet,
          \bfMarkedIndices{\loginIdx+1}\rangle}
      {\genericElmt \bfIn \langle
          \bfHashFnSet,
          \bfMarkedIndices{\loginIdx}\rangle} \\
& \approx (1-\remarkProb) + \remarkProb(\markProb)^{\expv{\setSize{\bfHashFnSet(\genericElmt)}}}
\end{align*}
it is prudent to measure
\honeyEntryProb{\breachingAttacker}{\cdot}{\cdot}
from \secref{sec:insight:brat} using
\begin{align}
  \honeyProbModified{\nmbrLoginsSeen}
  & = \honeyProb \times
  \left(\markProb\right)^{\expv{\setSize{\bfHashFnSet(\genericElmt)}}} \times \nonumber\\
  & \hspace{1em} \left(1-\remarkProb + \remarkProb(\markProb)^{\expv{\setSize{\bfHashFnSet(\genericElmt)}}}\right)^{\nmbrLoginsSeen}
  \label{eqn:honeyProbModified}
\end{align}
as its first argument.  (\expv{\setSize{\bfHashFnSet(\genericElmt)}}
can be instantiated using \eqnref{eqn:oneBFInsertion}.)  As discussed
above, this degradation in the probability with which a
\breachingAttacker inputs a honeyword (i.e., reflected in our use of
\honeyProbModified{\nmbrLoginsSeen} in lieu of \honeyProb in
\honeyEntryProb{\breachingAttacker}{\honeyProbModified{\nmbrLoginsSeen}}{\pwdRankSet})
is not an artifact of our construction, but rather an analogous
degradation is present in the original Amnesia
design~\cite{wang2021:amnesia}.

The entry of a honeyword \loginPassword by a \breachingAttacker is
necessary but not sufficient to detect the \breachingAttacker; in
addition, its doing so (after observing \nmbrLoginsSeen logins by the
legitimate user, and then himself logging in another $\nmbrLoginsTotal
- \nmbrLoginsSeen$ times) must leave $\pwdHashFn(\userPassword)
\bfNotIn \langle \bfHashFnSet,
\bfMarkedIndices{\nmbrLoginsTotal}\rangle$, for the user-chosen
password \userPassword.  Recall that by the analysis of
\secref{sec:insight:brat}, the \breachingAttacker can do no better
than attempting the most likely password in its rank ordering
\pwdList{\cdot} that its monitoring suggests is still viable.  To
provide a conservative analysis, suppose that the next most-likely
viable password \userPassword is indeed the only other possibility for
the user-chosen password, in the \breachingAttacker's view.  Denote
$\bfUserIndices = \bfHashFnSet(\pwdHashFn(\userPassword))$ and
$\bfLoginIndices = \bfHashFnSet(\pwdHashFn(\loginPassword))$.  Then,
\begin{align}
\prob{\pwdHashFn(\userPassword) \bfNotIn \langle
  \bfHashFnSet,
  \bfMarkedIndices{\nmbrLoginsSeen+1}\rangle}
\approx \remarkProb \left(1 - (\markProb)^{\expv{\setSize{\bfUserIndices \setminus \bfLoginIndices}}}\right)
\label{eqn:userPasswordRemoved}
\end{align}
where \expv{\setSize{\bfUserIndices \setminus \bfLoginIndices}} can be
evaluated as in \eqnref{eqn:bfSetDifference}, and for $\loginIdx >
\nmbrLoginsSeen$,
\begin{align}
& \cprob{\Big}{\pwdHashFn(\userPassword) \bfNotIn \langle
  \bfHashFnSet,
  \bfMarkedIndices{\loginIdx+1}\rangle}
  {\pwdHashFn(\userPassword) \bfNotIn \langle
  \bfHashFnSet,
  \bfMarkedIndices{\loginIdx}\rangle} \nonumber\\
  & \approx (1 - \remarkProb) +
  \remarkProb \left(1 - (\markProb)^{\expv{\setSize{\bfUserIndices \setminus \bfLoginIndices}}}\right)
  \label{eqn:userPasswordStillRemoved}
\end{align}
The true-detection probability for a \breachingAttacker who attacks
a \pwdRankSet-account, then, is
\begin{align*}
\trueAlarmProb{\pwdRankSet}(\nmbrLoginsSeen,\nmbrLoginsTotal)
& \approx \honeyEntryProb{\breachingAttacker}{\honeyProbModified{\nmbrLoginsSeen}}{\pwdRankSet} \times
    \prob{\pwdHashFn(\userPassword) \bfNotIn \langle
    \bfHashFnSet,
    \bfMarkedIndices{\nmbrLoginsSeen+1}\rangle} \\
& \hspace{0.5em} \times \prod_{\loginIdx=\nmbrLoginsSeen+1}^{\nmbrLoginsTotal-1}\cprob{\Big}{\pwdHashFn(\userPassword) \bfNotIn \langle
    \bfHashFnSet, \bfMarkedIndices{\loginIdx + 1}\rangle}
    {\pwdHashFn(\userPassword) \bfNotIn \langle \bfHashFnSet,
  \bfMarkedIndices{\loginIdx}\rangle}
\end{align*}
for any $\nmbrLoginsTotal \ge \nmbrLoginsSeen$.  The factors can be
plugged in from \eqnref{eqn:honeyProbModified},
\eqnref{eqn:userPasswordRemoved}, and
\eqnref{eqn:userPasswordStillRemoved}.  Similar to
\secref{sec:honeychecker:brat}, the true-detection probability for a
\breachingAttacker who attacks accounts $\accountsSubset \subseteq
\accountsSet$ is
\begin{align}
  \trueAlarmProb(\nmbrLoginsSeen,\nmbrLoginsTotal,\accountsSubset)
  = 1- \prod_{\accountId \in \accountsSubset}
  (1 - \trueAlarmProb{\pwdRankSet{\accountId}}(\nmbrLoginsSeen,\nmbrLoginsTotal))
\label{eqn:amnesia:trueAlarmProb}
\end{align}
and the \textit{minimum} true-detection probability for
\breachingAttacker who attacks \nmbrAccountsAttempted accounts is
\begin{align}
  \trueAlarmProb(\nmbrLoginsSeen,\nmbrLoginsTotal,\nmbrAccountsAttempted)
  = \min_{\accountsSubset \subseteq \accountsSet:\setSize{\accountsSubset} = \nmbrAccountsAttempted}
  \trueAlarmProb(\nmbrLoginsSeen,\nmbrLoginsTotal,\accountsSubset)
\label{eqn:amnesia:minTrueAlarmProb}
\end{align}

\subsubsection{Security evaluation}
\label{sec:amnesia:local:eval}

We now provide results for the Amnesia integration given in
\secref{sec:amnesia:local:design}, analogous to those of
\secref{sec:honeychecker:eval} but informed by the analysis in
\secsref{sec:amnesia:local:raat}{sec:amnesia:local:brat}.  We again
take the requirement \eqnref{eqn:fdpBounds} on \falseAlarmProb and
estimate \prob{\rvPwdRankChosen \le \pwdRank} using
\eqnref{eqn:wangCracked-TGI}, \eqnref{eqn:mazurekCracked},
\eqnref{eqn:xuCracked}, or \eqnref{eqn:wangCracked} with some
blocklisting algorithm.

\input{figures/fig_roc_ba_cmp.tex}

Analogous to \figref{fig:roc_hc_cmp:tdp}, in
\figref{fig:roc_ba_cmp:tdp} we plot $\trueAlarmProb(\nmbrLoginsSeen,
\nmbrLoginsTotal, \nmbrAccountsAttempted)$ as a function of the
fraction $\nmbrAccountsAttempted / \nmbrAccounts$ of accounts accessed
by the \breachingAttacker, produced using $10{,}000$ simulations with
$\pwdCount = 1000$ and $\fdpBound = 10^{-1}$.  For these plots, we set
$\nmbrLoginsSeen = 10$ and $\nmbrLoginsTotal = 31$ (i.e., the
\breachingAttacker logs in up to 20 additional times after its first);
we will illustrate the effects of varying these parameters below.  We
again set $\bfNmbrHashFns = 20$ and $\bfHashFnRange = 128$ and chose
$\markProb = 0.95$ and $\remarkProb = 0.065$.
\figref{fig:roc_ba_cmp:tdp} again highlights the utility of
blocklisting; e.g., blocklisting using \TGIPP reduces the fraction of
accounts accessed by the \breachingAttacker at which the
true-detection probability reaches $0.5$ from $\approx 32\%$ with no
blocklisting to $\approx 8.6\%$.  At the same time, comparing
\figref{fig:roc_ba_cmp:tdp} with \figref{fig:roc_hc_cmp:tdp}, we see
that the \breachingAttacker can access $\approx 34.4\%$ more accounts
than in the honeychecker design of \secref{sec:honeychecker:design}
before the true-detection probability reaches this value (in the case
of blocklisting with \TGIPP), at least for $\nmbrLoginsSeen = 10$ and
$\nmbrLoginsTotal = 31$.  As such, it is evident that the feature that
makes Amnesia symmetric in the sense of \secref{sec:related}, i.e.,
that the site has no data for which its secrecy survives the breach,
comes at a cost in detection power.  Also, \figref{fig:roc_ba_cmp:tdp}
again highlights the relative strengths of the CMU and CKL-PCFC
datasets, showing that our Amnesia integration will detect a
\breachingAttacker with certainty after it accesses only about $5\%$
and $0.0004\%$ of the accounts in these datasets, respectively.

\figref{fig:roc_ba_cmp:fdp} shows the impact of constraining the
false-positive probability to be $\fdpBound \le 10^{-1}$.  Comparing to
\figref{fig:roc_hc_cmp:fdp}, the cost of Amnesia's weak assumptions is
evident, in that when $\fdpBound = 10^{-5}$, there is minimal
true-detection power for a dataset as weak as
\eqnref{eqn:wangCracked-TGI}, even with blocklisting.  However,
datasets \eqnref{eqn:mazurekCracked} and \eqnref{eqn:xuCracked} retain
substantial true-detection power even at $\fdpBound = 10^{-5}$, in
that the \breachingAttacker will be detected with probability $\ge
0.5$ after it accesses $\approx 30\%$ of the accounts.

The impact of \nmbrLoginsSeen and \nmbrLoginsTotal are shown in
\tblref{tbl:logins}.  Recall that \nmbrLoginsSeen is the number of
logins by the legitimate user across which the \breachingAttacker
monitors the password database before accessing the account himself,
and $\nmbrLoginsTotal - \nmbrLoginsSeen$ is the number of logins by
the \breachingAttacker to first access the account and then to attempt
to return the system to a state in which the next login by the
legitimate user will not trigger an alarm.  \tblref{tbl:logins}
suggests the true-detection probability decays modestly when
\nmbrLoginsSeen and $\nmbrLoginsTotal - \nmbrLoginsSeen$ increase.

\input{figures/fig_roc_ba_heatmap.tex}

\subsection{Detecting a breach with remote help}
\label{sec:amnesia:remote}

\subsubsection{Background on remote monitoring in Amnesia}
\label{sec:amnesia:remote:background}

An aspect of the Amnesia design that requires a bit more adaptation to
accommodate our approach here is a target site's ability to solicit
help from other sites to monitor for the entry of the target's
honeywords at those monitoring sites.  Critically, Amnesia enables
this monitoring without the target \targetSite disclosing its
honeywords (or the user-chosen password) for accounts to the
monitoring site \monitorSite; without \monitorSite being able to
induce a breach alarm at \targetSite with any higher probability than
a \raisingAlarmAttacker could; and without placing \monitorSite's
accounts at risk.  Remote monitoring is useful because attempting
passwords breached from \targetSite at other sites is an effective way
to find the user-chosen password, since users tend to reuse passwords
across sites~\cite{das2014:tangled, pearman2017:habitat,
  wang2018:domino, mayer2022:managers}.

Amnesia achieves remote monitoring via a protocol it calls ``private
containment retrieval,'' denoted \pcr{}.  To request that \monitorSite
monitor for an account at \targetSite, \targetSite sends to
\monitorSite a data structure that contains the set of hashes of
passwords (one real, and the others honeywords) for that account at
\targetSite.  This data structure is encrypted under a public key
\pubKey whose private key \privKey is known only to \targetSite.

Upon receiving a login attempt for the same user's account at
\monitorSite for which the submitted password \loginPassword is
incorrect (even accounting for typos,
e.g.,~\cite{chatterjee2016:typo}), \monitorSite inputs
$\pwdHashFn(\loginPassword)$ as the \textit{test plaintext} and a
specific value \plaintext as the \textit{response plaintext} to a
local \textit{response computation}, together with the encrypted data
structure received from \targetSite.  If the test plaintext
$\pwdHashFn(\loginPassword)$ matches a hash value in the (plaintext of
the) encrypted data structure, then this computation produces a
ciphertext of \plaintext. Otherwise, it produces a ciphertext of a
random plaintext.  Even though \monitorSite knows the ciphertext
produced is either of \plaintext or of a uniformly random plaintext,
\monitorSite nevertheless cannot tell which type of ciphertext it
produced (see~\cite{wang2021:amnesia}).

\monitorSite returns this ciphertext to \targetSite, who decrypts it
using \privKey.  \targetSite can then test whether the resulting
plaintext is the response plaintext \plaintext corresponding to any
\loginPassword in its set of passwords for this account.  If so,
\targetSite acts (for the purposes of breach detection) as if
\loginPassword had been input in a local login attempt to this
account.

\subsubsection{Adapting remote monitoring in Amnesia}
\label{sec:amnesia:remote:protocol}

We describe a way in \appref{app:pcr-k} to use the \pcr{} protocol
summarized above to enable remote monitoring for entry of Bernoulli
honeywords in the Amnesia integration of \secref{sec:amnesia:local}.
However, this adaptation costs \bfNmbrHashFns \pcr{} responses per
incorrect login attempt at the monitoring site \monitorSite, yielding
substantially greater costs.  Here we instead provide a more efficient
protocol to convey \loginPassword to \targetSite iff
$\pwdHashFn(\loginPassword) \bfIn \langle \bfHashFnSet, \bfIndices
\rangle$.  Our design improves on that of \appref{app:pcr-k} by
roughly an order of magnitude for the common operations, namely the
response generation by \monitorSite and the response processing by
\targetSite, as we will show.  

\myparagraph{Cryptographic primitives}
\label{sec:amnesia:remote:protocol:crypto}
Our protocol builds on an encryption scheme \encScheme with algorithms
\keygen, \encrypt{}, \decrypt{}, and \encMult{[\cdot]}.
\begin{itemize}[nosep,leftmargin=1em,labelwidth=*,align=left]
\item \keygen is a randomized algorithm that
  outputs a public-key/private-key pair $\langle\pubKey,
  \privKey\rangle \gets \keygen()$.  The value of \pubKey
  determines a plaintext space that is a cyclic group
  $(\plaintextGroup, \plaintextGroupOperator)$ of prime order
  \plaintextGroupOrder, with generator \plaintextGroupGenerator and
  identity element \plaintextGroupIdentity.  We assume below that a
  password \loginPassword can be encoded as a plaintext in
  \plaintextGroup.  For any $\plaintext \in \plaintextGroup$, we use
  $\plaintext^{-1}$ to denote the inverse of \plaintext and
  $\plaintext^{\genericNat} = \plaintext{1} \plaintextGroupOperator
  \ldots \plaintextGroupOperator \plaintext{\genericNat}$ where each
  $\plaintext{\plaintextIdx} = \plaintext$.  The value of \pubKey also
  determines a ciphertext space $\ciphertextSpace{\pubKey} =
  \bigcup_{\plaintext \in
    \plaintextGroup} \ciphertextSpace{\pubKey}(\plaintext)$, where
  $\ciphertextSpace{\pubKey}(\plaintext)$ denotes the ciphertexts for
  plaintext \plaintext.  Below our discussion implicitly assumes that
  $\setSize{\ciphertextSpace{\pubKey}(\plaintext)} =
  \setSize{\ciphertextSpace{\pubKey}(\plaintextAlt)}$ for any
  $\plaintext, \plaintextAlt \in \plaintextGroup$; this is not
  necessary, but simplifies the discussion below and holds true in our
  implementation.
\item \encrypt{} is a randomized algorithm that on input public key
  \pubKey and a plaintext $\plaintext \in \plaintextGroup$, outputs a
  ciphertext $\ciphertext{} \gets \encrypt{\pubKey}(\plaintext)$
  chosen uniformly at random from
  $\ciphertextSpace{\pubKey}(\plaintext)$.
\item \decrypt{} is a deterministic algorithm that on input a private
  key \privKey and ciphertext $\ciphertext{}
  \in \ciphertextSpace{\pubKey}$ for the \pubKey corresponding to
  \privKey, outputs the plaintext $\plaintext \gets
  \decrypt{\privKey}(\ciphertext{})$ such that $\ciphertext{}
  \in \ciphertextSpace{\pubKey}(\plaintext)$.  If $\ciphertext{}
  \not\in \ciphertextSpace{\pubKey}$, then
  $\decrypt{\privKey}(\ciphertext{})$ returns \failureSymbol.
\item \encMult{[\cdot]} is a randomized algorithm that, on input a
  public key \pubKey and ciphertexts $\ciphertext{1}
  \in \ciphertextSpace{\pubKey}(\plaintext{1})$ and $\ciphertext{2}
  \in \ciphertextSpace{\pubKey}(\plaintext{2})$, outputs a ciphertext
  $\ciphertext{} \gets \ciphertext{1} \encMult{\pubKey}
  \ciphertext{2}$ chosen uniformly at random from
  $\ciphertextSpace{\pubKey}(\plaintext{1} \plaintextGroupOperator
  \plaintext{2})$.
\end{itemize}

Given this functionality, it will be convenient to define two
additional operators.  Below, ``$\genericRV \distEqual
\genericRVAlt$'' denotes that random variables \genericRV and
\genericRVAlt are distributed identically.
\begin{itemize}[nosep,leftmargin=1em,labelwidth=*,align=left]
\item The \encProd{\pubKey} operator denotes repetition of
  \encMult{\pubKey}, i.e.,
  \[
  \encProd{\pubKey}{\ciphertextIdx=1}{\genericNat} \ciphertext{\ciphertextIdx}
  \distEqual \ciphertext{1} \encMult{\pubKey} \ciphertext{2} \encMult{\pubKey}
  \ldots \encMult{\pubKey} \ciphertext{\genericNat}
  \]
\item The \encRand{\pubKey} operator produces a random ciphertext of
  \plaintextGroupIdentity if its argument is a ciphertext of
  \plaintextGroupIdentity, and otherwise produces a ciphertext of a
  random element of $\plaintextGroup \setminus
  \{\plaintextGroupIdentity\}$.  Specifically,
  \[
  \encRand{\pubKey}(\ciphertext) \distEqual
  \encProd{\pubKey}{\ciphertextIdx=1}{\genericNat} \ciphertext
  \mbox{~~where $\genericNat \getsr \relPrimeResidues{\plaintextGroupOrder}$}
  \]
\end{itemize}

\myparagraph{Protocol description}
\label{sec:amnesia:remote:protocol:design}
The protocol for \targetSite to deploy a monitor for a selected
account to site \monitorSite is shown in \figref{fig:monitor:deploy},
and the protocol for \monitorSite to send a monitoring response (upon
an attempted login to that account) to \targetSite is shown in
\figref{fig:monitor:respond}.  Deployment begins by \targetSite
creating a public/private key pair $\langle \pubKey, \privKey\rangle$
(\lineref{prot:deploy:keygen}) that it will save for processing
monitoring responses later (\ref{prot:deploy:targetSave}).  The
monitoring request itself includes \bfHashFnRange ciphertexts
$\{\ciphertext{\ciphertextIdx}\}_{\ciphertextIdx=1}^{\bfHashFnRange}$
that encode which indices \ciphertextIdx are in the account's Bloom
filter indices \bfIndices (encoded as $\ciphertext{\ciphertextIdx}
\in \ciphertextSpace{\pubKey}(\plaintextGroupGenerator)$) and which
are not (encoded as $\ciphertext{\ciphertextIdx}
\in \ciphertextSpace{\pubKey}(\plaintextGroupGenerator^{-1})$); see
\lineref{prot:deploy:ctexts}.  In addition, the monitoring request
(\msgref{prot:deploy:request}) includes \pubKey; the uniform hash
functions \bfHashFnSet for the account's Bloom filter; the number
\bfNmbrIndices of indices in \bfIndices
(\lineref{prot:deploy:nmbrIndices}); and a noninteractive
zero-knowledge proof \zkp that
$\{\ciphertext{\ciphertextIdx}\}_{\ciphertextIdx=1}^{\bfHashFnRange}$
$\subseteq$ $\ciphertextSpace{\pubKey}(\plaintextGroupGenerator) \cup
\ciphertextSpace{\pubKey}(\plaintextGroupGenerator^{-1})$ (generated
using \zkpGen in \lineref{prot:deploy:zkp}).\footnote{As in
Amnesia, if the password-hashing function \pwdHashFn in use at
\targetSite is salted, then the salt can be sent in
\ref{prot:deploy:request} to \monitorSite.  Or, \pwdHashFn could be
implemented as an oblivious pseudorandom function
(e.g.,~\cite{freedman2005:oprf}) keyed with the salt, which
\monitorSite would evaluate on \loginPassword with an extra
interaction with \targetSite in \figref{fig:monitor:respond}.}

\setcounter{requesterLineNmbr}{0}
\setcounter{responderLineNmbr}{0}
\setcounter{messageNmbr}{0}
\newcolumntype{R}{>{$}p{0.5in}<{$}}
\newcolumntype{M}{>{$}p{0.3in}<{$}}
\begin{figure}[t]
  \begin{oframed}
    \vspace{-1ex}
     $\begin{array}{@{}R@{}r@{}M@{}l@{}} 
      \mathrlap{\parbox{0.4\columnwidth}{\centering $\targetSite(\langle \bfHashFnSet, \bfIndices\rangle)$}}
      & & & \parbox{0.65\columnwidth}{\centering $\monitorSite(\cdot)$} \\[10pt]
      \requesterLabel{prot:deploy:keygen}~\mathrlap{\langle\pubKey,\privKey\rangle \gets \keygen()} \\
      \requesterLabel{prot:deploy:nmbrIndices}~\mathrlap{\bfNmbrIndices \gets \setSize{\bfIndices}} \\
      \requesterLabel{prot:deploy:ctexts}~\mathrlap{\forall \ciphertextIdx \in \{1, \ldots, \bfHashFnRange\}:} \\
      \mathrlap{\hspace{2em}\ciphertext{\ciphertextIdx} \gets
        \left\{\begin{array}{ll}
        \encrypt{\pubKey}(\plaintextGroupGenerator^{-1}) & \mbox{if $\ciphertextIdx \not\in \bfIndices$} \\
        \encrypt{\pubKey}(\plaintextGroupGenerator) & \mbox{if $\ciphertextIdx \in \bfIndices$}
        \end{array}\right.} \\
      \requesterLabel{prot:deploy:zkp}~\mathrlap{\zkp \gets \zkpGen(\langle\pubKey,
      \{\ciphertext{\ciphertextIdx}\}_{\ciphertextIdx=1}^{\bfHashFnRange}\rangle)}\\
      \requesterLabel{prot:deploy:targetSave}~\mathrlap{\mbox{save $\langle\pubKey,\privKey\rangle$}}
      \\[10pt]
      & \messageLabel{prot:deploy:request}
      & \mathrlap{\xrightarrow{\makebox[9.75em]{$\pubKey, \bfHashFnSet, \bfNmbrIndices, \{\ciphertext{\ciphertextIdx}\}_{\ciphertextIdx=1}^{\bfHashFnRange}$, \zkp}}}
      \\[6pt]
      & & & \responderLabel{prot:deploy:zkpVerify}~\mbox{abort if $\neg\zkpVerify(\langle\pubKey, \{\ciphertext{\ciphertextIdx}\}_{\ciphertextIdx=1}^{\bfHashFnRange}\rangle, \zkp)$} \\
      & & & \responderLabel{prot:deploy:count}~\ciphertext \gets \encProd{\pubKey}{\ciphertextIdx=1}{\bfHashFnRange} \ciphertext{\ciphertextIdx}  \\
      & & & \responderLabel{prot:deploy:check}~\derivedCiphertext{0} \gets \ciphertext \encMult{\pubKey} \encrypt{\pubKey}(\plaintextGroupGenerator^{\bfHashFnRange-2\bfNmbrIndices})\\
      & & & \responderLabel{prot:deploy:monitorSave}~\mbox{save $\langle \pubKey, \bfHashFnSet, \{\ciphertext{\ciphertextIdx}\}_{\ciphertextIdx=1}^{\bfHashFnRange}, \derivedCiphertext{0} \rangle$}
    \end{array}$
    \vspace{-1ex}
  \end{oframed}
  \vspace{-1ex}
  \caption{Monitor deployment}
  \label{fig:monitor:deploy}
\end{figure}

Upon receiving a well-formed monitoring request
\ref{prot:deploy:request} (in particular, where \pubKey is a valid
public key), \monitorSite checks the zero-knowledge proof
\zkp using \zkpVerify (\lineref{prot:deploy:zkpVerify}) and aborts if the
check returns \boolFalse.  If this check returns \boolTrue and so
$\{\ciphertext{\ciphertextIdx}\}_{\ciphertextIdx=1}^{\bfHashFnRange}
\subseteq \ciphertextSpace{\pubKey}(\plaintextGroupGenerator) \cup
\ciphertextSpace{\pubKey}(\plaintextGroupGenerator^{-1})$ (except with
probability the soundness error of \zkp), then \monitorSite calculates
\derivedCiphertext{0} to be a ciphertext of \plaintextGroupIdentity if
and only if the claimed number \bfNmbrIndices is accurate
(\linesref{prot:deploy:count}{prot:deploy:check}).  That is,
\begin{align*}
  \setSize{\{\ciphertext{\ciphertextIdx}\}_{\ciphertextIdx=1}^{\bfHashFnRange}
    \cap \ciphertextSpace{\pubKey}(\plaintextGroupGenerator)} =
  \bfNmbrIndices
  & \Leftrightarrow \ciphertext
  \in \ciphertextSpace{\pubKey}(\plaintextGroupGenerator^{\bfNmbrIndices}
  \plaintextGroupGenerator^{-(\bfHashFnRange-\bfNmbrIndices)})
  & \mbox{in \lineref{prot:deploy:count}} \\
  & \Leftrightarrow \derivedCiphertext{0} \in
  \ciphertextSpace{\pubKey}(\plaintextGroupIdentity)
  & \mbox{in \lineref{prot:deploy:check}}
\end{align*}
Finally, \monitorSite saves \pubKey, \bfHashFnSet,
$\{\ciphertext{\ciphertextIdx}\}_{\ciphertextIdx=1}^{\bfHashFnRange}$,
and \derivedCiphertext{0} in \lineref{prot:deploy:monitorSave}.

As we will see, $\derivedCiphertext{0} \not\in
\ciphertextSpace{\pubKey}(\plaintextGroupIdentity)$ ensures that
\targetSite learns nothing from \monitorSite; i.e., \targetSite cannot
gain any information about logins at \monitorSite if it reports an
incorrect value of \bfNmbrIndices.  \bfNmbrIndices is reported in
\ref{prot:deploy:request} primarily to permit \monitorSite to refuse
the monitoring request if \bfNmbrIndices is larger than \monitorSite
deems appropriate.  That is, using \bfNmbrIndices, \monitorSite can
calculate \honeyProb using \eqnref{eqn:honeyProb} with
$\setSize{\bfIndices} = \bfNmbrIndices$ and accept the monitoring
request only if \honeyProb is acceptably small.  If not, \monitorSite
can just drop this request (not shown in \figref{fig:monitor:deploy}).

\newcolumntype{R}{>{$}p{0.5in}<{$}}
\newcolumntype{M}{>{$}p{0.4in}<{$}}
\begin{figure}[t]
  \begin{oframed}
    \vspace{-1ex}
     $\begin{array}{@{}R@{}r@{}M@{}l@{}} 
      \mathrlap{\parbox{0.4\columnwidth}{\centering $\targetSite(\langle \bfHashFnSet, \bfIndices\rangle, \langle\pubKey,\privKey\rangle)$}}
      & & & \parbox{0.6\columnwidth}{\centering $\monitorSite(\langle \pubKey, \bfHashFnSet, \{\ciphertext{\ciphertextIdx}\}_{\ciphertextIdx=1}^{\bfHashFnRange}, \derivedCiphertext{0}\rangle)$} \\[10pt]
      & & & \parbox{1.85in}{Upon login attempt with incorrect password \loginPassword:} \\[8pt]
      & & & \responderLabel{prot:respond:indices}~
      \bfLoginIndices \gets \bfHashFnSet(\pwdHashFn(\loginPassword)) \\
      & & & \responderLabel{prot:respond:derived}~
      \derivedCiphertext{1} \gets \encProd{\pubKey}{\ciphertextIdx \in \bfLoginIndices}{} \left(\ciphertext{\ciphertextIdx} \encMult{\pubKey} \encrypt{\pubKey}(\plaintextGroupGenerator^{-1})\right) \\
      & & & \responderLabel{prot:respond:retvalZero}~
      \responseCiphertext{0} \gets \encRand{\pubKey}(\derivedCiphertext{0}) \encMult{\pubKey} \encRand{\pubKey}(\derivedCiphertext{1}) \\
      & & & \responderLabel{prot:respond:retvalOne}~
      \responseCiphertext{1} \gets \encRand{\pubKey}(\responseCiphertext{0}) \encMult{\pubKey} \encrypt{\pubKey}(\loginPassword)
      \\[6pt]
      & \messageLabel{prot:respond:response}
      & \mathrlap{\xleftarrow{\makebox[9.75em]{$\responseCiphertext{0}, \responseCiphertext{1}$}}}
      \\[6pt]
      \requesterLabel{prot:respond:testCtext}~\mathrlap{\mbox{abort if $\decrypt{\privKey}(\responseCiphertext{0}) \neq \plaintextGroupIdentity$}} \\
      \requesterLabel{prot:respond:retval}~\mathrlap{\loginPassword \gets \decrypt{\privKey}(\responseCiphertext{1})} \\
      \requesterLabel{prot:respond:testMon}~\mathrlap{\mbox{abort if $\loginPassword = \failureSymbol$ or $\pwdHashFn(\loginPassword) \bfNotIn \langle \bfHashFnSet, \bfIndices\rangle$}} \\
      \requesterLabel{prot:respond:return}~\mathrlap{\mbox{return \loginPassword}}
      \\
    \end{array}$
    \vspace{-1ex}
  \end{oframed}
  \vspace{-1ex}
  \caption{Monitor response}
  \label{fig:monitor:respond}
\end{figure}

The monitor site \monitorSite generates a response using the protocol
shown in \figref{fig:monitor:respond}.  The provided (incorrect)
password \loginPassword in a login attempt to the monitored account is
used to generate its set $\bfLoginIndices \gets
\bfHashFnSet(\pwdHashFn(\loginPassword))$ of Bloom-filter indices, in
\lineref{prot:respond:indices}.  For each such index \ciphertextIdx,
the ciphertext $\ciphertext{\ciphertextIdx} \encMult{\pubKey}
\encrypt{\pubKey}(\plaintextGroupGenerator^{-1})$ is calculated in
\lineref{prot:respond:derived}, yielding a ciphertext of
\plaintextGroupIdentity if $\ciphertext{\ciphertextIdx}
\in \ciphertextSpace{\pubKey}(\plaintextGroupGenerator)$ and a
ciphertext of $\plaintextGroupGenerator^{-2}$ otherwise.  These
ciphertexts are combined using \encMult{\pubKey}, ensuring that
\derivedCiphertext{1} is a ciphertext of \plaintextGroupIdentity if
and only if all of them are.  The ciphertext \responseCiphertext{0} is
then set to be a ciphertext of \plaintextGroupIdentity if and only if
both \derivedCiphertext{0} and \derivedCiphertext{1} are (with
overwhelming probability; \lineref{prot:respond:retvalZero}), and then
\responseCiphertext{1} is set to be a ciphertext of \loginPassword in
that case (and only that case; \lineref{prot:respond:retvalOne}).

Upon receiving \responseCiphertext{0}, \responseCiphertext{1} in
\msgref{prot:respond:response}, \targetSite tests whether
$\responseCiphertext{0}
\in \ciphertextSpace{\pubKey}(\plaintextGroupIdentity)$ and aborts if
not (\lineref{prot:respond:testCtext}); aborting here indicates that
$\pwdHashFn(\loginPassword) \bfNotIn \langle \bfHashFnSet, \bfIndices
\rangle$.  As such, if the protocol does not abort in
\lineref{prot:respond:testCtext} but $\pwdHashFn(\loginPassword)
\bfNotIn \langle \bfHashFnSet, \bfIndices \rangle$ in
\lineref{prot:respond:testMon}, then this reveals that \monitorSite
has misbehaved (and so the protocol again aborts).  Otherwise,
\loginPassword is returned and \targetSite treats it as if it were
entered in a local login attempt for this account, for the purposes of
breach detection.

We prove the cryptographic security of this protocol in
\appref{app:analysis}.  Below we sketch the cryptographic arguments
and additionally discuss other factors that bear on its security for
our purposes.

\subsubsection{Security against a malicious \monitorSite}
\label{sec:amnesia:remote:securityForTarget}

The first threat that we address relative to a malicious \monitorSite
is the possibility that \monitorSite learns information about the
honeywords or user-chosen password for an account at \targetSite for
which it is asked to monitor.  Of the values \monitorSite receives in
the protocol (see \msgref{prot:deploy:request}), \pubKey and
\bfHashFnSet are chosen independently from these passwords;
$\{\ciphertext{\ciphertextIdx}\}_{\ciphertextIdx=1}^{\bfHashFnRange}$
are ciphertexts encrypted using \pubKey; and \zkp is a
zero-knowledge proof that
$\{\ciphertext{\ciphertextIdx}\}_{\ciphertextIdx=1}^{\bfHashFnRange}$
are well-formed.  This protocol does disclose $\setSize{\bfIndices} =
\bfNmbrIndices$ to \monitorSite, though \bfNmbrIndices is a tunable
parameter chosen by \targetSite even prior to user password
registration and, combined with $\bfNmbrHashFns =
\setSize{\bfHashFnSet}$, merely reveals to \monitorSite the value of
\honeyProb (i.e., $\honeyProb =
(\bfNmbrIndices/\bfHashFnRange)^{\bfNmbrHashFns}$) in use at
\targetSite.  It does not, however, provide the adversary any ability
to distinguish $\targetSite(\langle \bfHashFnSet, \bfIndices \rangle)$
from $\targetSite(\langle \bfHashFnSet, \bfIndicesAlt \rangle)$ for
distinct \bfIndices and \bfIndicesAlt, provided that
$\setSize{\bfIndices} = \setSize{\bfIndicesAlt} = \bfNmbrIndices$.
Thus, if a malicious \monitorSite has a non-negligible advantage in
distinguishing between $\targetSite(\langle \bfHashFnSet, \bfIndices
\rangle)$ and $\targetSite(\langle \bfHashFnSet, \bfIndicesAlt
\rangle)$ on the basis of \msgref{prot:deploy:request}, even for
\bfIndices and \bfIndicesAlt of \monitorSite's own choosing
(satisfying $\setSize{\bfIndices} = \setSize{\bfIndicesAlt} =
\bfNmbrIndices$), then there is an
IND-CPA~\cite{bellare1998:relations} adversary with non-negligible
advantage against \encScheme.  This claim is in the
random oracle model~\cite{bellare1993:oracles}---in particular, the
non-interactive proof \zkp built using the Fiat-Shamir
heuristic~\cite{fiat1986:fiatshamir} implies random oracles.  Below we
will instantiate the encryption scheme \encScheme and
zero-knowledge proof \zkp in our implementation.

\iftrue

Given that the protocol does not leak information about the honeywords
to \monitorSite, we now consider the threat of \monitorSite using its
vantage point to induce a false breach alarm at \targetSite.
Interestingly, because \monitorSite learns \bfHashFnSet and might know
the user-chosen password \userPassword at \targetSite (e.g., if it was
reused at \monitorSite), \monitorSite \textit{does} gain an advantage
in inducing a false alarm at \targetSite.  Specifically, while
\monitorSite does not know \bfMarkedIndices or \bfIndices at
\targetSite, it knows that $\setSize{\bfIndices} = \bfNmbrIndices$ and
$\bfHashFnSet(\pwdHashFn(\userPassword)) \subseteq \bfMarkedIndices$.
So, to attempt to induce a false alarm at \targetSite, it can select a
password \loginPassword to maximize
\[
\cprob{\Bigg}{\begin{array}{@{}l@{}}
    \bfHashFnSet(\pwdHashFn(\loginPassword)) \subseteq \bfIndices~\wedge\\
    \bfHashFnSet(\pwdHashFn(\loginPassword)) \cap (\bfIndices \setminus
    \bfMarkedIndices) \neq \emptyset
    \end{array}}{\begin{array}{@{}l@{}}
    \setSize{\bfIndices} = \bfNmbrIndices~\wedge\\
    \bfHashFnSet(\pwdHashFn(\userPassword)) \subseteq \bfMarkedIndices
    \end{array}}
\]
where the probability is with respect to choice of \bfIndices and
\bfMarkedIndices by \targetSite.  Returning $\responseCiphertext{0}
\in \ciphertextSpace{\pubKey}(\plaintextGroupIdentity)$ and
$\responseCiphertext{1} \in \ciphertextSpace{\pubKey}(\loginPassword)$
will then induce a false alarm with higher probability than our
calculations in previous sections would suggest.  Since we model
\pwdHashFn as a random oracle, \monitorSite must search for such a
\loginPassword in a brute-force manner, but it can do so
\textit{offline}, whereas a \raisingAlarmAttacker as discussed
previously can attempt to induce a false alarm only as an online
attack.  Note that this attack is equally relevant for our adaptation
of the Amnesia protocol discussed in \appref{app:pcr-k}, since
\monitorSite obtains \bfHashFnSet there, as well.

To defend against this added risk of false alarms, \targetSite can use
two distinct Bloom filters per account for (both local and remote)
breach detection: one private Bloom filter $\langle \bfHashFnSetPriv,
\bfIndicesPriv, \bfMarkedIndicesPriv\rangle$ with marks and one
``public'' one $\langle \bfHashFnSetPub, \bfIndicesPub\rangle$ without
marks.  The detection rule in \secref{sec:amnesia:local:design} can
be modified to be
\[
  \loginResult{\loginPassword} \!=\!
  \left\{\begin{array}{@{\hspace{0.15em}}l@{\hspace{0.25em}}l}
  \loginFailure & \mbox{if $\pwdHashFn(\loginPassword)\!\bfNotIn\!\!\langle
    \bfHashFnSetPriv, \bfIndicesPriv\rangle \vee \pwdHashFn(\loginPassword)
    \!\bfNotIn\!\! \langle\bfHashFnSetPub, \bfIndicesPub\rangle$} \\
  \loginSuccess & \mbox{if $\pwdHashFn(\loginPassword) \!\bfIn\!\! \langle
    \bfHashFnSetPriv, \bfMarkedIndicesPriv\rangle$} \wedge
    \pwdHashFn(\loginPassword) \!\bfIn\!\! \langle\bfHashFnSetPub,
    \bfIndicesPub\rangle\\
  \loginAlarm & \mbox{otherwise}
  \end{array}\right.
\]
Here, $\bfHashFnSetPriv$ and $\bfHashFnSetPub$ are independently
chosen.  Then, \targetSite can use $\langle \bfHashFnSetPub,
\bfIndicesPub \rangle$ in the protocol of
\secref{sec:amnesia:remote:protocol:design} while keeping $\langle
\bfHashFnSetPriv, \bfIndicesPriv, \bfMarkedIndicesPriv\rangle$
private.  To limit the false detection probability against
\monitorSite to at most that provided in
\secref{sec:amnesia:local:raat} against a local \raisingAlarmAttacker
without knowledge of $\langle \bfHashFnSet, \bfIndices,
\bfMarkedIndices\rangle$, \targetSite can configure $\langle
\bfHashFnSetPriv, \bfIndicesPriv, \bfMarkedIndicesPriv \rangle$ as
prescribed previously. In this case, even with knowledge of
\bfHashFnSetPub and \userPassword, \monitorSite can do no better in
triggering a false alarm than a local \raisingAlarmAttacker could
have, due to the privacy of $\langle \bfHashFnSetPriv, \bfIndicesPriv,
\bfMarkedIndicesPriv\rangle$ at the \emph{unbreached} \targetSite.

In this design, increasing \setSize{\bfHashFnSetPub} reduces the true
detection probability.  Decreasing \setSize{\bfHashFnSetPub}, on the
other hand, increases the number of failed login passwords at
\monitorSite that are transmitted to \targetSite via the protocol of
\secref{sec:amnesia:remote:protocol:design}; as such, \monitorSite
might refuse a monitoring request for which
$\setSize{\bfHashFnSetPub}$ is too small.  Such impacts can be
quantified using the principles we developed previously.

\subsubsection{Security against a malicious \targetSite}
\label{sec:amnesia:remote:securityForMonitor}

The security risk that a malicious \targetSite poses to a monitor
\monitorSite is that \targetSite will learn more about the passwords
entered in login attempts at \monitorSite than it should.  The login
passwords \loginPassword on which \monitorSite executes the response
protocol (\figref{fig:monitor:respond}) are at \monitorSite's
discretion.  The key property that we show about this protocol is that
if \monitorSite executes the response protocol on a password
\loginPassword, then no information about \loginPassword is conveyed
to \targetSite unless the monitoring request is well-formed and
$\pwdHashFn(\loginPassword) \bfIn \langle\bfHashFnSet,
\bfIndicesAlt\rangle$ for the \bfIndicesAlt represented by the
monitoring request, no matter how \targetSite misbehaves.  This
security argument follows from three facts.
\begin{itemize}[nosep,leftmargin=1em,labelwidth=*,align=left]
\item If a monitor \monitorSite accepts a monitor request from
  \targetSite (\lineref{prot:deploy:monitorSave}), then
  $\{\ciphertext{\ciphertextIdx}\}_{\ciphertextIdx=1}^{\bfHashFnRange}
  \subseteq \ciphertextSpace{\pubKey}(\plaintextGroupGenerator)
  \cup \ciphertextSpace{\pubKey}(\plaintextGroupGenerator^{-1})$
  except with probability the soundness error of \zkp.

\item Let $\bfIndicesAlt = \{\ciphertextIdx :
  \ciphertext{\ciphertextIdx}
  \in \ciphertextSpace{\pubKey}(\plaintextGroupGenerator)\}$.  If
  $\{\ciphertext{\ciphertextIdx}\}_{\ciphertextIdx=1}^{\bfHashFnRange}
  \subseteq \ciphertextSpace{\pubKey}(\plaintextGroupGenerator)
  \cup \ciphertextSpace{\pubKey}(\plaintextGroupGenerator^{-1})$ but
  $\setSize{\bfIndicesAlt} \neq \bfNmbrIndices$, then $\derivedCiphertext{0}
  \in \ciphertextSpace{\pubKey} \setminus
  \ciphertextSpace{\pubKey}(\plaintextGroupIdentity)$ in
  \lineref{prot:deploy:monitorSave}.  Therefore,
  $\responseCiphertext{0}
  \in \ciphertextSpace{\pubKey}(\plaintextGroupIdentity)$ with
  probability at most $1/(\plaintextGroupOrder-1)$ and otherwise is
  uniformly distributed in $\ciphertextSpace{\pubKey} \setminus
  \ciphertextSpace{\pubKey}(\plaintextGroupIdentity)$.  When
  $\responseCiphertext{0} \in \ciphertextSpace{\pubKey} \setminus
  \ciphertextSpace{\pubKey}(\plaintextGroupIdentity)$, \responseCiphertext{1} is uniformly distributed in
  $\ciphertextSpace{\pubKey}\setminus\ciphertextSpace{\pubKey}(\loginPassword)$
  (\lineref{prot:respond:retvalOne}).

\item If
  $\{\ciphertext{\ciphertextIdx}\}_{\ciphertextIdx=1}^{\bfHashFnRange}
  \subseteq \ciphertextSpace{\pubKey}(\plaintextGroupGenerator)
  \cup \ciphertextSpace{\pubKey}(\plaintextGroupGenerator^{-1})$ and
  $\setSize{\bfIndicesAlt} = \bfNmbrIndices$ but in a run of the
  response protocol (\figref{fig:monitor:respond}) on \loginPassword,
  it is the case that $\pwdHashFn(\loginPassword) \bfNotIn \langle
  \bfHashFnSet, \bfIndicesAlt \rangle$, then the value
  \responseCiphertext{0} returned from \monitorSite is uniformly
  distributed in $\ciphertextSpace{\pubKey}
  \setminus \ciphertextSpace{\pubKey}(\plaintextGroupIdentity)$, and
  so \responseCiphertext{1} is uniformly distributed in
  $\ciphertextSpace{\pubKey}
  \setminus \ciphertextSpace{\pubKey}(\loginPassword)$.  This is
  immediate since in \lineref{prot:respond:derived},
  \derivedCiphertext{1} will be generated in
  $\ciphertextSpace{\pubKey}(\plaintextGroupGenerator^{-2
    \setSize{\bfHashFnSet(\pwdHashFn(\loginPassword)) \setminus
      \bfIndicesAlt}})$.  Therefore, \responseCiphertext{0} is
  uniformly distributed in $\ciphertextSpace{\pubKey}
  \setminus \ciphertextSpace{\pubKey}(\plaintextGroupIdentity)$
  (\ref{prot:respond:retvalZero}).
\end{itemize}

Given these facts, \monitorSite conveys information to \targetSite in
a run of the response protocol (\figref{fig:monitor:respond}) only if
$\{\ciphertext{\ciphertextIdx}\}_{\ciphertextIdx=1}^{\bfHashFnRange}
\subseteq \ciphertextSpace{\pubKey}(\plaintextGroupGenerator)
\cup \ciphertextSpace{\pubKey}(\plaintextGroupGenerator^{-1})$,
$\setSize{\bfIndicesAlt} = \bfNmbrIndices$, and
$\pwdHashFn(\loginPassword) \bfIn \langle \bfHashFnSet, \bfIndicesAlt
\rangle$.

\subsubsection{Performance}
\label{sec:amnesia:remote:performance}

We implemented the protocol of \secref{sec:amnesia:remote:protocol}
and compared its performance to Amnesia's \pcr{} protocol, retrieved
from \url{https://github.com/k3coby/pcr-go}.  To facilitate comparison
between \pcr{} and ours, we implemented ours in Go (as \pcr{} is) and
used the same encryption scheme \encScheme, namely ElGamal
encryption~\cite{elgamal1985:public-key} with \plaintextGroup being
the elliptic-curve group secp256r1~\cite{certicom2000:sec2v1}.  The
construction of the zero-knowledge proof \zkp is discussed in
\appref{app:analysis:primitives:zkp}.  Our implementation can be
retrieved from \url{https://github.com/k3coby/bhwmonitoring-go}.  In
our experiments, \targetSite and \monitorSite executed on separate AWS
EC2 instances having a single 2.50\gigahertz vCPU running Ubuntu
20.04.4.  All datapoints are the means of 50 runs; we report relative
standard deviations (\relstddev) in each caption.

\input{figures/fig_pcr_cmp2.tex}

\figref{fig:comparison2} provides a comparison of costs.  Note that
some vertical axes are log-scale.  In \figref{fig:comparison2}, we
denote by \nmbrExplicitHoneywords the number of explicit honeywords
with which \pcr{}'s monitoring request is configured, up to
$\nmbrExplicitHoneywords = 2^{12}$; this value is not excessive, but
rather even larger numbers are suggested in the Amnesia design to
offset certain threats~\cite[\secrefstatic{5.4}]{wang2021:amnesia}.
``Request generation by \targetSite''
(\figref{fig:cmp2:requestGenTime}) refers to execution at \targetSite
preceding the request message (i.e.,
\linesref{prot:deploy:keygen}{prot:deploy:targetSave} in
\figref{fig:monitor:deploy}, and the analogous instructions for
\pcr{}); ``Request validation by \monitorSite''
(\figref{fig:cmp2:requestValidateTime}) refers to execution at
\monitorSite following receipt of a well-formed request message
(\refs{prot:deploy:zkpVerify}{prot:deploy:monitorSave}); ``Response
generation by \monitorSite'' (\figref{fig:cmp2:responseTime}) refers
to execution at \monitorSite preceding the response message
(\refs{prot:respond:indices}{prot:respond:retvalOne} in
\figref{fig:monitor:respond}); and ``Response processing by
\targetSite'' (\figref{fig:cmp2:retrievalTime}) refers to execution at
\targetSite following receipt of a well-formed response message
(\refs{prot:respond:testCtext}{prot:respond:return}).  In the last
case, ``Negative'' refers to the case $\pwdHashFn(\loginPassword)
\bfNotIn \langle \bfHashFnSet, \bfIndices \rangle$, and ``Positive''
refers to the case $\pwdHashFn(\loginPassword) \bfIn \langle
\bfHashFnSet, \bfIndices \rangle$.

As shown in \figref{fig:comparison2}, in all measures except for
response computation by \monitorSite (\figref{fig:cmp2:responseTime}),
our protocol eventually outperforms the original \pcr{} protocol for
Amnesia as the number of honeywords is increased.  Response generation
by \monitorSite is more expensive by a mere $5\millisecs$, which could
be eliminated by reducing \bfNmbrHashFns from $\bfNmbrHashFns = 20$ to
$\bfNmbrHashFns = 10$.  We have used $\bfNmbrHashFns = 20$ throughout
this paper primarily to produce true-detection curves containing more
points and so that are smoother.  Reducing to $\bfNmbrHashFns = 10$
would have little practical effect.

Though we arrived at our remote-monitoring protocol through its
support for Bernoulli honeywords, it will work with any Bloom filter
provided by \targetSite.  So, it presents an alternative for remote
monitoring in an adaptation of Amnesia using a Bloom filter (vs.\ a
Cuckoo filter~\cite{fan2014:cuckoo}), even one populated with explicit
honeywords.

\section{Discussion}
\label{sec:discussion}

\myparagraph{Online password guessing} Our primary threat models
considered in this paper, namely \raisingAlarmAttackers and
\breachingAttackers, leave one additional threat model to consider:
The risk that an attacker \textit{not} knowing the user-chosen
password succeeds in an online password guessing attack to access an
account---versus to induce a false breach alarm, as a
\raisingAlarmAttacker does \textit{with} knowledge of the user-chosen
password.  Denote by $\guessingProb(\nmbrLoginsPerAccount)$ the
probability with which an online attacker would succeed in
\underline{g}uessing the \underline{p}assword for a particular account
within \nmbrLoginsPerAccount tries.  Then, Bernoulli honeywords
increase his probability of accessing this account to no more than
$\guessingProb(\nmbrLoginsPerAccount) +
\falseAlarmProb(\nmbrLoginsPerAccount, 1)$, since entry of either the
user-chosen password or a honeyword is necessary (though not
sufficient in our specific designs\footnote{Our honeychecker design
(\secref{sec:honeychecker:design}) also requires
$\bfHashFnSet(\pwdHashFn(\loginPassword)) =
\bfHashFnSet(\pwdHashFn(\userPassword))$ and our Amnesia adaptation
(\secref{sec:amnesia:local:design}) also requires
$\pwdHashFn(\loginPassword) \bfIn \langle \bfHashFnSet,
\bfMarkedIndices\rangle$.}) to access the account.  As discussed in
\secref{sec:honeychecker:eval}, Flor\^{e}ncio et
al.~\cite{florencio2014:guide} recommend resisting up to
$\nmbrLoginsPerAccount = 10^6$ online guesses in a prolonged
depth-first campaign, which we interpreted to require
$\falseAlarmProb(10^6, 10) \le 10^{-1}$ (and so $\falseAlarmProb(10^6,
1) < 10^{-2}$).  Still, if the user-chosen password satisfies
$\guessingProb(10^6) \ll 10^{-2}$, then it is conceivable that
Bernoulli honeywords could theoretically weaken the account to access
by an online guessing attack.  That said, any such weakening is of
little practical importance, for online guessing attacks or others.
Flor\^{e}ncio et al.\ put it this way: ``... consider two passwords
which withstand $10^6$ and $10^{12}$ guesses respectively ... there is
no apparent scenario in which the extra guess-resistance of the second
password helps.  ... both will survive online guessing, but neither
will survive offline attack''~\cite[p.\ 43]{florencio2014:guide}.  We
therefore conclude that the advantages brought by Bernoulli honeywords
far outweigh any additional risk of account access by an online
guessing attack.  In those rare cases of a user-chosen password
capable of withstanding even an offline attack, a site can simply
exempt this account from using Bernoulli honeywords.

\myparagraph{Space efficiency}
Though not our primary motivation for using them, Bloom filters are
very space-efficient data structures, which has additional benefits
for our designs.  For example, Wang et
al.~\cite[\secrefstatic{V.B}]{wang2022:honeywords} estimates the
honeyword storage for $10^7$ accounts costs $12.8\gigabytes$ for 40
honeywords per user.  In our design, storing Bloom filters with
$\bfHashFnRange = 128$ (the parameter we chose for our security
evaluations) for the same number of users would cost $160\megabytes$
only.

This space efficiency also has benefits for remote monitoring.  For
example, with $\bfHashFnRange = 128$ our protocol in
\secref{sec:amnesia:remote:protocol} would require \monitorSite to
store $<1\gigabytes$ for $10^5$ monitoring requests.  In contrast, the
authors of Amnesia show that if 4096 \textit{explicit} honeywords are
deployed for each account, it requires $32\gigabytes$ to store the
same number of monitoring requests.

A direction for future research is more space-efficient methods for
implementing Bernoulli honeywords.  Certain natural candidates,
e.g., simply storing $\pwdHashFn(\userPassword) \bmod \pwdHashBins$
for the user-chosen password \userPassword, would implement
$\honeyProb = 1/\pwdHashBins$.  However, a similarly efficient
construction is possible using Bloom filters, by setting
$\bfHashFnRange = \pwdHashBins$, $\bfNmbrHashFns = 1$, and
$\setSize{\bfIndices} = 1$.

\section{Conclusion}
\label{sec:conclusion}

In this paper we have made the case for choosing honeywords from all
possible passwords as a Bernoulli process, in contrast to previous
proposals to generate a small number of honeywords per account using
heuristics.  We have shown how to realize this idea within existing
honeyword system designs, namely the original honeychecker-based
design of Juels \& Rivest and the more recent Amnesia proposal.
Moreover, we have shown that our design enables even greater
efficiency than the previous Amnesia proposal for remotely monitoring
for the entry of a site's honeywords elsewhere.  Most critically,
though, we have shown that Bernoulli honeywords permit analytic
estimation of true and false breach-detection rates, which we provide
for our realizations.  In particular, when evaluated against realistic
threats, Bernoulli honeywords enable detection of credential database
breaches with a risk of false alarms that is quantifiable, tunable,
and independent of the adversary's knowledge of the site's users.

\noindent\textit{Acknowledgment}:
This research was supported by grant W911NF-17-1-0370 from the
U.S.\ Army Research Office.

\bibliographystyle{IEEEtranS}
\bibliography{full,main}

\appendices

\section{A Remote-Monitoring Alternative}
\label{app:pcr-k}

A fairly direct method to adapt the \pcr{} protocol of Amnesia to
accommodate the local-detection design in \secref{sec:amnesia:local}
using Bloom filter $\langle \bfHashFnSet, \bfIndices \rangle$ is for
\targetSite to include the elements of \bfIndices in the encrypted
data structure it sends to \monitorSite, instead of the (hashes of
the) account passwords themselves.  Upon receiving a login attempt
with password \loginPassword, \monitorSite can then execute the
response computation \bfNmbrHashFns times, once using
$\bfHashFn{\bfHashFnIdx}(\pwdHashFn(\loginPassword))$ as the test
plaintext for $\bfHashFnIdx = 1 \ldots \bfNmbrHashFns$.  As long as
the response plaintexts in these executions were chosen to combine to
produce \loginPassword (say, by a \bfNmbrHashFns-out-of-\bfNmbrHashFns
secret sharing), \targetSite would obtain \loginPassword if and only
if $\pwdHashFn(\loginPassword) \bfIn \langle \bfHashFnSet, \bfIndices
\rangle$.  We refer to this adaptation using \bfNmbrHashFns executions
of the \pcr{} protocol as \pcr{\bfNmbrHashFns}.

\input{figures/fig_pcr_cmp.tex}

We implemented \pcr{\bfNmbrHashFns} as a simple modification to the Go
implementation of the original Amnesia remote-monitoring protocol.
The comparison between our protocol of \secref{sec:amnesia:remote} and
\pcr{\bfNmbrHashFns} is given in \figref{fig:comparison}; note that
all vertical axes are log-scale.  As before, all datapoints are the
means of 50 runs; we report relative standard deviations (\relstddev)
in each caption.

Monitor deployment is a rare cost compared to monitor responses, as
one monitor deployment could produce thousands of monitor responses
during its lifetime at \monitorSite.  So, while the monitor request is
larger in our design than in \pcr{\bfNmbrHashFns}
(\figref{fig:cmp:requestSize}) and the monitor request is an
order-of-magnitude more costly for \targetSite to create
(\figref{fig:cmp:requestGenTime}) and \monitorSite to validate
(\figref{fig:cmp:requestValidateTime}), these costs are still modest
($\le 200\millisecs$ in all cases shown) and concern us little.
Moreover, $\approx 79\%$ of the monitor request size is consumed by
the proof \zkp, which is not saved once it is checked
(\lineref{prot:deploy:targetSave}); as such, the storage consumed by
the saved monitor request at \monitorSite is similar to that in
\pcr{\bfNmbrHashFns}.  The far more important costs are response
generation by \monitorSite (\figref{fig:cmp:responseTime}), response
processing by \targetSite (\figref{fig:cmp:retrievalTime}), and the
response message size (\figref{fig:cmp:responseSize}), since these are
incurred per unsuccessful login at \monitorSite.  As we can see, our
protocol outperforms \pcr{\bfNmbrHashFns} by an order of magnitude in
all of these measures.

\input{proofs.tex}

\end{document}

%% file: macros.tex
\newcommand{\myparagraph}[1]{\smallskip\noindent\textit{#1}:\xspace}
\newcounter{note}[section]

\newcommand{\refs}[2]{\mbox{\ref{#1}--\ref{#2}}\xspace}
\newcommand{\secref}[1]{\mbox{Sec.~\ref{#1}}\xspace}

\newcommand{\lineref}[1]{\mbox{line~\ref{#1}}\xspace}
\newcommand{\linesref}[2]{\mbox{lines~\ref{#1}--\ref{#2}}\xspace}
\newcommand{\secsref}[2]{\mbox{Secs.~\ref{#1}--\ref{#2}}\xspace}
\newcommand{\secrefstatic}[1]{\mbox{Sec.~{#1}}}
\newcommand{\figref}[1]{\mbox{Fig.~\ref{#1}}\xspace}

\newcommand{\subfigsref}[2]{\mbox{Figs.~\ref{#1}--\subref{#2}}\xspace}
\newcommand{\figrefstatic}[1]{\mbox{Fig.~{#1}}}

\newcommand{\tblref}[1]{\mbox{Table~\ref{#1}}\xspace}
\newcommand{\tblrefstatic}[1]{\mbox{Table~{#1}}\xspace}

\newcommand{\appref}[1]{\mbox{App.~\ref{#1}}\xspace}
\newcommand{\eqnref}[1]{\mbox{(\ref{#1})}\xspace}

\newcommand{\propref}[1]{\mbox{Prop.~\ref{#1}}\xspace}

\newcommand{\msgref}[1]{\mbox{message~\ref{#1}}\xspace}

\newtheorem{prop}{Proposition}
\newtheorem{defn}{Definition}

\newcommand{\kilobytes}{\ensuremath{\mathrm{KB}}\xspace}

\newcommand{\megabytes}{\ensuremath{\mathrm{MB}}\xspace}
\newcommand{\gigabytes}{\ensuremath{\mathrm{GB}}\xspace}

\newcommand{\gigahertz}{\ensuremath{\mathrm{GHz}}\xspace}

\newcommand{\millisecs}{\ensuremath{\mathrm{ms}}\xspace}

\newcommand{\relstddev}{\ensuremath{\chi}\xspace}

\newcounter{requesterLineNmbr}
\renewcommand{\therequesterLineNmbr}{\ensuremath{\mathsf{t\arabic{requesterLineNmbr}}}}
\newcommand{\requesterLabel}[1]{\refstepcounter{requesterLineNmbr}\label{#1}\therequesterLineNmbr.}

\newcounter{responderLineNmbr}
\renewcommand{\theresponderLineNmbr}{\ensuremath{\mathsf{m\arabic{responderLineNmbr}}}}
\newcommand{\responderLabel}[1]{\refstepcounter{responderLineNmbr}\label{#1}\theresponderLineNmbr.}

\newcounter{messageNmbr}
\renewcommand{\themessageNmbr}{\ensuremath{\mathsf{r\arabic{messageNmbr}}}}
\newcommand{\messageLabel}[1]{\refstepcounter{messageNmbr}\label{#1}\themessageNmbr.~~}

\newlength{\figureheight}
\newcommand{\figurewidth}{\columnwidth}

\definecolor{curve_color}{rgb}{0.129411764705882,0.380392156862745,0.549019607843137}

\newcommand{\raisingAlarmAttacker}{\textit{raat}\xspace}
\newcommand{\breachingAttacker}{\textit{brat}\xspace}
\newcommand{\raisingAlarmAttackers}{\textit{raats}\xspace}
\newcommand{\breachingAttackers}{\textit{brats}\xspace}

\newcommand{\genericRV}{\ensuremath{\rvNotation{Y}}\xspace}
\newcommand{\genericRVAlt}{\ensuremath{\rvNotation{Y}'}\xspace}
\NewDocumentCommand{\genericNat}{ g }{\ensuremath{z\IfNoValueF{#1}{_{#1}}{}}\xspace}
\newcommand{\residues}[1]{\ensuremath{\mathbb{Z}_{#1}}\xspace}
\newcommand{\relPrimeResidues}[1]{\ensuremath{\mathbb{Z}_{#1}^{\ast}}\xspace}
\newcommand{\setSize}[1]{\ensuremath{\left|{#1}\right|}\xspace}

\newcommand{\getsr}{\ensuremath{\overset{\scriptscriptstyle\$}{\leftarrow}}\xspace}
\newcommand{\defined}{\ensuremath{\stackrel{\text{\tiny def}}{=}}\xspace}
\newcommand{\boolTrue}{\ensuremath{\mathit{true}}\xspace}
\newcommand{\boolFalse}{\ensuremath{\mathit{false}}\xspace}

\newcommand{\distEqual}{\ensuremath{\mathrel{\;\overset{d}{=}\;}}\xspace}

\newcommand{\prob}[1]{\ensuremath{\mathop{\mathbb{P}}\left({#1}\right)}\xspace}
\newcommand{\expv}[1]{\ensuremath{\mathop{\mathbb{E}}\left({#1}\right)}\xspace}
\newcommand{\cset}[3]{\ensuremath{{#1\{}}{#2}\ensuremath{\;{#1|}} \ifmmode{\;}\fi {#3}\ensuremath{{#1\}}}\xspace}
\newcommand{\cprob}[3]{\ensuremath{\mathop{\mathbb{P}}{#1(}}{#2}\ensuremath{\;{#1|}} \ifmmode{\;}\fi {#3}\ensuremath{{#1)}}\xspace}
\newcommand{\cexpv}[3]{\ensuremath{\mathop{\mathbb{E}}{#1(}}{#2}\ensuremath{\;{#1|}} \ifmmode{\;}\fi {#3}\ensuremath{{#1)}}\xspace}
\newcommand{\setNotation}[1]{\ensuremath{\mathcal{#1}}\xspace}
\newcommand{\rvNotation}[1]{\ensuremath{\varmathbb{#1}}\xspace}

\makeatletter
\newcommand{\vast}{\bBigg@{4}}
\newcommand{\Vast}{\bBigg@{5}}
\makeatother

\newcommand{\encScheme}{\ensuremath{\mathcal{E}}\xspace}
\newcommand{\keygen}{\ensuremath{\mathsf{Gen}}\xspace}
\newcommand{\encrypt}[1]{\ensuremath{\mathsf{Enc}_{#1}}\xspace}
\newcommand{\decrypt}[1]{\ensuremath{\mathsf{Dec}_{#1}}\xspace}
\newcommand{\encMult}[1]{\ensuremath{\times_{#1}}\xspace}
\NewDocumentCommand{\plaintext}{ g g }{\ensuremath{m\IfNoValueF{#1}{\IfNoValueTF{#2}{_{#1}}{_{{#1},{#2}}}}}\xspace}
\NewDocumentCommand{\plaintextAlt}{ g g }{\ensuremath{m\IfNoValueF{#1}{\IfNoValueTF{#2}{_{#1}}{_{{#1},{#2}}}}'}\xspace}
\NewDocumentCommand{\plaintextAltAlt}{ g g }{\ensuremath{\hat{m}\IfNoValueF{#1}{\IfNoValueTF{#2}{_{#1}}{_{{#1},{#2}}}}}\xspace}
\newcommand{\plaintextIdx}{\ensuremath{j}\xspace}
\NewDocumentCommand{\ciphertext}{ g }{\ensuremath{c\IfNoValueF{#1}{_{#1}}{}}\xspace}
\NewDocumentCommand{\derivedCiphertext}{ g }{\ensuremath{d\IfNoValueF{#1}{_{#1}}{}}\xspace}
\NewDocumentCommand{\derivedCiphertextAlt}{ g }{\ensuremath{d'\IfNoValueF{#1}{_{#1}}{}}\xspace}
\newcommand{\responseCiphertext}[1]{\ensuremath{\hat{c}_{#1}}\xspace}
\NewDocumentCommand{\ciphertextIdx}{ g }{\ensuremath{j\IfNoValueF{#1}{_{#1}}{}}\xspace}
\newcommand{\ciphertextSpace}[1]{\ensuremath{C_{#1}}\xspace}
\newcommand{\privKey}{\ensuremath{\mathit{sk}}\xspace}
\newcommand{\pubKey}{\ensuremath{\mathit{pk}}\xspace}
\NewDocumentCommand{\encProd}{ g g g }{\ensuremath{\displaystyle\operatorname*{\textstyle\prod_{\mathrlap{#1}}}\IfNoValueF{#2}{_{#2}}\IfNoValueF{#3}{^{#3}}\hphantom{_{#1}}}\xspace}
\newcommand{\encRand}[1]{\ensuremath{\$_{#1}}\xspace}
\newcommand{\failureSymbol}{\ensuremath{\bot}\xspace}
\newcommand{\plaintextGroup}{\ensuremath{\setNotation{G}}\xspace}
\newcommand{\plaintextGroupOrder}{\ensuremath{q}\xspace}
\newcommand{\plaintextGroupGenerator}{\ensuremath{g}\xspace}
\newcommand{\plaintextGroupOperator}{\ensuremath{\circ}\xspace}
\newcommand{\plaintextGroupIdentity}{\ensuremath{\mathbf{1}}\xspace}

\NewDocumentCommand{\niwipGen}{ o g }{\ensuremath{\mathsf{wipGen}\IfNoValueF{#1}{_{#1}}\IfNoValueF{#2}{\left[{#2}\right]}}\xspace}
\NewDocumentCommand{\niwipGenAlt}{ o g }{\ensuremath{\mathsf{wipGen}'\IfNoValueF{#1}{_{#1}}\IfNoValueF{#2}{\left[{#2}\right]}}\xspace}

\newcommand{\zkp}{\ensuremath{\Psi}\xspace}
\NewDocumentCommand{\zkpGen}{ o }{\ensuremath{\mathsf{zkpGen}\IfNoValueF{#1}{_{#1}}}\xspace}
\NewDocumentCommand{\zkpGenAlt}{ o }{\ensuremath{\widehat{\mathsf{zkpGen}}\IfNoValueF{#1}{_{#1}}}\xspace}
\newcommand{\zkpVerify}{\ensuremath{\mathsf{zkpVerify}}\xspace}
\newcommand{\zkpSim}{\ensuremath{\mathsf{zkpSim}}\xspace}
\newcommand{\zkpSimHash}{\ensuremath{\zkpSim.\mathsf{hash}}\xspace}
\newcommand{\zkpSimProve}{\ensuremath{\zkpSim.\mathsf{prove}}\xspace}
\newcommand{\pcr}[1]{\ensuremath{\text{PCR}^{#1}}\xspace}

\newcommand{\pwdCountIdx}{\ensuremath{j}\xspace}
\newcommand{\pwdCountIdxAlt}{\ensuremath{j'}\xspace}
\NewDocumentCommand{\pwdRank}{ g }{\ensuremath{r\IfNoValueF{#1}{_{#1}}}\xspace}
\NewDocumentCommand{\pwdRankAlt}{ g }{\ensuremath{r\IfNoValueF{#1}{_{#1}}'}\xspace}
\NewDocumentCommand{\pwdRankSet}{ g }{\ensuremath{\setNotation{R}\IfNoValueF{#1}{_{#1}}}\xspace}
\newcommand{\pwdCount}{\ensuremath{\theta}\xspace}
\newcommand{\pwdRankChosen}{\ensuremath{\hat{r}}\xspace}
\newcommand{\pwdRankChosenAlt}{\ensuremath{\check{r}}\xspace}
\newcommand{\pwdRankMax}{\ensuremath{r^{\ast}}\xspace}
\NewDocumentCommand{\rvPwdRankChosen}{ g }{\ensuremath{\rvNotation{U}\IfNoValueF{#1}{_{\text{\tiny #1}}}{}}\xspace}
\NewDocumentCommand{\rvPwdRank}{ g }{\ensuremath{\rvNotation{R}\IfNoValueF{#1}{_{#1}}}\xspace}
\newcommand{\honeywordInd}[1]{\ensuremath{s[#1]}\xspace}
\newcommand{\rvHoneywordInd}[1]{\ensuremath{\rvNotation{S}[#1]}\xspace}
\newcommand{\pwdList}[1]{\ensuremath{\mathsf{pwd}[#1]}\xspace}
\newcommand{\honeyProb}{\ensuremath{p_{\mathsf{h}}}\xspace}
\newcommand{\honeyProbModified}[1]{\ensuremath{\hat{p}_{\mathsf{h}}({#1})}\xspace}
\NewDocumentCommand{\nmbrAccountsAttempted}{ g }{\ensuremath{n\IfNoValueF{#1}{_{#1}}}\xspace}
\NewDocumentCommand{\nmbrAccounts}{ g }{\ensuremath{N\IfNoValueF{#1}{_{#1}}}\xspace}
\NewDocumentCommand{\accountId}{ g }{\ensuremath{a\IfNoValueF{#1}{_{#1}}}\xspace}
\newcommand{\accountsSet}{\ensuremath{\setNotation{A}}\xspace}
\newcommand{\accountsSubset}{\ensuremath{\setNotation{A}'}\xspace}
\NewDocumentCommand{\honeyEntryProb}{ g g g }{\ensuremath{\textsc{hwin}^{#1}\IfNoValueF{#2}{\left({#2},{#3}\right)}}\xspace}
\NewDocumentCommand{\trueAlarmProb}{ g }{\ensuremath{\textsc{tdp}\IfNoValueF{#1}{_{#1}}}\xspace}
\NewDocumentCommand{\trueAlarmProbAlt}{ g }{\ensuremath{\widetilde{\textsc{tdp}}\IfNoValueF{#1}{_{#1}}}\xspace}
\newcommand{\falseAlarmProb}{\ensuremath{\textsc{fdp}}\xspace}
\newcommand{\guessingProb}{\ensuremath{\textsc{gp}}\xspace}
\newcommand{\loginResult}[1]{\ensuremath{\mathsf{login}({#1})}\xspace}
\newcommand{\loginSuccess}{\ensuremath{\mathit{success}}\xspace}
\newcommand{\loginFailure}{\ensuremath{\mathit{failure}}\xspace}
\newcommand{\loginAlarm}{\ensuremath{\mathit{alarm}}\xspace}
\newcommand{\nmbrLoginsSeen}{\ensuremath{L}\xspace}
\newcommand{\nmbrLoginsTotal}{\ensuremath{L'}\xspace}
\newcommand{\nmbrLoginsPerAccount}{\ensuremath{\ell}\xspace}
\newcommand{\loginIdx}{\ensuremath{l}\xspace}
\newcommand{\nmbrExplicitHoneywords}{\ensuremath{h}\xspace}

\newcommand{\bfNmbrHashFns}{\ensuremath{k}\xspace}
\newcommand{\bfHashFn}[1]{\ensuremath{f_{#1}}\xspace}
\NewDocumentCommand{\bfHashFnSet}{ g }{\ensuremath{\setNotation{F}\IfNoValueF{#1}{_{#1}}}\xspace}
\newcommand{\bfHashFnSetPriv}{\ensuremath{\bfHashFnSet_{\mathsf{pr}}}\xspace}
\newcommand{\bfHashFnSetPub}{\ensuremath{\bfHashFnSet_{\mathsf{pu}}}\xspace}
\newcommand{\bfHashFnIdx}{\ensuremath{i}\xspace}
\newcommand{\bfHashFnIdxAlt}{\ensuremath{i'}\xspace}
\newcommand{\bfHashFnDomainBits}{\ensuremath{v}\xspace}
\newcommand{\bfHashFnRange}{\ensuremath{b}\xspace}
\NewDocumentCommand{\bfIndices}{ g }{\ensuremath{\setNotation{B}\IfNoValueF{#1}{_{#1}}}\xspace}
\newcommand{\bfIndicesPriv}{\ensuremath{\bfIndices_{\mathsf{pr}}}\xspace}
\newcommand{\bfIndicesPub}{\ensuremath{\bfIndices_{\mathsf{pu}}}\xspace}
\newcommand{\bfIndicesAlt}{\ensuremath{\setNotation{B}'}\xspace}
\newcommand{\bfNmbrIndices}{\ensuremath{b'}\xspace}
\NewDocumentCommand{\bfMarkedIndices}{ g }{\ensuremath{\setNotation{M}\IfNoValueF{#1}{_{#1}}{}}\xspace}
\newcommand{\bfMarkedIndicesPriv}{\ensuremath{\bfMarkedIndices_{\mathsf{pr}}}\xspace}

\newcommand{\genericElmt}{\ensuremath{e}\xspace}
\newcommand{\genericElmtAlt}{\ensuremath{e'}\xspace}
\newcommand{\bfTestIn}{\ensuremath{\mathrel{\;\overset{?}{\in}_{\textsc{b}}\;}}\xspace}
\newcommand{\bfIn}{\ensuremath{\in_{\textsc{b}}}\xspace}
\newcommand{\bfNotIn}{\ensuremath{\not\in_{\textsc{b}}}\xspace}
\newcommand{\pwdHashFn}{\ensuremath{H}\xspace}
\newcommand{\pwdHashBins}{\ensuremath{w}\xspace}

\newcommand{\userPassword}{\ensuremath{\hat{\pi}}\xspace}
\newcommand{\loginPassword}{\ensuremath{\pi}\xspace}

\newcommand{\bfUserIndices}{\ensuremath{\hat{\setNotation{S}}}\xspace}
\newcommand{\bfLoginIndices}{\ensuremath{\setNotation{S}}\xspace}

\newcommand{\markProb}{\ensuremath{p_{\mathsf{mark}}}\xspace}
\newcommand{\remarkProb}{\ensuremath{p_{\mathsf{remark}}}\xspace}
\newcommand{\monitorSite}{\ensuremath{\mathsf{M}}\xspace}
\newcommand{\targetSite}{\ensuremath{\mathsf{T}}\xspace}

\newcommand{\passwordGuesses}{\ensuremath{x}\xspace}
\newcommand{\wangCracked}{\ensuremath{y}\xspace}

\newcommand{\TGI}{\ensuremath{\text{TG-I}}\xspace}
\newcommand{\TGIPP}{\ensuremath{\text{TG-I}''}\xspace}
\newcommand{\TGIPPP}{\ensuremath{\text{TG-I}'''}\xspace}
\newcommand{\PCFG}{\ensuremath{\text{PCFG}}\xspace}
\newcommand{\blistAlg}{\ensuremath{A}\xspace}

\NewDocumentCommand{\tdpTolerance}{ g }{\ensuremath{\delta\IfNoValueF{#1}{_{#1}}}\xspace}

\ExplSyntaxOn
\cs_new:Npn \intensity #1
   {\fp_eval:n {\MinIntensity + (1.0-\MinIntensity) * ((({#1}*1000-\value{MinNumber})/
   (\value{MaxNumber}-\value{MinNumber})))}
   }
\ExplSyntaxOff
\newcommand{\MinIntensity}{0.5}   
\newcounter{MinNumber}
\newcounter{MaxNumber}
\newcommand{\ApplyGradientX}[1]{\cellcolor[gray]{\intensity{#1}}{\parbox{2em}{\centering \raggedleft{#1}}}}
\newcommand{\ApplyGradientY}[1]{\cellcolor[gray]{\intensity{#1}}{\parbox{1.5em}{\centering \raggedleft{#1}}}}
\newcolumntype{X}{>{\collectcell\ApplyGradientX}c<{\endcollectcell}}
\newcolumntype{Y}{>{\collectcell\ApplyGradientY}c<{\endcollectcell}}

\newcommand{\yIntercept}{\ensuremath{\alpha}\xspace}
\newcommand{\slope}{\ensuremath{\beta}\xspace}

\newcommand{\fdpBound}{\ensuremath{\epsilon}\xspace}

\newcommand{\codeExpt}{\ensuremath{\mathtt{Experiment}~}}

\newcommand{\codeIf}{\ensuremath{\mathtt{if}~}}
\newcommand{\codeThen}{\ensuremath{\mathtt{then}~}}

\newcommand{\codeReturn}{\ensuremath{\mathtt{return}~}}

\newcommand{\elgPrivKey}{\ensuremath{u}\xspace}
\newcommand{\elgPubKey}{\ensuremath{U}\xspace}
\newcommand{\elgEphemeralPrivKey}[1]{\ensuremath{v_{#1}}\xspace}
\newcommand{\elgEphemeralPubKey}[1]{\ensuremath{V_{#1}}\xspace}
\newcommand{\elgCiphertext}[1]{\ensuremath{W_{#1}}\xspace}
\newcommand{\elgGroup}{\ensuremath{G}\xspace}
\newcommand{\elgGroupGenerator}{\ensuremath{g}\xspace}
\newcommand{\elgGroupExponent}{\ensuremath{y}\xspace}

\newcommand{\lang}{\ensuremath{\mathcal{L}}\xspace}
\newcommand{\relation}{\ensuremath{\mathcal{R}_{\lang}}\xspace}
\newcommand{\statement}{\ensuremath{\textsc{x}}\xspace}
\NewDocumentCommand{\witness}{ g }{\ensuremath{w\IfNoValueF{#1}{_{#1}}}\xspace}
\newcommand{\zkpName}{\ensuremath{\Pi}\xspace}
\newcommand{\zkpSoundnessAdvantage}[1]{\ensuremath{\mathsf{Adv}^{\textsc{Snd}}_{#1}}\xspace}
\newcommand{\zkpExperiment}[2]{\ensuremath{\mathsf{Expt}^{\textsc{zkp}\mbox{,}
{#1}}_{#2}}\xspace}
\newcommand{\zkpAdversary}{\ensuremath{\mathfrak{D}_\textsc{zkp}}\xspace}
\newcommand{\zkpAdversaryAlt}{\ensuremath{\hat{\mathfrak{D}}_\textsc{zkp}}\xspace}

\newcommand{\zkpAdvantage}[1]{\ensuremath{\mathsf{Adv}^{\textsc{zkp}}_{#1}}\xspace}
\newcommand{\zkpBit}{\ensuremath{\textsf{b}''}\xspace}
\newcommand{\zkpAdversaryBit}{\ensuremath{\hat{\textsf{b}''}}\xspace}

\newcommand{\randomOracle}{\ensuremath{\mathsf{hash}}\xspace}
\newcommand{\randomOracles}{\ensuremath{\mathcal{H}}\xspace}

\newcommand{\roQueryBoundZKP}{\ensuremath{q_\textsc{zkp}}\xspace}
\newcommand{\timeBound}{\ensuremath{t}\xspace}
\newcommand{\timeBoundZKP}{\ensuremath{t_\textsc{zkp}}\xspace}
\newcommand{\timeBoundSound}{\ensuremath{t_\textsc{snd}}\xspace}

\newcommand{\indcpaExperiment}[2]{\ensuremath{\mathsf{Expt}^{\textsc{cpa}
\mbox{,}{#1}}_{#2}}\xspace}
\newcommand{\indcpaAdversary}{\ensuremath{\mathfrak{D}_\textsc{cpa}}\xspace}
\newcommand{\indcpaAdvantage}[1]{\ensuremath{\mathsf{Adv}^{\textsc{cpa}}_{#1}}\xspace}
\newcommand{\indcpaLROracle}[3]{\ensuremath{\encrypt{#1}(\textrm{LR}({#2},{#3},\indcpaBit))}\xspace}
\newcommand{\indcpaBit}{\ensuremath{\textsf{b}'}\xspace}
\newcommand{\indcpaAdversaryBit}{\ensuremath{\hat{\textsf{b}'}}\xspace}
\newcommand{\hybridIndex}{\ensuremath{i}\xspace}
\newcommand{\hybridCPA}[1]{\ensuremath{\mathsf{H}{\left(#1\right)}}\xspace}
\newcommand{\lrQueryBound}{\ensuremath{q_\textsc{cpa}}\xspace}
\newcommand{\timeBoundCPA}{\ensuremath{t_\textsc{cpa}}\xspace}
\newcommand{\timeEnc}{\ensuremath{t_\textsc{enc}}\xspace}
\newcommand{\timeHash}{\ensuremath{t_\textsc{hash}}\xspace}

\newcommand{\targetAdversary}{\ensuremath{\mathfrak{T}}\xspace}

\newcommand{\protocolName}{}

\newcommand{\targetAlgorithms}[2]{\ensuremath{\targetSite_{{#1}\mbox{-}{#2}}}\xspace}
\newcommand{\monitorExperiment}[2]{\ensuremath{\mathsf{Expt}^{\targetSite
\mbox{,}{#1}}_{#2}}\xspace}
\NewDocumentCommand{\monitorAdversary}{ o }{\ensuremath{\mathfrak{M}_{\IfNoValueF{#1}{#1}}}\xspace}
\newcommand{\monitorAdversaryState}{\ensuremath{\phi}\xspace}
\newcommand{\monitorAdvantage}[1]{\ensuremath{\mathsf{Adv}^{\targetSite}_
{#1}}\xspace}
\newcommand{\monitorExptBit}{\ensuremath{\textsf{b}}\xspace}
\newcommand{\monitorAdversaryBit}{\ensuremath{\hat{\textsf{b}}}\xspace}
\newcommand{\worldBitOne}{\ensuremath{\alpha}\xspace}
\newcommand{\worldBitTwo}{\ensuremath{\beta}\xspace}
\newcommand{\probHybrid}[1]{\ensuremath{\mathcal{P}}{\left(#1\right)}\xspace}

\newcommand{\roQueryBound}{\ensuremath{q_{\textsc{ro}}}\xspace}

%% file: figures/fig_fitting.tex
\begin{figure}[t]
\captionsetup[subfigure]{font=small,labelfont=small}
\begin{subfigure}[b]{\columnwidth}
  \newcommand{\stepequation}{\refstepcounter{equation}(\theequation)}
  {\footnotesize
  \begin{tabular}{@{}r@{\hspace{0.9em}}c@{\hspace{0.9em}}c@{\hspace{0.9em}}c@{\hspace{0.9em}}c@{\hspace{0.9em}}c@{}}
    \toprule
    \multicolumn{1}{c}{\multirow{2}{*}{Data source}} & \multicolumn{2}{c}{Best fit} & \multirow{2}{*}{$R^2$} & \multirow{2}{*}{RMSE} & \\
                & \multicolumn{2}{c}{$\prob{\rvPwdRankChosen \le \pwdRank} = \yIntercept + \slope\log_{10}(\pwdRank)$} & & \\
    \midrule
    \TGI~\cite[\figrefstatic{8}]{wang2016:targeted} & $\yIntercept = 0.0419$ & $\slope = 0.0771$ & $0.99$ & $0.004$ & \stepequation\label{eqn:wangCracked-TGI} \\
    \TGIPP~\cite[\figrefstatic{8}]{wang2016:targeted} & $\yIntercept = 0.0118$ & $\slope = 0.0601$ & $0.99$ & $0.004$ & \stepequation\label{eqn:wangCracked-TGIPP} \\
    \TGIPPP~\cite[\figrefstatic{8}]{wang2016:targeted} & $\yIntercept = 0.0013$ & $\slope = 0.0310$ & $0.97$ & $0.005$ & \stepequation\label{eqn:wangCracked-TGIPPP}\\
    \PCFG~\cite[\figrefstatic{8}]{wang2016:targeted} & $\yIntercept = -0.0047$ & $\slope = 0.0172$ & $0.96$ & $0.003$ & \stepequation\label{eqn:wangCracked-PCFG}\\
    CMU~\cite[\figrefstatic{7}]{mazurek2013:guessability} & $\yIntercept = -0.4119$ & $\slope = 0.0602$ & $0.98$ & $0.022$ & \stepequation\label{eqn:mazurekCracked} \\
    CKL-PCFG~\cite[\figrefstatic{5}]{xu2021:chunk}& $\yIntercept = -0.6938$ & $\slope = 0.1240$ & $0.97$ & $0.053$ & \stepequation\label{eqn:xuCracked} \\
    \bottomrule
  \end{tabular}}
  \caption{Data sources and best-fit CDFs}
  \label{fig:fitting:data}
  \end{subfigure}
\\[12pt]
  \begin{subfigure}[b]{.48\columnwidth}
    \setlength\figureheight{2.4in}
    \begin{minipage}[b]{\textwidth}
      \centering
      \vspace*{0em}\resizebox{!}{13.9em}{\input{figures/fitting.tex}}
      \hspace*{1.25em} 
      \begin{minipage}[t]{11em}
       \vspace*{-0.65em}
        \caption{Original (thicker) CDFs and best-fit (thinner) CDFs for data sources in \figref{fig:fitting:data}}
      \label{fig:fitting:lines}
      \end{minipage}%
      \hspace*{1.25em} 
      \begin{minipage}[t]{10em}
       \vspace*{-0.65em}
        \caption{CDF \eqnref{eqn:wangCracked-TGI} and CDF \eqnref{eqn:wangCracked} with various indicated blocklisting algorithms}
        \label{fig:fitting:blocklist}
      \end{minipage}%
    \end{minipage}
  \end{subfigure}%
  \caption{Data sources used in our evaluations}
  \label{fig:fitting}
\end{figure}

%% file: figures/fitting.tex
\begin{tikzpicture}
\pgfplotsset{every axis/.append style={
					ylabel=\mbox{$\prob{\rvPwdRankChosen \le \pwdRank} $},
					xlabel=\mbox{$\log_{10}(\pwdRank)$},
					compat=1.8,
                    label style={font=\normalsize},
                    tick label style={font=\small}  
                    }}

\begin{axis}[
ylabel style={align=center}, 
xmin=0, xmax=24,
ymin=0, ymax=1,
xtick={0, 6, 12, 18, 24},
scaled ticks=base 10:0,
ytick={0, 0.2, 0.4, 0.6, 0.8, 1.0},
yticklabels={0, 0.2, 0.4, 0.6, 0.8, 1.0},
width=1.1\figurewidth,
height=1.1\figurewidth,
tick align=outside,
tick pos=left,
xmajorgrids,
xminorgrids,
minor x tick num=1,
x grid style={lightgray!92.026143790849673!black},
minor y tick num=1,
ymajorgrids,
yminorgrids,
y grid style={lightgray!92.026143790849673!black},
legend entries={{~\eqnref{eqn:wangCracked-TGI}$~$},{~\eqnref{eqn:wangCracked-TGIPP}$~$},
{~\eqnref{eqn:wangCracked-TGIPPP}},{~\eqnref{eqn:wangCracked-PCFG}}, {~
\eqnref{eqn:mazurekCracked}}, {~\eqnref{eqn:xuCracked}}},
legend style={{font={\fontsize{17pt}{12}\selectfont}},{draw=none}},
legend cell align={left},
legend columns=3,
legend pos={outer north east},
legend style={at={(1.10, 1.32)}, anchor=north east, draw=none, nodes={scale=0.458,
transform shape}}
]
\addlegendimage{line width=1pt, solid, curve_color}
\addlegendimage{line width=1pt, densely dotted, curve_color}
\addlegendimage{line width=1pt, dotted, curve_color}
\addlegendimage{line width=1pt, loosely dashdotted, curve_color}
\addlegendimage{line width=1pt, densely dashdotted, curve_color}
\addlegendimage{line width=1pt, dashed, curve_color}
\addplot [line width=0.8pt, curve_color]
table {%
0 	 0.04187
1 	 0.11901
2 	 0.19615
3 	 0.27329
4 	 0.35043
5 	 0.42757
6 	 0.50471
7 	 0.58185
8 	 0.65899
9 	 0.73613
10 	 0.81327
11 	 0.89041
12 	 0.96755
13 	 1.04469
};
\addplot [line width=0.8pt, curve_color, densely dotted]
table {%
0 	 0.01180
1 	 0.07190
2 	 0.13200
3 	 0.19210
4 	 0.25220
5 	 0.31230
6 	 0.37240
7 	 0.43250
8 	 0.49260
9 	 0.55270
10	 0.61280
11	 0.67290
12	 0.73300
13	 0.79310
14	 0.85320
15	 0.91330
16	 0.97340
17	 1.03350
};
\addplot [line width=0.8pt, curve_color, dotted]
table {%
0	 0.00127
1	 0.03224
2	 0.06321
3	 0.09418
4	 0.12515
5	 0.15612
6	 0.18709
7	 0.21806
8	 0.24903
9	 0.28000
10	 0.31097
11	 0.34194
12	 0.37291
13	 0.40388
14	 0.43485
15	 0.46582
16	 0.49679
17	 0.52776
18	 0.55873
19	 0.58970
20	 0.62067
21	 0.65164
22	 0.68261
23	 0.71358
24	 0.74455
25	 0.77552
26	 0.80649
27	 0.83746
28	 0.86843
29	 0.89940
30	 0.93037
31	 0.96134
32	 0.99231
33	 1.02328
};
\addplot [line width=0.8pt, curve_color, loosely dashdotted]
table {%
0	 -0.00466
1	 0.01249
2	 0.02964
3	 0.04679
4	 0.06394
5	 0.08109
6	 0.09824
7	 0.11539
8	 0.13254
9	 0.14969
10	 0.16684
11	 0.18399
12	 0.20114
13	 0.21829
14	 0.23544
15	 0.25259
16	 0.26974
17	 0.28689
18	 0.30404
19	 0.32119
20	 0.33834
21	 0.35549
22	 0.37264
23	 0.38979
24	 0.40694
25	 0.42409
26	 0.44124
27	 0.45839
28	 0.47554
29	 0.49269
30	 0.50984
31	 0.52699
32	 0.54414
33	 0.56129
34	 0.57844
35	 0.59559
36	 0.61274
37	 0.62989
38	 0.64704
39	 0.66419
40	 0.68134
41	 0.69849
42	 0.71564
43	 0.73279
44	 0.74994
45	 0.76709
46	 0.78424
47	 0.80139
48	 0.81854
49	 0.83569
50	 0.85284
51	 0.86999
52	 0.88714
53	 0.90429
54	 0.92144
55	 0.93859
56	 0.95574
57	 0.97289
58	 0.99004
59	 1.00719
};
\addplot [line width=0.8pt, curve_color, dashdotted]
table {%
0 	 -0.41190
1 	 -0.35173
2 	 -0.29156
3 	 -0.23139
4 	 -0.17122
5 	 -0.11105
6 	 -0.05088
7 	 0.00929
8 	 0.06946
9 	 0.12963
10 	 0.18980
11 	 0.24997
12 	 0.31014
13 	 0.37031
14 	 0.43048
15 	 0.49065
16 	 0.55082
17 	 0.61099
18 	 0.67116
19 	 0.73133
20 	 0.79150
21 	 0.85167
22 	 0.91184
23 	 0.97201
24 	 1.03218
};
\addplot [line width=2pt, curve_color, dashdotted]
table {
5.7478	0.0017
5.7688	0.0017
5.7900	0.0017
5.8112	0.0017
5.8289	0.0026
5.8682	0.0026
5.8897	0.0026
5.9113	0.0026
5.9329	0.0027
5.9547	0.0030
5.9765	0.0030
5.9984	0.0030
6.0204	0.0030
6.0425	0.0030
6.0646	0.0030
6.0868	0.0030
6.1091	0.0030
6.1315	0.0030
6.1540	0.0033
6.1766	0.0034
6.1992	0.0034
6.2219	0.0034
6.2447	0.0034
6.2676	0.0037
6.2906	0.0039
6.3136	0.0040
6.3368	0.0043
6.3600	0.0043
6.3833	0.0045
6.4067	0.0048
6.4302	0.0048
6.4538	0.0050
6.4774	0.0052
6.5012	0.0052
6.5250	0.0052
6.5489	0.0055
6.5729	0.0061
6.5970	0.0061
6.6212	0.0061
6.6455	0.0065
6.6698	0.0068
6.6943	0.0069
6.7106	0.0069
6.7434	0.0046
6.7681	0.0049
6.7929	0.0055
6.8178	0.0061
6.8428	0.0100
6.8679	0.0104
6.8889	0.0117
6.9183	0.0117
6.9183	0.0129
6.9437	0.0136
6.9692	0.0139
6.9947	0.0139
7.0203	0.0142
7.0461	0.0144
7.0719	0.0148
7.0935	0.0157
7.1238	0.0157
7.1674	0.0151
7.1937	0.0158
7.2200	0.0164
7.2465	0.0174
7.2819	0.0174
7.3086	0.0177
7.3354	0.0183
7.3623	0.0192
7.3893	0.0192
7.4118	0.0200
7.4435	0.0200
7.4708	0.0205
7.4982	0.0212
7.5257	0.0224
7.5533	0.0229
7.5810	0.0235
7.6087	0.0241
7.6366	0.0244
7.6646	0.0283
7.6833	0.0292
7.7115	0.0296
7.7398	0.0305
7.7634	0.0305
7.8156	0.0293
7.8443	0.0305
7.8826	0.0309
7.9115	0.0318
7.9405	0.0328
7.9696	0.0352
8.0086	0.0392
8.0135	0.0392
8.0576	0.0379
8.0871	0.0398
8.1267	0.0406
8.1565	0.0419
8.1864	0.0442
8.2364	0.0459
8.2666	0.0475
8.2969	0.0490
8.3273	0.0509
8.3578	0.0523
8.3885	0.0536
8.4192	0.0550
8.4501	0.0564
8.4811	0.0580
8.5121	0.0593
8.5381	0.0608
8.5747	0.0633
8.6061	0.0646
8.6271	0.0733
8.7117	0.0757
8.7489	0.0779
8.7757	0.0798
8.8078	0.0820
8.8347	0.0843
8.8617	0.0862
8.8942	0.0881
8.9213	0.0905
8.9486	0.0921
8.9814	0.0956
9.0143	0.0973
9.0473	0.1001
9.0722	0.1001
9.0971	0.1025
9.1049	0.1025
9.1249	0.1046
9.1360	0.1047
9.1583	0.1072
9.1606	0.1102
9.1807	0.1102
9.1907	0.1110
9.2031	0.1124
9.2054	0.1126
9.2256	0.1150
9.2278	0.1173
9.2504	0.1178
9.2707	0.1203
9.2735	0.1203
9.2933	0.1224
9.3070	0.1224
9.3274	0.1249
9.3411	0.1264
9.3730	0.1270
9.3959	0.1286
9.4097	0.1318
9.4326	0.1318
9.4418	0.1322
9.4534	0.1340
9.4672	0.1340
9.4880	0.1363
9.5019	0.1366
9.5228	0.1390
9.5344	0.1414
9.5600	0.1416
9.5869	0.1442
9.5927	0.1467
9.6185	0.1468
9.6455	0.1491
9.6538	0.1492
9.6750	0.1517
9.6773	0.1543
9.6986	0.1545
9.7223	0.1572
9.7247	0.1572
9.7460	0.1601
9.7484	0.1601
9.7698	0.1632
9.7722	0.1632
9.7937	0.1653
9.8080	0.1679
9.8320	0.1689
9.8536	0.1713
9.8566	0.1713
9.8837	0.1735
9.8921	0.1763
9.9163	0.1767
9.9381	0.1788
9.9526	0.1814
9.9775	0.1814
9.9867	0.1807
10.0110	0.1831
10.0135	0.1859
10.0379	0.1859
10.0600	0.1872
10.0723	0.1903
10.0846	0.1903
10.0969	0.1927
10.0999	0.1927
10.1215	0.1946
10.1363	0.1951
10.1586	0.1975
10.1710	0.1999
10.1983	0.2017
10.2332	0.2021
10.2582	0.2044
10.2732	0.2047
10.2958	0.2064
10.3109	0.2064
10.3335	0.2090
10.3361	0.2090
10.3588	0.2121
10.3613	0.2138
10.3967	0.2139
10.4285	0.2168
10.4374	0.2173
10.4603	0.2194
10.4757	0.2229
10.5012	0.2229
10.5051	0.2230
10.5243	0.2254
10.5397	0.2263
10.5726	0.2296
10.5784	0.2296
10.6016	0.2322
10.6171	0.2335
10.6508	0.2335
10.6795	0.2365
10.6821	0.2368
10.7099	0.2398
10.7219	0.2398
10.7448	0.2417
10.7605	0.2430
10.7973	0.2438
10.8336	0.2470
10.8395	0.2470
10.8634	0.2505
10.8793	0.2510
10.9032	0.2530
10.9191	0.2536
10.9431	0.2558
10.9592	0.2560
10.9877	0.2591
11.0000	0.2595
11.0302	0.2621
11.0403	0.2621
11.0639	0.2643
11.0774	0.2671
11.1072	0.2673
11.1361	0.2709
11.1479	0.2710
11.1724	0.2728
11.1887	0.2728
11.2133	0.2749
11.2270	0.2776
11.2572	0.2780
11.2819	0.2801
11.2957	0.2826
11.3260	0.2826
11.3509	0.2847
11.3675	0.2850
11.3995	0.2871
11.4099	0.2871
11.4343	0.2888
11.4482	0.2916
11.4790	0.2916
11.5042	0.2940
11.5211	0.2940
11.5464	0.2960
11.5633	0.2960
11.5887	0.2987
11.6057	0.2990
11.6312	0.3007
11.6482	0.3007
11.6738	0.3038
11.6766	0.3038
11.7023	0.3063
11.7166	0.3064
11.7452	0.3088
11.7481	0.3111
11.7767	0.3111
11.8026	0.3137
11.8055	0.3137
11.8315	0.3163
11.8343	0.3195
11.8632	0.3195
11.8893	0.3221
11.9067	0.3221
11.9329	0.3249
11.9474	0.3272
11.9803	0.3272
11.9912	0.3293
12.0234	0.3295
12.0499	0.3314
12.0675	0.3314
12.0940	0.3336
12.1117	0.3339
12.1384	0.3360
12.1561	0.3374
12.1947	0.3383
12.2275	0.3403
12.2312	0.3407
12.2648	0.3430
12.2761	0.3430
12.3023	0.3449
12.3203	0.3452
12.3474	0.3476
12.3655	0.3479
12.3926	0.3500
12.4108	0.3505
12.4381	0.3528
12.4563	0.3535
12.4837	0.3562
12.5019	0.3566
12.5294	0.3588
12.5478	0.3592
12.5753	0.3615
12.5937	0.3615
12.6291	0.3644
12.6399	0.3649
12.6677	0.3675
12.6862	0.3684
12.7141	0.3707
12.7327	0.3707
12.7607	0.3730
12.7763	0.3752
12.8106	0.3752
12.8388	0.3770
12.8544	0.3796
12.8890	0.3796
12.9015	0.3824
12.9204	0.3828
12.9567	0.3849
12.9686	0.3849
13.0042	0.3863
13.0161	0.3874
13.0598	0.3876
13.0917	0.3901
13.1077	0.3928
13.1429	0.3933
13.1718	0.3956
13.1911	0.3959
13.2201	0.3980
13.2362	0.4006
13.2726	0.4006
13.2847	0.4026
13.3204	0.4029
13.3497	0.4052
13.3692	0.4066
13.4150	0.4082
13.4641	0.4082
13.4970	0.4099
13.5168	0.4099
13.5465	0.4116
13.5663	0.4120
13.5961	0.4150
13.5995	0.4151
13.6293	0.4174
13.6493	0.4174
13.6876	0.4201
13.6960	0.4225
13.7328	0.4225
13.7630	0.4252
13.7831	0.4268
13.8303	0.4273
13.8640	0.4297
13.8843	0.4312
13.9318	0.4317
13.9658	0.4339
13.9863	0.4355
14.0341	0.4356
14.0684	0.4384
14.0856	0.4408
14.1234	0.4408
14.1545	0.4433
14.1752	0.4433
14.2063	0.4449
14.2271	0.4449
14.2584	0.4465
14.2793	0.4465
14.3107	0.4489
14.3316	0.4491
14.3631	0.4511
14.3806	0.4537
14.4193	0.4537
14.4510	0.4552
14.4721	0.4552
14.5039	0.4576
14.5252	0.4579
14.5571	0.4602
14.5784	0.4605
14.6194	0.4635
14.6318	0.4635
14.6640	0.4655
14.6855	0.4657
14.7177	0.4676
14.7393	0.4676
14.7717	0.4699
14.7897	0.4724
14.8294	0.4727
14.8620	0.4751
14.8838	0.4753
14.9165	0.4777
14.9384	0.4777
14.9803	0.4800
14.9940	0.4806
15.0444	0.4806
15.0811	0.4829
15.1032	0.4829
15.1364	0.4841
15.1549	0.4841
};
\addplot [line width=2pt, curve_color]
table {%
0.0024	0.0180
0.3010	0.0277
0.4771	0.0464
0.6021	0.0648
0.6990	0.0725
0.7782	0.0789
0.8451	0.0909
0.9031	0.1015
0.9542	0.1085
1.0000	0.1094
1.0414	0.1142
1.0792	0.1158
1.1139	0.1188
1.1461	0.1209
1.1761	0.1228
1.2041	0.1241
1.2304	0.1269
1.2553	0.1304
1.2788	0.1336
1.3010	0.1357
1.3222	0.1375
1.3424	0.1394
1.3617	0.1401
1.3802	0.1412
1.3979	0.1435
1.4150	0.1445
1.4314	0.1461
1.4472	0.1479
1.4624	0.1498
1.4771	0.1509
1.4914	0.1523
1.5051	0.1528
1.5185	0.1537
1.5315	0.1544
1.5441	0.1553
1.5563	0.1567
1.5682	0.1585
1.5798	0.1611
1.5911	0.1622
1.6021	0.1636
1.6128	0.1643
1.6232	0.1650
1.6335	0.1655
1.6435	0.1661
1.6532	0.1671
1.6628	0.1675
1.6721	0.1689
1.6812	0.1705
1.6902	0.1717
1.6990	0.1721
1.7076	0.1733
1.7160	0.1738
1.7243	0.1749
1.7324	0.1761
1.7404	0.1768
1.7482	0.1779
1.7559	0.1788
1.7634	0.1805
1.7709	0.1809
1.7782	0.1816
1.7853	0.1825
1.7924	0.1830
1.7993	0.1832
1.8062	0.1837
1.8129	0.1846
1.8195	0.1855
1.8261	0.1862
1.8325	0.1867
1.8388	0.1871
1.8451	0.1874
1.8513	0.1878
1.8573	0.1885
1.8633	0.1895
1.8692	0.1904
1.8751	0.1913
1.8808	0.1920
1.8865	0.1927
1.8921	0.1932
1.8976	0.1936
1.9031	0.1938
1.9085	0.1943
1.9138	0.1950
1.9191	0.1952
1.9243	0.1964
1.9294	0.1971
1.9345	0.1973
1.9395	0.1980
1.9445	0.1985
1.9494	0.1989
1.9542	0.1992
1.9590	0.1992
1.9638	0.1996
1.9685	0.1998
1.9731	0.2003
1.9777	0.2005
1.9823	0.2012
1.9868	0.2017
1.9912	0.2019
1.9956	0.2019
2.0000	0.2024
2.0043	0.2024
2.0086	0.2026
2.0128	0.2026
2.0170	0.2031
2.0212	0.2035
2.0253	0.2035
2.0294	0.2040
2.0334	0.2042
2.0374	0.2045
2.0414	0.2049
2.0453	0.2052
2.0492	0.2056
2.0531	0.2056
2.0569	0.2063
2.0607	0.2063
2.0645	0.2063
2.0682	0.2065
2.0719	0.2068
2.0755	0.2068
2.0792	0.2072
2.0828	0.2075
2.0864	0.2077
2.0899	0.2077
2.0934	0.2082
2.0969	0.2082
2.1004	0.2086
2.1038	0.2086
2.1072	0.2091
2.1106	0.2091
2.1139	0.2095
2.1173	0.2095
2.1206	0.2100
2.1239	0.2100
2.1271	0.2105
2.1303	0.2107
2.1335	0.2109
2.1367	0.2109
2.1399	0.2114
2.1430	0.2114
2.1461	0.2118
2.1492	0.2118
2.1523	0.2123
2.1553	0.2130
2.1584	0.2130
2.1614	0.2135
2.1644	0.2135
2.1673	0.2142
2.1703	0.2142
2.1732	0.2144
2.1761	0.2146
2.1790	0.2148
2.1818	0.2148
2.1847	0.2153
2.1875	0.2153
2.1903	0.2155
2.1931	0.2155
2.1959	0.2160
2.1987	0.2162
2.2014	0.2165
2.2041	0.2167
2.2068	0.2169
2.2095	0.2169
2.2122	0.2174
2.2148	0.2176
2.2175	0.2178
2.2201	0.2181
2.2227	0.2183
2.2253	0.2183
2.2279	0.2183
2.2304	0.2188
2.2330	0.2188
2.2355	0.2190
2.2380	0.2192
2.2405	0.2192
2.2430	0.2195
2.2455	0.2197
2.2480	0.2199
2.2504	0.2204
2.2529	0.2204
2.2553	0.2206
2.2577	0.2206
2.2601	0.2208
2.2625	0.2208
2.2648	0.2211
2.2672	0.2211
2.2695	0.2211
2.2718	0.2211
2.2742	0.2215
2.2765	0.2215
2.2788	0.2220
2.2810	0.2220
2.2833	0.2220
2.2856	0.2225
2.2878	0.2225
2.2900	0.2225
2.2923	0.2229
2.2945	0.2229
2.2967	0.2229
2.2989	0.2232
2.3010	0.2234
2.3032	0.2234
2.3054	0.2238
2.3075	0.2241
2.3096	0.2241
2.3118	0.2241
2.3139	0.2243
2.3160	0.2245
2.3181	0.2245
2.3201	0.2248
2.3222	0.2248
2.3243	0.2248
2.3263	0.2248
2.3284	0.2250
2.3304	0.2250
2.3324	0.2252
2.3345	0.2252
2.3365	0.2252
2.3385	0.2255
2.3404	0.2255
2.3424	0.2255
2.3444	0.2257
2.3464	0.2257
2.3483	0.2259
2.3502	0.2262
2.3522	0.2262
2.3541	0.2264
2.3560	0.2264
2.3579	0.2268
2.3598	0.2271
2.3617	0.2271
2.3636	0.2273
2.3655	0.2273
2.3674	0.2275
2.3692	0.2278
2.3711	0.2278
2.3729	0.2282
2.3747	0.2282
2.3766	0.2285
2.3784	0.2289
2.3802	0.2289
2.3820	0.2292
2.3838	0.2292
2.3856	0.2294
2.3874	0.2294
2.3892	0.2296
2.3909	0.2298
2.3927	0.2298
2.3945	0.2298
2.3962	0.2298
2.3979	0.2301
2.3997	0.2301
2.4014	0.2303
2.4031	0.2303
2.4048	0.2303
2.4065	0.2305
2.4082	0.2305
2.4099	0.2308
2.4116	0.2308
2.4133	0.2308
2.4150	0.2308
2.4166	0.2312
2.4183	0.2312
2.4200	0.2317
2.4216	0.2317
2.4232	0.2319
2.4249	0.2319
2.4265	0.2322
2.4281	0.2322
2.4298	0.2324
2.4314	0.2326
2.4330	0.2326
2.4346	0.2328
2.4362	0.2328
2.4378	0.2331
2.4393	0.2331
2.4409	0.2331
2.4425	0.2331
2.4440	0.2333
2.4456	0.2333
2.4472	0.2335
2.4487	0.2335
2.4502	0.2338
2.4518	0.2338
2.4533	0.2342
2.4548	0.2342
2.4564	0.2342
2.4579	0.2342
2.4594	0.2342
2.4609	0.2347
2.4624	0.2347
2.4639	0.2349
2.4654	0.2349
2.4669	0.2352
2.4683	0.2352
2.4698	0.2354
2.4713	0.2354
2.4728	0.2354
2.4742	0.2354
2.4757	0.2358
2.4771	0.2358
2.4786	0.2361
2.4800	0.2361
2.4814	0.2363
2.4829	0.2363
2.4843	0.2363
2.4857	0.2363
2.4871	0.2363
2.4886	0.2368
2.4900	0.2368
2.4914	0.2370
2.4928	0.2370
2.4942	0.2372
2.4955	0.2372
2.4969	0.2375
2.4983	0.2375
2.4997	0.2375
2.5011	0.2375
2.5024	0.2375
2.5038	0.2377
2.5051	0.2377
2.5065	0.2377
2.5079	0.2377
2.5092	0.2377
2.5105	0.2379
2.5119	0.2379
2.5132	0.2382
2.5145	0.2382
2.5159	0.2382
2.5172	0.2382
2.5185	0.2382
2.5198	0.2384
2.5211	0.2384
2.5224	0.2386
2.5237	0.2386
2.5250	0.2388
2.5263	0.2388
2.5276	0.2388
2.5289	0.2391
2.5302	0.2391
2.5315	0.2393
2.5328	0.2393
2.5340	0.2393
2.5353	0.2395
2.5366	0.2395
2.5378	0.2398
2.5391	0.2398
2.5403	0.2398
2.5416	0.2398
2.5428	0.2398
2.5441	0.2400
2.5453	0.2400
2.5465	0.2400
2.5478	0.2402
2.5490	0.2402
2.5502	0.2405
2.5514	0.2405
2.5527	0.2405
2.5539	0.2405
2.5551	0.2405
2.5563	0.2407
2.5575	0.2407
2.5587	0.2407
2.5599	0.2412
2.5611	0.2412
2.5623	0.2412
2.5635	0.2414
2.5647	0.2414
2.5658	0.2416
2.5670	0.2416
2.5682	0.2416
2.5694	0.2418
2.5705	0.2418
2.5717	0.2418
2.5729	0.2418
2.5740	0.2418
2.5752	0.2423
2.5763	0.2423
2.5775	0.2423
2.5786	0.2423
2.5798	0.2423
2.5809	0.2423
2.5821	0.2425
2.5832	0.2425
2.5843	0.2425
2.5855	0.2428
2.5866	0.2428
2.5877	0.2428
2.5888	0.2428
2.5899	0.2428
2.5911	0.2430
2.5922	0.2430
2.5933	0.2430
2.5944	0.2432
2.5955	0.2432
2.5966	0.2432
2.5977	0.2432
2.5988	0.2432
2.5999	0.2432
2.6010	0.2435
2.6021	0.2435
2.6031	0.2437
2.6042	0.2437
2.6053	0.2437
2.6064	0.2439
2.6075	0.2439
2.6085	0.2439
2.6096	0.2442
2.6107	0.2442
2.6117	0.2442
2.6128	0.2442
2.6138	0.2442
2.6149	0.2442
2.6160	0.2446
2.6170	0.2446
2.6180	0.2446
2.6191	0.2448
2.6201	0.2448
2.6212	0.2448
2.6222	0.2448
2.6232	0.2448
2.6243	0.2448
2.6253	0.2451
2.6263	0.2451
2.6274	0.2451
2.6284	0.2453
2.6294	0.2453
2.6304	0.2453
2.6314	0.2455
2.6325	0.2455
2.6335	0.2455
2.6345	0.2458
2.6355	0.2458
2.6365	0.2458
2.6375	0.2460
2.6385	0.2460
2.6395	0.2460
2.6405	0.2460
2.6415	0.2465
2.6425	0.2465
2.6435	0.2465
2.6444	0.2467
2.6454	0.2467
2.6464	0.2467
2.6474	0.2467
2.6484	0.2467
2.6493	0.2467
2.6503	0.2472
2.6513	0.2472
2.6522	0.2472
2.6532	0.2472
2.6542	0.2472
2.6551	0.2472
2.6561	0.2472
2.6571	0.2472
2.6580	0.2472
2.6590	0.2472
2.6599	0.2478
2.6609	0.2478
2.6618	0.2478
2.6628	0.2478
2.6637	0.2478
2.6646	0.2478
2.6656	0.2483
2.6665	0.2483
2.6675	0.2483
2.6684	0.2483
2.6693	0.2483
2.6702	0.2483
2.6712	0.2483
2.6721	0.2485
2.6730	0.2485
2.6739	0.2485
2.6749	0.2485
2.6758	0.2488
2.6767	0.2488
2.6776	0.2488
2.6785	0.2490
2.6794	0.2490
2.6803	0.2490
2.6812	0.2492
2.6821	0.2492
2.6830	0.2492
2.6839	0.2492
2.6848	0.2497
2.6857	0.2497
2.6866	0.2497
2.6875	0.2502
2.6884	0.2502
2.6893	0.2502
2.6902	0.2502
2.6911	0.2502
2.6920	0.2502
2.6928	0.2502
2.6937	0.2504
2.6946	0.2504
2.6955	0.2504
2.6964	0.2504
2.6972	0.2506
2.6981	0.2506
2.6990	0.2506
2.6998	0.2506
2.7007	0.2506
2.7016	0.2506
2.7024	0.2506
2.7033	0.2506
2.7042	0.2506
2.7050	0.2506
2.7059	0.2506
2.7067	0.2506
2.7076	0.2506
2.7084	0.2506
2.7093	0.2506
2.7101	0.2506
2.7110	0.2506
2.7118	0.2506
2.7126	0.2508
2.7135	0.2508
2.7143	0.2508
2.7152	0.2508
2.7160	0.2511
2.7168	0.2511
2.7177	0.2511
2.7185	0.2511
2.7193	0.2513
2.7202	0.2513
2.7210	0.2513
2.7218	0.2515
2.7226	0.2515
2.7235	0.2515
2.7243	0.2515
2.7251	0.2515
2.7259	0.2515
2.7267	0.2515
2.7275	0.2515
2.7284	0.2520
2.7292	0.2520
2.7300	0.2520
2.7308	0.2520
2.7316	0.2522
2.7324	0.2522
2.7332	0.2522
2.7340	0.2522
2.7348	0.2525
2.7356	0.2525
2.7364	0.2525
2.7372	0.2525
2.7380	0.2527
2.7388	0.2527
2.7396	0.2527
2.7404	0.2527
2.7412	0.2527
2.7419	0.2527
2.7427	0.2527
2.7435	0.2527
2.7443	0.2532
2.7451	0.2532
2.7459	0.2532
2.7466	0.2532
2.7474	0.2534
2.7482	0.2534
2.7490	0.2534
2.7497	0.2534
2.7505	0.2536
2.7513	0.2536
2.7520	0.2536
2.7528	0.2536
2.7536	0.2541
2.7543	0.2541
2.7551	0.2541
2.7559	0.2541
2.7566	0.2541
2.7574	0.2541
2.7582	0.2541
2.7589	0.2541
2.7597	0.2545
2.7604	0.2545
2.7612	0.2545
2.7619	0.2545
2.7627	0.2548
2.7634	0.2548
2.7642	0.2548
2.7649	0.2548
2.7657	0.2548
2.7664	0.2548
2.7672	0.2548
2.7679	0.2548
2.7686	0.2550
2.7694	0.2550
2.7701	0.2550
2.7709	0.2550
2.7716	0.2550
2.7723	0.2552
2.7731	0.2552
2.7738	0.2552
2.7745	0.2552
2.7752	0.2552
2.7760	0.2552
2.7767	0.2552
2.7774	0.2552
2.7782	0.2552
2.7789	0.2552
2.7796	0.2552
2.7803	0.2552
2.7810	0.2552
2.7818	0.2557
2.7825	0.2557
2.7832	0.2557
2.7839	0.2557
2.7846	0.2557
2.7853	0.2557
2.7860	0.2557
2.7868	0.2557
2.7875	0.2557
2.7882	0.2557
2.7889	0.2557
2.7896	0.2557
2.7903	0.2557
2.7910	0.2559
2.7917	0.2559
2.7924	0.2559
2.7931	0.2559
2.7938	0.2562
2.7945	0.2562
2.7952	0.2562
2.7959	0.2562
2.7966	0.2562
2.7973	0.2564
2.7980	0.2564
2.7987	0.2564
2.7993	0.2564
2.8000	0.2566
2.8007	0.2566
2.8014	0.2566
2.8021	0.2566
2.8028	0.2566
2.8035	0.2566
2.8041	0.2566
2.8048	0.2566
2.8055	0.2566
2.8062	0.2566
2.8069	0.2566
2.8075	0.2566
2.8082	0.2566
2.8089	0.2566
2.8096	0.2573
2.8102	0.2573
2.8109	0.2573
2.8116	0.2573
2.8122	0.2573
2.8129	0.2575
2.8136	0.2575
2.8142	0.2575
2.8149	0.2575
2.8156	0.2578
2.8162	0.2578
2.8169	0.2578
2.8176	0.2578
2.8182	0.2578
2.8189	0.2580
2.8195	0.2580
2.8202	0.2580
2.8209	0.2580
2.8215	0.2580
2.8222	0.2582
2.8228	0.2582
2.8235	0.2582
2.8241	0.2582
2.8248	0.2582
2.8254	0.2587
2.8261	0.2587
2.8267	0.2587
2.8274	0.2587
2.8280	0.2592
2.8287	0.2592
2.8293	0.2592
2.8299	0.2592
2.8306	0.2592
2.8312	0.2594
2.8319	0.2594
2.8325	0.2594
2.8331	0.2594
2.8338	0.2594
2.8344	0.2594
2.8351	0.2594
2.8357	0.2594
2.8363	0.2594
2.8370	0.2594
2.8376	0.2596
2.8382	0.2596
2.8388	0.2596
2.8395	0.2596
2.8401	0.2596
2.8407	0.2598
2.8414	0.2598
2.8420	0.2598
2.8426	0.2598
2.8432	0.2598
2.8439	0.2601
2.8445	0.2601
2.8451	0.2601
2.8457	0.2601
2.8463	0.2601
2.8470	0.2603
2.8476	0.2603
2.8482	0.2603
2.8488	0.2603
2.8494	0.2603
2.8500	0.2603
2.8506	0.2603
2.8513	0.2603
2.8519	0.2603
2.8525	0.2603
2.8531	0.2605
2.8537	0.2605
2.8543	0.2605
2.8549	0.2605
2.8555	0.2605
2.8561	0.2610
2.8567	0.2610
2.8573	0.2610
2.8579	0.2610
2.8585	0.2610
2.8591	0.2610
2.8597	0.2610
2.8603	0.2610
2.8609	0.2610
2.8615	0.2610
2.8621	0.2610
2.8627	0.2615
2.8633	0.2615
2.8639	0.2615
2.8645	0.2615
2.8651	0.2615
2.8657	0.2615
2.8663	0.2615
2.8669	0.2615
2.8675	0.2615
2.8681	0.2615
2.8686	0.2617
2.8692	0.2617
2.8698	0.2617
2.8704	0.2617
2.8710	0.2617
2.8716	0.2617
2.8722	0.2619
2.8727	0.2619
2.8733	0.2619
2.8739	0.2619
2.8745	0.2619
2.8751	0.2622
2.8756	0.2622
2.8762	0.2622
2.8768	0.2622
2.8774	0.2622
2.8779	0.2626
2.8785	0.2626
2.8791	0.2626
2.8797	0.2626
2.8802	0.2626
2.8808	0.2626
2.8814	0.2628
2.8820	0.2628
2.8825	0.2628
2.8831	0.2628
2.8837	0.2628
2.8842	0.2631
2.8848	0.2631
2.8854	0.2631
2.8859	0.2631
2.8865	0.2631
2.8871	0.2631
2.8876	0.2631
2.8882	0.2631
2.8887	0.2631
2.8893	0.2631
2.8899	0.2631
2.8904	0.2631
2.8910	0.2633
2.8915	0.2633
2.8921	0.2633
2.8927	0.2633
2.8932	0.2633
2.8938	0.2633
2.8943	0.2633
2.8949	0.2633
2.8954	0.2633
2.8960	0.2633
2.8965	0.2633
2.8971	0.2635
2.8976	0.2635
2.8982	0.2635
2.8987	0.2635
2.8993	0.2635
2.8998	0.2638
2.9004	0.2638
2.9009	0.2638
2.9015	0.2638
2.9020	0.2638
2.9025	0.2638
2.9031	0.2640
2.9036	0.2640
2.9042	0.2640
2.9047	0.2640
2.9053	0.2640
2.9058	0.2640
2.9063	0.2640
2.9069	0.2640
2.9074	0.2640
2.9079	0.2640
2.9085	0.2640
2.9090	0.2640
2.9096	0.2645
2.9101	0.2645
2.9106	0.2645
2.9112	0.2645
2.9117	0.2645
2.9122	0.2645
2.9128	0.2647
2.9133	0.2647
2.9138	0.2647
2.9143	0.2647
2.9149	0.2647
2.9154	0.2647
2.9159	0.2647
2.9165	0.2647
2.9170	0.2647
2.9175	0.2647
2.9180	0.2647
2.9186	0.2649
2.9191	0.2649
2.9196	0.2649
2.9201	0.2649
2.9206	0.2649
2.9212	0.2649
2.9217	0.2652
2.9222	0.2652
2.9227	0.2652
2.9232	0.2652
2.9238	0.2652
2.9243	0.2652
2.9248	0.2654
2.9253	0.2654
2.9258	0.2654
2.9263	0.2654
2.9269	0.2654
2.9274	0.2654
2.9279	0.2654
2.9284	0.2656
2.9289	0.2656
2.9294	0.2656
2.9299	0.2656
2.9304	0.2656
2.9309	0.2656
2.9315	0.2661
2.9320	0.2661
2.9325	0.2661
2.9330	0.2661
2.9335	0.2661
2.9340	0.2661
2.9345	0.2661
2.9350	0.2661
2.9355	0.2661
2.9360	0.2661
2.9365	0.2661
2.9370	0.2661
2.9375	0.2663
2.9380	0.2663
2.9385	0.2663
2.9390	0.2663
2.9395	0.2663
2.9400	0.2663
2.9405	0.2668
2.9410	0.2668
2.9415	0.2668
2.9420	0.2668
2.9425	0.2668
2.9430	0.2668
2.9435	0.2668
2.9440	0.2668
2.9445	0.2668
2.9450	0.2668
2.9455	0.2668
2.9460	0.2668
2.9465	0.2668
2.9469	0.2672
2.9474	0.2672
2.9479	0.2672
2.9484	0.2672
2.9489	0.2672
2.9494	0.2672
2.9499	0.2672
2.9504	0.2672
2.9509	0.2672
2.9513	0.2672
2.9518	0.2672
2.9523	0.2672
2.9528	0.2672
2.9533	0.2675
2.9538	0.2675
2.9542	0.2675
2.9547	0.2675
2.9552	0.2675
2.9557	0.2675
2.9562	0.2679
2.9566	0.2679
2.9571	0.2679
2.9576	0.2679
2.9581	0.2679
2.9586	0.2679
2.9590	0.2679
2.9595	0.2679
2.9600	0.2679
2.9605	0.2679
2.9609	0.2679
2.9614	0.2679
2.9619	0.2679
2.9624	0.2682
2.9628	0.2682
2.9633	0.2682
2.9638	0.2682
2.9643	0.2682
2.9647	0.2682
2.9652	0.2682
2.9657	0.2684
2.9661	0.2684
2.9666	0.2684
2.9671	0.2684
2.9675	0.2684
2.9680	0.2684
2.9685	0.2684
2.9689	0.2686
2.9694	0.2686
2.9699	0.2686
2.9703	0.2686
2.9708	0.2686
2.9713	0.2686
2.9717	0.2686
2.9722	0.2686
2.9727	0.2686
2.9731	0.2686
2.9736	0.2686
2.9741	0.2686
2.9745	0.2686
2.9750	0.2688
2.9754	0.2688
2.9759	0.2688
2.9763	0.2688
2.9768	0.2688
2.9773	0.2688
2.9777	0.2688
2.9782	0.2688
2.9786	0.2688
2.9791	0.2688
2.9795	0.2688
2.9800	0.2688
2.9805	0.2688
2.9809	0.2688
2.9814	0.2691
2.9818	0.2691
2.9823	0.2691
2.9827	0.2691
2.9832	0.2691
2.9836	0.2691
2.9841	0.2691
2.9845	0.2691
2.9850	0.2691
2.9854	0.2691
2.9859	0.2691
2.9863	0.2691
2.9868	0.2691
2.9872	0.2691
2.9877	0.2691
2.9881	0.2691
2.9886	0.2691
2.9890	0.2691
2.9894	0.2691
2.9899	0.2691
2.9903	0.2691
2.9908	0.2693
2.9912	0.2693
2.9917	0.2693
2.9921	0.2693
2.9926	0.2693
2.9930	0.2693
2.9934	0.2693
2.9939	0.2693
2.9943	0.2693
2.9948	0.2693
2.9952	0.2693
2.9956	0.2693
2.9961	0.2693
2.9965	0.2693
2.9969	0.2695
2.9974	0.2695
2.9978	0.2695
2.9983	0.2695
2.9987	0.2695
2.9991	0.2695
2.9996	0.2695
3.0000	0.2698
};
\addplot [line width=2pt, curve_color, densely dotted]
table {%
0.00209	0.01786
0.00846	0.01750
0.01484	0.01774
0.02121	0.01797
0.02759	0.01821
0.03396	0.01821
0.04034	0.01856
0.04671	0.01868
0.05309	0.01880
0.05946	0.01916
0.06584	0.01927
0.07221	0.01939
0.07859	0.01963
0.08496	0.01963
0.09134	0.01998
0.09771	0.02010
0.10408	0.02034
0.11046	0.02057
0.11683	0.02057
0.12321	0.02105
0.12958	0.02105
0.13596	0.02128
0.14233	0.02152
0.14871	0.02152
0.15508	0.02176
0.16146	0.02199
0.16783	0.02211
0.17421	0.02247
0.18058	0.02247
0.18696	0.02282
0.19333	0.02294
0.19971	0.02306
0.20608	0.02341
0.21246	0.02341
0.21883	0.02353
0.22521	0.02388
0.23158	0.02388
0.23637	0.02388
0.26027	0.02495
0.26665	0.02507
0.27302	0.02530
0.29852	0.02613
0.30490	0.02625
0.31127	0.02648
0.31765	0.02672
0.32402	0.02707
0.33040	0.02719
0.33677	0.02755
0.34315	0.02767
0.34952	0.02814
0.35590	0.02849
0.36227	0.02861
0.36865	0.02897
0.37502	0.02944
0.38140	0.02956
0.38777	0.02991
0.39415	0.03015
0.40052	0.03038
0.40690	0.03062
0.41327	0.03098
0.41964	0.03133
0.42602	0.03157
0.43239	0.03192
0.43877	0.03228
0.44514	0.03239
0.45152	0.03275
0.45789	0.03287
0.46427	0.03346
0.47064	0.03369
0.47702	0.03393
0.48339	0.03428
0.48977	0.03452
0.49614	0.03499
0.50252	0.03535
0.50889	0.03570
0.51527	0.03618
0.52164	0.03629
0.55033	0.03807
0.55671	0.03854
0.56149	0.03830
0.58539	0.04043
0.59177	0.04055
0.59814	0.04090
0.60452	0.04138
0.61089	0.04232
0.61727	0.04303
0.62364	0.04374
0.63002	0.04469
0.63639	0.04516
0.64277	0.04610
0.64914	0.04693
0.65552	0.04752
0.66189	0.04847
0.66827	0.04918
0.67464	0.04989
0.68102	0.05083
0.68739	0.05142
0.69377	0.05225
0.70014	0.05308
0.70652	0.05355
0.71289	0.05426
0.71927	0.05497
0.72564	0.05532
0.73202	0.05615
0.73839	0.05674
0.74477	0.05733
0.75114	0.05792
0.75752	0.05851
0.76389	0.05922
0.77027	0.05970
0.77664	0.05981
0.80214	0.06064
0.80852	0.06076
0.81489	0.06064
0.84358	0.06088
0.84995	0.06135
0.85633	0.06218
0.86270	0.06277
0.86908	0.06348
0.87545	0.06419
0.88183	0.06478
0.88820	0.06549
0.89458	0.06608
0.90095	0.06679
0.90733	0.06714
0.91370	0.06738
0.92008	0.06750
0.92645	0.06773
0.93283	0.06785
0.93920	0.06809
0.94558	0.06809
0.95195	0.06833
0.95833	0.06868
0.96470	0.06915
0.97108	0.06951
0.97745	0.06986
0.98383	0.07033
0.99020	0.07069
0.99658	0.07116
1.00295	0.07152
1.00933	0.07163
1.01570	0.07199
1.02208	0.07211
1.02845	0.07234
1.03483	0.07258
1.04120	0.07282
1.04758	0.07317
1.05395	0.07329
1.08264	0.07542
1.08901	0.07577
1.09380	0.07565
1.11133	0.07932
1.11770	0.07967
1.12408	0.08014
1.13045	0.08062
1.13683	0.08109
1.14320	0.08156
1.14958	0.08204
1.15595	0.08239
1.16233	0.08274
1.16870	0.08310
1.17508	0.08345
1.18145	0.08393
1.18783	0.08428
1.19420	0.08464
1.20058	0.08511
1.20695	0.08558
1.21333	0.08617
1.21970	0.08665
1.22608	0.08724
1.23245	0.08771
1.23883	0.08818
1.24520	0.08830
1.25158	0.08877
1.25795	0.08913
1.26433	0.08925
1.27070	0.08936
1.27708	0.08960
1.28345	0.09007
1.28983	0.09043
1.29620	0.09078
1.30258	0.09126
1.30895	0.09161
1.31533	0.09196
1.32170	0.09232
1.34401	0.09563
1.35039	0.09586
1.35676	0.09598
1.37907	0.09835
1.38545	0.09858
1.39182	0.09894
1.39820	0.09917
1.40457	0.09953
1.41095	0.09977
1.41732	0.10012
1.42370	0.10059
1.43007	0.10107
1.43645	0.10142
1.44282	0.10177
1.44920	0.10213
1.45557	0.10272
1.46195	0.10319
1.46832	0.10331
1.47470	0.10355
1.48107	0.10402
1.48745	0.10473
1.49382	0.10532
1.50020	0.10591
1.50657	0.10638
1.51295	0.10662
1.51932	0.10698
1.52570	0.10733
1.53207	0.10768
1.53845	0.10792
1.54482	0.10828
1.55120	0.10863
1.55757	0.10898
1.56395	0.10934
1.57032	0.10993
1.57670	0.11076
1.58307	0.11099
1.58785	0.11135
1.61495	0.11359
1.62132	0.11383
1.62610	0.11395
1.65001	0.11655
1.65639	0.11690
1.66276	0.11738
1.66914	0.11773
1.67551	0.11820
1.68188	0.11868
1.68826	0.11915
1.69463	0.11939
1.70101	0.11974
1.70738	0.11986
1.71376	0.12021
1.72013	0.12057
1.72651	0.12104
1.73288	0.12175
1.73926	0.12234
1.74510	0.12309
1.75201	0.12423
1.75838	0.12482
1.76476	0.12506
1.77113	0.12541
1.77751	0.12553
1.78388	0.12589
1.79026	0.12600
1.79663	0.12600
1.80301	0.12636
1.80938	0.12648
1.81576	0.12660
1.82213	0.12695
1.82851	0.12742
1.83488	0.12790
1.84126	0.12801
1.84763	0.12837
1.85401	0.12860
1.85879	0.12860
1.88907	0.13026
1.89545	0.13026
1.92732	0.13168
1.93370	0.13215
1.94007	0.13215
1.94645	0.13262
1.95282	0.13298
1.95920	0.13310
1.96557	0.13345
1.97195	0.13357
1.97832	0.13416
1.98523	0.13428
1.99107	0.13534
1.99744	0.13605
2.00382	0.13629
2.01019	0.13652
2.01657	0.13712
2.02294	0.13759
2.02932	0.13782
2.03569	0.13794
2.04207	0.13830
2.04844	0.13842
2.05482	0.13889
2.06119	0.13924
2.06757	0.13960
2.07394	0.13983
2.08032	0.14031
2.08669	0.14066
2.09307	0.14102
2.09944	0.14125
2.10582	0.14161
2.11219	0.14208
2.11857	0.14220
2.12494	0.14255
2.13132	0.14255
2.13769	0.14267
2.16319	0.14456
2.16957	0.14444
2.17435	0.14468
2.20144	0.14622
2.20782	0.14633
2.21419	0.14645
2.22057	0.14681
2.22694	0.14716
2.23332	0.14763
2.23969	0.14787
2.24607	0.14834
2.25244	0.14870
2.25882	0.14870
2.26519	0.14905
2.27157	0.14964
2.27794	0.14976
2.28432	0.15000
2.29069	0.15023
2.29707	0.15059
2.30344	0.15071
2.30982	0.15106
2.31619	0.15130
2.32257	0.15165
2.32894	0.15201
2.33532	0.15224
2.34169	0.15260
2.34807	0.15284
2.35444	0.15307
2.36082	0.15343
2.36719	0.15390
2.37357	0.15402
2.37994	0.15449
2.38632	0.15473
2.39269	0.15520
2.39907	0.15555
2.40544	0.15603
2.41182	0.15626
2.41819	0.15591
2.44369	0.15768
2.45007	0.15804
2.45485	0.15815
2.48194	0.15945
2.48832	0.15969
2.49469	0.15993
2.50107	0.16016
2.50744	0.16064
2.51382	0.16099
2.52019	0.16099
2.52657	0.16135
2.53294	0.16158
2.53932	0.16205
2.54569	0.16241
2.55207	0.16288
2.55844	0.16324
2.56482	0.16335
2.57119	0.16395
2.57757	0.16430
2.58394	0.16465
2.59032	0.16489
2.59669	0.16513
2.60307	0.16536
2.60944	0.16560
2.61581	0.16584
2.62219	0.16619
2.62856	0.16643
2.63494	0.16666
2.64131	0.16714
2.64769	0.16761
2.65406	0.16773
2.66044	0.16844
2.66681	0.16867
2.67319	0.16903
2.67956	0.16926
2.68594	0.16950
2.69231	0.16950
2.71781	0.17151
2.72419	0.17175
2.75288	0.17364
2.75925	0.17387
2.76563	0.17423
2.77200	0.17458
2.77838	0.17482
2.78475	0.17517
2.79113	0.17565
2.79750	0.17577
2.80388	0.17612
2.81025	0.17636
2.81663	0.17671
2.82300	0.17707
2.82938	0.17777
2.83575	0.17789
2.84213	0.17825
2.84850	0.17848
2.85488	0.17884
2.86125	0.17919
2.86763	0.17955
2.87400	0.18002
2.88038	0.18026
2.88675	0.18073
2.89313	0.18097
2.89950	0.18144
2.90588	0.18179
2.91225	0.18227
2.91863	0.18250
2.92500	0.18286
2.93137	0.18321
2.93775	0.18368
2.94412	0.18392
2.95050	0.18439
2.95687	0.18439
2.96166	0.18439
2.98556	0.18628
2.99194	0.18628
};
\addplot [line width=2pt, curve_color, dotted]
table {%
0.00527	0.01112
0.01165	0.01112
0.01802	0.01112
0.02440	0.01135
0.03077	0.01135
0.03715	0.01135
0.04352	0.01159
0.04990	0.01159
0.05627	0.01159
0.06265	0.01159
0.06902	0.01159
0.07540	0.01171
0.08177	0.01206
0.08815	0.01206
0.09452	0.01206
0.10090	0.01206
0.10727	0.01206
0.11365	0.01230
0.12002	0.01230
0.12640	0.01254
0.13277	0.01254
0.13915	0.01254
0.14552	0.01254
0.15190	0.01254
0.15827	0.01254
0.16465	0.01301
0.17102	0.01301
0.17740	0.01301
0.18377	0.01301
0.19015	0.01301
0.19652	0.01301
0.20290	0.01325
0.20927	0.01325
0.21565	0.01325
0.22202	0.01348
0.22840	0.01348
0.23477	0.01348
0.23955	0.01348
0.26665	0.01384
0.27302	0.01396
0.27940	0.01396
0.28577	0.01396
0.29215	0.01396
0.29852	0.01407
0.30490	0.01431
0.31127	0.01443
0.31765	0.01455
0.32402	0.01490
0.33040	0.01502
0.33677	0.01502
0.36546	0.01585
0.37183	0.01608
0.37821	0.01632
0.38458	0.01632
0.39096	0.01667
0.39733	0.01679
0.40371	0.01703
0.41008	0.01726
0.41646	0.01726
0.42283	0.01750
0.42921	0.01774
0.43558	0.01774
0.44196	0.01797
0.44833	0.01821
0.45471	0.01821
0.46108	0.01856
0.46746	0.01868
0.47383	0.01868
0.48021	0.01916
0.48658	0.01916
0.49296	0.01927
0.49933	0.01939
0.50571	0.01963
0.51208	0.01963
0.51846	0.01963
0.52483	0.01975
0.53121	0.01998
0.53758	0.02010
0.54396	0.02010
0.55033	0.02046
0.55671	0.02057
0.56308	0.02057
0.56946	0.02057
0.57583	0.02105
0.58221	0.02105
0.58858	0.02116
0.59496	0.02105
0.62046	0.02164
0.62683	0.02152
0.63321	0.02176
0.63958	0.02176
0.64596	0.02187
0.65233	0.02199
0.65871	0.02199
0.66508	0.02199
0.67146	0.02199
0.67783	0.02223
0.68421	0.02247
0.69058	0.02247
0.69536	0.02247
0.72246	0.02329
0.72883	0.02341
0.73520	0.02341
0.74158	0.02353
0.74795	0.02377
0.75433	0.02388
0.76070	0.02388
0.76708	0.02436
0.77345	0.02436
0.77983	0.02436
0.78620	0.02459
0.79258	0.02459
0.79895	0.02459
0.80533	0.02459
0.81170	0.02483
0.81808	0.02483
0.82445	0.02483
0.83083	0.02483
0.83720	0.02483
0.84358	0.02483
0.84995	0.02483
0.85633	0.02495
0.86270	0.02530
0.86908	0.02530
0.87545	0.02530
0.88183	0.02530
0.88820	0.02542
0.89458	0.02554
0.90095	0.02577
0.90733	0.02577
0.91370	0.02577
0.92008	0.02601
0.92645	0.02625
0.93283	0.02625
0.93920	0.02672
0.94558	0.02672
0.95195	0.02684
0.98064	0.02778
0.98702	0.02790
0.99339	0.02814
0.99977	0.02837
1.00614	0.02861
1.01252	0.02861
1.01889	0.02897
1.02527	0.02908
1.03164	0.02908
1.03802	0.02956
1.04439	0.02956
1.05076	0.02991
1.07945	0.03121
1.08583	0.03145
1.09220	0.03180
1.09858	0.03192
1.10495	0.03228
1.11133	0.03239
1.11770	0.03263
1.12408	0.03275
1.13045	0.03287
1.13683	0.03298
1.14320	0.03322
1.14958	0.03334
1.15595	0.03334
1.16233	0.03358
1.16870	0.03381
1.17508	0.03381
1.18145	0.03381
1.18783	0.03405
1.19420	0.03428
1.20058	0.03428
1.20695	0.03428
1.21333	0.03464
1.21970	0.03476
1.22608	0.03476
1.23245	0.03476
1.23883	0.03476
1.24520	0.03511
1.25158	0.03523
1.25795	0.03523
1.26433	0.03558
1.27070	0.03570
1.27708	0.03570
1.28345	0.03594
1.28983	0.03594
1.29620	0.03594
1.30258	0.03618
1.30895	0.03594
1.33764	0.03700
1.34401	0.03712
1.35039	0.03724
1.35676	0.03759
1.36314	0.03759
1.36951	0.03771
1.37589	0.03795
1.38226	0.03807
1.38864	0.03807
1.39501	0.03842
1.40139	0.03854
1.40776	0.03842
1.43645	0.03913
1.44282	0.03937
1.44920	0.03949
1.45557	0.03949
1.46195	0.03984
1.46832	0.04067
1.47470	0.04149
1.48107	0.04185
1.48745	0.04220
1.49382	0.04232
1.50020	0.04232
1.50657	0.04232
1.51295	0.04232
1.51932	0.04244
1.52570	0.04279
1.53207	0.04279
1.53845	0.04279
1.54482	0.04303
1.55120	0.04315
1.55757	0.04327
1.56395	0.04362
1.57032	0.04374
1.57670	0.04421
1.58307	0.04421
1.58945	0.04457
1.59582	0.04469
1.60220	0.04469
1.60857	0.04480
1.61495	0.04492
1.62132	0.04516
1.62770	0.04516
1.63407	0.04528
1.64045	0.04551
1.64682	0.04563
1.65320	0.04599
1.65957	0.04610
1.68826	0.04705
1.69463	0.04717
1.70101	0.04740
1.70738	0.04776
1.71376	0.04800
1.72013	0.04835
1.72651	0.04859
1.73288	0.04906
1.73926	0.04930
1.74563	0.05000
1.75201	0.05060
1.77751	0.05190
1.78388	0.05178
1.79026	0.05190
1.79663	0.05225
1.80301	0.05225
1.80938	0.05272
1.81576	0.05296
1.82213	0.05320
1.82851	0.05343
1.83488	0.05367
1.84126	0.05367
1.84763	0.05414
1.85401	0.05438
1.86038	0.05438
1.86676	0.05473
1.87313	0.05497
1.87951	0.05521
1.88588	0.05544
1.89226	0.05580
1.89863	0.05639
1.90501	0.05662
1.91138	0.05674
1.91776	0.05698
1.92413	0.05721
1.93051	0.05745
1.93688	0.05792
1.94326	0.05792
1.94963	0.05840
1.95601	0.05851
1.96238	0.05875
1.96876	0.05899
1.97513	0.05911
1.98151	0.05934
1.98788	0.05970
1.99426	0.06017
2.00063	0.06041
2.02932	0.06123
2.03569	0.06159
2.04207	0.06194
2.04844	0.06218
2.05482	0.06265
2.06119	0.06277
2.06757	0.06301
2.07394	0.06324
2.08032	0.06360
2.08669	0.06360
2.09307	0.06383
2.09785	0.06407
2.12494	0.06502
2.13132	0.06525
2.13769	0.06537
2.14407	0.06561
2.15044	0.06596
2.15682	0.06596
2.16319	0.06643
2.16957	0.06655
2.17594	0.06679
2.18232	0.06714
2.18869	0.06738
2.19507	0.06785
2.20144	0.06785
2.20782	0.06809
2.21419	0.06833
2.22057	0.06868
2.22694	0.06903
2.23332	0.06927
2.23969	0.06963
2.24607	0.06974
2.25244	0.07010
2.25882	0.07057
2.26519	0.07081
2.27157	0.07104
2.27794	0.07140
2.28432	0.07163
2.29069	0.07211
2.29707	0.07258
2.30344	0.07282
2.30982	0.07353
2.31619	0.07364
2.32257	0.07400
2.32894	0.07412
2.33532	0.07447
2.34169	0.07459
2.36082	0.07636
2.36719	0.07648
2.37357	0.07660
2.37994	0.07707
2.38632	0.07719
2.39269	0.07754
2.39907	0.07790
2.40544	0.07814
2.41182	0.07825
2.41819	0.07837
2.42457	0.07873
2.43094	0.07884
2.45963	0.08014
2.46600	0.08014
2.47238	0.08014
2.47875	0.08050
2.48513	0.08074
2.49150	0.08109
2.49788	0.08133
2.50425	0.08156
2.51063	0.08192
2.51700	0.08227
2.52338	0.08239
2.52975	0.08263
2.53613	0.08286
2.54250	0.08310
2.54888	0.08357
2.55525	0.08405
2.56163	0.08428
2.56800	0.08440
2.57438	0.08475
2.58075	0.08499
2.58713	0.08535
2.59350	0.08558
2.59988	0.08605
2.60625	0.08629
2.61263	0.08653
2.61900	0.08676
2.62538	0.08688
2.63175	0.08724
2.63813	0.08759
2.64450	0.08806
2.65088	0.08830
2.65725	0.08854
2.66363	0.08877
2.67000	0.08901
2.67638	0.08913
2.70188	0.09066
2.70825	0.09078
2.71463	0.09102
2.72100	0.09102
2.72738	0.09126
2.73375	0.09149
2.74013	0.09149
2.74650	0.09196
2.75288	0.09196
2.75925	0.09220
2.76563	0.09232
2.77200	0.09232
2.79750	0.09421
2.80388	0.09409
2.81025	0.09445
2.81663	0.09456
2.82300	0.09480
2.82938	0.09504
2.83575	0.09527
2.84213	0.09527
2.84850	0.09575
2.85488	0.09575
2.86125	0.09575
2.86763	0.09598
2.87400	0.09622
2.88038	0.09634
2.88675	0.09646
2.89313	0.09681
2.89950	0.09705
2.90588	0.09728
2.91225	0.09752
2.91863	0.09787
2.92500	0.09811
2.93137	0.09847
2.93775	0.09870
2.94412	0.09894
2.95050	0.09929
2.95687	0.09977
2.96325	0.10012
2.96962	0.10059
2.97600	0.10071
2.98237	0.10095
2.98875	0.10118
2.99512	0.10130
};
\addplot [line width=2pt, curve_color, loosely dashdotted]
table {%
0.00049	0.00237
0.02121	0.00237
0.02759	0.00273
0.03396	0.00284
0.04034	0.00284
0.04946	0.00284
0.08496	0.00308
0.09134	0.00308
0.09771	0.00308
0.10408	0.00308
0.11046	0.00308
0.11683	0.00308
0.12321	0.00320
0.12958	0.00324
0.13596	0.00332
0.16783	0.00332
0.17421	0.00332
0.18058	0.00332
0.18696	0.00344
0.19333	0.00344
0.19971	0.00351
0.20608	0.00355
0.21246	0.00355
0.21724	0.00355
0.24433	0.00355
0.25071	0.00355
0.25708	0.00355
0.26346	0.00371
0.26983	0.00379
0.27621	0.00391
0.28258	0.00403
0.28896	0.00403
0.29533	0.00403
0.30011	0.00403
0.32402	0.00403
0.33040	0.00403
0.33677	0.00414
0.34315	0.00426
0.34952	0.00426
0.35590	0.00426
0.36227	0.00426
0.36865	0.00450
0.37502	0.00450
0.37980	0.00450
0.40371	0.00450
0.41008	0.00450
0.41646	0.00462
0.42283	0.00474
0.42921	0.00474
0.43558	0.00474
0.44196	0.00474
0.44833	0.00474
0.45471	0.00474
0.45949	0.00474
0.48658	0.00474
0.49296	0.00497
0.49933	0.00497
0.50571	0.00497
0.51208	0.00497
0.51846	0.00497
0.52483	0.00497
0.53121	0.00509
0.53758	0.00509
0.56627	0.00521
0.57264	0.00521
0.57902	0.00521
0.58539	0.00521
0.59177	0.00521
0.59814	0.00533
0.60452	0.00533
0.61089	0.00556
0.64596	0.00604
0.65233	0.00615
0.65871	0.00615
0.66508	0.00627
0.67146	0.00639
0.67783	0.00639
0.68421	0.00663
0.69058	0.00663
0.69696	0.00663
0.72564	0.00675
0.73202	0.00686
0.73839	0.00686
0.74477	0.00686
0.75114	0.00686
0.75752	0.00686
0.76389	0.00686
0.77027	0.00686
0.77664	0.00686
0.78142	0.00686
0.80852	0.00710
0.81489	0.00710
0.82127	0.00710
0.82764	0.00710
0.83402	0.00710
0.84039	0.00722
0.84677	0.00734
0.85314	0.00734
0.85952	0.00734
0.88820	0.00781
0.89458	0.00781
0.90095	0.00781
0.90733	0.00805
0.91370	0.00828
0.92008	0.00828
0.92645	0.00828
0.93283	0.00828
0.93920	0.00828
0.96470	0.00946
0.97108	0.00946
0.97745	0.01017
0.98383	0.01029
0.99020	0.01053
0.99658	0.01088
1.00295	0.01100
1.00933	0.01112
1.01570	0.01112
1.04439	0.01135
1.05076	0.01135
1.05714	0.01147
1.06351	0.01159
1.06989	0.01159
1.07626	0.01159
1.08264	0.01183
1.08901	0.01206
1.09539	0.01206
1.10017	0.01218
1.12726	0.01218
1.13364	0.01218
1.14001	0.01230
1.14639	0.01242
1.15276	0.01254
1.15914	0.01254
1.16551	0.01254
1.17189	0.01254
1.17826	0.01254
1.20695	0.01325
1.21333	0.01325
1.21970	0.01348
1.22608	0.01348
1.23245	0.01360
1.23883	0.01365
1.24520	0.01384
1.25158	0.01384
1.25795	0.01396
1.28345	0.01466
1.28983	0.01490
1.29620	0.01514
1.30258	0.01537
1.30895	0.01537
1.31533	0.01537
1.32170	0.01549
1.32808	0.01585
1.33445	0.01585
1.36632	0.01601
1.37270	0.01601
1.37907	0.01608
1.38545	0.01608
1.39182	0.01620
1.39820	0.01632
1.40457	0.01632
1.41095	0.01632
1.41732	0.01644
1.44601	0.01679
1.45239	0.01679
1.45876	0.01679
1.46514	0.01703
1.47151	0.01726
1.47789	0.01726
1.48426	0.01750
1.49064	0.01750
1.49701	0.01762
1.52251	0.01821
1.52889	0.01821
1.53526	0.01821
1.54164	0.01821
1.54801	0.01821
1.55439	0.01821
1.56076	0.01868
1.56714	0.01868
1.57351	0.01892
1.60539	0.01939
1.61176	0.01939
1.61814	0.01951
1.62451	0.01963
1.63089	0.01963
1.63726	0.01963
1.64364	0.01975
1.65001	0.01986
1.65639	0.01986
1.68188	0.02057
1.68826	0.02057
1.69463	0.02057
1.70101	0.02069
1.70738	0.02081
1.71376	0.02105
1.72013	0.02105
1.72651	0.02128
1.73288	0.02140
1.73767	0.02152
1.76157	0.02211
1.76795	0.02235
1.77432	0.02247
1.78070	0.02258
1.78707	0.02294
1.79345	0.02294
1.79982	0.02294
1.80620	0.02294
1.81257	0.02294
1.84126	0.02388
1.84763	0.02388
1.85401	0.02412
1.86038	0.02436
1.86676	0.02436
1.87313	0.02447
1.87951	0.02471
1.88588	0.02507
1.89226	0.02542
1.91457	0.02672
1.92095	0.02672
1.92732	0.02672
1.93370	0.02684
1.94007	0.02696
1.94645	0.02707
1.95282	0.02719
1.95920	0.02731
1.96557	0.02755
1.97035	0.02767
1.99426	0.02849
2.00063	0.02861
2.00701	0.02861
2.01338	0.02861
2.01976	0.02873
2.02613	0.02897
2.03251	0.02908
2.03888	0.02920
2.04526	0.02932
2.07394	0.02956
2.08032	0.02956
2.08669	0.02968
2.09307	0.02979
2.09944	0.03003
2.10582	0.03003
2.11219	0.03003
2.11857	0.03003
2.12494	0.03027
2.15363	0.03086
2.16001	0.03098
2.16638	0.03098
2.17276	0.03109
2.17913	0.03133
2.18551	0.03145
2.19188	0.03157
2.19826	0.03168
2.20463	0.03192
2.23332	0.03263
2.23969	0.03275
2.24607	0.03287
2.25244	0.03287
2.25882	0.03310
2.26519	0.03322
2.27157	0.03334
2.27794	0.03358
2.28432	0.03369
2.30982	0.03476
2.31619	0.03476
2.32257	0.03499
2.32894	0.03511
2.33532	0.03511
2.34169	0.03523
2.34807	0.03547
2.35444	0.03570
2.36082	0.03594
2.38950	0.03677
2.39588	0.03689
2.40225	0.03736
2.40863	0.03748
2.41500	0.03759
2.42138	0.03795
2.42775	0.03819
2.43413	0.03830
2.46600	0.03913
2.47238	0.03925
2.47875	0.03949
2.48513	0.03960
2.49150	0.03996
2.49788	0.03996
2.50425	0.03996
2.51063	0.04031
2.51700	0.04043
2.54569	0.04114
2.55207	0.04114
2.55844	0.04138
2.56482	0.04149
2.57119	0.04173
2.57757	0.04185
2.58394	0.04185
2.59032	0.04220
2.59669	0.04232
2.62219	0.04315
2.62856	0.04339
2.63494	0.04350
2.64131	0.04362
2.64769	0.04398
2.65406	0.04409
2.66044	0.04421
2.66681	0.04445
2.67319	0.04457
2.70188	0.04528
2.70825	0.04551
2.71463	0.04563
2.72100	0.04587
2.72738	0.04622
2.73375	0.04634
2.74013	0.04646
2.74650	0.04681
2.75288	0.04729
2.77838	0.04752
2.78475	0.04764
2.79113	0.04764
2.79750	0.04788
2.80388	0.04835
2.81025	0.04847
2.81663	0.04847
2.82300	0.04870
2.85806	0.04965
2.86444	0.04965
2.88038	0.05083
2.88675	0.05107
2.89313	0.05107
2.89950	0.05107
2.93137	0.05264
2.93775	0.05272
2.96325	0.05331
2.96962	0.05355
2.97600	0.05367
2.98078	0.05367
};
\addplot [line width=0.8pt, curve_color, densely dashed]
table {%
0	-0.6938
1	-0.5698
2	-0.4458
3	-0.3218
4	-0.1978
5	-0.0738
6	0.0502
7	0.1742
8	0.2982
9	0.4222
10	0.5462
11	0.6702
12	0.7942
13	0.9182
14	1.0422
};
\addplot [line width=1.8pt, curve_color, densely dashed]
table {%
5.66096	0.00519
5.73376	0.00536
5.82470	0.00519
5.86836	0.00519
5.91203	0.00689
5.95569	0.00859
6.02119	0.01029
6.05393	0.01369
6.10851	0.01709
6.14126	0.01879
6.19236	0.02321
6.22675	0.02679
6.26115	0.03036
6.31847	0.03571
6.37049	0.03919
6.40324	0.03919
6.46873	0.04599
6.52331	0.05449
6.55924	0.05893
6.60510	0.06071
6.63949	0.06786
6.67614	0.07150
6.70828	0.07679
6.74268	0.08036
6.77707	0.08571
6.80713	0.09105
6.84586	0.09286
6.88025	0.09821
6.90318	0.10536
6.93758	0.10893
6.97197	0.11429
7.02544	0.11995
7.05223	0.12679
7.08662	0.13393
7.13460	0.14120
7.17826	0.14885
7.22193	0.15650
7.26559	0.16246
7.30925	0.17096
7.35292	0.17861
7.39658	0.18541
7.42933	0.18881
7.46208	0.20071
7.50574	0.20581
7.54940	0.21261
7.59307	0.22111
7.63673	0.22791
7.68403	0.23528
7.72406	0.24194
7.76772	0.25427
7.81138	0.26192
7.85504	0.27127
7.89871	0.28062
7.94237	0.28827
7.98603	0.29762
8.02970	0.30612
8.07336	0.31462
8.11702	0.32397
8.16069	0.33247
8.20435	0.34097
8.24801	0.34863
8.29168	0.35713
8.33534	0.36563
8.37900	0.37328
8.42267	0.38178
8.46633	0.38943
8.50999	0.39623
8.55366	0.40303
8.59732	0.40983
8.64098	0.41663
8.68829	0.42060
8.73195	0.42683
8.77197	0.43193
8.81564	0.44214
8.86294	0.45290
8.89205	0.45574
8.94663	0.46424
8.99029	0.47104
9.03395	0.47784
9.07762	0.48464
9.12128	0.49144
9.16494	0.49824
9.20861	0.50504
9.25227	0.51184
9.29593	0.51864
9.33960	0.52289
9.38326	0.52969
9.42692	0.53565
9.47059	0.54245
9.51425	0.54925
9.55791	0.55605
9.60158	0.56030
9.64524	0.56625
9.68890	0.57305
9.73257	0.57815
9.77623	0.58325
9.81989	0.58920
9.84968	0.59643
9.90722	0.60195
9.93997	0.60365
9.97580	0.60893
10.03821 0.61725
10.08187 0.62150
10.12554 0.62746
10.16920 0.63171
10.21286 0.63766
10.25653 0.64361
10.30019 0.64786
10.34385 0.65381
10.38752 0.65806
10.43118 0.66316
10.47484 0.66826
10.51851 0.67251
10.56217 0.67846
10.60583 0.68271
10.64950 0.68781
10.69316 0.69206
10.73682 0.69631
10.78049 0.70311
10.82415 0.70651
10.86781 0.71161
10.91148 0.71757
10.95514 0.72012
10.99880 0.72522
11.04247 0.72947
11.08613 0.73287
11.12979 0.73712
11.17346 0.74222
11.21712 0.74647
11.26078 0.74987
11.30445 0.75412
11.34811 0.75837
11.39177 0.76262
11.43544 0.76687
11.47910 0.77027
11.52276 0.77367
11.56643 0.77792
11.61009 0.78132
11.65375 0.78557
11.69742 0.78982
11.74108 0.79322
11.78474 0.79662
11.82841 0.80002
11.87207 0.80342
11.91573 0.80682
11.95940 0.81023
11.99214 0.81278
12.03581 0.81788
12.09039 0.82128
12.13405 0.82298
12.17771 0.82808
12.22138 0.82978
12.26504 0.83233
12.30870 0.83573
12.35237 0.83658
12.39603 0.83998
12.43969 0.84168
12.48336 0.84423
12.52702 0.84763
12.57068 0.85018
12.61435 0.85358
12.65801 0.85613
12.70167 0.85868
12.74534 0.86123
12.78900 0.86378
12.83266 0.86633
12.87633 0.86888
12.91999 0.87058
12.96365 0.87398
13.00732 0.87568
13.05098 0.87738
13.09464 0.88078
13.13831 0.88248
13.18197 0.88418
13.22563 0.88758
13.26930 0.88928
13.31296 0.89098
13.35662 0.89353
13.40029 0.89523
13.44395 0.89693
13.48761 0.89863
13.53128 0.90033
13.57494 0.90289
13.61860 0.90459
13.66227 0.90629
13.70593 0.90799
13.74959 0.90969
13.79326 0.91054
13.83692 0.91309
13.88058 0.91309
13.92425 0.91479
13.96791 0.91649
14.00066 0.91989
};
\end{axis}
\end{tikzpicture}

\begin{tikzpicture}
\pgfplotsset{every axis/.append style={
					xlabel=\mbox{$\log_{10}(\pwdRank)$},
					compat=1.8,
                    label style={font=\normalsize},
                    tick label style={font=\small}  
                    }}

\begin{axis}[
ylabel style={align=center}, 
xmin=0, xmax=12.5,
ymin=0, ymax=1,
xtick={0, 2.5, 5, 7.5, 10, 12.5},
xticklabels={0, 2.5, 5, 7.5, 10, 12.5},
scaled ticks=base 10:0,
ytick={0, 0.2, 0.4, 0.6, 0.8, 1.0},
yticklabels={0, 0.2, 0.4, 0.6, 0.8, 1.0},
width=1.1\figurewidth,
height=1.1\figurewidth,
tick align=outside,
tick pos=left,
xmajorgrids,
x grid style={lightgray!92.026143790849673!black},
minor y tick num=1,
ymajorgrids,
yminorgrids,
y grid style={lightgray!92.026143790849673!black},
legend entries={{~\eqnref{eqn:wangCracked-TGI}$~~~$}, {~\TGIPP},{~\TGIPPP$~~~$},{~\PCFG}},
legend style={{font={\fontsize{17pt}{12}\selectfont}},{draw=none}},
legend cell align={left},
legend columns=2,
legend pos={outer north east},
legend style={at={(1.03, 1.32)}, anchor=north east, draw=none, nodes={scale=0.458,
transform shape}}
]
\addlegendimage{line width=1pt, solid, curve_color}
\addlegendimage{line width=1pt, densely dotted, curve_color}
\addlegendimage{line width=1pt, dotted, curve_color}
\addlegendimage{line width=1pt, loosely dashdotted, curve_color}
\addplot [line width=1.0pt, curve_color]
table {%
0 	 0.04187
1 	 0.11901
2 	 0.19615
3 	 0.27329
4 	 0.35043
5 	 0.42757
6 	 0.50471
7 	 0.58185
8 	 0.65899
9 	 0.73613
10 	 0.81327
11 	 0.89041
12 	 0.96755
13 	 1.04469
};
\addplot [line width=1.0pt, curve_color, densely dotted]
table {%
0 	 0.03007
1 	 0.04711
2 	 0.06415
3 	 0.08119
4 	 0.09823
5 	 0.11527
6 	 0.13231
7 	 0.26745
8 	 0.40259
9 	 0.53773
10 	 0.67287
11 	 0.80801
12 	 0.94315
13 	 1.07829
};
\addplot [line width=1.0pt, curve_color, dotted]
table {%
0 	 0.04060
1 	 0.08677
2 	 0.13294
3 	 0.17911
4 	 0.22528
5 	 0.27145
6 	 0.31763
7 	 0.42391
8 	 0.53019
9 	 0.63647
10 	 0.74275
11 	 0.84903
12 	 0.95531
13 	 1.06159
};
\addplot [line width=1.0pt, curve_color, loosely dashdotted]
table {%
0 	 0.04187
1 	 0.10652
2 	 0.16651
3 	 0.22650
4 	 0.28649
5 	 0.34648
6 	 0.40647
7 	 0.49891
8 	 0.59135
9 	 0.68379
10 	 0.77623
11 	 0.86867
12 	 0.96111
13 	 1.05355
};
\end{axis}
\end{tikzpicture}

%% file: figures/fig_roc_hc_cmp.tex
\begin{figure}[t]
\captionsetup[subfigure]{font=small,labelfont=small}
  \begin{subfigure}[b]{.1\columnwidth}
    \setlength\figureheight{2in}
    \centering
    \hspace*{1.35em}
    \resizebox{!}{2.05em}{\input{figures/roc/roc_legend.tex}}
  \end{subfigure}
  \vspace*{0.8em}

  \hspace*{-0.8em}
  \begin{subfigure}[b]{.48\columnwidth}
    \setlength\figureheight{2.4in}
    \begin{minipage}[b]{1.05\textwidth}
      \centering
      \vspace*{0em}\resizebox{!}{11.6em}{\input{figures/roc/roc_hc_cmp.tex}}
      \hspace*{3.0em} 
      \begin{minipage}[t]{11em}
       \vspace*{-0.65em}
        \caption{\parbox[t]{5.3em}{$\fdpBound = 10^{-1}$}}
        \label{fig:roc_hc_cmp:tdp}
      \end{minipage}%
      \hspace*{0.2em} 
      \begin{minipage}[t]{11em}
       \vspace*{-0.65em}
        \caption{\parbox[t]{5.3em}{$\trueAlarmProb(\nmbrAccountsAttempted)
        \ge 0.5$}}
        \label{fig:roc_hc_cmp:fdp}
      \end{minipage}%
    \end{minipage}
  \end{subfigure}%
  \caption{True detection probability for honeychecker integration
    (\secref{sec:honeychecker:design}).  \figref{fig:roc_hc_cmp:tdp}
    shows $\trueAlarmProb(\nmbrAccountsAttempted)$ per fraction
    $\nmbrAccountsAttempted / \nmbrAccounts$ of accounts accessed by
    the \breachingAttacker.  \figref{fig:roc_hc_cmp:fdp} shows the
    fraction $\nmbrAccountsAttempted / \nmbrAccounts$ of accounts
    attempted by the \breachingAttacker at which
    $\trueAlarmProb(\nmbrAccountsAttempted) \ge 0.5$, per bound
    \fdpBound in \eqnref{eqn:fdpBounds}.  \TGIPP, \TGIPPP, and \PCFG
    represent different blocklists adopted for
    \eqnref{eqn:wangCracked}; see
    \secref{sec:honeychecker:eval:rank}.}
  \label{fig:roc_hc_cmp}
\end{figure}

%% file: figures/roc/roc_legend.tex
\newenvironment{customlegend}[1][]{%
    \begingroup
    \csname pgfplots@init@cleared@structures\endcsname
    \pgfplotsset{#1}%
}{%
    \csname pgfplots@createlegend\endcsname
    \endgroup
}%

\def\addlegendimage{\csname pgfplots@addlegendimage\endcsname}

\begin{tikzpicture}

\begin{customlegend}[
    legend style={{font={\fontsize{12pt}{12}\selectfont}},{draw=none}},
    legend columns=3,
    legend cell align={left},
    legend entries={~\eqnref{eqn:mazurekCracked}$\quad$, ~\eqnref{eqn:xuCracked}$\qquad$,
    ~\eqnref{eqn:wangCracked} with \TGIPP$\quad$,
~\eqnref{eqn:wangCracked} with \TGIPPP$\quad$, ~\eqnref{eqn:wangCracked} with
\PCFG$\quad$,
~\eqnref{eqn:wangCracked-TGI}$\quad$}]
\addlegendimage{line width=2.4pt, curve_color, densely dotted}
\addlegendimage{line width=2.4pt, curve_color, dashed}
\addlegendimage{line width=1.6pt, curve_color}
\addlegendimage{line width=1.6pt, curve_color, dotted}
\addlegendimage{line width=1.6pt, curve_color, dash pattern=on 1pt off 3pt on 3pt off 3pt}
\addlegendimage{line width=1.6pt, curve_color, dashed}

\end{customlegend}

\end{tikzpicture}

%% file: figures/roc/roc_hc_cmp.tex
\begin{tikzpicture}
\pgfplotsset{every axis/.append style={
					ylabel=\mbox{$\trueAlarmProb(\nmbrAccountsAttempted)$},
					xlabel=\mbox{$\nmbrAccountsAttempted/\nmbrAccounts$},
					compat=1.8,
                    label style={font=\normalsize},
                    tick label style={font=\small}  
                    }}

\begin{axis}[
ylabel style={align=center}, 
xmin=-0.02, xmax=0.5,
ymin=0, ymax=1.03,
xtick={0, 0.1, 0.2, 0.3, 0.4, 0.5},
ytick={0, 0.2, 0.4, 0.6, 0.8, 1.0},
yticklabels={0, 0.2, 0.4, 0.6, 0.8, 1.0},
width=1.2\figurewidth,
height=1.1\figurewidth,
tick align=outside,
tick pos=left,
xmajorgrids,
minor x tick num=1,
x grid style={lightgray!92.026143790849673!black}, minor y tick num=1,
ymajorgrids, yminorgrids, y grid style={lightgray!92.026143790849673!black},
]
\addplot [line width=1pt, curve_color]
table {%
0.0000e+00	0.0000
4.5398e-02	0.0102
4.8507e-02	0.0203
5.0765e-02	0.0303
5.2172e-02	0.0404
5.3333e-02	0.0504
5.4322e-02	0.0606
5.5158e-02	0.0709
5.5916e-02	0.0813
5.6612e-02	0.0913
5.7261e-02	0.1015
5.7741e-02	0.1119
5.8174e-02	0.1222
5.8514e-02	0.1322
5.8854e-02	0.1423
5.9195e-02	0.1525
5.9458e-02	0.1628
5.9721e-02	0.1733
6.0015e-02	0.1838
6.0247e-02	0.1939
6.0479e-02	0.2041
6.0680e-02	0.2144
6.0881e-02	0.2246
6.1097e-02	0.2348
6.1267e-02	0.2454
6.1407e-02	0.2555
6.1546e-02	0.2659
6.1716e-02	0.2760
6.1855e-02	0.2871
6.1994e-02	0.2981
6.2149e-02	0.3083
6.2288e-02	0.3187
6.2412e-02	0.3292
6.2536e-02	0.3396
6.2675e-02	0.3501
6.2814e-02	0.3618
6.2938e-02	0.3725
6.3077e-02	0.3833
6.3216e-02	0.3941
6.3340e-02	0.4046
6.3464e-02	0.4153
6.3588e-02	0.4256
6.3696e-02	0.4356
6.3804e-02	0.4460
6.3943e-02	0.4571
6.4052e-02	0.4691
6.4160e-02	0.4791
6.4284e-02	0.4902
6.4392e-02	0.5013
6.4500e-02	0.5126
6.4608e-02	0.5235
6.4717e-02	0.5344
6.4825e-02	0.5445
6.4933e-02	0.5553
6.5057e-02	0.5663
6.5165e-02	0.5766
6.5289e-02	0.5878
6.5397e-02	0.5986
6.5506e-02	0.6098
6.5614e-02	0.6213
6.5738e-02	0.6319
6.5861e-02	0.6422
6.5970e-02	0.6524
6.6093e-02	0.6636
6.6202e-02	0.6744
6.6310e-02	0.6849
6.6418e-02	0.6952
6.6542e-02	0.7073
6.6666e-02	0.7176
6.6789e-02	0.7280
6.6882e-02	0.7388
6.7006e-02	0.7500
6.7130e-02	0.7601
6.7253e-02	0.7710
6.7377e-02	0.7820
6.7501e-02	0.7934
6.7625e-02	0.8034
6.7764e-02	0.8145
6.7903e-02	0.8255
6.8042e-02	0.8372
6.8197e-02	0.8474
6.8336e-02	0.8580
6.8491e-02	0.8683
6.8645e-02	0.8788
6.8831e-02	0.8889
6.9017e-02	0.8990
6.9187e-02	0.9091
6.9419e-02	0.9196
6.9604e-02	0.9298
6.9883e-02	0.9402
7.0208e-02	0.9505
7.0548e-02	0.9606
7.0950e-02	0.9707
7.1461e-02	0.9807
7.2249e-02	0.9907
7.6673e-02	1.0000
1.0000e+00	1.0000
};
\addplot [line width=1pt, curve_color, dotted]
table {%
0.0000e+00	0.0000
1.0247e-01	0.0101
1.0713e-01	0.0202
1.0998e-01	0.0302
1.1223e-01	0.0403
1.1392e-01	0.0504
1.1556e-01	0.0604
1.1667e-01	0.0705
1.1765e-01	0.0805
1.1867e-01	0.0907
1.1947e-01	0.1008
1.2028e-01	0.1111
1.2105e-01	0.1213
1.2170e-01	0.1314
1.2232e-01	0.1414
1.2291e-01	0.1516
1.2348e-01	0.1621
1.2413e-01	0.1721
1.2468e-01	0.1821
1.2518e-01	0.1922
1.2563e-01	0.2026
1.2605e-01	0.2126
1.2646e-01	0.2227
1.2688e-01	0.2331
1.2730e-01	0.2432
1.2762e-01	0.2534
1.2799e-01	0.2634
1.2843e-01	0.2737
1.2881e-01	0.2837
1.2923e-01	0.2943
1.2959e-01	0.3046
1.2994e-01	0.3146
1.3030e-01	0.3248
1.3066e-01	0.3351
1.3098e-01	0.3454
1.3129e-01	0.3555
1.3160e-01	0.3656
1.3194e-01	0.3762
1.3228e-01	0.3864
1.3260e-01	0.3969
1.3293e-01	0.4073
1.3324e-01	0.4179
1.3349e-01	0.4281
1.3380e-01	0.4384
1.3412e-01	0.4484
1.3441e-01	0.4590
1.3469e-01	0.4691
1.3500e-01	0.4791
1.3528e-01	0.4898
1.3559e-01	0.5003
1.3581e-01	0.5107
1.3605e-01	0.5210
1.3636e-01	0.5317
1.3670e-01	0.5422
1.3700e-01	0.5523
1.3729e-01	0.5625
1.3757e-01	0.5726
1.3786e-01	0.5834
1.3816e-01	0.5934
1.3844e-01	0.6038
1.3865e-01	0.6138
1.3895e-01	0.6239
1.3921e-01	0.6340
1.3946e-01	0.6447
1.3972e-01	0.6549
1.4003e-01	0.6649
1.4032e-01	0.6752
1.4062e-01	0.6854
1.4089e-01	0.6954
1.4122e-01	0.7057
1.4153e-01	0.7157
1.4182e-01	0.7261
1.4212e-01	0.7361
1.4241e-01	0.7464
1.4274e-01	0.7567
1.4308e-01	0.7667
1.4337e-01	0.7769
1.4369e-01	0.7873
1.4400e-01	0.7975
1.4431e-01	0.8076
1.4475e-01	0.8179
1.4512e-01	0.8281
1.4549e-01	0.8381
1.4589e-01	0.8487
1.4629e-01	0.8590
1.4666e-01	0.8693
1.4708e-01	0.8797
1.4753e-01	0.8897
1.4807e-01	0.8998
1.4860e-01	0.9099
1.4912e-01	0.9202
1.4968e-01	0.9303
1.5036e-01	0.9403
1.5104e-01	0.9505
1.5189e-01	0.9605
1.5307e-01	0.9705
1.5438e-01	0.9807
1.5602e-01	0.9907
1.6308e-01	1.0000
1.0000e+00	1.0000
};
\addplot [line width=1pt, curve_color, dash pattern=on 1pt off 3pt on 3pt off 3pt]
table {%
0.0000e+00	0.0000
1.2772e-01	0.0101
1.3465e-01	0.0202
1.3800e-01	0.0303
1.4077e-01	0.0403
1.4277e-01	0.0504
1.4465e-01	0.0604
1.4622e-01	0.0704
1.4747e-01	0.0805
1.4881e-01	0.0907
1.4985e-01	0.1008
1.5078e-01	0.1109
1.5185e-01	0.1210
1.5262e-01	0.1313
1.5347e-01	0.1413
1.5421e-01	0.1515
1.5505e-01	0.1615
1.5576e-01	0.1718
1.5636e-01	0.1819
1.5693e-01	0.1921
1.5762e-01	0.2025
1.5836e-01	0.2129
1.5898e-01	0.2230
1.5958e-01	0.2331
1.6020e-01	0.2431
1.6079e-01	0.2534
1.6136e-01	0.2634
1.6184e-01	0.2738
1.6236e-01	0.2840
1.6289e-01	0.2941
1.6343e-01	0.3042
1.6383e-01	0.3143
1.6430e-01	0.3248
1.6472e-01	0.3351
1.6521e-01	0.3452
1.6567e-01	0.3559
1.6614e-01	0.3659
1.6654e-01	0.3761
1.6694e-01	0.3867
1.6744e-01	0.3967
1.6784e-01	0.4068
1.6821e-01	0.4170
1.6858e-01	0.4272
1.6889e-01	0.4374
1.6920e-01	0.4476
1.6960e-01	0.4579
1.6997e-01	0.4682
1.7033e-01	0.4785
1.7072e-01	0.4885
1.7112e-01	0.4986
1.7146e-01	0.5088
1.7183e-01	0.5189
1.7222e-01	0.5290
1.7256e-01	0.5392
1.7297e-01	0.5492
1.7332e-01	0.5592
1.7373e-01	0.5694
1.7404e-01	0.5797
1.7440e-01	0.5901
1.7478e-01	0.6004
1.7522e-01	0.6108
1.7556e-01	0.6209
1.7590e-01	0.6315
1.7627e-01	0.6422
1.7663e-01	0.6523
1.7701e-01	0.6627
1.7738e-01	0.6730
1.7777e-01	0.6830
1.7819e-01	0.6937
1.7854e-01	0.7039
1.7891e-01	0.7139
1.7932e-01	0.7239
1.7969e-01	0.7346
1.8011e-01	0.7446
1.8051e-01	0.7549
1.8088e-01	0.7649
1.8134e-01	0.7751
1.8178e-01	0.7852
1.8226e-01	0.7956
1.8272e-01	0.8061
1.8315e-01	0.8164
1.8365e-01	0.8264
1.8413e-01	0.8365
1.8461e-01	0.8465
1.8513e-01	0.8567
1.8563e-01	0.8667
1.8611e-01	0.8771
1.8671e-01	0.8871
1.8725e-01	0.8973
1.8792e-01	0.9074
1.8875e-01	0.9174
1.8946e-01	0.9274
1.9014e-01	0.9375
1.9086e-01	0.9476
1.9177e-01	0.9576
1.9296e-01	0.9676
1.9429e-01	0.9776
1.9666e-01	0.9876
2.0209e-01	0.9976
2.0946e-01	1.0000
1.0000e+00	1.0000
};
\addplot [line width=1pt, curve_color, dashed]
table {%
0.0000e+00	0.0000
1.4874e-01	0.0101
1.5670e-01	0.0202
1.6147e-01	0.0304
1.6482e-01	0.0405
1.6716e-01	0.0505
1.6945e-01	0.0606
1.7120e-01	0.0707
1.7321e-01	0.0809
1.7499e-01	0.0911
1.7656e-01	0.1012
1.7794e-01	0.1114
1.7932e-01	0.1216
1.8048e-01	0.1317
1.8165e-01	0.1418
1.8275e-01	0.1519
1.8385e-01	0.1620
1.8464e-01	0.1722
1.8561e-01	0.1824
1.8659e-01	0.1926
1.8734e-01	0.2029
1.8810e-01	0.2130
1.8883e-01	0.2232
1.8953e-01	0.2334
1.9033e-01	0.2435
1.9109e-01	0.2536
1.9172e-01	0.2636
1.9231e-01	0.2737
1.9296e-01	0.2838
1.9362e-01	0.2940
1.9417e-01	0.3042
1.9482e-01	0.3145
1.9548e-01	0.3245
1.9613e-01	0.3346
1.9670e-01	0.3447
1.9727e-01	0.3548
1.9780e-01	0.3649
1.9839e-01	0.3749
1.9885e-01	0.3851
1.9936e-01	0.3951
1.9990e-01	0.4051
2.0034e-01	0.4152
2.0085e-01	0.4253
2.0137e-01	0.4354
2.0178e-01	0.4454
2.0226e-01	0.4556
2.0270e-01	0.4658
2.0318e-01	0.4758
2.0365e-01	0.4859
2.0413e-01	0.4959
2.0462e-01	0.5061
2.0509e-01	0.5163
2.0550e-01	0.5266
2.0592e-01	0.5370
2.0639e-01	0.5477
2.0688e-01	0.5577
2.0733e-01	0.5681
2.0779e-01	0.5783
2.0824e-01	0.5883
2.0872e-01	0.5984
2.0917e-01	0.6086
2.0962e-01	0.6190
2.1007e-01	0.6291
2.1055e-01	0.6393
2.1109e-01	0.6494
2.1155e-01	0.6595
2.1203e-01	0.6695
2.1253e-01	0.6795
2.1305e-01	0.6897
2.1348e-01	0.6998
2.1393e-01	0.7099
2.1446e-01	0.7201
2.1491e-01	0.7301
2.1537e-01	0.7401
2.1582e-01	0.7504
2.1633e-01	0.7606
2.1687e-01	0.7709
2.1755e-01	0.7813
2.1802e-01	0.7913
2.1860e-01	0.8013
2.1922e-01	0.8114
2.1990e-01	0.8215
2.2057e-01	0.8315
2.2116e-01	0.8420
2.2195e-01	0.8520
2.2275e-01	0.8620
2.2349e-01	0.8720
2.2419e-01	0.8820
2.2487e-01	0.8921
2.2567e-01	0.9023
2.2657e-01	0.9123
2.2742e-01	0.9226
2.2838e-01	0.9326
2.2943e-01	0.9426
2.3064e-01	0.9526
2.3223e-01	0.9626
2.3421e-01	0.9727
2.3658e-01	0.9828
2.4003e-01	0.9928
2.6119e-01	1.0000
1.0000e+00	1.0000
};
\addplot [line width=1.8pt, curve_color, densely dotted]
table {%
0.0000e+00	0.0000
1.8318e-04	1.0000
1.0000e+00	1.0000
};
\addplot [line width=1.8pt, curve_color, dashed]
table {%
0.0000e+00	0.0000
1.4777e-08	1.0000
1.0000e+00	1.0000
};
\end{axis}
\end{tikzpicture}

\begin{tikzpicture}
\pgfplotsset{every axis/.append style={
					ylabel=\mbox{\small $\frac{1}{\nmbrAccounts}\min \cset{\normalsize}{\nmbrAccountsAttempted}{\trueAlarmProb(\nmbrAccountsAttempted) \ge 0.5}$},
					xlabel=\mbox{$\fdpBound$},
					compat=1.8,
                    label style={font=\normalsize},
                    tick label style={font=\small},
                    xticklabel style = {font=\small,yshift=-0.5ex}    
                    }}

\begin{axis}[
ylabel style={align=center}, 
xmin=1e-5, xmax=1e-1,
ymin=0, ymax=1.03,
xmode=log,
log basis x={10},
xtick={1e-5, 1e-4, 1e-3, 1e-2, 1e-1},
xticklabels={$10^{-5}$,,$10^{-3}$,,$10^{-1}$},
yticklabels={},
ytick={0, 0.2, 0.4, 0.6, 0.8, 1.0},
yticklabels={0, 0.2, 0.4, 0.6, 0.8, 1.0},
width=0.85\figurewidth,
height=1.1\figurewidth,
tick align=outside,
tick pos=left,
xmajorgrids,
minor x tick num=1,
x grid style={lightgray!92.026143790849673!black},
minor y tick num=1,
ymajorgrids,
yminorgrids,
y grid style={lightgray!92.026143790849673!black},
]
\addplot [line width=1pt, curve_color]
table {%
1.0000e-05	3.6355e-01
1.0000e-04	1.0807e-01
1.0000e-03	8.6634e-02
1.0000e-02	7.4833e-02
1.0000e-01	6.4392e-02
};
\addplot [line width=1pt, curve_color, dotted]
table {%
1.0000e-05  5.1690e-01
1.0000e-04  2.5908e-01
1.0000e-03  2.0354e-01
1.0000e-02  1.6479e-01
1.0000e-01	1.3559e-01
};
\addplot [line width=1pt, curve_color, dash pattern=on 1pt off 3pt on 3pt off 3pt]
table {%
1.0000e-05	5.8712e-01
1.0000e-04	3.4389e-01
1.0000e-03	2.6137e-01
1.0000e-02	2.0977e-01
1.0000e-01	1.7146e-01
};
\addplot [line width=1pt, curve_color, dashed]
table {%
1.0000e-05	6.6341e-01
1.0000e-04	4.1787e-01
1.0000e-03	3.2312e-01
1.0000e-02	2.5556e-01
1.0000e-01	2.0462e-01
};
\addplot [line width=1.8pt, curve_color, densely dotted]
table {%
1.0000e-05	7.3273e-04
1.0000e-04	1.8318e-04
1.0000e-03	1.8318e-04
1.0000e-02	1.8318e-04
1.0000e-01	1.8318e-04
};
\addplot [line width=1.8pt, curve_color, dashed]
table {%
1.0000e-05	1.4777e-08
1.0000e-04	1.4777e-08
1.0000e-03	1.4777e-08
1.0000e-02	1.4777e-08
1.0000e-01	1.4777e-08
};
\end{axis}
\end{tikzpicture}

%% file: figures/fig_pwdcount_hc.tex
\begin{wrapfigure}{r}{0.46\columnwidth} 
  \vspace{-2ex}
  \hspace*{1.8em}
  \begin{subfigure}[b]{.1\columnwidth}
    \setlength\figureheight{2in}
    \centering
    \hspace*{1.85em}
    \resizebox{!}{2.7em}{\input{figures/roc/pwdcount_legend_hc.tex}}
  \end{subfigure}

  \hspace*{-1.2em}
  \begin{subfigure}[b]{.48\columnwidth}
  \captionsetup[subfigure]{font=small,labelfont=small}
    \setlength\figureheight{2.4in}
      \centering
      \vspace*{0em}\resizebox{!}{10.2em}{\input{figures/roc/pwdcount_hc.tex}}
  \end{subfigure}%
  \caption{True-detection probability for honeychecker integration
    (\secref{sec:honeychecker:design}) as function of \pwdCount for
    varying fractions $\nmbrAccountsAttempted / \nmbrAccounts$ of
    accounts accessed by the \breachingAttacker, based on
    \eqnref{eqn:wangCracked-TGI}; see
    \secref{sec:honeychecker:eval:rank}.}
  \label{fig:pwdcount:hc}
  \vspace{-2ex}
\end{wrapfigure}

%% file: figures/roc/pwdcount_legend_hc.tex
\hspace*{-1.0em}
\newenvironment{customlegend}[1][]{%
    \begingroup
    \csname pgfplots@init@cleared@structures\endcsname
    \pgfplotsset{#1}%
}{%
    \csname pgfplots@createlegend\endcsname
    \endgroup
}%

\def\addlegendimage{\csname pgfplots@addlegendimage\endcsname}

\begin{tikzpicture}
\begin{customlegend}[
    legend style={{font={\fontsize{12pt}{12}\selectfont}},{draw=none}},
    legend columns=1,
    legend cell align={left},
    legend entries={~$\nmbrAccountsAttempted/\nmbrAccounts = 0.23\qquad$,
    ~$\nmbrAccountsAttempted/\nmbrAccounts = 0.21\qquad$,
    ~$\nmbrAccountsAttempted/\nmbrAccounts = 0.19\qquad$}]
\addlegendimage{line width=1.6pt, curve_color}
\addlegendimage{line width=1.6pt, curve_color, dotted}
\addlegendimage{line width=1.6pt, curve_color, dash pattern=on 1pt off 3pt on 3pt off 3pt}
\end{customlegend}
\end{tikzpicture}

%% file: figures/roc/pwdcount_hc.tex
\begin{tikzpicture}
\pgfplotsset{every axis/.append style={
					ylabel=\mbox{$\trueAlarmProb(\nmbrAccountsAttempted)$},
					xlabel=\mbox{$\pwdCount$},
					compat=1.8,
                    label style={font=\normalsize},
                    tick label style={font=\normalsize},
                    xticklabel style = {font=\normalsize,yshift=-0.5ex} 
                    }}

\begin{axis}[
ylabel style={align=center}, 
xmin=1, xmax=3162,
ymin=0, ymax=1.03,
xmode=log,
log basis x={10},
ytick={0, 0.2, 0.4, 0.6, 0.8, 1.0},
yticklabels={0, 0.2, 0.4, 0.6, 0.8, 1.0},
width=\figurewidth,
height=\figurewidth,
tick align=outside,
tick pos=left,
xmajorgrids,
minor x tick num=1,
x grid style={lightgray!92.026143790849673!black}, minor y tick num=1,
ymajorgrids, yminorgrids, y grid style={lightgray!92.026143790849673!black},
]
\addplot [line width=1pt, curve_color]
table {%
1 		0.9785
2 		0.9533
3 		0.9544
10		0.9534
32		0.9545
100		0.9517
316		0.9527
1000	0.9526
3162	0.9518
};
\addplot [line width=1pt, curve_color, dotted]
table {%
1 		0.7304
2 		0.6284
3 		0.6294
10		0.6283
32		0.6289
100		0.6179
316		0.6185
1000	0.6291
3162	0.6185	
};
\addplot [line width=1pt, curve_color, dash pattern=on 1pt off 3pt on 3pt off 3pt]
table {%
1 		0.3235
2 		0.2494
3 		0.2501
10		0.2528
32		0.2508
100		0.2422
316		0.2426
1000	0.2435
3162	0.2427
};



\end{axis}
\end{tikzpicture}

%% file: figures/fig_roc_ba_cmp.tex
\begin{figure}[t]
\captionsetup[subfigure]{font=small,labelfont=small}
  \begin{subfigure}[b]{.1\columnwidth}
    \setlength\figureheight{2in}
    \centering
    \hspace*{1.35em}
    \resizebox{!}{2.05em}{\input{figures/roc/roc_legend.tex}}
  \end{subfigure}
  \vspace*{0.5em}

  \hspace*{-0.5em}
  \begin{subfigure}[b]{.48\columnwidth}
    \setlength\figureheight{2.4in}
    \begin{minipage}[b]{1.05\textwidth}
      \centering
      \vspace*{0em}\resizebox{!}{11.6em}{\input{figures/roc/roc_ba_cmp.tex}}
      \hspace*{3.2em} 
      \begin{minipage}[t]{11em}
       \vspace*{-0.65em}
        \caption{\parbox[t]{5.3em}{$\fdpBound = 10^{-1}$}}
        \label{fig:roc_ba_cmp:tdp}
      \end{minipage}%
      \hspace*{0.0em} 
      \begin{minipage}[t]{11em}
       \vspace*{-0.65em}
        \caption{\parbox[t]{8.3em}{$\trueAlarmProb(10, 31, \nmbrAccountsAttempted)
        \ge 0.5$}}
        \label{fig:roc_ba_cmp:fdp}
      \end{minipage}%
    \end{minipage}
  \end{subfigure}%
  \caption{True detection probability for Amnesia integration
    (\secref{sec:amnesia:local:design}).  \figref{fig:roc_ba_cmp:tdp}
    shows $\trueAlarmProb(10, 31, \nmbrAccountsAttempted)$ per
    fraction $\nmbrAccountsAttempted / \nmbrAccounts$ of accounts
    accessed by the \breachingAttacker.  \figref{fig:roc_ba_cmp:fdp}
    shows the fraction $\nmbrAccountsAttempted / \nmbrAccounts$ of
    accounts attempted by the \breachingAttacker at which
    $\trueAlarmProb(10,31,\nmbrAccountsAttempted) \ge 0.5$, per bound
    \fdpBound in \eqnref{eqn:fdpBounds}.  \TGIPP, \TGIPPP, and \PCFG
    represent different blocklists adopted for
    \eqnref{eqn:wangCracked}; see
    \secref{sec:honeychecker:eval:rank}.}
  \label{fig:roc_ba_cmp}
\end{figure}

%% file: figures/roc/roc_ba_cmp.tex
\begin{tikzpicture}
\pgfplotsset{every axis/.append style={
					ylabel=\mbox{$\trueAlarmProb(10, 31, \nmbrAccountsAttempted)$},
                    xlabel=\mbox{$\nmbrAccountsAttempted/\nmbrAccounts$},
					compat=1.8,
                    label style={font=\normalsize},
                    tick label style={font=\small}
                    }}

\begin{axis}[
ylabel style={align=center}, 
xmin=-0.02, xmax=0.5,
ymin=0, ymax=1.03,
xtick={0, 0.1, 0.2, 0.3, 0.4, 0.5},
ytick={0, 0.2, 0.4, 0.6, 0.8, 1.0},
yticklabels={0, 0.2, 0.4, 0.6, 0.8, 1.0},
width=1.2\figurewidth,
height=1.1\figurewidth,
tick align=outside,
tick pos=left,
xmajorgrids,
minor x tick num=1,
x grid style={lightgray!92.026143790849673!black}, minor y tick num=1,
ymajorgrids, yminorgrids, y grid style={lightgray!92.026143790849673!black},
]
\addplot [line width=1pt, curve_color]
table {%
0.0000e+00	0.0000
6.7795e-02	0.0101
7.1105e-02	0.0202
7.2853e-02	0.0304
7.4152e-02	0.0407
7.5018e-02	0.0510
7.5822e-02	0.0612
7.6658e-02	0.0712
7.7307e-02	0.0815
7.7833e-02	0.0916
7.8344e-02	0.1018
7.8885e-02	0.1119
7.9303e-02	0.1221
7.9736e-02	0.1322
8.0184e-02	0.1423
8.0556e-02	0.1524
8.0927e-02	0.1626
8.1252e-02	0.1726
8.1623e-02	0.1826
8.1932e-02	0.1930
8.2195e-02	0.2030
8.2458e-02	0.2133
8.2675e-02	0.2235
8.2860e-02	0.2336
8.3077e-02	0.2443
8.3232e-02	0.2545
8.3402e-02	0.2649
8.3603e-02	0.2755
8.3757e-02	0.2861
8.3928e-02	0.2978
8.4082e-02	0.3084
8.4237e-02	0.3193
8.4392e-02	0.3303
8.4515e-02	0.3407
8.4608e-02	0.3512
8.4732e-02	0.3631
8.4840e-02	0.3734
8.4964e-02	0.3840
8.5088e-02	0.3951
8.5196e-02	0.4062
8.5304e-02	0.4175
8.5397e-02	0.4278
8.5490e-02	0.4383
8.5598e-02	0.4485
8.5722e-02	0.4592
8.5846e-02	0.4692
8.5954e-02	0.4809
8.6078e-02	0.4917
8.6186e-02	0.5027
8.6310e-02	0.5140
8.6418e-02	0.5240
8.6542e-02	0.5350
8.6650e-02	0.5454
8.6774e-02	0.5562
8.6913e-02	0.5666
8.7037e-02	0.5769
8.7160e-02	0.5876
8.7284e-02	0.5980
8.7423e-02	0.6085
8.7547e-02	0.6199
8.7671e-02	0.6312
8.7794e-02	0.6414
8.7934e-02	0.6522
8.8057e-02	0.6630
8.8212e-02	0.6744
8.8351e-02	0.6858
8.8491e-02	0.6967
8.8661e-02	0.7072
8.8815e-02	0.7174
8.8955e-02	0.7275
8.9109e-02	0.7377
8.9264e-02	0.7482
8.9434e-02	0.7589
8.9620e-02	0.7696
8.9821e-02	0.7799
9.0037e-02	0.7902
9.0223e-02	0.8008
9.0409e-02	0.8121
9.0641e-02	0.8222
9.0842e-02	0.8324
9.1043e-02	0.8425
9.1290e-02	0.8532
9.1553e-02	0.8641
9.1816e-02	0.8744
9.2110e-02	0.8847
9.2435e-02	0.8950
9.2791e-02	0.9050
9.3224e-02	0.9151
9.3548e-02	0.9251
9.4012e-02	0.9351
9.4523e-02	0.9451
9.5080e-02	0.9555
9.5791e-02	0.9655
9.6828e-02	0.9758
9.8312e-02	0.9858
1.0148e-01	0.9959
1.0994e-01	1.0000
1.0000e+00	1.0000
};
\addplot [line width=1pt, curve_color, dotted]
table {%
0.0000e+00	0.0000
1.4673e-01	0.0101
1.5372e-01	0.0202
1.5861e-01	0.0302
1.6227e-01	0.0403
1.6527e-01	0.0504
1.6776e-01	0.0605
1.6974e-01	0.0706
1.7178e-01	0.0806
1.7364e-01	0.0907
1.7514e-01	0.1009
1.7664e-01	0.1113
1.7791e-01	0.1216
1.7921e-01	0.1317
1.8020e-01	0.1421
1.8105e-01	0.1521
1.8209e-01	0.1623
1.8303e-01	0.1725
1.8383e-01	0.1826
1.8475e-01	0.1927
1.8569e-01	0.2028
1.8645e-01	0.2129
1.8717e-01	0.2230
1.8793e-01	0.2332
1.8872e-01	0.2433
1.8957e-01	0.2535
1.9024e-01	0.2635
1.9082e-01	0.2736
1.9151e-01	0.2839
1.9223e-01	0.2942
1.9282e-01	0.3044
1.9349e-01	0.3144
1.9415e-01	0.3246
1.9468e-01	0.3347
1.9528e-01	0.3448
1.9581e-01	0.3548
1.9632e-01	0.3649
1.9690e-01	0.3753
1.9737e-01	0.3854
1.9788e-01	0.3958
1.9839e-01	0.4058
1.9895e-01	0.4160
1.9944e-01	0.4263
1.9989e-01	0.4363
2.0046e-01	0.4466
2.0097e-01	0.4567
2.0151e-01	0.4669
2.0201e-01	0.4772
2.0250e-01	0.4873
2.0300e-01	0.4976
2.0334e-01	0.5076
2.0376e-01	0.5178
2.0424e-01	0.5280
2.0470e-01	0.5383
2.0509e-01	0.5487
2.0547e-01	0.5589
2.0594e-01	0.5690
2.0639e-01	0.5790
2.0690e-01	0.5893
2.0727e-01	0.5996
2.0765e-01	0.6097
2.0801e-01	0.6199
2.0835e-01	0.6303
2.0878e-01	0.6406
2.0911e-01	0.6507
2.0948e-01	0.6609
2.0990e-01	0.6713
2.1028e-01	0.6814
2.1065e-01	0.6915
2.1103e-01	0.7015
2.1144e-01	0.7118
2.1188e-01	0.7223
2.1225e-01	0.7326
2.1270e-01	0.7429
2.1314e-01	0.7530
2.1353e-01	0.7633
2.1396e-01	0.7734
2.1444e-01	0.7835
2.1494e-01	0.7936
2.1542e-01	0.8037
2.1591e-01	0.8137
2.1647e-01	0.8237
2.1707e-01	0.8337
2.1772e-01	0.8437
2.1851e-01	0.8537
2.1924e-01	0.8641
2.1990e-01	0.8743
2.2071e-01	0.8844
2.2153e-01	0.8944
2.2233e-01	0.9044
2.2323e-01	0.9144
2.2444e-01	0.9245
2.2563e-01	0.9347
2.2691e-01	0.9447
2.2835e-01	0.9547
2.3024e-01	0.9647
2.3200e-01	0.9749
2.3554e-01	0.9851
2.4286e-01	0.9952
2.6355e-01	1.0000
1.0000e+00	1.0000
};
\addplot [line width=1pt, curve_color, dash pattern=on 1pt off 3pt on 3pt off 3pt]
table {%
0.0000e+00	0.0000
1.8725e-01	0.0101
1.9678e-01	0.0202
2.0307e-01	0.0303
2.0719e-01	0.0403
2.1053e-01	0.0504
2.1319e-01	0.0604
2.1559e-01	0.0704
2.1876e-01	0.0805
2.2097e-01	0.0906
2.2286e-01	0.1008
2.2475e-01	0.1109
2.2668e-01	0.1211
2.2836e-01	0.1312
2.2962e-01	0.1414
2.3075e-01	0.1515
2.3184e-01	0.1615
2.3319e-01	0.1715
2.3441e-01	0.1817
2.3571e-01	0.1917
2.3666e-01	0.2018
2.3783e-01	0.2120
2.3890e-01	0.2222
2.3997e-01	0.2324
2.4100e-01	0.2424
2.4190e-01	0.2525
2.4286e-01	0.2625
2.4377e-01	0.2726
2.4467e-01	0.2826
2.4552e-01	0.2927
2.4637e-01	0.3028
2.4730e-01	0.3129
2.4810e-01	0.3229
2.4880e-01	0.3332
2.4968e-01	0.3433
2.5047e-01	0.3534
2.5099e-01	0.3635
2.5186e-01	0.3735
2.5257e-01	0.3840
2.5334e-01	0.3940
2.5409e-01	0.4040
2.5481e-01	0.4141
2.5554e-01	0.4243
2.5614e-01	0.4343
2.5686e-01	0.4444
2.5760e-01	0.4544
2.5845e-01	0.4646
2.5902e-01	0.4746
2.5958e-01	0.4849
2.6015e-01	0.4950
2.6077e-01	0.5050
2.6133e-01	0.5151
2.6193e-01	0.5252
2.6253e-01	0.5352
2.6309e-01	0.5453
2.6363e-01	0.5557
2.6447e-01	0.5658
2.6498e-01	0.5758
2.6561e-01	0.5858
2.6621e-01	0.5958
2.6686e-01	0.6060
2.6745e-01	0.6160
2.6815e-01	0.6262
2.6869e-01	0.6363
2.6926e-01	0.6465
2.6979e-01	0.6565
2.7036e-01	0.6667
2.7101e-01	0.6769
2.7153e-01	0.6871
2.7217e-01	0.6971
2.7274e-01	0.7072
2.7339e-01	0.7177
2.7396e-01	0.7282
2.7457e-01	0.7383
2.7511e-01	0.7484
2.7588e-01	0.7584
2.7658e-01	0.7686
2.7732e-01	0.7789
2.7816e-01	0.7889
2.7893e-01	0.7990
2.7976e-01	0.8090
2.8057e-01	0.8191
2.8133e-01	0.8295
2.8208e-01	0.8397
2.8287e-01	0.8497
2.8382e-01	0.8598
2.8485e-01	0.8698
2.8600e-01	0.8799
2.8702e-01	0.8899
2.8818e-01	0.9000
2.8945e-01	0.9100
2.9048e-01	0.9200
2.9218e-01	0.9300
2.9372e-01	0.9401
2.9545e-01	0.9503
2.9747e-01	0.9603
3.0012e-01	0.9703
3.0343e-01	0.9806
3.0863e-01	0.9906
3.3000e-01	1.0000
1.0000e+00	1.0000
};
\addplot [line width=1pt, curve_color, dashed]
table {%
0.0000e+00	0.0000
2.2731e-01	0.0101
2.3785e-01	0.0202
2.4451e-01	0.0302
2.5132e-01	0.0405
2.5594e-01	0.0505
2.5969e-01	0.0606
2.6304e-01	0.0707
2.6609e-01	0.0808
2.6906e-01	0.0908
2.7152e-01	0.1009
2.7382e-01	0.1110
2.7628e-01	0.1212
2.7798e-01	0.1312
2.7998e-01	0.1413
2.8233e-01	0.1514
2.8382e-01	0.1615
2.8535e-01	0.1715
2.8686e-01	0.1816
2.8838e-01	0.1917
2.8996e-01	0.2019
2.9130e-01	0.2119
2.9279e-01	0.2221
2.9399e-01	0.2322
2.9498e-01	0.2422
2.9642e-01	0.2525
2.9766e-01	0.2625
2.9871e-01	0.2725
3.0012e-01	0.2826
3.0103e-01	0.2926
3.0225e-01	0.3027
3.0329e-01	0.3128
3.0425e-01	0.3229
3.0542e-01	0.3330
3.0646e-01	0.3430
3.0727e-01	0.3532
3.0822e-01	0.3633
3.0914e-01	0.3733
3.0994e-01	0.3834
3.1093e-01	0.3935
3.1180e-01	0.4035
3.1257e-01	0.4138
3.1339e-01	0.4239
3.1421e-01	0.4339
3.1505e-01	0.4440
3.1585e-01	0.4540
3.1668e-01	0.4642
3.1744e-01	0.4743
3.1832e-01	0.4844
3.1919e-01	0.4945
3.2010e-01	0.5046
3.2105e-01	0.5148
3.2193e-01	0.5248
3.2278e-01	0.5349
3.2355e-01	0.5449
3.2434e-01	0.5550
3.2507e-01	0.5651
3.2598e-01	0.5753
3.2677e-01	0.5853
3.2753e-01	0.5954
3.2844e-01	0.6054
3.2924e-01	0.6156
3.3016e-01	0.6257
3.3101e-01	0.6359
3.3174e-01	0.6459
3.3257e-01	0.6561
3.3333e-01	0.6662
3.3426e-01	0.6762
3.3511e-01	0.6862
3.3586e-01	0.6964
3.3665e-01	0.7064
3.3749e-01	0.7168
3.3834e-01	0.7270
3.3953e-01	0.7371
3.4055e-01	0.7471
3.4140e-01	0.7571
3.4227e-01	0.7673
3.4304e-01	0.7774
3.4395e-01	0.7874
3.4498e-01	0.7974
3.4583e-01	0.8074
3.4686e-01	0.8175
3.4773e-01	0.8276
3.4883e-01	0.8376
3.5013e-01	0.8478
3.5127e-01	0.8578
3.5255e-01	0.8678
3.5392e-01	0.8778
3.5514e-01	0.8879
3.5641e-01	0.8980
3.5784e-01	0.9080
3.5934e-01	0.9181
3.6132e-01	0.9281
3.6332e-01	0.9381
3.6581e-01	0.9482
3.6796e-01	0.9582
3.7112e-01	0.9683
3.7521e-01	0.9784
3.8131e-01	0.9885
3.9305e-01	0.9985
4.1065e-01	1.0000
1.0000e+00	1.0000
};
\addplot [line width=1.8pt, curve_color, densely dotted]
table {%
0.0000e+00  0.0000
1.8318e-04  0.0416
3.6637e-04  0.0615
5.4955e-04  0.0799
7.3273e-04  0.1001
9.1592e-04  0.1201
1.0991e-03  0.1369
1.2823e-03  0.1548
1.4655e-03  0.1742
1.6487e-03  0.1901
1.8318e-03  0.2060
2.0150e-03  0.2242
2.1982e-03  0.2382
2.3814e-03  0.2537
2.5646e-03  0.2692
2.7478e-03  0.2818
2.9309e-03  0.2963
3.1141e-03  0.3123
3.2973e-03  0.3272
3.4805e-03  0.3406
3.6637e-03  0.3569
3.8469e-03  0.3682
4.0300e-03  0.3822
4.2132e-03  0.3940
4.3964e-03  0.4053
4.5796e-03  0.4165
4.7628e-03  0.4303
4.9460e-03  0.4432
5.1291e-03  0.4550
5.4955e-03  0.4759
5.6787e-03  0.4862
6.0451e-03  0.5065
6.4114e-03  0.5252
6.7778e-03  0.5442
7.1442e-03  0.5627
7.5105e-03  0.5777
7.8769e-03  0.5951
8.2433e-03  0.6127
8.6096e-03  0.6303
8.9760e-03  0.6449
9.3424e-03  0.6591
9.7087e-03  0.6738
1.0075e-02  0.6844
1.0441e-02  0.6977
1.0808e-02  0.7111
1.1174e-02  0.7240
1.1541e-02  0.7358
1.1907e-02  0.7464
1.2456e-02  0.7611
1.3006e-02  0.7748
1.3556e-02  0.7886
1.4105e-02  0.8004
1.4655e-02  0.8123
1.5387e-02  0.8267
1.5937e-02  0.8375
1.6670e-02  0.8513
1.7402e-02  0.8614
1.8135e-02  0.8725
1.8868e-02  0.8828
1.9601e-02  0.8931
2.0517e-02  0.9043
2.1432e-02  0.9146
2.2715e-02  0.9247
2.3997e-02  0.9347
2.5646e-02  0.9449
2.7661e-02  0.9557
2.9859e-02  0.9660
3.3156e-02  0.9762
3.7186e-02  0.9863
5.0742e-02  0.9964
9.1958e-02  1.0000
1.0000e+00  1.0000
};
\addplot [line width=1.8pt, curve_color, dashed]
table {%
0.0000e+00  0.0000
1.4777e-08  0.0396
2.9554e-08  0.0595
4.4331e-08  0.0758
5.9108e-08  0.0940
7.3886e-08  0.1122
8.8663e-08  0.1300
1.0344e-07  0.1483
1.1822e-07  0.1643
1.3299e-07  0.1812
1.4777e-07  0.1966
1.6255e-07  0.2102
1.7733e-07  0.2272
1.9210e-07  0.2420
2.0688e-07  0.2576
2.2166e-07  0.2712
2.3643e-07  0.2862
2.5121e-07  0.3003
2.6599e-07  0.3141
2.8077e-07  0.3267
2.9554e-07  0.3391
3.1032e-07  0.3528
3.2510e-07  0.3660
3.3987e-07  0.3789
3.5465e-07  0.3905
3.6943e-07  0.4039
3.8421e-07  0.4149
4.1376e-07  0.4336
4.2854e-07  0.4447
4.4331e-07  0.4559
4.5809e-07  0.4678
4.8764e-07  0.4899
5.1720e-07  0.5098
5.4675e-07  0.5285
5.7631e-07  0.5483
5.9108e-07  0.5586
6.2064e-07  0.5758
6.5019e-07  0.5919
6.7975e-07  0.6104
7.0930e-07  0.6263
7.3886e-07  0.6437
7.6841e-07  0.6587
7.9796e-07  0.6728
8.2752e-07  0.6884
8.5707e-07  0.6995
8.8663e-07  0.7109
9.1618e-07  0.7210
9.4574e-07  0.7319
9.9007e-07  0.7459
1.0196e-06  0.7561
1.0640e-06  0.7707
1.1083e-06  0.7817
1.1526e-06  0.7958
1.1969e-06  0.8075
1.2413e-06  0.8210
1.2856e-06  0.8324
1.3447e-06  0.8458
1.3890e-06  0.8559
1.4334e-06  0.8659
1.4925e-06  0.8769
1.5516e-06  0.8869
1.6403e-06  0.8993
1.7289e-06  0.9110
1.8324e-06  0.9219
1.9358e-06  0.9328
2.0540e-06  0.9435
2.1722e-06  0.9537
2.3643e-06  0.9645
2.6008e-06  0.9749
2.9850e-06  0.9851
4.0046e-06  0.9951
7.1373e-06  1.0000
1.0000e+00  1.0000
};
\end{axis}
\end{tikzpicture}

\begin{tikzpicture}
\pgfplotsset{every axis/.append style={
                    ylabel=\mbox{\scriptsize $\frac{1}{\nmbrAccounts}\min\cset{\normalsize}{\nmbrAccountsAttempted}{\trueAlarmProb(10,31,\nmbrAccountsAttempted) \ge 0.5}$},
                    xlabel=\mbox{$\fdpBound$},
                    compat=1.8,
                    label style={font=\normalsize},
                    tick label style={font=\small},
                    xticklabel style = {font=\small,yshift=-0.5ex}    
                    }}

\begin{axis}[
ylabel style={align=center}, 
xmin=1e-5, xmax=1e-1,
ymin=0, ymax=1.03,
xmode=log,
log basis x={10},
xtick={1e-5, 1e-4, 1e-3, 1e-2, 1e-1},
xticklabels={$10^{-5}$,,$10^{-3}$,,$10^{-1}$},
yticklabels={},
ytick={0, 0.2, 0.4, 0.6, 0.8, 1.0},
yticklabels={0, 0.2, 0.4, 0.6, 0.8, 1.0},
width=0.85\figurewidth,
height=1.1\figurewidth,
tick align=outside,
tick pos=left,
xmajorgrids,
minor x tick num=1,
x grid style={lightgray!92.026143790849673!black},
minor y tick num=1,
ymajorgrids,
yminorgrids,
y grid style={lightgray!92.026143790849673!black},
]
\addplot [line width=1pt, curve_color]
table {%
1.0000e-05  9.8118e-01
1.0000e-04  5.0145e-01
1.0000e-03  1.6043e-01
1.0000e-02  1.0478e-01
1.0000e-01  8.6186e-02
};
\addplot [line width=1pt, curve_color, dotted]
table {%
1.0000e-05  9.8900e-01
1.0000e-04  6.3376e-01
1.0000e-03  3.4796e-01
1.0000e-02  2.5472e-01
1.0000e-01  2.0334e-01
};
\addplot [line width=1pt, curve_color, dash pattern=on 1pt off 3pt on 3pt off 3pt]
table {%
1.0000e-05  9.9174e-01
1.0000e-04  6.8975e-01
1.0000e-03  4.3756e-01
1.0000e-02  3.2878e-01
1.0000e-01  2.6077e-01
};
\addplot [line width=1pt, curve_color, dashed]
table {%
1.0000e-05  9.9439e-01
1.0000e-04  7.5323e-01
1.0000e-03  5.3979e-01
1.0000e-02  4.1088e-01
1.0000e-01  3.2010e-01
};
\addplot [line width=1.8pt, curve_color, densely dotted]
table {%
1.0000e-05  2.9071e-01
1.0000e-04  1.3208e-01
1.0000e-03  2.5279e-02
1.0000e-02  6.0451e-03
1.0000e-01  6.0451e-03 
};
\addplot [line width=1.8pt, curve_color, dashed]
table {%
1.0000e-05  2.5911e-02
1.0000e-04  6.7975e-07
1.0000e-03  6.0586e-07
1.0000e-02  5.4675e-07
1.0000e-01  5.1720e-07
};
\end{axis}
\end{tikzpicture}

%% file: figures/fig_roc_ba_heatmap.tex
\begin{table}
  \hspace*{0.8em}
  \begin{subfigure}[b]{0.425\columnwidth}
  \setcounter{MinNumber}{850}%
  \setcounter{MaxNumber}{1000}%
  \centering
      {\footnotesize
        \setlength{\tabcolsep}{.16667em}
        \begin{tabular}{cr|@{\hspace{.16667em}}|@{\hspace{1.5pt}}*{5}{X}@{\hspace{1.5pt}}}
          & & \multicolumn{5}{c}{$\nmbrLoginsTotal-\nmbrLoginsSeen$} \\
          & ~\nmbrLoginsSeen~
          & \multicolumn{1}{c}{1}
          & \multicolumn{1}{c}{6}
          & \multicolumn{1}{c}{11}
          & \multicolumn{1}{c}{16}
          & \multicolumn{1}{c}{21}\\
          \cline{2-7} \\[-2.35ex]
          &  0 & .998 & .995 & .990 & .981 & .967  \\
          &  5 & .997 & .995 & .989 & .980 & .967  \\
          & 10 & .997 & .994 & .989 & .980 & .966  \\
          & 15 & .997 & .994 & .989 & .979 & .965  \\
          & 20 & .997 & .994 & .988 & .977 & .965
        \end{tabular}
      }
      \hspace*{-0.35em}\begin{minipage}[t]{17em}
      \caption{${\nmbrAccountsAttempted/5{,}459} \approx 0.03$}
      \label{tbl:login:3pct}
      \end{minipage}
  \end{subfigure}
  \hspace{1.5em}
  \begin{subfigure}[b]{0.425\columnwidth}
  \setcounter{MinNumber}{850}%
  \setcounter{MaxNumber}{1000}%
  \centering
      {\footnotesize
        \setlength{\tabcolsep}{.16667em}
        \begin{tabular}{cr|@{\hspace{.16667em}}|@{\hspace{1.5pt}}*{5}{X}@{\hspace{1.5pt}}}
          & & \multicolumn{5}{c}{$\nmbrLoginsTotal-\nmbrLoginsSeen$} \\
          & ~\nmbrLoginsSeen~
          & \multicolumn{1}{c}{1}
          & \multicolumn{1}{c}{6}
          & \multicolumn{1}{c}{11}
          & \multicolumn{1}{c}{16}
          & \multicolumn{1}{c}{21}\\
          \cline{2-7} \\[-2.35ex]
          &  0 & 1.00 & 1.00 & 1.00 & .999 & .997  \\
          &  5 & 1.00 & 1.00 & 1.00 & .998 & .997  \\
          & 10 & 1.00 & 1.00 & .999 & .998 & .997  \\
          & 15 & 1.00 & 1.00 & .999 & .998 & .997  \\
          & 20 & 1.00 & 1.00 & .999 & .997 & .996
        \end{tabular}
      }
      \hspace*{-0.35em}\begin{minipage}[t]{17em}
      \caption{${\nmbrAccountsAttempted/5{,}459} \approx 0.05$}
      \label{tbl:login:5pct}
      \end{minipage}
  \end{subfigure}
  \caption{$\trueAlarmProb(\nmbrLoginsSeen, \nmbrLoginsTotal,
\nmbrAccountsAttempted)$ with varying $\nmbrLoginsSeen$ and $\nmbrLoginsTotal-\nmbrLoginsSeen$
based on estimate \eqnref{eqn:mazurekCracked} for \prob{\rvPwdRankChosen
\le \pwdRank}}
  \label{tbl:logins}
\end{table}

%% file: figures/fig_pcr_cmp2.tex
\begin{figure}[ht]
  \hspace*{6em}
  \begin{subfigure}[b]{.1\columnwidth}
    \setlength\figureheight{1.8in}
    \resizebox{!}{1.15em}{\input{figures/pcr/cmp2_legend.tex}}
  \end{subfigure}\vspace*{0.8em}

  \captionsetup[subfigure]{font=small,labelfont=small, oneside,margin={-0.5in,0in}}
  \begin{subfigure}[b]{.9\columnwidth}
  \setlength\figureheight{1.5in}
    \centering
    \hspace*{-0.4em}
    \resizebox{!}{8.35em}{\input{figures/pcr/cmp2_fig_1.tex}}
     \hspace*{-1.95em}\begin{minipage}[t]{17em}
      \vspace*{-1.5em}\caption{\parbox[t]{6.5em}{Request \\generation by \targetSite\\
      \hspace{1em}($\relstddev < 0.10$)}}
      \label{fig:cmp2:requestGenTime}
    \end{minipage}%
    \hspace*{-8.25em}\begin{minipage}[t]{17em}
      \vspace*{-1.5em}\caption{\parbox[t]{6.5em}{Request\\ validation by \monitorSite\\
      \hspace{1em}($\relstddev < 0.10$)}}
      \label{fig:cmp2:requestValidateTime}
    \end{minipage}%
    \hspace*{-6.95em}\begin{minipage}[t]{17em}
      \vspace*{-1.5em}\caption{\parbox[t]{6.5em}{Size of \\request \\ 
      \hspace{1em}($\relstddev < 0.01$)}}
      \label{fig:cmp2:requestSize}
    \end{minipage}%
  \end{subfigure}

  \begin{subfigure}[b]{.9\columnwidth}
  \setlength\figureheight{1.5in}
    \centering
    \hspace*{-0.4em}
    \resizebox{!}{8.35em}{\input{figures/pcr/cmp2_fig_2.tex}}
     \hspace*{-1.95em}\begin{minipage}[t]{17em}
      \vspace*{-1.5em}\caption{\parbox[t]{6.5em}{Response \\generation by \monitorSite\\
      \hspace{1em}($\relstddev < 0.12$)}}
      \label{fig:cmp2:responseTime}
    \end{minipage}%
    \hspace*{-8.25em}\begin{minipage}[t]{17em}
      \vspace*{-1.5em}\caption{\parbox[t]{6.5em}{Response \\processing by \targetSite\\
      \hspace{1em}($\relstddev < 0.60$)}}
      \label{fig:cmp2:retrievalTime}
    \end{minipage}%
    \hspace*{-6.95em}\begin{minipage}[t]{17em}
      \vspace*{-1.5em}\caption{\parbox[t]{6.5em}{Size of \\response \\
      \hspace{1em}($\relstddev < 0.01$)}}
      \label{fig:cmp2:responseSize}
    \end{minipage}%
  \end{subfigure}

  \caption{Costs for our protocol
    (\secref{sec:amnesia:remote:protocol}) and
    \pcr{}~\cite{wang2021:amnesia}.  One-time costs for \targetSite to
    deploy a monitoring request to \monitorSite are shown in
    (\subref{fig:cmp2:requestGenTime})--(\subref{fig:cmp2:requestSize}),
    and costs for \monitorSite to return a response to \targetSite,
    once per incorrect login attempt, are shown in
    (\subref{fig:cmp2:responseTime})--(\subref{fig:cmp2:responseSize}).
    In (\subref{fig:cmp2:retrievalTime}), ``negative'' and
    ``positive'' refer to the cases in our protocol when
    $\pwdHashFn(\loginPassword) \bfNotIn \langle \bfHashFnSet,
    \bfIndices \rangle$ and $\pwdHashFn(\loginPassword) \bfIn \langle
    \bfHashFnSet, \bfIndices \rangle$, respectively; \pcr{} has
    analogous cases.  Numbers shown are for $\bfHashFnRange = 128$ and
    $\bfNmbrHashFns = 20$ for our protocol.  \nmbrExplicitHoneywords
    is the number of explicit honeywords with which \pcr{}'s
    monitoring request was configured.}
  \label{fig:comparison2}
\end{figure}

%% file: figures/pcr/cmp2_legend.tex
\newenvironment{customlegend}[1][]{%
    \begingroup
    \csname pgfplots@init@cleared@structures\endcsname
    \pgfplotsset{#1}%
}{%
    \csname pgfplots@createlegend\endcsname
    \endgroup
}%

\def\addlegendimage{\csname pgfplots@addlegendimage\endcsname}

\begin{tikzpicture}

\begin{customlegend}[
    legend style={{font={\fontsize{12pt}{12}\selectfont}},{draw=none}},
    legend columns=3,
    legend cell align={left},
    legend entries={{~Our protocol$\quad$}, {~\pcr{}~\cite{wang2021:amnesia}}}]
\addlegendimage{ybar,ybar legend,draw=curve_color,fill=white,line width=1.5pt,postaction={pattern=crosshatch, pattern color=curve_color}, legend image
code/.code={\draw (0cm,-0.1cm) rectangle (0.5cm,0.15cm);}}

\addlegendimage{ybar,ybar legend,draw=curve_color,fill=white,line width=1.5pt,postaction={pattern=crosshatch dots, pattern color=curve_color}, legend image
code/.code={\draw (0cm,-0.1cm) rectangle (0.5cm,0.15cm);}}

\end{customlegend}

\end{tikzpicture}

%% file: figures/pcr/cmp2_fig_1.tex
\newcommand{\offSet}{0.4}
\newcommand{\textWidth}{3cm}

\newcommand{\startX}{-0.3}
\newcommand{\barWidth}{0.2}
\newcommand{\gapWidth}{0.2}

\definecolor{darkgray176}{RGB}{176,176,176}

\hspace*{-2.6em}
\begin{tikzpicture}
\pgfplotsset{every axis/.append style={
                    xlabel={},
                    compat=1.12,
                    label style={font=\large},
                    tick label style={font=\large}  
                    }}

\begin{axis}[
width = 0.65\columnwidth,
height = 0.65\columnwidth,
tick align=outside,
tick pos=left,
x grid style={darkgray176},
xmin=-0.49, xmax=1.29,
xtick style={color=black},
xtick={-0.49},
xticklabels={},
y grid style={darkgray176},
ylabel=\mbox{Time (\millisecs)},
y label style={at={(axis description cs:-0.26,.5)}},
ymin=1, ymax=1000,
ytick={1, 10, 100, 1000},
ytick style={color=black},
ymode = log,
ymajorgrids,
yminorgrids,
minor y tick num=1,
y grid style={lightgray!92.026143790849673!black},
]

\draw[draw=curve_color,fill=white,line width=1.5pt,postaction={pattern=crosshatch,
pattern color=curve_color}] (axis cs:\startX,0.0001) rectangle (axis cs:\startX+\barWidth,68.21);

\draw[draw=curve_color,fill=white,line width=1.5pt,postaction={pattern=crosshatch
dots, pattern color=curve_color}] (axis cs:\startX+\barWidth+\gapWidth,0.0001)
rectangle (axis
cs:\startX+2*\barWidth+\gapWidth,2.75);
\draw[draw=curve_color,fill=white,line width=1.5pt,postaction={pattern=crosshatch
dots, pattern color=curve_color}] (axis cs:\startX+2*\barWidth+\gapWidth,0.0001)
rectangle (axis cs:\startX+3*\barWidth+\gapWidth,9.48);
\draw[draw=curve_color,fill=white,line width=1.5pt,postaction={pattern=crosshatch dots,
pattern color=curve_color}] (axis cs:\startX+3*\barWidth+\gapWidth,0.0001)
rectangle (axis cs:\startX+4*\barWidth+\gapWidth,35.41);
\draw[draw=curve_color,fill=white,line width=1.5pt,postaction={pattern=crosshatch dots, pattern color=curve_color}] (axis cs:\startX+4*\barWidth+\gapWidth,0.0001)
rectangle (axis cs:\startX+5*\barWidth+\gapWidth,140.29);
\draw[draw=curve_color,fill=white,line width=1.5pt,postaction={pattern=crosshatch dots, pattern color=curve_color}] (axis cs:\startX+5*\barWidth+\gapWidth,0.0001)
rectangle (axis cs:\startX+6*\barWidth+\gapWidth,558.67);

\end{axis}

\node[text width=\textWidth, 
      scale=1.2,
      anchor=south, 
      right,
      text=black,
      rotate=45
] at (\startX+2*\barWidth+\gapWidth+\offSet,-1.075){\small$\boldsymbol{\nmbrExplicitHoneywords = 2^4}$};
\node[text width=\textWidth, 
      scale=1.2,
      anchor=south, 
      right,
      text=black,
      rotate=45
] at (\startX+6*\barWidth+\gapWidth+\offSet,-1.075){\small$\boldsymbol{\nmbrExplicitHoneywords
= 2^8}$};
\node[text width=\textWidth, 
      scale=1.2,
      anchor=south, 
      right,
      text=black,
      rotate=45
] at (\startX+9.7*\barWidth+\gapWidth+\offSet,-1.175){\small$\boldsymbol{\nmbrExplicitHoneywords
= 2^{12}}$};

\end{tikzpicture}

\hspace*{-4.5em}
\begin{tikzpicture}
\pgfplotsset{every axis/.append style={
                    xlabel={},
                    compat=1.12,
                    label style={font=\large},
                    tick label style={font=\large}  
                    }}

\begin{axis}[
width = 0.65\columnwidth,
height = 0.65\columnwidth,
tick align=outside,
tick pos=left,
x grid style={darkgray176},
xmin=-0.49, xmax=1.29,
xtick style={color=black},
xtick={-0.49},
xticklabels={},
y grid style={darkgray176},
ylabel=\mbox{Time (\millisecs)},
y label style={at={(axis description cs:-0.26,.5)}},
ymin=1, ymax=1000,
ytick={1, 10, 100, 1000},
ytick style={color=black},
ymode = log,
ymajorgrids,
yminorgrids,
minor y tick num=1,
y grid style={lightgray!92.026143790849673!black},
]

\draw[draw=curve_color,fill=white,line width=1.5pt,postaction={pattern=crosshatch,
pattern color=curve_color}] (axis cs:\startX,0.0001) rectangle (axis cs:\startX+\barWidth,123.08);

\draw[draw=curve_color,fill=white,line width=1.5pt,postaction={pattern=crosshatch
dots, pattern color=curve_color}] (axis cs:\startX+\barWidth+\gapWidth,0.0001)
rectangle (axis
cs:\startX+2*\barWidth+\gapWidth,2.81);
\draw[draw=curve_color,fill=white,line width=1.5pt,postaction={pattern=crosshatch
dots, pattern color=curve_color}] (axis cs:\startX+2*\barWidth+\gapWidth,0.0001)
rectangle (axis cs:\startX+3*\barWidth+\gapWidth,9.59);
\draw[draw=curve_color,fill=white,line width=1.5pt,postaction={pattern=crosshatch dots,
pattern color=curve_color}] (axis cs:\startX+3*\barWidth+\gapWidth,0.0001)
rectangle (axis cs:\startX+4*\barWidth+\gapWidth,36.27);
\draw[draw=curve_color,fill=white,line width=1.5pt,postaction={pattern=crosshatch dots, pattern color=curve_color}] (axis cs:\startX+4*\barWidth+\gapWidth,0.0001)
rectangle (axis cs:\startX+5*\barWidth+\gapWidth,142.43);
\draw[draw=curve_color,fill=white,line width=1.5pt,postaction={pattern=crosshatch dots, pattern color=curve_color}] (axis cs:\startX+5*\barWidth+\gapWidth,0.0001)
rectangle (axis cs:\startX+6*\barWidth+\gapWidth,568.38);

\end{axis}

\node[text width=\textWidth, 
      scale=1.2,
      anchor=south, 
      right,
      text=black,
      rotate=45
] at (\startX+2*\barWidth+\gapWidth+\offSet,-1.075){\small$\boldsymbol{\nmbrExplicitHoneywords = 2^4}$};
\node[text width=\textWidth, 
      scale=1.2,
      anchor=south, 
      right,
      text=black,
      rotate=45
] at (\startX+6*\barWidth+\gapWidth+\offSet,-1.075){\small$\boldsymbol{\nmbrExplicitHoneywords
= 2^8}$};
\node[text width=\textWidth, 
      scale=1.2,
      anchor=south, 
      right,
      text=black,
      rotate=45
] at (\startX+9.7*\barWidth+\gapWidth+\offSet,-1.175){\small$\boldsymbol{\nmbrExplicitHoneywords
= 2^{12}}$};

\end{tikzpicture}

\hspace*{-4.1em}
\begin{tikzpicture}
\pgfplotsset{every axis/.append style={
                    xlabel={},
                    compat=1.12,
                    label style={font=\large},
                    tick label style={font=\large}  
                    }}

\begin{axis}[
width = 0.65\columnwidth,
height = 0.65\columnwidth,
tick align=outside,
tick pos=left,
x grid style={darkgray176},
xmin=-0.49, xmax=1.29,
xtick style={color=black},
xtick={-0.49},
xticklabels={},
y grid style={darkgray176},
ylabel=\mbox{Message size (\kilobytes)},
y label style={at={(axis description cs:-0.26,.5)}},
ymin=1, ymax=1000,
ytick={1, 10, 100, 1000},
ytick style={color=black},
ymode = log,
ymajorgrids,
yminorgrids,
minor y tick num=1,
y grid style={lightgray!92.026143790849673!black},
]

\draw[draw=curve_color,fill=white,line width=1.5pt,postaction={pattern=crosshatch,
pattern color=curve_color}] (axis cs:\startX,0.0001) rectangle (axis cs:\startX+\barWidth,44.87);

\draw[draw=curve_color,fill=white,line width=1.5pt,postaction={pattern=crosshatch
dots, pattern color=curve_color}] (axis cs:\startX+\barWidth+\gapWidth,0.0001)
rectangle (axis
cs:\startX+2*\barWidth+\gapWidth,1.92);
\draw[draw=curve_color,fill=white,line width=1.5pt,postaction={pattern=crosshatch
dots, pattern color=curve_color}] (axis cs:\startX+2*\barWidth+\gapWidth,0.0001)
rectangle (axis cs:\startX+3*\barWidth+\gapWidth,5.80);
\draw[draw=curve_color,fill=white,line width=1.5pt,postaction={pattern=crosshatch dots,
pattern color=curve_color}] (axis cs:\startX+3*\barWidth+\gapWidth,0.0001)
rectangle (axis cs:\startX+4*\barWidth+\gapWidth,20.42);
\draw[draw=curve_color,fill=white,line width=1.5pt,postaction={pattern=crosshatch dots, pattern color=curve_color}] (axis cs:\startX+4*\barWidth+\gapWidth,0.0001)
rectangle (axis cs:\startX+5*\barWidth+\gapWidth,78.97);
\draw[draw=curve_color,fill=white,line width=1.5pt,postaction={pattern=crosshatch dots, pattern color=curve_color}] (axis cs:\startX+5*\barWidth+\gapWidth,0.0001)
rectangle (axis cs:\startX+6*\barWidth+\gapWidth,313.86);

\end{axis}

\node[text width=\textWidth, 
      scale=1.2,
      anchor=south, 
      right,
      text=black,
      rotate=45
] at (\startX+2*\barWidth+\gapWidth+\offSet,-1.075){\small$\boldsymbol{\nmbrExplicitHoneywords = 2^4}$};
\node[text width=\textWidth, 
      scale=1.2,
      anchor=south, 
      right,
      text=black,
      rotate=45
] at (\startX+6*\barWidth+\gapWidth+\offSet,-1.075){\small$\boldsymbol{\nmbrExplicitHoneywords
= 2^8}$};
\node[text width=\textWidth, 
      scale=1.2,
      anchor=south, 
      right,
      text=black,
      rotate=45
] at (\startX+9.7*\barWidth+\gapWidth+\offSet,-1.175){\small$\boldsymbol{\nmbrExplicitHoneywords
= 2^{12}}$};

\end{tikzpicture}

%% file: figures/pcr/cmp2_fig_2.tex
\newcommand{\offSet}{0.4}
\newcommand{\textWidth}{3cm}

\newcommand{\startX}{-0.3}
\newcommand{\barWidth}{0.2}
\newcommand{\gapWidth}{0.2}

\definecolor{darkgray176}{RGB}{176,176,176}

\hspace*{-2.0em}
\begin{tikzpicture}
\pgfplotsset{every axis/.append style={
                    xlabel={},
                    compat=1.12,
                    label style={font=\large},
                    tick label style={font=\large}  
                    }}

\begin{axis}[
width = 0.665\columnwidth,
height = 0.665\columnwidth,
tick align=outside,
tick pos=left,
x grid style={darkgray176},
xmin=-0.49, xmax=1.29,
xtick style={color=black},
xtick={-0.49},
xticklabels={},
y grid style={darkgray176},
ylabel=\mbox{Time (\millisecs)},
y label style={at={(axis description cs:-0.23,.5)}},
ymin=0, ymax=20,
ytick style={color=black},
ymajorgrids,
yminorgrids,
minor y tick num=1,
y grid style={lightgray!92.026143790849673!black},
]

\draw[draw=curve_color,fill=white,line width=1.5pt,postaction={pattern=crosshatch,
pattern color=curve_color}] (axis cs:\startX,0.0001) rectangle (axis cs:\startX+\barWidth,9.78);

\draw[draw=curve_color,fill=white,line width=1.5pt,postaction={pattern=crosshatch
dots, pattern color=curve_color}] (axis cs:\startX+\barWidth+\gapWidth,0.0001)
rectangle (axis
cs:\startX+2*\barWidth+\gapWidth,4.31);
\draw[draw=curve_color,fill=white,line width=1.5pt,postaction={pattern=crosshatch
dots, pattern color=curve_color}] (axis cs:\startX+2*\barWidth+\gapWidth,0.0001)
rectangle (axis cs:\startX+3*\barWidth+\gapWidth,4.97);
\draw[draw=curve_color,fill=white,line width=1.5pt,postaction={pattern=crosshatch dots,
pattern color=curve_color}] (axis cs:\startX+3*\barWidth+\gapWidth,0.0001)
rectangle (axis cs:\startX+4*\barWidth+\gapWidth,4.65);
\draw[draw=curve_color,fill=white,line width=1.5pt,postaction={pattern=crosshatch dots, pattern color=curve_color}] (axis cs:\startX+4*\barWidth+\gapWidth,0.0001)
rectangle (axis cs:\startX+5*\barWidth+\gapWidth,5.04);
\draw[draw=curve_color,fill=white,line width=1.5pt,postaction={pattern=crosshatch dots, pattern color=curve_color}] (axis cs:\startX+5*\barWidth+\gapWidth,0.0001)
rectangle (axis cs:\startX+6*\barWidth+\gapWidth,5.11);

\end{axis}


\node[text width=\textWidth, 
      scale=1.2,
      anchor=south, 
      right,
      text=white,
      rotate=45
] at (\startX+\barWidth-0.8*\offSet,-1.275){\small\textbf{Negative}};

\node[text width=\textWidth, 
      scale=1.2,
      anchor=south, 
      right,
      text=white,
      rotate=45
] at (\startX+2*\barWidth+\offSet,-1.175){\small\textbf{Positive}};


\node[text width=\textWidth, 
      scale=1.2,
      anchor=south, 
      right,
      text=black,
      rotate=45
] at (\startX+2*\barWidth+\gapWidth+\offSet,-1.075){\small$\boldsymbol{\nmbrExplicitHoneywords = 2^4}$};
\node[text width=\textWidth, 
      scale=1.2,
      anchor=south, 
      right,
      text=black,
      rotate=45
] at (\startX+6.5*\barWidth+\gapWidth+\offSet,-1.075){\small$\boldsymbol{\nmbrExplicitHoneywords
= 2^8}$};
\node[text width=\textWidth, 
      scale=1.2,
      anchor=south, 
      right,
      text=black,
      rotate=45
] at (\startX+10.2*\barWidth+\gapWidth+\offSet,-1.175){\small$\boldsymbol{\nmbrExplicitHoneywords
= 2^{12}}$};

\end{tikzpicture}

\hspace*{-4.2em}
\begin{tikzpicture}
\pgfplotsset{every axis/.append style={
                              xlabel={},
                              compat=1.12,
                    label style={font=\large},
                    tick label style={font=\large}  
                    }}

\begin{axis}[
width = 0.665\columnwidth,
height = 0.665\columnwidth,
tick align=outside,
tick pos=left,
x grid style={darkgray176},
xmin=-0.49, xmax=1.69,
xtick style={color=black},
xtick={-0.49},
xticklabels={},
y grid style={darkgray176},
ylabel=\mbox{Time (\millisecs)},
y label style={at={(axis description cs:-0.25,.5)}},
ymin=0.1, ymax=1000,
ytick={0.1, 1, 10, 100, 1000},
ytick style={color=black},
ymode = log,
ymajorgrids,
yminorgrids,
minor y tick num=1,
y grid style={lightgray!92.026143790849673!black},
]

\draw[draw=curve_color,fill=white,line width=1.5pt,postaction={pattern=crosshatch,
pattern color=curve_color}] (axis cs:\startX,0.0001) rectangle (axis cs:\startX+\barWidth,0.26);

\draw[draw=curve_color,fill=white,line width=1.5pt,postaction={pattern=crosshatch dots, pattern color=curve_color}] (axis cs:\startX+\barWidth,0.0001)
rectangle (axis
cs:\startX+2*\barWidth,1.80);

\draw[draw=curve_color,fill=white,line width=1.5pt,postaction={pattern=crosshatch,
pattern color=curve_color}] (axis cs:\startX+\barWidth+2*\gapWidth,0.0001)
rectangle (axis cs:\startX+2*\barWidth+2*\gapWidth,0.49);

\draw[draw=curve_color,fill=white,line width=1.5pt,postaction={pattern=crosshatch dots, pattern color=curve_color}] (axis cs:\startX+2*\barWidth+2*\gapWidth,0.0001)
rectangle (axis
cs:\startX+3*\barWidth+2*\gapWidth,2.08);
\draw[draw=curve_color,fill=white,line width=1.5pt,postaction={pattern=crosshatch dots, pattern color=curve_color}] (axis cs:\startX+3*\barWidth+2*\gapWidth,0.0001)
rectangle (axis cs:\startX+4*\barWidth+2*\gapWidth,4.81);
\draw[draw=curve_color,fill=white,line width=1.5pt,postaction={pattern=crosshatch dots,
pattern color=curve_color}] (axis cs:\startX+4*\barWidth+2*\gapWidth,0.0001)
rectangle (axis cs:\startX+5*\barWidth+2*\gapWidth,13.69);
\draw[draw=curve_color,fill=white,line width=1.5pt,postaction={pattern=crosshatch dots, pattern color=curve_color}] (axis cs:\startX+5*\barWidth+2*\gapWidth,0.0001)
rectangle (axis cs:\startX+6*\barWidth+2*\gapWidth,45.79);
\draw[draw=curve_color,fill=white,line width=1.5pt,postaction={pattern=crosshatch dots, pattern color=curve_color}] (axis cs:\startX+6*\barWidth+2*\gapWidth,0.0001)
rectangle (axis cs:\startX+7*\barWidth+2*\gapWidth,171.18);

\end{axis}

\node[text width=\textWidth, 
      scale=1.2,
      anchor=south, 
      right,
      text=black,
      rotate=45
] at (\startX+\barWidth-0.8*\offSet,-1.275){\small\textbf{Negative}};

\node[text width=\textWidth, 
      scale=1.2,
      anchor=south, 
      right,
      text=black,
      rotate=45
] at (\startX+2*\barWidth+\offSet,-1.175){\small\textbf{Positive}};
\node[text width=\textWidth, 
      scale=1.2,
      anchor=south, 
      right,
      text=black,
      rotate=45
] at (\startX+6.5*\barWidth+2*\gapWidth+0.325*\offSet,-1.120){\small$\boldsymbol{\nmbrExplicitHoneywords
= 2^6}$};
\node[text width=\textWidth, 
      scale=1.2,
      anchor=south, 
      right,
      text=black,
      rotate=45
] at (\startX+9.4*\barWidth+2*\gapWidth+0.375*\offSet,-1.200){\small$\boldsymbol{\nmbrExplicitHoneywords
= 2^{10}}$};
\end{tikzpicture}

\hspace*{-2.85em}
\begin{tikzpicture}
\pgfplotsset{every axis/.append style={
                    xlabel={},
                    compat=1.12,
                    label style={font=\large},
                    tick label style={font=\large}  
                    }}

\begin{axis}[
width = 0.665\columnwidth,
height = 0.665\columnwidth,
tick align=outside,
tick pos=left,
x grid style={darkgray176},
xmin=-0.49, xmax=1.29,
xtick style={color=black},
xtick={-0.49},
xticklabels={},
y grid style={darkgray176},
ylabel=\mbox{Message size (\kilobytes)},
y label style={at={(axis description cs:-0.23,.5)}},
ymin=0, ymax=2,
ytick={0, 0.5, 1.0, 1.5, 2.0},
yticklabels={0, 0.5, 1.0, 1.5, 2.0},
ytick style={color=black},
ymajorgrids,
yminorgrids,
minor y tick num=1,
y grid style={lightgray!92.026143790849673!black},
]

\draw[draw=curve_color,fill=white,line width=1.5pt,postaction={pattern=crosshatch,
pattern color=curve_color}] (axis cs:\startX,0.0001) rectangle (axis cs:\startX+\barWidth,0.22);

\draw[draw=curve_color,fill=white,line width=1.5pt,postaction={pattern=crosshatch
dots, pattern color=curve_color}] (axis cs:\startX+\barWidth+\gapWidth,0.0001)
rectangle (axis
cs:\startX+2*\barWidth+\gapWidth,1.21);
\draw[draw=curve_color,fill=white,line width=1.5pt,postaction={pattern=crosshatch
dots, pattern color=curve_color}] (axis cs:\startX+2*\barWidth+\gapWidth,0.0001)
rectangle (axis cs:\startX+3*\barWidth+\gapWidth,1.21);
\draw[draw=curve_color,fill=white,line width=1.5pt,postaction={pattern=crosshatch dots,
pattern color=curve_color}] (axis cs:\startX+3*\barWidth+\gapWidth,0.0001)
rectangle (axis cs:\startX+4*\barWidth+\gapWidth,1.21);
\draw[draw=curve_color,fill=white,line width=1.5pt,postaction={pattern=crosshatch dots, pattern color=curve_color}] (axis cs:\startX+4*\barWidth+\gapWidth,0.0001)
rectangle (axis cs:\startX+5*\barWidth+\gapWidth,1.21);
\draw[draw=curve_color,fill=white,line width=1.5pt,postaction={pattern=crosshatch dots, pattern color=curve_color}] (axis cs:\startX+5*\barWidth+\gapWidth,0.0001)
rectangle (axis cs:\startX+6*\barWidth+\gapWidth,1.21);

\end{axis}


\node[text width=\textWidth, 
      scale=1.2,
      anchor=south, 
      right,
      text=white,
      rotate=45
] at (\startX+\barWidth-0.8*\offSet,-1.275){\small\textbf{Negative}};

\node[text width=\textWidth, 
      scale=1.2,
      anchor=south, 
      right,
      text=white,
      rotate=45
] at (\startX+2*\barWidth+\offSet,-1.175){\small\textbf{Positive}};


\node[text width=\textWidth, 
      scale=1.2,
      anchor=south, 
      right,
      text=black,
      rotate=45
] at (\startX+2*\barWidth+\gapWidth+\offSet,-1.075){\small$\boldsymbol{\nmbrExplicitHoneywords = 2^4}$};
\node[text width=\textWidth, 
      scale=1.2,
      anchor=south, 
      right,
      text=black,
      rotate=45
] at (\startX+6.5*\barWidth+\gapWidth+\offSet,-1.075){\small$\boldsymbol{\nmbrExplicitHoneywords
= 2^8}$};
\node[text width=\textWidth, 
      scale=1.2,
      anchor=south, 
      right,
      text=black,
      rotate=45
] at (\startX+10.2*\barWidth+\gapWidth+\offSet,-1.175){\small$\boldsymbol{\nmbrExplicitHoneywords
= 2^{12}}$};

\end{tikzpicture}

%% file: figures/fig_pcr_cmp.tex
\begin{figure}[ht]
  \hspace*{4.5em}
  \begin{subfigure}[b]{.1\columnwidth}
    \setlength\figureheight{1.8in}
    \resizebox{!}{1.15em}{\input{figures/pcr/cmp_legend.tex}}
  \end{subfigure}\vspace*{0.8em}

  \captionsetup[subfigure]{font=small,labelfont=small, oneside,margin={-0.5in,0in}}
  \begin{subfigure}[b]{.9\columnwidth}
  \setlength\figureheight{1.5in}
    \centering
    \hspace*{-0.4em}
    \resizebox{!}{8.35em}{\input{figures/pcr/cmp_fig_1.tex}}
     \hspace*{-1.95em}\begin{minipage}[t]{17em}
      \vspace*{-1.5em}\caption{\parbox[t]{6.5em}{Request \\generation by \targetSite\\
      \hspace{1em}($\relstddev < 0.10$)}}
      \label{fig:cmp:requestGenTime}
    \end{minipage}%
    \hspace*{-8.25em}\begin{minipage}[t]{17em}
      \vspace*{-1.5em}\caption{\parbox[t]{6.5em}{Request\\ validation by \monitorSite\\
      \hspace{1em}($\relstddev < 0.10$)}}
      \label{fig:cmp:requestValidateTime}
    \end{minipage}%
    \hspace*{-6.95em}\begin{minipage}[t]{17em}
      \vspace*{-1.5em}\caption{\parbox[t]{6.5em}{Size of \\request \\ 
      \hspace{1em}($\relstddev < 0.01$)}}
      \label{fig:cmp:requestSize}
    \end{minipage}%
  \end{subfigure}

  \begin{subfigure}[b]{.9\columnwidth}
  \setlength\figureheight{1.5in}
    \centering
    \hspace*{-0.4em}
    \resizebox{!}{8.35em}{\input{figures/pcr/cmp_fig_2.tex}}
     \hspace*{-1.95em}\begin{minipage}[t]{17em}
      \vspace*{-1.5em}\caption{\parbox[t]{6.5em}{Response \\generation by \monitorSite\\
      \hspace{1em}($\relstddev < 0.12$)}}
      \label{fig:cmp:responseTime}
    \end{minipage}%
    \hspace*{-8.25em}\begin{minipage}[t]{17em}
      \vspace*{-1.5em}\caption{\parbox[t]{6.5em}{Response \\processing by \targetSite\\
      \hspace{1em}($\relstddev < 0.60$)}}
      \label{fig:cmp:retrievalTime}
    \end{minipage}%
    \hspace*{-6.95em}\begin{minipage}[t]{17em}
      \vspace*{-1.5em}\caption{\parbox[t]{6.5em}{Size of \\response \\
      \hspace{1em}($\relstddev < 0.01$)}}
      \label{fig:cmp:responseSize}
    \end{minipage}%
  \end{subfigure}

  \caption{Performance comparison between our protocol
    (\secref{sec:amnesia:remote:protocol}) and \pcr{\bfNmbrHashFns}.
    One-time costs for \targetSite to deploy a monitoring request to
    \monitorSite are shown in
    (\subref{fig:cmp:requestGenTime})--(\subref{fig:cmp:requestSize}),
    and costs for \monitorSite to return a response to \targetSite,
    once per incorrect login attempt, are shown in
    (\subref{fig:cmp:responseTime})--(\subref{fig:cmp:responseSize}).
    In (\subref{fig:cmp:retrievalTime}), ``negative'' refers to the
    case $\pwdHashFn(\loginPassword) \bfNotIn \langle \bfHashFnSet,
    \bfIndices \rangle$, and ``positive'' refers to the case
    $\pwdHashFn(\loginPassword) \bfIn \langle \bfHashFnSet, \bfIndices
    \rangle$.  In \pcr{\bfNmbrHashFns}, the ``positive'' cost depends
    on \bfNmbrHashFns.  Request generation, validation, and size in
    \pcr{\bfNmbrHashFns} also depends on $\bfNmbrIndices =
    \setSize{\bfIndices}$; numbers shown are for $\bfNmbrIndices =
    \bfHashFnRange/2$ in
    (\subref{fig:cmp:requestGenTime})--(\subref{fig:cmp:requestSize}).}
  \label{fig:comparison}
\end{figure}

%% file: figures/pcr/cmp_legend.tex
\newenvironment{customlegend}[1][]{%
    \begingroup
    \csname pgfplots@init@cleared@structures\endcsname
    \pgfplotsset{#1}%
}{%
    \csname pgfplots@createlegend\endcsname
    \endgroup
}%

\def\addlegendimage{\csname pgfplots@addlegendimage\endcsname}

\begin{tikzpicture}

\begin{customlegend}[
    legend style={{font={\fontsize{12pt}{12}\selectfont}},{draw=none}},
    legend columns=3,
    legend cell align={left},
    legend entries={{~Our protocol$\quad$}, {~\pcr{\bfNmbrHashFns}, adapted from Amnesia~\cite{wang2021:amnesia}}}]
\addlegendimage{ybar,ybar legend,draw=curve_color,fill=white,line width=1.5pt,postaction={pattern=crosshatch, pattern color=curve_color}, legend image
code/.code={\draw (0cm,-0.1cm) rectangle (0.5cm,0.15cm);}}

\addlegendimage{ybar,ybar legend,draw=curve_color,fill=white,line width=1.5pt,postaction={pattern=crosshatch dots, pattern color=curve_color}, legend image
code/.code={\draw (0cm,-0.1cm) rectangle (0.5cm,0.15cm);}}

\end{customlegend}

\end{tikzpicture}

%% file: figures/pcr/cmp_fig_1.tex
\newcommand{\offSet}{-0.3}
\newcommand{\textWidth}{3cm}

\newcommand{\startX}{-0.4}
\newcommand{\barWidth}{0.2}
\newcommand{\gapWidth}{0.2}

\definecolor{darkgray176}{RGB}{176,176,176}

\hspace*{-2.6em}
\begin{tikzpicture}
\pgfplotsset{every axis/.append style={
                              xlabel={},
                              compat=1.12,
                    label style={font=\large},
                    tick label style={font=\large}  
                    }}

\begin{axis}[
width = 0.65\columnwidth,
height = 0.65\columnwidth,
tick align=outside,
tick pos=left,
x grid style={darkgray176},
xmin=-0.49, xmax=1.89,
xtick style={color=black},
xtick={-0.49},
xticklabels={},
y grid style={darkgray176},
ylabel={Time (\millisecs)},
y label style={at={(axis description cs:-0.26,.5)}},
ymin=1, ymax=1000,
ytick={1, 10, 100, 1000},
ytick style={color=black},
ymode = log,
ymajorgrids,
yminorgrids,
minor y tick num=1,
y grid style={lightgray!92.026143790849673!black},
]

\draw[draw=curve_color,fill=white,line width=1.5pt,postaction={pattern=crosshatch, pattern color=curve_color}] (axis cs:\startX,0.0001)
rectangle (axis
cs:\startX+\barWidth,33.8);
\draw[draw=curve_color,fill=white,line width=1.5pt,postaction={pattern=crosshatch dots, pattern color=curve_color}] (axis cs:\startX+\barWidth,0.0001)
rectangle (axis
cs:\startX+2*\barWidth,4.93);

\draw[draw=curve_color,fill=white,line width=1.5pt,postaction={pattern=crosshatch, pattern color=curve_color}] (axis cs:\startX+2*\barWidth+\gapWidth,0.0001) rectangle (axis cs:\startX+3*\barWidth+\gapWidth,68.02);
\draw[draw=curve_color,fill=white,line width=1.5pt,postaction={pattern=crosshatch dots, pattern color=curve_color}] (axis cs:\startX+3*\barWidth+\gapWidth,0.0001)
rectangle (axis cs:\startX+4*\barWidth+\gapWidth,9.27);

\draw[draw=curve_color,fill=white,line width=1.5pt,postaction={pattern=crosshatch,
pattern color=curve_color}] (axis cs:\startX+4*\barWidth+2*\gapWidth,0.0001)
rectangle (axis cs:\startX+5*\barWidth+2*\gapWidth,102.42);
\draw[draw=curve_color,fill=white,line width=1.5pt,postaction={pattern=crosshatch dots,
pattern color=curve_color}] (axis cs:\startX+5*\barWidth+2*\gapWidth,0.0001)
rectangle (axis cs:\startX+6*\barWidth+2*\gapWidth,13.56);

\draw[draw=curve_color,fill=white,line width=1.5pt,postaction={pattern=crosshatch, pattern color=curve_color}] (axis cs:\startX+6*\barWidth+3*\gapWidth,0.0001)
rectangle (axis cs:\startX+7*\barWidth+3*\gapWidth,137.31);
\draw[draw=curve_color,fill=white,line width=1.5pt,postaction={pattern=crosshatch dots, pattern color=curve_color}] (axis cs:\startX+7*\barWidth+3*\gapWidth,0.0001) rectangle (axis cs:\startX+8*\barWidth+3*\gapWidth,17.85);

\end{axis}


\node[text width=\textWidth, 
      scale=1.2,
      anchor=south, 
      right,
      text=white,
      rotate=45
] at (\startX+0.78*\offSet,-1.275){\small\textbf{Negative}};

\node[text width=\textWidth,
      scale=1.2,
      anchor=south, 
      right,
      text=white,
      rotate=45
] at (\startX+6*\barWidth+1.5*\offSet,-1.175){\small\textbf{Positive}};


\node[text width=\textWidth, 
      scale=1.2,
      anchor=south, 
      right,
      text=black,
      rotate=45
] at (\offSet,-1.1){\small$\boldsymbol{\bfHashFnRange = 64}$};
\node[text width=\textWidth, 
      scale=1.2,
      anchor=south, 
      right,
      text=black,
      rotate=45
] at (\offSet+3.8*\barWidth,-1.25){\small$\boldsymbol{\bfHashFnRange = 128}$};
\node[text width=\textWidth, 
      scale=1.2,
      anchor=south, 
      right,
      text=black,
      rotate=45
] at (\offSet+8.5*\barWidth,-1.25){\small$\boldsymbol{\bfHashFnRange = 192}$};
\node[text width=\textWidth, 
      scale=1.2,
      anchor=south, 
      right,
      text=black,
      rotate=45
] at (\offSet+12.7*\barWidth,-1.25){\small$\boldsymbol{\bfHashFnRange = 256}$};
\end{tikzpicture}

\hspace*{-4.2em}
\begin{tikzpicture}
\pgfplotsset{every axis/.append style={
                              xlabel={},
                              compat=1.12,
                    label style={font=\large},
                    tick label style={font=\large}  
                    }}

\begin{axis}[
width = 0.65\columnwidth,
height = 0.65\columnwidth,
tick align=outside,
tick pos=left,
x grid style={darkgray176},
xmin=-0.49, xmax=1.89,
xtick style={color=black},
xtick={-0.49},
xticklabels={},
y grid style={darkgray176},
ylabel={Time (\millisecs)},
y label style={at={(axis description cs:-0.26,.5)}},
ymin=1, ymax=1000,
ytick={1, 10, 100, 1000},
ytick style={color=black},
ymode = log,
ymajorgrids,
yminorgrids,
minor y tick num=1,
y grid style={lightgray!92.026143790849673!black},
]

\draw[draw=curve_color,fill=white,line width=1.5pt,postaction={pattern=crosshatch, pattern color=curve_color}] (axis cs:\startX,0.0001)
rectangle (axis
cs:\startX+\barWidth,61.6);
\draw[draw=curve_color,fill=white,line width=1.5pt,postaction={pattern=crosshatch dots, pattern color=curve_color}] (axis cs:\startX+\barWidth,0.0001)
rectangle (axis
cs:\startX+2*\barWidth,4.82);

\draw[draw=curve_color,fill=white,line width=1.5pt,postaction={pattern=crosshatch, pattern color=curve_color}] (axis cs:\startX+2*\barWidth+\gapWidth,0.0001) rectangle (axis cs:\startX+3*\barWidth+\gapWidth,122.97);
\draw[draw=curve_color,fill=white,line width=1.5pt,postaction={pattern=crosshatch dots, pattern color=curve_color}] (axis cs:\startX+3*\barWidth+\gapWidth,0.0001)
rectangle (axis cs:\startX+4*\barWidth+\gapWidth,9.44);

\draw[draw=curve_color,fill=white,line width=1.5pt,postaction={pattern=crosshatch,
pattern color=curve_color}] (axis cs:\startX+4*\barWidth+2*\gapWidth,0.0001)
rectangle (axis cs:\startX+5*\barWidth+2*\gapWidth,183.37);
\draw[draw=curve_color,fill=white,line width=1.5pt,postaction={pattern=crosshatch dots,
pattern color=curve_color}] (axis cs:\startX+5*\barWidth+2*\gapWidth,0.0001)
rectangle (axis cs:\startX+6*\barWidth+2*\gapWidth,14.03);

\draw[draw=curve_color,fill=white,line width=1.5pt,postaction={pattern=crosshatch, pattern color=curve_color}] (axis cs:\startX+6*\barWidth+3*\gapWidth,0.0001)
rectangle (axis cs:\startX+7*\barWidth+3*\gapWidth,244.84);
\draw[draw=curve_color,fill=white,line width=1.5pt,postaction={pattern=crosshatch dots, pattern color=curve_color}] (axis cs:\startX+7*\barWidth+3*\gapWidth,0.0001) rectangle (axis cs:\startX+8*\barWidth+3*\gapWidth,18.89);

\end{axis}


\node[text width=\textWidth, 
      scale=1.2,
      anchor=south, 
      right,
      text=white,
      rotate=45
] at (\startX+0.78*\offSet,-1.275){\small\textbf{Negative}};

\node[text width=\textWidth,
      scale=1.2,
      anchor=south, 
      right,
      text=white,
      rotate=45
] at (\startX+6*\barWidth+1.5*\offSet,-1.175){\small\textbf{Positive}};


\node[text width=\textWidth, 
      scale=1.2,
      anchor=south, 
      right,
      text=black,
      rotate=45
] at (\offSet,-1.1){\small$\boldsymbol{\bfHashFnRange = 64}$};
\node[text width=\textWidth, 
      scale=1.2,
      anchor=south, 
      right,
      text=black,
      rotate=45
] at (\offSet+3.8*\barWidth,-1.25){\small$\boldsymbol{\bfHashFnRange = 128}$};
\node[text width=\textWidth, 
      scale=1.2,
      anchor=south, 
      right,
      text=black,
      rotate=45
] at (\offSet+8.5*\barWidth,-1.25){\small$\boldsymbol{\bfHashFnRange = 192}$};
\node[text width=\textWidth, 
      scale=1.2,
      anchor=south, 
      right,
      text=black,
      rotate=45
] at (\offSet+12.7*\barWidth,-1.25){\small$\boldsymbol{\bfHashFnRange = 256}$};
\end{tikzpicture}

\hspace*{-3.9em}
\begin{tikzpicture}
\pgfplotsset{every axis/.append style={
                              xlabel={},
                              compat=1.12,
                    label style={font=\large},
                    tick label style={font=\large}  
                    }}

\begin{axis}[
width = 0.65\columnwidth,
height = 0.65\columnwidth,
tick align=outside,
tick pos=left,
x grid style={darkgray176},
xmin=-0.49, xmax=1.89,
xtick style={color=black},
xtick={-0.49},
xticklabels={},
y grid style={darkgray176},
ylabel={Message size (\kilobytes)},
y label style={at={(axis description cs:-0.26,.5)}},
ymin=1, ymax=100,
ytick={1, 10, 100},
ytick style={color=black},
ymode = log,
ymajorgrids,
yminorgrids,
minor y tick num=1,
y grid style={lightgray!92.026143790849673!black},
]

\draw[draw=curve_color,fill=white,line width=1.5pt,postaction={pattern=crosshatch, pattern color=curve_color}] (axis cs:\startX,0.0001)
rectangle (axis
cs:\startX+\barWidth,22.78);
\draw[draw=curve_color,fill=white,line width=1.5pt,postaction={pattern=crosshatch dots, pattern color=curve_color}] (axis cs:\startX+\barWidth,0.0001)
rectangle (axis
cs:\startX+2*\barWidth,3.10);

\draw[draw=curve_color,fill=white,line width=1.5pt,postaction={pattern=crosshatch, pattern color=curve_color}] (axis cs:\startX+2*\barWidth+\gapWidth,0.0001) rectangle (axis cs:\startX+3*\barWidth+\gapWidth,44.86);
\draw[draw=curve_color,fill=white,line width=1.5pt,postaction={pattern=crosshatch dots, pattern color=curve_color}] (axis cs:\startX+3*\barWidth+\gapWidth,0.0001)
rectangle (axis cs:\startX+4*\barWidth+\gapWidth,5.59);

\draw[draw=curve_color,fill=white,line width=1.5pt,postaction={pattern=crosshatch,
pattern color=curve_color}] (axis cs:\startX+4*\barWidth+2*\gapWidth,0.0001)
rectangle (axis cs:\startX+5*\barWidth+2*\gapWidth,66.94);
\draw[draw=curve_color,fill=white,line width=1.5pt,postaction={pattern=crosshatch dots,
pattern color=curve_color}] (axis cs:\startX+5*\barWidth+2*\gapWidth,0.0001)
rectangle (axis cs:\startX+6*\barWidth+2*\gapWidth,8.07);

\draw[draw=curve_color,fill=white,line width=1.5pt,postaction={pattern=crosshatch, pattern color=curve_color}] (axis cs:\startX+6*\barWidth+3*\gapWidth,0.0001)
rectangle (axis cs:\startX+7*\barWidth+3*\gapWidth,88.99);
\draw[draw=curve_color,fill=white,line width=1.5pt,postaction={pattern=crosshatch dots, pattern color=curve_color}] (axis cs:\startX+7*\barWidth+3*\gapWidth,0.0001) rectangle (axis cs:\startX+8*\barWidth+3*\gapWidth,10.54);

\end{axis}


\node[text width=\textWidth, 
      scale=1.2,
      anchor=south, 
      right,
      text=white,
      rotate=45
] at (\startX+0.78*\offSet,-1.275){\small\textbf{Negative}};

\node[text width=\textWidth,
      scale=1.2,
      anchor=south, 
      right,
      text=white,
      rotate=45
] at (\startX+6*\barWidth+1.5*\offSet,-1.175){\small\textbf{Positive}};


\node[text width=\textWidth, 
      scale=1.2,
      anchor=south, 
      right,
      text=black,
      rotate=45
] at (\offSet,-1.1){\small$\boldsymbol{\bfHashFnRange = 64}$};
\node[text width=\textWidth, 
      scale=1.2,
      anchor=south, 
      right,
      text=black,
      rotate=45
] at (\offSet+3.8*\barWidth,-1.25){\small$\boldsymbol{\bfHashFnRange = 128}$};
\node[text width=\textWidth, 
      scale=1.2,
      anchor=south, 
      right,
      text=black,
      rotate=45
] at (\offSet+8.5*\barWidth,-1.25){\small$\boldsymbol{\bfHashFnRange = 192}$};
\node[text width=\textWidth, 
      scale=1.2,
      anchor=south, 
      right,
      text=black,
      rotate=45
] at (\offSet+12.7*\barWidth,-1.25){\small$\boldsymbol{\bfHashFnRange = 256}$};
\end{tikzpicture}

%% file: figures/pcr/cmp_fig_2.tex
\newcommand{\offSet}{-0.3}
\newcommand{\textWidth}{3cm}

\newcommand{\startX}{-0.4}
\newcommand{\barWidth}{0.2}
\newcommand{\gapWidth}{0.2}

\definecolor{darkgray176}{RGB}{176,176,176}

\hspace*{-2.6em}
\begin{tikzpicture}
\pgfplotsset{every axis/.append style={
                              xlabel={},
                              compat=1.12,
                    label style={font=\large},
                    tick label style={font=\large}  
                    }}

\begin{axis}[
width = 0.65\columnwidth,
height = 0.65\columnwidth,
tick align=outside,
tick pos=left,
x grid style={darkgray176},
xmin=-0.49, xmax=3.09,
xtick style={color=black},
xtick={-0.49},
xticklabels={},
y grid style={darkgray176},
ylabel={Time (\millisecs)},
y label style={at={(axis description cs:-0.26,.5)}},
ymin=1, ymax=150,
ytick={1, 10, 100},
ytick style={color=black},
ymode = log,
ymajorgrids,
yminorgrids,
minor y tick num=1,
y grid style={lightgray!92.026143790849673!black},
]

\draw[draw=curve_color,fill=white,line width=1.5pt,postaction={pattern=crosshatch, pattern color=curve_color}] (axis cs:\startX,0.0001)
rectangle (axis
cs:\startX+\barWidth,3.47);
\draw[draw=curve_color,fill=white,line width=1.5pt,postaction={pattern=crosshatch dots, pattern color=curve_color}] (axis cs:\startX+\barWidth,0.0001)
rectangle (axis
cs:\startX+2*\barWidth,19.44);

\draw[draw=curve_color,fill=white,line width=1.5pt,postaction={pattern=crosshatch, pattern color=curve_color}] (axis cs:\startX+2*\barWidth+\gapWidth,0.0001) rectangle (axis cs:\startX+3*\barWidth+\gapWidth,5.60);
\draw[draw=curve_color,fill=white,line width=1.5pt,postaction={pattern=crosshatch dots, pattern color=curve_color}] (axis cs:\startX+3*\barWidth+\gapWidth,0.0001)
rectangle (axis cs:\startX+4*\barWidth+\gapWidth,38.97);

\draw[draw=curve_color,fill=white,line width=1.5pt,postaction={pattern=crosshatch,
pattern color=curve_color}] (axis cs:\startX+4*\barWidth+2*\gapWidth,0.0001)
rectangle (axis cs:\startX+5*\barWidth+2*\gapWidth,7.65);
\draw[draw=curve_color,fill=white,line width=1.5pt,postaction={pattern=crosshatch dots,
pattern color=curve_color}] (axis cs:\startX+5*\barWidth+2*\gapWidth,0.0001)
rectangle (axis cs:\startX+6*\barWidth+2*\gapWidth,58.28);

\draw[draw=curve_color,fill=white,line width=1.5pt,postaction={pattern=crosshatch, pattern color=curve_color}] (axis cs:\startX+6*\barWidth+3*\gapWidth,0.0001)
rectangle (axis cs:\startX+7*\barWidth+3*\gapWidth,9.75);
\draw[draw=curve_color,fill=white,line width=1.5pt,postaction={pattern=crosshatch dots, pattern color=curve_color}] (axis cs:\startX+7*\barWidth+3*\gapWidth,0.0001) rectangle (axis cs:\startX+8*\barWidth+3*\gapWidth,77.89);

\draw[draw=curve_color,fill=white,line width=1.5pt,postaction={pattern=crosshatch, pattern color=curve_color}] (axis cs:\startX+8*\barWidth+4*\gapWidth,0.0001)
rectangle (axis cs:\startX+9*\barWidth+4*\gapWidth,11.00);
\draw[draw=curve_color,fill=white,line width=1.5pt,postaction={pattern=crosshatch dots, pattern color=curve_color}] (axis cs:\startX+9*\barWidth+4*\gapWidth,0.0001)
rectangle (axis cs:\startX+10*\barWidth+4*\gapWidth,96.96);

\draw[draw=curve_color,fill=white,line width=1.5pt,postaction={pattern=crosshatch, pattern color=curve_color}] (axis cs:\startX+10*\barWidth+5*\gapWidth,0.0001)
rectangle (axis
cs:\startX+11*\barWidth+5*\gapWidth,12.56);
\draw[draw=curve_color,fill=white,line width=1.5pt,postaction={pattern=crosshatch dots, pattern color=curve_color}] (axis cs:\startX+11*\barWidth+5*\gapWidth,0.0001) rectangle (axis
cs:\startX+12*\barWidth+5*\gapWidth,116.56);

\end{axis}


\node[text width=\textWidth, 
      scale=1.2,
      anchor=south, 
      right,
      text=white,
      rotate=45
] at (\startX+0.78*\offSet,-1.275){\small\textbf{Negative}};

\node[text width=\textWidth,
      scale=1.2,
      anchor=south, 
      right,
      text=white,
      rotate=45
] at (\startX+6*\barWidth+1.5*\offSet,-1.175){\small\textbf{Positive}};


\node[text width=\textWidth, 
      scale=1.2,
      anchor=south, 
      right,
      text=black,
      rotate=45
] at (\offSet,-0.975){\small$\boldsymbol{\bfNmbrHashFns = 5}$};
\node[text width=\textWidth, 
      scale=1.2,
      anchor=south, 
      right,
      text=black,
      rotate=45
] at (\offSet+5.5*\barWidth,-1.15){\small$\boldsymbol{\bfNmbrHashFns = 15}$};
\node[text width=\textWidth, 
      scale=1.2,
      anchor=south, 
      right,
      text=black,
      rotate=45
] at (\offSet+11*\barWidth,-1.15){\small$\boldsymbol{\bfNmbrHashFns = 25}$};

\end{tikzpicture}

\hspace*{-3.2em}
\begin{tikzpicture}
\pgfplotsset{every axis/.append style={
                              xlabel={},
                              compat=1.12,
                    label style={font=\large},
                    tick label style={font=\large}  
                    }}

\begin{axis}[
width = 0.65\columnwidth,
height = 0.65\columnwidth,
tick align=outside,
tick pos=left,
x grid style={darkgray176},
xmin=-0.49, xmax=1.69,
xtick style={color=black},
xtick={-0.49},
xticklabels={},
y grid style={darkgray176},
ylabel={Time (\millisecs)},
y label style={at={(axis description cs:-0.26,.5)}},
ymin=0.1, ymax=200,
ytick={0.1, 1, 10, 100},
ytick style={color=black},
ymode = log,
ymajorgrids,
yminorgrids,
minor y tick num=1,
y grid style={lightgray!92.026143790849673!black},
]

\draw[draw=curve_color,fill=white,line width=1.5pt,postaction={pattern=crosshatch,
pattern color=curve_color}] (axis cs:\startX,0.0001) rectangle (axis cs:\startX+\barWidth,0.28);

\draw[draw=curve_color,fill=white,line width=1.5pt,postaction={pattern=crosshatch dots, pattern color=curve_color}] (axis cs:\startX+\barWidth,0.0001)
rectangle (axis
cs:\startX+2*\barWidth,2.00);

\draw[draw=curve_color,fill=white,line width=1.5pt,postaction={pattern=crosshatch,
pattern color=curve_color}] (axis cs:\startX+\barWidth+2*\gapWidth,0.0001)
rectangle (axis cs:\startX+2*\barWidth+2*\gapWidth,0.49);

\draw[draw=curve_color,fill=white,line width=1.5pt,postaction={pattern=crosshatch dots, pattern color=curve_color}] (axis cs:\startX+2*\barWidth+2*\gapWidth,0.0001)
rectangle (axis
cs:\startX+3*\barWidth+2*\gapWidth,4.24);
\draw[draw=curve_color,fill=white,line width=1.5pt,postaction={pattern=crosshatch dots, pattern color=curve_color}] (axis cs:\startX+3*\barWidth+2*\gapWidth,0.0001)
rectangle (axis cs:\startX+4*\barWidth+2*\gapWidth,8.94);
\draw[draw=curve_color,fill=white,line width=1.5pt,postaction={pattern=crosshatch dots,
pattern color=curve_color}] (axis cs:\startX+4*\barWidth+2*\gapWidth,0.0001)
rectangle (axis cs:\startX+5*\barWidth+2*\gapWidth,13.55);
\draw[draw=curve_color,fill=white,line width=1.5pt,postaction={pattern=crosshatch dots, pattern color=curve_color}] (axis cs:\startX+5*\barWidth+2*\gapWidth,0.0001)
rectangle (axis cs:\startX+6*\barWidth+2*\gapWidth,17.12);
\draw[draw=curve_color,fill=white,line width=1.5pt,postaction={pattern=crosshatch dots, pattern color=curve_color}] (axis cs:\startX+6*\barWidth+2*\gapWidth,0.0001)
rectangle (axis cs:\startX+7*\barWidth+2*\gapWidth,20.98);
\draw[draw=curve_color,fill=white,line width=1.5pt,postaction={pattern=crosshatch dots, pattern color=curve_color}] (axis cs:\startX+7*\barWidth+2*\gapWidth,0.0001) rectangle (axis
cs:\startX+8*\barWidth+2*\gapWidth,25.83);

\end{axis}

\node[text width=\textWidth, 
      scale=1.2,
      anchor=south, 
      right,
      text=black,
      rotate=45
] at (\startX+0.78*\offSet,-1.275){\small\textbf{Negative}};

\node[text width=\textWidth,
      scale=1.2,
      anchor=south, 
      right,
      text=black,
      rotate=45
] at (\startX+6*\barWidth+1.5*\offSet,-1.175){\small\textbf{Positive}};
\node[text width=\textWidth, 
      scale=1.2,
      anchor=south, 
      right,
      text=black,
      rotate=45
] at (\startX+7.5*\barWidth+2*\gapWidth+\offSet,-1.110){\small$\boldsymbol{\bfNmbrHashFns = 10}$};
\node[text width=\textWidth, 
      scale=1.2,
      anchor=south, 
      right,
      text=black,
      rotate=45
] at (\startX+10.8*\barWidth+2*\gapWidth+\offSet,-1.100){\small$\boldsymbol{\bfNmbrHashFns
= 20}$};
\node[text width=\textWidth, 
      scale=1.2,
      anchor=south, 
      right,
      text=black,
      rotate=45
] at (\startX+13.25*\barWidth+2*\gapWidth+0.375*\offSet,-1.100){\small$\boldsymbol{\bfNmbrHashFns
= 30}$};
\end{tikzpicture}

\hspace*{-4.8em}
\begin{tikzpicture}
\pgfplotsset{every axis/.append style={
                              xlabel={},
                              compat=1.12,
                    label style={font=\large},
                    tick label style={font=\large}  
                    }}

\begin{axis}[
width = 0.65\columnwidth,
height = 0.65\columnwidth,
tick align=outside,
tick pos=left,
x grid style={darkgray176},
xmin=-0.49, xmax=1.29,
xtick style={color=black},
xtick={-0.49},
xticklabels={},
y grid style={darkgray176},
ylabel={Message size (\kilobytes)},
y label style={at={(axis description cs:-0.26,.5)}},
ymin=0.1, ymax=100,
ytick={0.1, 1, 10, 100},
ytick style={color=black},
ymode = log,
ymajorgrids,
yminorgrids,
minor y tick num=1,
y grid style={lightgray!92.026143790849673!black},
]

\draw[draw=curve_color,fill=white,line width=1.5pt,postaction={pattern=crosshatch,
pattern color=curve_color}] (axis cs:\startX,0.0001) rectangle (axis cs:\startX+\barWidth,0.22);

\draw[draw=curve_color,fill=white,line width=1.5pt,postaction={pattern=crosshatch
dots, pattern color=curve_color}] (axis cs:\startX+\barWidth+\gapWidth,0.0001)
rectangle (axis
cs:\startX+2*\barWidth+\gapWidth,2.98);
\draw[draw=curve_color,fill=white,line width=1.5pt,postaction={pattern=crosshatch
dots, pattern color=curve_color}] (axis cs:\startX+2*\barWidth+\gapWidth,0.0001)
rectangle (axis cs:\startX+3*\barWidth+\gapWidth,6.07);
\draw[draw=curve_color,fill=white,line width=1.5pt,postaction={pattern=crosshatch dots,
pattern color=curve_color}] (axis cs:\startX+3*\barWidth+\gapWidth,0.0001)
rectangle (axis cs:\startX+4*\barWidth+\gapWidth,9.52);
\draw[draw=curve_color,fill=white,line width=1.5pt,postaction={pattern=crosshatch dots, pattern color=curve_color}] (axis cs:\startX+4*\barWidth+\gapWidth,0.0001)
rectangle (axis cs:\startX+5*\barWidth+\gapWidth,14.78);
\draw[draw=curve_color,fill=white,line width=1.5pt,postaction={pattern=crosshatch dots, pattern color=curve_color}] (axis cs:\startX+5*\barWidth+\gapWidth,0.0001)
rectangle (axis cs:\startX+6*\barWidth+\gapWidth,19.77);
\draw[draw=curve_color,fill=white,line width=1.5pt,postaction={pattern=crosshatch
dots, pattern color=curve_color}] (axis cs:\startX+6*\barWidth+\gapWidth,0.0001) rectangle (axis
cs:\startX+7*\barWidth+\gapWidth,25.83);

\end{axis}


\node[text width=\textWidth, 
      scale=1.2,
      anchor=south, 
      right,
      text=white,
      rotate=45
] at (\startX+0.78*\offSet,-1.275){\small\textbf{Negative}};

\node[text width=\textWidth,
      scale=1.2,
      anchor=south, 
      right,
      text=white,
      rotate=45
] at (\startX+6*\barWidth+1.5*\offSet,-1.175){\small\textbf{Positive}};


\node[text width=\textWidth, 
      scale=1.2,
      anchor=south, 
      right,
      text=black,
      rotate=45
] at (\offSet+4.5*\barWidth,-0.975){\small$\boldsymbol{\bfNmbrHashFns = 5}$};
\node[text width=\textWidth, 
      scale=1.2,
      anchor=south, 
      right,
      text=black,
      rotate=45
] at (\offSet+7.7*\barWidth,-1.15){\small$\boldsymbol{\bfNmbrHashFns = 15}$};
\node[text width=\textWidth, 
      scale=1.2,
      anchor=south, 
      right,
      text=black,
      rotate=45
] at (\offSet+11.9*\barWidth,-1.15){\small$\boldsymbol{\bfNmbrHashFns = 25}$};

\end{tikzpicture}

%% file: proofs.tex
\section{Security Analysis of Remote Monitoring Protocol}
\label{app:analysis}

In this appendix, we prove security for the protocol described in
\secref{sec:amnesia:remote:protocol:design}.  We start by defining the
primitives on which security relies, in
\secref{app:analysis:primitives}.  We then prove security against a
malicious target site \targetSite in \secref{app:analysis:target}, and
against a malicious monitor site \monitorSite in
\secref{app:analysis:monitor}.

\subsection{Primitives}
\label{app:analysis:primitives}

In this section we provide definitions for the primitives used in our
cryptographic protocol described in \secref{sec:amnesia:remote}.

\subsubsection{IND-CPA secure encryption}
\label{app:analysis:primitives:enc}
Our protocol relies on a partially homomorphic encryption
scheme \encScheme achieving indistinguishability under chosen
plaintext attack (IND-CPA) security~\cite{bellare1998:relations}.
We define the IND-CPA
experiment \indcpaExperiment{\indcpaBit}{\encScheme} as:
\begin{center}
\begin{minipage}{0.5\columnwidth}
\begin{tabbing}
\codeExpt $\indcpaExperiment{\indcpaBit}{\encScheme}(\indcpaAdversary)$ \\
\hspace{1em} \= \kill
\> $\langle \pubKey, \privKey \rangle \gets \keygen()$ \\
\> $\indcpaAdversaryBit \gets \indcpaAdversary^{\indcpaLROracle{\pubKey}{\cdot}{\cdot}}(\pubKey)$ \\
\> \codeReturn \indcpaAdversaryBit
\end{tabbing}
\end{minipage}
\end{center}
The IND-CPA adversary \indcpaAdversary is given access to a
``left-or-right'' oracle \indcpaLROracle{\pubKey}{\cdot}{\cdot} that
takes two plaintexts $\plaintext{0}, \plaintext{1}$ as inputs and
returns $\encrypt{\pubKey}(\plaintext{\indcpaBit})$. Finally,
\indcpaAdversary returns a bit \indcpaAdversaryBit, which the
experiment returns. We define
\begin{align*}
  &\indcpaAdvantage{\encScheme}(\indcpaAdversary) = \prob{\indcpaExperiment{0}{\encScheme}(\indcpaAdversary) = 0} - \prob{\indcpaExperiment{1}{\encScheme}(\indcpaAdversary) = 0} \\
  &\indcpaAdvantage{\encScheme}(\timeBoundCPA, \lrQueryBound) =
  \max_{\indcpaAdversary} \indcpaAdvantage{\encScheme}(\indcpaAdversary)
\end{align*}
where the maximum is taken over all IND-CPA adversaries
\indcpaAdversary running in time \timeBoundCPA and making up to
\lrQueryBound oracle queries.

ElGamal encryption~\cite{elgamal1985:public-key} can be used to
instantiate the homomorphic encryption scheme \encScheme
in~\secref{sec:amnesia:remote:protocol:crypto}. Given are a cyclic
group \elgGroup of prime order
\plaintextGroupOrder and a generator \elgGroupGenerator of \elgGroup.
\begin{itemize}[nosep,leftmargin=1em,labelwidth=*,align=left]
\item $\keygen()$ returns a key pair $\langle \pubKey, \privKey
\rangle$, including a private key $\privKey =
  \langle \elgPrivKey\rangle$ and a public key $\pubKey
  = \langle \elgGroup, \elgGroupGenerator, \elgPubKey \rangle$, where
  $\elgPrivKey \getsr \residues{\plaintextGroupOrder}$ and
  $\elgPubKey \gets \elgGroupGenerator^{\elgPrivKey}$.

\item $\encrypt{\langle\elgGroup,
  \elgGroupGenerator,\elgPubKey\rangle}(\plaintext)$ returns
  $\langle\elgEphemeralPubKey{}, \elgCiphertext{} \rangle$ where
  $\elgEphemeralPubKey{} \gets
  \elgGroupGenerator^{\elgEphemeralPrivKey{}}$,
  $\elgEphemeralPrivKey{} \getsr \residues{\plaintextGroupOrder}$, and
  $\elgCiphertext{} \gets \plaintext
  \elgPubKey^{\elgEphemeralPrivKey{}}$ for a plaintext $\plaintext \in \elgGroup$.

\item $\decrypt{\langle \elgPrivKey \rangle}(\langle\elgEphemeralPubKey{}, \elgCiphertext{} \rangle)$ returns
  $\plaintext \gets \elgCiphertext{}\elgEphemeralPubKey{}^{-\elgPrivKey}$.

\item $\langle \elgEphemeralPubKey{1}, \elgCiphertext{1} \rangle
  \encMult{\langle \elgGroup, \elgGroupGenerator, \elgPubKey\rangle}
  \langle \elgEphemeralPubKey{2}, \elgCiphertext{2} \rangle$ returns
  $\langle \elgEphemeralPubKey{1} \elgEphemeralPubKey{2}
  \elgGroupGenerator^\elgGroupExponent, \elgCiphertext{1}
  \elgCiphertext{2} \elgPubKey^\elgGroupExponent\rangle$ for
  $\elgGroupExponent \getsr \residues{\plaintextGroupOrder}$ if
  $\{\elgEphemeralPubKey{1}, \elgCiphertext{1},
  \elgEphemeralPubKey{2}, \elgCiphertext{2}\} \subseteq \elgGroup$ and
  returns $\bot$ otherwise.
\end{itemize}

The IND-CPA security of ElGamal encryption was proved by Tsiounis and
Yung~\cite{tsiounis1998:elgamalcpa}.  Our protocols descriptions leave
implicit the checks needed to determine whether ciphertexts are
well-formed, but \propref{prop:ciphertextSpace} indicates that these
are trivial.

\begin{prop}
  For ElGamal encryption,
  $\ciphertextSpace{\langle \elgGroup, \elgGroupGenerator,
    \elgPubKey\rangle}$ $=$ $\elgGroup \times \elgGroup$.
    \label{prop:ciphertextSpace}
\end{prop}

\begin{proof}
$\ciphertextSpace{\langle \elgGroup, \elgGroupGenerator, \elgPubKey\rangle}$
$\subseteq$ $\elgGroup \times \elgGroup$ follows from the fact
that \elgGroup is a cyclic group. Also, given that for any
$\langle \elgEphemeralPubKey{}, \elgCiphertext{}\rangle
\in \elgGroup \times \elgGroup$,
there exists $ \plaintext \in \elgGroup$ such that $\plaintext
= \elgCiphertext{}\elgEphemeralPubKey{}^{-\elgPrivKey}$ for
$\elgPrivKey \in \residues{\plaintextGroupOrder}$. Therefore
$\langle \elgEphemeralPubKey{}, \elgCiphertext{}\rangle \in
\ciphertextSpace{\langle \elgGroup, \elgGroupGenerator,
\elgPubKey\rangle}(\plaintext)$.
\end{proof}

In addition, ElGamal is well-known to be multiplicatively homomorphic,
which is confirmed in \propref{prop:homomorphic}.

\begin{prop}
For ElGamal encryption, if
$\ciphertext{1} \in \ciphertextSpace{\langle \elgGroup, \elgGroupGenerator, \elgPubKey\rangle}(\plaintext{1})$
and
$\ciphertext{2} \in \ciphertextSpace{\langle \elgGroup, \elgGroupGenerator, \elgPubKey\rangle}(\plaintext{2})$, then
$\ciphertext{1} \encMult{\langle \elgGroup, \elgGroupGenerator, \elgPubKey\rangle} \ciphertext{2}$
is uniformly distributed in
$\ciphertextSpace{\langle \elgGroup, \elgGroupGenerator, \elgPubKey\rangle}(\plaintext{1}\plaintext{2})$.
\label{prop:homomorphic}
\end{prop}

\begin{proof}
Let $\ciphertext{1}
= \langle \elgEphemeralPubKey{1}, \elgCiphertext{1} \rangle$ and
$\ciphertext{2}
= \langle \elgEphemeralPubKey{2}, \elgCiphertext{2} \rangle$. For
$\elgEphemeralPubKey{} = \elgEphemeralPubKey{1} \elgEphemeralPubKey{2}
= \elgGroupGenerator^{\elgEphemeralPrivKey{1}+\elgEphemeralPrivKey{2}}$,
$\elgCiphertext{}
= \plaintext{1}\plaintext{2} \elgPubKey^{\elgEphemeralPrivKey
{1}+\elgEphemeralPrivKey{2}}$, and for any
$\elgGroupExponent \getsr \residues{\plaintextGroupOrder}$, $$\langle \elgEphemeralPubKey{1}, \elgCiphertext{1} \rangle \encMult{\langle \elgGroup, \elgGroupGenerator, \elgPubKey\rangle} \langle \elgEphemeralPubKey{2}, \elgCiphertext{2} \rangle
= \langle \elgEphemeralPubKey{}\elgGroupGenerator^\elgGroupExponent, \elgCiphertext{}\elgPubKey^\elgGroupExponent\rangle$$
which is a re-randomization of
$\langle \elgEphemeralPubKey{}, \elgCiphertext{}\rangle$ and is
uniformly distributed in
$\ciphertextSpace{\langle \elgGroup, \elgGroupGenerator, \elgPubKey\rangle}(\plaintext{1}\plaintext{2})$.
\end{proof}

\subsubsection{Noninteractive zero-knowledge proofs}
\label{app:analysis:primitives:zkp}

Our protocol in \secref{sec:amnesia:remote} additionally leverages a noninteractive
zero-knowledge proof of membership for an NP language \lang,
implemented by scheme $\zkpName = (\zkpGen$, \zkpVerify, $\zkpSim)$,
in the random oracle model~\cite{bellare1993:oracles}.
\zkpSim offers two interfaces, denoted \zkpSimHash and \zkpSimProve,
that share state between them.  Let \relation be the witness relation
for \lang, and let \randomOracles denote the set of all functions from
$\{0, 1\}^{\ast}$ to $\{0,1\}^{\infty}$.  On input
$(\statement, \witness) \in \relation$ and with access to a random
oracle $\randomOracle \getsr \randomOracles$,
$\zkpGen[\witness]^{\randomOracle}(\statement)$ produces a proof \zkp
(using the witness \witness) that $\statement \in \lang$, so that if
$\zkp \gets \zkpGen[\witness]^{\randomOracle}(\statement)$ then
$\zkpVerify^{\randomOracle}(\statement, \zkp)$ returns true.  We
reduce the security of our protocol to the following adversary
advantages against \zkpName.

\paragraph{Soundness advantage}
\label{app:analysis:primitives:zkp:soundness}
For any \zkpGenAlt, the soundness advantage is
\[
\zkpSoundnessAdvantage{\zkpName}(\zkpGenAlt) =
\max_{\statement \not\in \lang}
\prob{\zkpVerify^{\randomOracle}(\statement, \zkpGenAlt[\bot]^{\randomOracle}(\statement))}
\]
For any time \timeBoundSound, the soundness advantage is
\[
\zkpSoundnessAdvantage{\zkpName}(\timeBoundSound) =
\max_{\zkpGenAlt} \zkpSoundnessAdvantage{\zkpName}(\zkpGenAlt)
\]
where the maximum is taken over all algorithms \zkpGenAlt running in
time \timeBoundSound.

\paragraph{Distinguishing advantage}
\label{app:analysis:primitives:zkp:zeroknowledge}
We define a distinguishing adversary to be an algorithm \zkpAdversary
that can participate in either of the experiments
\zkpExperiment{\zkpBit}{\zkpName} defined below:

\smallskip

\begin{minipage}[t]{0.4\columnwidth}
\begin{tabbing}
\hspace{1em} \= \hspace{1em} \= \kill
$\codeExpt \zkpExperiment{0}{\zkpName}(\zkpAdversary)$ \\
\> $(\statement, \witness) \getsr \relation$ \\
\> $\randomOracle \getsr \randomOracles$ \\
\> $\zkp \gets \zkpGen[\witness]^{\randomOracle}(\statement)$ \\
\> $\zkpAdversaryBit \gets \zkpAdversary^{\randomOracle}(\statement, \zkp)$ \\
\> \codeReturn \zkpAdversaryBit
\end{tabbing}
\end{minipage}
\hfill
\begin{minipage}[t]{0.4\columnwidth}
\begin{tabbing}
\hspace{1em} \= \hspace{1em} \= \kill
$\codeExpt \zkpExperiment{1}{\zkpName}(\zkpAdversary)$ \\
\> $(\statement, \witness) \getsr \relation$ \\
\> $\zkp \gets \zkpSimProve(\statement)$ \\
\> $\zkpAdversaryBit \gets \zkpAdversary^{\zkpSimHash}(\statement, \zkp)$ \\
\> \codeReturn \zkpAdversaryBit
\end{tabbing}
\end{minipage}

\smallskip

In words, the adversary \zkpAdversary must distinguish between a real
proof output from $\zkpGen[\witness]^{\randomOracle}(\statement)$ and a
proof output from the simulator $\zkpSimProve(\statement)$ without
knowledge of the witness \witness but with the ability to implement
the hash function \zkpSimHash.  This permits \zkpSim to leverage the
standard technique of ``backpatching'' the random oracle outputs on
inputs that \zkpAdversary has not yet queried.  We define
\begin{align*}
  &\zkpAdvantage{\zkpName}(\zkpAdversary) = \prob{\zkpExperiment{1}{\zkpName}
  (\zkpAdversary) = 1} - \prob{\zkpExperiment{0}{\zkpName}(\zkpAdversary) = 1} \\
  &\zkpAdvantage{\zkpName}(\timeBoundZKP, \roQueryBound) =
  \max_{\zkpAdversary} \zkpAdvantage{\zkpName}(\zkpAdversary)
\end{align*}
where the maximum is taken over all adversaries
\zkpAdversary making \roQueryBound random oracle queries and running in
time \timeBoundZKP.

Our implementation leverages a zero-knowledge proof of the equality of
discrete logarithms, due to Chaum and
Pedersen~\cite{chaum1992:wallet}.  More specifically, this technique
demonstrates that an ElGamal ciphertext
$\langle \elgEphemeralPubKey{}, \elgCiphertext{} \rangle$ satisfies
$\langle \elgEphemeralPubKey{}, \elgCiphertext{} \rangle \in \ciphertextSpace{\langle \elgGroup, \elgGroupGenerator, \elgPubKey\rangle}(\elgGroupGenerator)$
by proving $\log_{\elgGroupGenerator} (\elgEphemeralPubKey{})
= \log_{\elgPubKey{}} (\elgCiphertext{}\elgGroupGenerator^{-1})$ in
zero knowledge, and similarly for a ciphertext
$\langle \elgEphemeralPubKey{}, \elgCiphertext{} \rangle$ satisfying
$\langle \elgEphemeralPubKey{}, \elgCiphertext{} \rangle \in \ciphertextSpace{\langle \elgGroup, \elgGroupGenerator, \elgPubKey\rangle}(\elgGroupGenerator^{-1})$.
We combine these zero-knowledge proof techniques to demonstrate only
that
$\langle \elgEphemeralPubKey{}, \elgCiphertext{} \rangle \in \ciphertextSpace{\langle \elgGroup, \elgGroupGenerator, \elgPubKey\rangle}(\elgGroupGenerator)
\cup \ciphertextSpace{\langle \elgGroup, \elgGroupGenerator, \elgPubKey\rangle}(\elgGroupGenerator^{-1})$
in zero knowledge using a technique due to Cramer et
al.~\cite{cramer1994:proofs}.

\subsection{Security against a Malicious \targetSite}
\label{app:analysis:target}

In this section we prove that a malicious \targetSite learns
nothing about \loginPassword from the response computed by an honest
\monitorSite unless \targetSite already guessed
$\pwdHashFn(\loginPassword)$, in the sense that $\pwdHashFn(\loginPassword) \bfIn \langle \bfHashFnSet, \bfIndicesAlt
\rangle$ for the Bloom filter $\langle \bfHashFnSet, \bfIndicesAlt
\rangle$ encoded in its request.  An important premise here is that
\pubKey is a valid public key of the underlying cryptosystem, which is
implicitly assumed to be verified by \monitorSite upon receiving
\msgref{prot:deploy:request}, and that the ciphertexts received in
\msgref{prot:deploy:request} are valid for the cryptosystem;
for the cryptosystem used in our implementation, this can be easily
verified (see \propref{prop:ciphertextSpace}).  Then, the following
propositions show that a malicious \targetSite learns nothing
about \loginPassword if any of the following three conditions is not
satisfied:
$\{\ciphertext{\ciphertextIdx}\}_{\ciphertextIdx=1}^{\bfHashFnRange}
\subseteq \ciphertextSpace{\pubKey}(\plaintextGroupGenerator)
\cup \ciphertextSpace{\pubKey}(\plaintextGroupGenerator^{-1})$,
$\setSize{\bfIndicesAlt} = \bfNmbrIndices$, or
$\pwdHashFn(\loginPassword) \bfIn \langle \bfHashFnSet, \bfIndicesAlt
\rangle$.

\begin{prop}
If $\langle\pubKey,
\bfHashFnSet, \bfNmbrIndices,
\{\ciphertext{\ciphertextIdx}\}_{\ciphertextIdx=1}^{\bfHashFnRange},
\zkp\rangle$, where
$\{\ciphertext{\ciphertextIdx}\}_{\ciphertextIdx=1}^{\bfHashFnRange} \not\subseteq
\left(\ciphertextSpace{\pubKey}(\plaintextGroupGenerator)
\cup \ciphertextSpace{\pubKey}(\plaintextGroupGenerator^{-1})\right)$, is produced by a \targetSite-adversary \targetAdversary running
in time \timeBoundSound, then
\monitorSite fails to abort in \lineref{prot:deploy:zkpVerify}
with probability at most
$\zkpSoundnessAdvantage{\zkpName}(\timeBoundSound)$.
\end{prop}

\begin{proof}
This is immediate from the definition of soundness advantage.
\end{proof}

\begin{prop}
Suppose \monitorSite receives
$\langle\pubKey, \bfHashFnSet, \bfNmbrIndices, \{\ciphertext{\ciphertextIdx}\}_{\ciphertextIdx=1}^{\bfHashFnRange}, \zkp\rangle$
where
$\{\ciphertext{\ciphertextIdx}\}_{\ciphertextIdx=1}^{\bfHashFnRange} \subseteq \ciphertextSpace{\pubKey}(\plaintextGroupGenerator) \cup \ciphertextSpace{\pubKey}(\plaintextGroupGenerator^{-1})$, and let $\bfIndicesAlt = \{\ciphertextIdx :
\ciphertext{\ciphertextIdx}
\in \ciphertextSpace{\pubKey}(\plaintextGroupGenerator)\}$. 
If $\setSize{\bfIndicesAlt} \neq \bfNmbrIndices$, then for any
$\plaintext, \plaintextAlt \in \plaintextGroup$,
\begin{align*}
\cprob{\Bigg}{\begin{array}{@{}r@{}}\responseCiphertext{0} \in \ciphertextSpace{\pubKey}(\plaintext)\\
\wedge~\responseCiphertext{1} \in \ciphertextSpace{\pubKey}(\plaintextAlt)\end{array}}
  {\begin{array}{@{}r@{}}\{\ciphertext{\ciphertextIdx}\}_{\ciphertextIdx=1}^{\bfHashFnRange} \subseteq \ciphertextSpace{\pubKey}(\plaintextGroupGenerator)
\cup \ciphertextSpace{\pubKey}(\plaintextGroupGenerator^{-1}) \\
\wedge~\setSize{\bfIndicesAlt} \neq \bfNmbrIndices\end{array}} \le \frac{1}{\plaintextGroupOrder-1}
\end{align*}
\label{prop:bfNmbrIndices}
\end{prop}

\begin{proof}
First note that $\responseCiphertext{0}
\in \ciphertextSpace{\pubKey}(\plaintextGroupIdentity) \Leftrightarrow
\responseCiphertext{1} \in \ciphertextSpace{\pubKey}(\loginPassword)$.
So, it suffices to quantify the probability in the proposition for the
cases $\plaintext = \plaintextGroupIdentity \wedge \plaintextAlt =
\loginPassword$ and $\plaintext \neq \plaintextGroupIdentity \wedge
\plaintextAlt \neq \loginPassword$.  If
$\{\ciphertext{\ciphertextIdx}\}_{\ciphertextIdx=1}^{\bfHashFnRange}
\subseteq \ciphertextSpace{\pubKey}(\plaintextGroupGenerator)
\cup \ciphertextSpace{\pubKey}(\plaintextGroupGenerator^{-1})$ but
$\setSize{\bfIndicesAlt} \neq \bfNmbrIndices$, then
$\derivedCiphertext{0} \in \ciphertextSpace{\pubKey}
\setminus \ciphertextSpace{\pubKey}(\plaintextGroupIdentity)$ in
\lineref{prot:deploy:monitorSave}.  We consider two cases.
\begin{itemize}[nosep,leftmargin=1em,labelwidth=*,align=left]
\item First suppose $\derivedCiphertext{1}
  \in \ciphertextSpace{\pubKey}
  \setminus \ciphertextSpace{\pubKey}(\plaintextGroupIdentity)$.  For
  the case $\plaintext = \plaintextGroupIdentity$ and $\plaintextAlt =
  \loginPassword$,
\begin{align*}
    &\cprob{\Bigg}{\begin{array}{@{}r@{}}\responseCiphertext{0} \in \ciphertextSpace{\pubKey}(\plaintextGroupIdentity)\\
\wedge~\responseCiphertext{1} \in \ciphertextSpace{\pubKey}(\loginPassword)\end{array}}
  {\begin{array}{@{}r@{}}
      \derivedCiphertext{0} \in \ciphertextSpace{\pubKey}
      \setminus \ciphertextSpace{\pubKey}(\plaintextGroupIdentity) \\
      \wedge~\derivedCiphertext{1} \in \ciphertextSpace{\pubKey}
      \setminus \ciphertextSpace{\pubKey}(\plaintextGroupIdentity)
      \end{array}} \\
  & = \sum_{\plaintextAltAlt \in \plaintextGroup\setminus\{\plaintextGroupIdentity\}}
  \cprob{\Bigg}{\begin{array}{@{}r@{}}
      \encRand{\pubKey}(\derivedCiphertext{0}) \in \ciphertextSpace{\pubKey}(\plaintextAltAlt)\\
      \wedge~\encRand{\pubKey}(\derivedCiphertext{1}) \in \ciphertextSpace{\pubKey}(\plaintextAltAlt^{-1})
      \end{array}}
        {\begin{array}{@{}r@{}}
            \derivedCiphertext{0} \in \ciphertextSpace{\pubKey}
            \setminus \ciphertextSpace{\pubKey}(\plaintextGroupIdentity) \\
            \wedge~\derivedCiphertext{1} \in \ciphertextSpace{\pubKey}
            \setminus \ciphertextSpace{\pubKey}(\plaintextGroupIdentity)
        \end{array}} \\
        & = (\plaintextGroupOrder-1) \frac{1}{\plaintextGroupOrder-1} \frac{1}{\plaintextGroupOrder-1} = \frac{1}{\plaintextGroupOrder-1}
\end{align*}
And for any $\plaintext \neq \plaintextGroupIdentity$ and
$\plaintextAlt \neq \loginPassword$,
\begin{align*}
    &\cprob{\Bigg}{\begin{array}{@{}r@{}}\responseCiphertext{0} \in \ciphertextSpace{\pubKey}(\plaintext)\\
\wedge~\responseCiphertext{1} \in \ciphertextSpace{\pubKey}(\plaintextAlt)\end{array}}
  {\begin{array}{@{}r@{}}
      \derivedCiphertext{0} \in \ciphertextSpace{\pubKey}
      \setminus \ciphertextSpace{\pubKey}(\plaintextGroupIdentity) \\
      \wedge~\derivedCiphertext{1} \in \ciphertextSpace{\pubKey}
      \setminus \ciphertextSpace{\pubKey}(\plaintextGroupIdentity)
      \end{array}} \\
  & = \sum_{\plaintextAltAlt \in \plaintextGroup \setminus \{\plaintextGroupIdentity, \plaintext\}} \!\!
  \cprob{\vast}{\begin{array}{@{}r@{}}
      \encRand{\pubKey}(\derivedCiphertext{0}) \in
      \ciphertextSpace{\pubKey}(\plaintextAltAlt)\\
      \wedge~\encRand{\pubKey}(\derivedCiphertext{1})
      \in \ciphertextSpace{\pubKey}(\plaintext \plaintextGroupOperator \plaintextAltAlt^{-1})\\
      \wedge~\encRand{\pubKey}(\responseCiphertext{0}) \in \ciphertextSpace{\pubKey}(\plaintextAlt \plaintextGroupOperator \loginPassword^{-1}) \end{array}}
        {\begin{array}{@{}r@{}}
            \derivedCiphertext{0} \in \ciphertextSpace{\pubKey}
            \setminus \ciphertextSpace{\pubKey}(\plaintextGroupIdentity) \\
            \wedge~\derivedCiphertext{1} \in \ciphertextSpace{\pubKey}
            \setminus \ciphertextSpace{\pubKey}(\plaintextGroupIdentity)
        \end{array}} \\
     & = (\plaintextGroupOrder-2) \frac{1}{\plaintextGroupOrder-1} \frac{1}{\plaintextGroupOrder-1} \frac{1}{\plaintextGroupOrder-1} = \frac{\plaintextGroupOrder-2}{(\plaintextGroupOrder-1)^3}
\end{align*}

\item Now suppose $\derivedCiphertext{1}
  \in \ciphertextSpace{\pubKey}(\plaintextGroupIdentity)$.  In this
  case $\plaintext = \plaintextGroupIdentity$ is not possible, since
  $\encRand{\pubKey}(\derivedCiphertext{0}) \encMult{\pubKey}
  \encRand{\pubKey}(\derivedCiphertext{1}) \in
  \ciphertextSpace{\pubKey} \setminus
  \ciphertextSpace{\pubKey}(\plaintextGroupIdentity)$.  For any
  $\plaintext \neq \plaintextGroupIdentity$ and $\plaintextAlt \neq
  \loginPassword$,
\begin{align*}
  &\cprob{\Bigg}{\begin{array}{@{}r@{}}\responseCiphertext{0} \in \ciphertextSpace{\pubKey}(\plaintext)\\
\wedge~\responseCiphertext{1} \in \ciphertextSpace{\pubKey}(\plaintextAlt)\end{array}}
  {\begin{array}{@{}r@{}}
      \derivedCiphertext{0} \in \ciphertextSpace{\pubKey}
      \setminus \ciphertextSpace{\pubKey}(\plaintextGroupIdentity) \\
      \wedge~\derivedCiphertext{1} \in \ciphertextSpace{\pubKey}(\plaintextGroupIdentity)
  \end{array}} \\
  & = \cprob{\Bigg}{\begin{array}{@{}r@{}}
      \encRand{\pubKey}(\derivedCiphertext{0}) \in
      \ciphertextSpace{\pubKey}(\plaintext)\\
      \wedge~\encRand{\pubKey}(\responseCiphertext{0}) \in \ciphertextSpace{\pubKey}(\plaintextAlt \plaintextGroupOperator \loginPassword^{-1}) \end{array}}
        {\begin{array}{@{}r@{}}
            \derivedCiphertext{0} \in \ciphertextSpace{\pubKey}
            \setminus \ciphertextSpace{\pubKey}(\plaintextGroupIdentity) \\
            \wedge~\derivedCiphertext{1} \in \ciphertextSpace{\pubKey}(\plaintextGroupIdentity)
        \end{array}} \\
     & = \frac{1}{\plaintextGroupOrder-1} \frac{1}{\plaintextGroupOrder-1} = \frac{1}{(\plaintextGroupOrder-1)^{2}}
\end{align*}
\end{itemize}
\end{proof}

\begin{prop}
Suppose \monitorSite receives
$\langle\pubKey, \bfHashFnSet, \bfNmbrIndices, \{\ciphertext{\ciphertextIdx}\}_{\ciphertextIdx=1}^{\bfHashFnRange}, \zkp\rangle$
where
$\{\ciphertext{\ciphertextIdx}\}_{\ciphertextIdx=1}^{\bfHashFnRange} \subseteq \ciphertextSpace{\pubKey}(\plaintextGroupGenerator) \cup \ciphertextSpace{\pubKey}(\plaintextGroupGenerator^{-1})$,
and let $\bfIndicesAlt = \{\ciphertextIdx
: \ciphertext{\ciphertextIdx} \in \ciphertextSpace{\pubKey}(\plaintextGroupGenerator)\}$.
If $\setSize{\bfIndicesAlt} = \bfNmbrIndices$ but
$\pwdHashFn(\loginPassword) \bfNotIn \langle \bfHashFnSet, \bfIndicesAlt \rangle$,
then for any $\plaintext, \plaintextAlt \in \plaintextGroup$,
\[
\cprob{\Bigg}{\begin{array}{@{}r@{}}\responseCiphertext{0} \in \ciphertextSpace{\pubKey}(\plaintext)\\
\wedge~\responseCiphertext{1} \in \ciphertextSpace{\pubKey}(\plaintextAlt)\end{array}}
  {\begin{array}{@{}r@{}}\{\ciphertext{\ciphertextIdx}\}_{\ciphertextIdx=1}^{\bfHashFnRange} \subseteq \ciphertextSpace{\pubKey}(\plaintextGroupGenerator)
\cup \ciphertextSpace{\pubKey}(\plaintextGroupGenerator^{-1}) \\
\wedge~\setSize{\bfIndicesAlt} = \bfNmbrIndices \wedge
\pwdHashFn(\loginPassword) \bfNotIn \langle \bfHashFnSet, \bfIndicesAlt
\rangle\end{array}}
\le \frac{1}{(\plaintextGroupOrder-1)^2}
\]
\end{prop}

\begin{proof}
First note that $\responseCiphertext{0}
\in \ciphertextSpace{\pubKey}(\plaintextGroupIdentity) \Leftrightarrow
\responseCiphertext{1} \in \ciphertextSpace{\pubKey}(\loginPassword)$.
So, it suffices to quantify the probability in the proposition for the
cases $\plaintext = \plaintextGroupIdentity \wedge \plaintextAlt =
\loginPassword$ and $\plaintext \neq \plaintextGroupIdentity \wedge
\plaintextAlt \neq \loginPassword$. If
$\{\ciphertext{\ciphertextIdx}\}_{\ciphertextIdx=1}^{\bfHashFnRange}
\subseteq \ciphertextSpace{\pubKey}(\plaintextGroupGenerator)
\cup \ciphertextSpace{\pubKey}(\plaintextGroupGenerator^{-1})$ and
$\setSize{\bfIndicesAlt} = \bfNmbrIndices$ but
$\pwdHashFn(\loginPassword) \bfNotIn \langle \bfHashFnSet,
\bfIndicesAlt \rangle$, then $\derivedCiphertext{0}
\in \ciphertextSpace{\pubKey}(\plaintextGroupIdentity)$ but in
\lineref{prot:respond:derived}, $\derivedCiphertext{1}
\in \ciphertextSpace{\pubKey}(\plaintextGroupGenerator^{-2
  \setSize{\bfHashFnSet(\pwdHashFn(\loginPassword)) \setminus
    \bfIndicesAlt}})$ and so $\derivedCiphertext{1}
\in \ciphertextSpace{\pubKey}\setminus
\ciphertextSpace{\pubKey}(\plaintextGroupIdentity)$.  In this case
$\plaintext = \plaintextGroupIdentity$ is not possible, since
$\encRand{\pubKey}(\derivedCiphertext{0}) \encMult{\pubKey}
\encRand{\pubKey}(\derivedCiphertext{1}) \in
  \ciphertextSpace{\pubKey} \setminus
  \ciphertextSpace{\pubKey}(\plaintextGroupIdentity)$.  For
  any $\plaintext \neq \plaintextGroupIdentity$ and $\plaintextAlt
  \neq \loginPassword$,
\begin{align*}
  &\cprob{\Bigg}{\begin{array}{@{}r@{}}\responseCiphertext{0} \in \ciphertextSpace{\pubKey}(\plaintext)\\
\wedge~\responseCiphertext{1} \in \ciphertextSpace{\pubKey}(\plaintextAlt)\end{array}}
  {\begin{array}{@{}r@{}}
      \derivedCiphertext{0} \in \ciphertextSpace{\pubKey}(\plaintextGroupIdentity) \\
      \wedge~\derivedCiphertext{1} \in \ciphertextSpace{\pubKey} \setminus \ciphertextSpace{\pubKey}(\plaintextGroupIdentity)
  \end{array}} \\
  & = \cprob{\Bigg}{\begin{array}{@{}r@{}}
      \encRand{\pubKey}(\derivedCiphertext{1}) \in
      \ciphertextSpace{\pubKey}(\plaintext)\\
      \wedge~\encRand{\pubKey}(\responseCiphertext{0}) \in \ciphertextSpace{\pubKey}(\plaintextAlt \plaintextGroupOperator \loginPassword^{-1}) \end{array}}
        {\begin{array}{@{}r@{}}
            \derivedCiphertext{0} \in \ciphertextSpace{\pubKey}(\plaintextGroupIdentity) \\
            \wedge~\derivedCiphertext{1} \in \ciphertextSpace{\pubKey}
            \setminus \ciphertextSpace{\pubKey}(\plaintextGroupIdentity)
        \end{array}} \\
     & = \frac{1}{\plaintextGroupOrder-1} \frac{1}{\plaintextGroupOrder-1} = \frac{1}{(\plaintextGroupOrder-1)^{2}}
\end{align*}
\end{proof}

A malicious \targetSite without knowledge
of $\pwdHashFn(\loginPassword)$ can increase the probability of
$\pwdHashFn(\loginPassword) \bfIn \langle
\bfHashFnSet, \bfIndicesAlt \rangle$ by increasing \setSize{\bfIndicesAlt}
(or $\bfNmbrIndices$, as the malicious \targetSite must ensure
$\setSize {\bfIndicesAlt} = \bfNmbrIndices$,
by \propref{prop:bfNmbrIndices}). In practice, \monitorSite could
determine an acceptable threshold and drop monitoring requests for
which \bfNmbrIndices exceeds that threshold.

\subsection{Security against a malicious \monitorSite}
\label{app:analysis:monitor}

We need to show that \msgref{prot:deploy:request} does not leak
information about \targetSite's input \bfIndices (except its size
\bfNmbrIndices), assuming \targetSite is honest.  More precisely, we
consider the following experiment to characterize success of a
malicious \monitorSite in distinguishing between two Bloom filters
$\langle \bfHashFnSet, \bfIndices{0}\rangle$ and $\langle
\bfHashFnSet, \bfIndices{1}\rangle$ for \bfIndices{0}, \bfIndices{1}
of its own choosing (but of the same size, and each containing
$\bfHashFnSet(\userPassword)$ for the user-chosen password
\userPassword), based on \msgref{prot:deploy:request}.
\begin{center}
\begin{minipage}{0.6\columnwidth}
\begin{tabbing}
\codeExpt $\monitorExperiment{\monitorExptBit}{}(\langle\monitorAdversary[1], \monitorAdversary[2]\rangle)$ \\
\hspace{1em} \= \kill
\> $\randomOracle \getsr \randomOracles$ \\
\> $\langle \bfIndices{0}, \bfIndices{1}, \monitorAdversaryState \rangle \gets \monitorAdversary[1]^{\randomOracle}(\bfHashFnSet)$ \\
\> $\langle \pubKey, \bfHashFnSet, \bfNmbrIndices, \{\ciphertext{\ciphertextIdx}\}_
{\ciphertextIdx=1}^{\bfHashFnRange}, \zkp \rangle\gets \targetAlgorithms{\ref{prot:deploy:keygen}}{\ref{prot:deploy:zkp}}(\langle \bfHashFnSet, \bfIndices{\monitorExptBit} \rangle)$ \\
\> $\monitorAdversaryBit \gets \monitorAdversary[2]^{\randomOracle}(\langle\pubKey, \bfHashFnSet,
\bfNmbrIndices, \{\ciphertext{\ciphertextIdx}\}_{\ciphertextIdx=1}^{\bfHashFnRange}, \zkp\rangle, \monitorAdversaryState)$ \\
\> \codeReturn \monitorAdversaryBit
\end{tabbing}
\end{minipage}
\end{center}
Here, \targetAlgorithms{\ref{prot:deploy:keygen}}{\ref{prot:deploy:zkp}}
denotes \linesref{prot:deploy:keygen}{prot:deploy:zkp}
in \figref{fig:monitor:deploy}, and \bfHashFnSet is assumed to be
sampled according the Bloom filter algorithm's specification. We
define the \monitorSite-adversary advantage as
\begin{align*}
   \monitorAdvantage{\protocolName}(\monitorAdversary) &= \prob{\monitorExperiment{0}{}(\monitorAdversary) = 0} - \prob{\monitorExperiment{1}{}(\monitorAdversary) = 0} \\[6pt]
   \monitorAdvantage{\protocolName}(\timeBound, \roQueryBound) &= \max_{\monitorAdversary} \monitorAdvantage
  {}(\monitorAdversary)
\end{align*}
where the maximum is taken over all adversaries \monitorAdversary that make at
most \roQueryBound random oracle queries and execute in time \timeBound.
In proving time bounds on adversaries, we ignore constant terms.

\begin{prop}
  \label{prop:monitor}
  \[
  \monitorAdvantage{\protocolName}(\timeBound, \roQueryBound) \le
  \bfHashFnRange \cdot \indcpaAdvantage{\encScheme}(\timeBoundCPA,
  \lrQueryBound) + 2 \cdot \zkpAdvantage{\zkpName}(\timeBoundZKP, \roQueryBoundZKP)~,
  \]
  where $\lrQueryBound = 1$, $\roQueryBoundZKP \le \roQueryBound$,
  $\timeBoundCPA \le \timeBound + \bfHashFnRange\cdot\timeEnc +
  \roQueryBound\cdot\timeHash$, $\timeBoundZKP \le \timeBound +
  \bfHashFnRange\cdot\timeEnc + \roQueryBound\cdot\timeHash$, and
  \timeEnc and \timeHash are the times for one invocation of
  \encrypt{} and \randomOracle, respectively.
\end{prop}

\begin{proof}
Since \monitorSite
observes ciphertexts
$\{\ciphertext{\ciphertextIdx}\}_{\ciphertextIdx=1}^{\bfHashFnRange}$ and a
noninteractive zero-knowledge proof
\zkp, and other components including
\pubKey and \bfHashFnSet that do not depend on \bfIndices, 
we reduce the advantage of a \monitorSite-adversary \monitorAdversary
to the IND-CPA advantage of the encryption scheme \encScheme and the
distinguishing advantage of the noninteractive zero-knowledge
proof \zkp.

\begin{figure*}
\begin{center}
\begin{tabular}{ll}
\begin{minipage}[t]{0.25\textwidth}
\begin{tabbing}
\codeExpt $\monitorExperiment{00}{}(\langle\monitorAdversary[1], \monitorAdversary[2]\rangle)$ \\
\hspace{1em} \= \hspace{1em} \= \kill
\> $\randomOracle \getsr \randomOracles$ \\
\> $\langle \bfIndices{0}, \bfIndices{1}, \monitorAdversaryState \rangle \gets \monitorAdversary[1]^{\randomOracle}(\bfHashFnSet)$ \\
\> \codeIf $\setSize{\bfIndices{0}} \neq \setSize{\bfIndices{1}}$ \codeThen \codeReturn 0 \\
\> $\langle \pubKey, \bfHashFnSet, \bfNmbrIndices, \{\ciphertext{\ciphertextIdx}\}_{\ciphertextIdx=1}^{\bfHashFnRange} \rangle\gets \targetAlgorithms{\ref{prot:deploy:keygen}}{\ref{prot:deploy:ctexts}}(\langle \bfHashFnSet, \bfIndices{0} \rangle)$ \\
\> $\zkp \gets \zkpGen^{\randomOracle}(\langle\pubKey, \{\ciphertext{\ciphertextIdx}\}_{\ciphertextIdx=1}^{\bfHashFnRange}\rangle)$ \\
\> $\monitorAdversaryBit \gets \monitorAdversary[2]^{\randomOracle}(\langle\pubKey, \bfHashFnSet, \bfNmbrIndices, \{\ciphertext{\ciphertextIdx}\}_{\ciphertextIdx=1}^{\bfHashFnRange}, \zkp\rangle, \monitorAdversaryState)$ \\
\> \codeReturn \monitorAdversaryBit
\end{tabbing}
\end{minipage}
&
\begin{minipage}[t]{0.2\textwidth}
\begin{tabbing}
\codeExpt $\monitorExperiment{01}{}(\langle\monitorAdversary[1], \monitorAdversary[2]\rangle)$ \\
\hspace{1em} \= \hspace{1em} \= \kill
\> $\langle \bfIndices{0}, \bfIndices{1}, \monitorAdversaryState \rangle \gets \monitorAdversary[1]^{\zkpSimHash}(\bfHashFnSet)$ \\
\> \codeIf $\setSize{\bfIndices{0}} \neq \setSize{\bfIndices{1}}$ \codeThen \codeReturn 0 \\
\> $\langle \pubKey, \bfHashFnSet, \bfNmbrIndices, \{\ciphertext{\ciphertextIdx}\}_{\ciphertextIdx=1}^{\bfHashFnRange} \rangle\gets \targetAlgorithms{\ref{prot:deploy:keygen}}{\ref{prot:deploy:ctexts}}(\langle \bfHashFnSet, \bfIndices{0} \rangle)$ \\
\> $\zkp \gets \zkpSimProve(\langle\pubKey, \{\ciphertext{\ciphertextIdx}\}_
{\ciphertextIdx=1}^{\bfHashFnRange}\rangle)$  \\
\> $\monitorAdversaryBit \gets \monitorAdversary[2]^{\zkpSimHash}(\langle\pubKey, \bfHashFnSet, \bfNmbrIndices, \{\ciphertext{\ciphertextIdx}\}_{\ciphertextIdx=1}^{\bfHashFnRange}, \zkp\rangle, \monitorAdversaryState)$ \\
\> \codeReturn \monitorAdversaryBit
\end{tabbing}
\end{minipage}
\\
\\
\begin{minipage}[t]{0.2\textwidth}
\begin{tabbing}
\codeExpt $\monitorExperiment{10}{}(\langle\monitorAdversary[1], \monitorAdversary[2]\rangle)$ \\
\hspace{1em} \= \hspace{1em} \= \kill
\> $\randomOracle \getsr \randomOracles$ \\
\> $\langle \bfIndices{0}, \bfIndices{1}, \monitorAdversaryState \rangle \gets \monitorAdversary[1]^{\randomOracle}(\bfHashFnSet)$ \\
\> \codeIf $\setSize{\bfIndices{0}} \neq \setSize{\bfIndices{1}}$ \codeThen \codeReturn 0 \\
\> $\langle \pubKey, \bfHashFnSet, \bfNmbrIndices, \{\ciphertext{\ciphertextIdx}\}_{\ciphertextIdx=1}^{\bfHashFnRange} \rangle\gets \targetAlgorithms{\ref{prot:deploy:keygen}}{\ref{prot:deploy:ctexts}}(\langle \bfHashFnSet, \bfIndices{1} \rangle)$ \\
\> $\zkp \gets \zkpGen^{\randomOracle}(\langle\pubKey, \{\ciphertext{\ciphertextIdx}\}_{\ciphertextIdx=1}^{\bfHashFnRange}\rangle)$ \\
\> $\monitorAdversaryBit \gets \monitorAdversary[2]^{\randomOracle}(\langle\pubKey, \bfHashFnSet, \bfNmbrIndices, \{\ciphertext{\ciphertextIdx}\}_{\ciphertextIdx=1}^{\bfHashFnRange}, \zkp\rangle, \monitorAdversaryState)$ \\
\> \codeReturn \monitorAdversaryBit
\end{tabbing}
\end{minipage}
&
\begin{minipage}[t]{0.25\textwidth}
\begin{tabbing}
\codeExpt $\monitorExperiment{11}{}(\langle\monitorAdversary[1], \monitorAdversary[2]\rangle)$ \\
\hspace{1em} \= \hspace{1em} \= \kill
\> $\langle \bfIndices{0}, \bfIndices{1}, \monitorAdversaryState \rangle \gets \monitorAdversary[1]^{\zkpSimHash}(\bfHashFnSet)$ \\
\> \codeIf $\setSize{\bfIndices{0}} \neq \setSize{\bfIndices{1}}$ \codeThen \codeReturn 0 \\
\> $\langle \pubKey, \bfHashFnSet, \bfNmbrIndices, \{\ciphertext{\ciphertextIdx}\}_{\ciphertextIdx=1}^{\bfHashFnRange} \rangle\gets \targetAlgorithms{\ref{prot:deploy:keygen}}{\ref{prot:deploy:ctexts}}(\langle \bfHashFnSet, \bfIndices{1} \rangle)$ \\
\> $\zkp \gets \zkpSimProve(\langle\pubKey, \{\ciphertext{\ciphertextIdx}\}_
{\ciphertextIdx=1}^{\bfHashFnRange}\rangle)$  \\
\> $\monitorAdversaryBit \gets \monitorAdversary[2]^{\zkpSimHash}(\langle\pubKey, \bfHashFnSet, \bfNmbrIndices, \{\ciphertext{\ciphertextIdx}\}_{\ciphertextIdx=1}^{\bfHashFnRange}, \zkp\rangle, \monitorAdversaryState)$ \\
\> \codeReturn \monitorAdversaryBit
\end{tabbing}
\end{minipage}
\end{tabular}
\end{center}
\caption{Definition of $\monitorExperiment{\worldBitOne\worldBitTwo}{}(\monitorAdversary)$ for $\worldBitOne, \worldBitTwo \in \{0,1\}$}
\label{fig:hybrid}
\end{figure*}

To do this, let $\monitorAdversary
= \langle\monitorAdversary[1], \monitorAdversary[2]\rangle$ be
a \monitorSite-adversary and define a sequence of experiments
for \monitorAdversary
\[
\monitorExperiment{00}{}(\monitorAdversary),~~\monitorExperiment{01}{}(\monitorAdversary),~~\monitorExperiment{11}{}(\monitorAdversary),~~\monitorExperiment{10}{}(\monitorAdversary)
\]
where we associate each pair $\worldBitOne, \worldBitTwo$ of bits to a
$(\worldBitOne, \worldBitTwo)$ hybrid experiment as shown
in \figref{fig:hybrid}.  In words, hybrid experiment
$\monitorExperiment{\worldBitOne\worldBitTwo}{}(\monitorAdversary)$
produces ciphertexts
$\{\ciphertext{\ciphertextIdx}\}_{\ciphertextIdx=1}^{\bfHashFnRange}$
based on $\langle \bfHashFnSet, \bfIndices{\worldBitOne} \rangle$ and,
depending on whether $\worldBitTwo = 0$ or $\worldBitTwo = 1$,
generates a real\footnote{\zkpGen leverages the witness produced by
\targetAlgorithms{\ref{prot:deploy:keygen}}{\ref{prot:deploy:ctexts}}
to do so.} \zkp or a simulated \zkp for the statement that
$\{\ciphertext{\ciphertextIdx}\}_{\ciphertextIdx=1}^{\bfHashFnRange} \subseteq
\ciphertextSpace{\pubKey}(\plaintextGroupGenerator) \cup
\ciphertextSpace{\pubKey}(\plaintextGroupGenerator^{-1})$.
So if we let $\probHybrid{\worldBitOne,\worldBitTwo} =
\prob{\monitorExperiment{\worldBitOne\worldBitTwo}{}(\monitorAdversary)=0}$
for bits $\worldBitOne, \worldBitTwo \in \{0, 1\}$, then it will be
the case that
\begin{align*}
\probHybrid{0, 0} &= \prob{\monitorExperiment{00}{}(\monitorAdversary)=0} = \prob{\monitorExperiment{0}{}(\monitorAdversary)=0} \\
\probHybrid{1, 0} &= \prob{\monitorExperiment{10}{}(\monitorAdversary)=0}= \prob{\monitorExperiment{1}{}(\monitorAdversary)=0}
\end{align*}
and so we have:
\begin{align}
&\monitorAdvantage{\protocolName}(\monitorAdversary) \nonumber\\
&= \prob{\monitorExperiment{0}{}(\monitorAdversary) = 0} - \prob{\monitorExperiment{1}{}(\monitorAdversary) = 0} \nonumber \\
&= \probHybrid{0, 0} - \probHybrid{1, 0} \nonumber\\
&= \probHybrid{0, 0} \!-\! \probHybrid{0, 1} \!+\! \probHybrid{0, 1} \!-\! \probHybrid{1, 1} \!+\! \probHybrid{1, 1} \!-\! \probHybrid{1, 0} 
\label{eqn:monitoradv}
\end{align}

We now show that:
\begin{align}
\probHybrid{0, 1} - \probHybrid{1, 1} &\le \bfHashFnRange
\cdot \indcpaAdvantage{\encScheme}(\timeBoundCPA, \lrQueryBound) \label{eqn:cpa} \\
\probHybrid{0, 0} - \probHybrid{0, 1} &\le
\zkpAdvantage{\zkpName}(\timeBoundZKP, \roQueryBoundZKP) \label{eqn:zkp1} \\
\probHybrid{1, 1} - \probHybrid{1, 0} &\le
\zkpAdvantage{\zkpName}(\timeBoundZKP, \roQueryBoundZKP) \label{eqn:zkp2}~,
\end{align}
where, for all adversaries \monitorAdversary that make at most \roQueryBound
random oracle queries and execute in total time \timeBound, we have
$\lrQueryBound = 1$, $\roQueryBoundZKP \le \roQueryBound$,
$\timeBoundCPA \le \timeBound + \bfHashFnRange\cdot\timeEnc +
\roQueryBound\cdot\timeHash$, and $\timeBoundZKP \le \timeBound +
\bfHashFnRange\cdot\timeEnc + \roQueryBound\cdot\timeHash$. So,
combining \eqnref{eqn:cpa}, \eqnref{eqn:zkp1}, and \eqnref{eqn:zkp2} with
\eqnref{eqn:monitoradv}, we have $$\monitorAdvantage{\protocolName}(\timeBound,
\roQueryBound) \le \bfHashFnRange \cdot
\indcpaAdvantage{\encScheme}(\timeBoundCPA, \lrQueryBound) + 2 \cdot
\zkpAdvantage{\zkpName}(\timeBoundZKP, \roQueryBoundZKP)~.$$

\paragraph*{Justification of \eqnref{eqn:cpa}}

Given \monitorSite-adversary $\monitorAdversary
= \langle \monitorAdversary[1], \monitorAdversary[2] \rangle$, we
construct an IND-CPA adversary \indcpaAdversary attacking the IND-CPA
experiment \indcpaExperiment {\indcpaBit}{\encScheme} defined
in \secref{app:analysis:primitives:enc}. \indcpaAdversary first
invokes \monitorAdversary[1] with \bfHashFnSet, servicing its oracle
queries using \zkpSimHash, and receives \bfIndices{0}
and \bfIndices{1} (aborting if they are of unequal size)
from \monitorAdversary[1].  \indcpaAdversary then sets
$\plaintext{0\ciphertextIdx} \gets \plaintextGroupGenerator$ if
$\ciphertextIdx \in \bfIndices{0}$ and
$\plaintext{0\ciphertextIdx} \gets \plaintextGroupGenerator^{-1}$
otherwise, and
$\plaintext{1\ciphertextIdx} \gets \plaintextGroupGenerator$ if
$\ciphertextIdx \in \bfIndices{1}$ and
$\plaintext{1\ciphertextIdx} \gets \plaintextGroupGenerator^{-1}$
otherwise.  \indcpaAdversary chooses an index uniformly at random
$\hybridIndex \getsr \{1, ..., \bfHashFnRange\}$ and computes
$\{\ciphertext{\ciphertextIdx}\}_ {\ciphertextIdx=1}^{\bfHashFnRange}$
as follows:
\begin{itemize}[nosep,leftmargin=1em,labelwidth=*,align=left]
  \item For $\ciphertextIdx < \hybridIndex$, \indcpaAdversary computes
  $\ciphertext{\ciphertextIdx} \gets
  \encrypt{\pubKey}(\plaintext{0\ciphertextIdx})$
  \item For $\ciphertextIdx = \hybridIndex$, \indcpaAdversary queries
  its oracle and obtains $\ciphertext{\ciphertextIdx}
  \gets
  \indcpaLROracle{\pubKey}{\plaintext{0\ciphertextIdx}}{\plaintext{1\ciphertextIdx}}$
  \item For $\ciphertextIdx > \hybridIndex$, \indcpaAdversary computes
  $\ciphertext{\ciphertextIdx} \gets
  \encrypt{\pubKey}(\plaintext{1\ciphertextIdx})$
\end{itemize}
\indcpaAdversary also executes
$\zkpSimProve(\langle\pubKey, \{\ciphertext{\ciphertextIdx}\}_
{\ciphertextIdx=1}^{\bfHashFnRange}\rangle)$
and generates a simulated noninteractive zero-knowledge proof \zkp for
$\{\ciphertext{\ciphertextIdx}\}_{\ciphertextIdx=1}^{\bfHashFnRange} \subseteq \ciphertextSpace{\pubKey}(\plaintextGroupGenerator) \cup \ciphertextSpace{\pubKey}(\plaintextGroupGenerator^{-1})$. Then \indcpaAdversary
invokes \monitorAdversary[2] with
$\langle\pubKey$, \bfHashFnSet, \bfNmbrIndices,
$\{\ciphertext{\ciphertextIdx}\}_{\ciphertextIdx=1}^{\bfHashFnRange}$,
$\zkp \rangle$ and services its
oracle queries using \zkpSimHash.  Finally \indcpaAdversary returns
the bit \monitorAdversaryBit returned by \monitorAdversary[2] as its
guess \indcpaAdversaryBit. Here we use $\hybridIndex \in \{1,
..., \bfHashFnRange\}$ to index the experiment simulated
by \indcpaAdversary for $\monitorExperiment{\indcpaBit
1}{\hybridIndex}(\monitorAdversary)$ and we
let $\hybridCPA{\indcpaBit, \hybridIndex} =
\monitorExperiment{\indcpaBit 1}{\hybridIndex}(\monitorAdversary)$.
Then, $\hybridCPA{0, \bfHashFnRange} \distEqual
\monitorExperiment{01}{}(\monitorAdversary)$, $\hybridCPA{1, 1}
\distEqual \monitorExperiment{11}{}(\monitorAdversary)$ and, for $2
\le \hybridIndex \le \bfHashFnRange$, $\hybridCPA{0,\hybridIndex-1}
\distEqual \hybridCPA{1,\hybridIndex}$.
Given some fixed choice of $\hybridIndex =
\hybridIndex^*$, the simulation provided by
\indcpaAdversary to \monitorAdversary is perfectly
indistinguishable from a real execution in $\hybridCPA{\indcpaBit, \hybridIndex^*}$.
Since \indcpaAdversary outputs $0$ if \monitorAdversary outputs $0$, we
have:
\begin{align*}
  &\prob{\indcpaExperiment{\indcpaBit}{\encScheme}(\indcpaAdversary) = 0} \\
  &= \sum_{\hybridIndex^*=1}^{\bfHashFnRange} \prob{\hybridIndex = \hybridIndex^*} \cdot \prob{\indcpaExperiment{\indcpaBit}{\encScheme}(\indcpaAdversary) = 0 \wedge \hybridIndex = \hybridIndex^*} \\
  &= \sum_{\hybridIndex^*=1}^{\bfHashFnRange} \frac{1}{\bfHashFnRange}
  \cdot \prob{\hybridCPA{\indcpaBit, \hybridIndex^*} = 0}~.
\end{align*}
Combining the above, we have:
\begin{align*}
  &\indcpaAdvantage{\encScheme}(\indcpaAdversary) \\
  &= \prob{\indcpaExperiment{0}{\encScheme}(\indcpaAdversary) = 0} - \prob{\indcpaExperiment{1}{\encScheme}(\indcpaAdversary) = 0} \\
  &\ge \sum_{\hybridIndex^*=1}^{\bfHashFnRange} \frac{1}{\bfHashFnRange}
  \cdot \Big(\prob{\hybridCPA{0, \hybridIndex^*} = 0}
  - \prob{\hybridCPA{1, \hybridIndex^*} = 0}\Big) \\ 
  &\ge \frac{1}{\bfHashFnRange} \cdot \Big(\prob{\hybridCPA{0, \bfHashFnRange}
  = 0} - \prob{\hybridCPA{1, 1} = 0}\Big) \\
  &\quad+ \sum_{\hybridIndex^*=2}^{\bfHashFnRange} \frac{1}{\bfHashFnRange}
  \cdot \Big(\prob{\hybridCPA{0, \hybridIndex^*-1} = 0}
  - \prob{\hybridCPA{1, \hybridIndex^*} = 0}\Big) \\
  &\ge \frac{1}{\bfHashFnRange} \cdot \Big(\prob{\hybridCPA{0, \bfHashFnRange}
  = 0} - \prob{\hybridCPA{1, 1} = 0}\Big) + 0 \\
  &\ge \frac{1}{\bfHashFnRange} \cdot \bigg(\prob{\monitorExperiment{01}
  {}(\monitorAdversary)
  = 0} - \prob{\monitorExperiment{11}{}(\monitorAdversary) = 0}\bigg)\\
  &\ge \frac{1}{\bfHashFnRange} \cdot \Big(\probHybrid{0, 1} - \probHybrid{1, 1}\Big)~.
\end{align*}
Here \indcpaAdversary makes one ``left-or-right'' oracle query in
constructing
$\{\ciphertext{\ciphertextIdx}\}_{\ciphertextIdx=1}^{\bfHashFnRange}$,
and runs in time at most $\timeBound + \bfHashFnRange\cdot\timeEnc +
\roQueryBound\cdot\timeHash$ which is the time for
\monitorAdversary plus the time to produce \bfHashFnRange ciphertexts
and answer at most \roQueryBound random oracle queries from
\monitorAdversary.

\paragraph*{Justification of \eqnref{eqn:zkp1} and \eqnref{eqn:zkp2}} 

Given an
\monitorSite-adversary $\monitorAdversary = \langle
\monitorAdversary[1], \monitorAdversary[2]
\rangle$ for the experiments \monitorExperiment{0\zkpBit}{}, we construct an
adversary \zkpAdversary for noninteractive zero-knowledge experiment
\zkpExperiment{\zkpBit}{\zkpName} defined in
\secref{app:analysis:primitives:zkp:zeroknowledge}. \zkpAdversary first invokes
\monitorAdversary[1] with 
\bfHashFnSet and receives two distinct Bloom filters $\langle \bfHashFnSet,
\bfIndices{0}\rangle$ and $\langle
\bfHashFnSet, \bfIndices{1}\rangle$ (of equal size, \bfNmbrIndices) from \monitorAdversary[1].
\zkpAdversary receives, from either \zkpGen (if $\zkpBit = 0$) or
\zkpSimProve (if $\zkpBit = 1$), a proof \zkp for the statement that $\{\ciphertext{\ciphertextIdx}\}_{\ciphertextIdx=1}^{\bfHashFnRange}
\subseteq \ciphertextSpace{\pubKey}(\plaintextGroupGenerator) \cup
\ciphertextSpace{\pubKey}(\plaintextGroupGenerator^{-1})$, as well as
\pubKey and $\{\ciphertext{\ciphertextIdx}\}_{\ciphertextIdx=1}^{\bfHashFnRange}$ produced
based on \bfIndices{0} as part of the public information for \zkp. Then
\zkpAdversary invokes \monitorAdversary[2] with $\langle
\pubKey, \bfHashFnSet, \bfNmbrIndices,
\{\ciphertext{\ciphertextIdx}\}_{\ciphertextIdx=1}^{\bfHashFnRange}, \zkp
\rangle$ as its expected input. Also, \zkpAdversary uses its random
oracle to reply to all random oracle queries by
\monitorAdversary[2] to verify \zkp (as in
\lineref{prot:deploy:zkpVerify} in \figref{fig:monitor:deploy}).
Finally, \zkpAdversary outputs $0$ if
\monitorAdversary[2] outputs $0$. Since the simulation provided by
\zkpAdversary to \monitorAdversary is perfectly
indistinguishable from a real execution in $\monitorExperiment
{0\zkpBit}{}(\monitorAdversary)$, we have:
\begin{align*}
&\zkpAdvantage{\zkpName}(\zkpAdversary) \\
&= \prob{\zkpExperiment{1}{\zkpName}(\zkpAdversary) = 1} - \prob{\zkpExperiment
{0}{\zkpName}(\zkpAdversary) = 1} \\
&= \prob{\zkpExperiment{0}{\zkpName}(\zkpAdversary) = 0} - \prob{\zkpExperiment
{1}{\zkpName}(\zkpAdversary) = 0} \\
&\ge \prob{\monitorExperiment
{00}{}(\monitorAdversary) = 0} - \prob{\monitorExperiment
{01}{}(\monitorAdversary) = 0} \\
&\ge \probHybrid{0, 0} - \probHybrid{0, 1}~.
\end{align*}

Similarly, we can construct a \zkpAdversaryAlt like \zkpAdversary,
except that it receives from \zkpGen (if $\zkpBit = 0$) or
\zkpSimProve (if $\zkpBit = 1$) a proof \zkp corresponding to
\bfIndices{1}, instead of \bfIndices{0}, and that outputs $1$ if
\monitorAdversary outputs $0$. So:
\begin{align*}
&\zkpAdvantage{\zkpName}(\zkpAdversaryAlt) \\
&= \prob{\zkpExperiment{1}{\zkpName}(\zkpAdversaryAlt) = 1} - \prob{\zkpExperiment{0}{\zkpName}(\zkpAdversaryAlt) = 1} \\
&\ge \prob{\monitorExperiment
{11}{}(\monitorAdversary) = 0} - \prob{\monitorExperiment
{10}{}(\monitorAdversary) = 0} \\
&\ge \probHybrid{1, 1} - \probHybrid{1, 0}~.
\end{align*}

Here \zkpAdversary and \zkpAdversaryAlt make at most \roQueryBound random oracle queries and
run in time at most $\timeBound + \bfHashFnRange\cdot\timeEnc +
\roQueryBound\cdot\timeHash$ which is the time-complexity of
\monitorAdversary plus the time costs of producing \bfHashFnRange ciphertexts
and answering at most \roQueryBound random oracle queries from
\monitorAdversary.
\end{proof}

%% file: main.bbl
\begin{thebibliography}{10}
\providecommand{\url}[1]{#1}
\csname url@samestyle\endcsname
\providecommand{\newblock}{\relax}
\providecommand{\bibinfo}[2]{#2}
\providecommand{\BIBentrySTDinterwordspacing}{\spaceskip=0pt\relax}
\providecommand{\BIBentryALTinterwordstretchfactor}{4}
\providecommand{\BIBentryALTinterwordspacing}{\spaceskip=\fontdimen2\font plus
\BIBentryALTinterwordstretchfactor\fontdimen3\font minus
  \fontdimen4\font\relax}
\providecommand{\BIBforeignlanguage}[2]{{%
\expandafter\ifx\csname l@#1\endcsname\relax
\typeout{** WARNING: IEEEtranS.bst: No hyphenation pattern has been}%
\typeout{** loaded for the language `#1'. Using the pattern for}%
\typeout{** the default language instead.}%
\else
\language=\csname l@#1\endcsname
\fi
#2}}
\providecommand{\BIBdecl}{\relax}
\BIBdecl

\bibitem{akshima2019:honeywords}
Akshima, D.~Chang, A.~Goel, S.~Mishra, and S.~K. Sanadhya, ``Generation of
  secure and reliable honeywords, preventing false detection,'' \emph{IEEE
  Transactions on Dependable and Secure Computing}, vol.~16, no.~5, 2019.

\bibitem{almeshekah2015:ersatz}
M.~H. Almeshekah, C.~N. Gutierrez, M.~J. Atallah, and E.~H. Spafford,
  ``{ErsatzPasswords}: Ending password cracking and detecting password
  leakage,'' in \emph{31\textsuperscript{st} Annual Computer Security
  Applications Conference}, Dec. 2015.

\bibitem{bellare1998:relations}
M.~Bellare, A.~Desai, D.~Pointcheval, and P.~Rogaway, ``Relations among notions
  of security for public-key encryption schemes,'' in \emph{Advances in
  Cryptology -- CRYPTO 1998}, ser. LNCS, vol. 1462, Aug. 1998.

\bibitem{bellare1993:oracles}
M.~Bellare and P.~Rogaway, ``Random oracles are practical: A paradigm for
  designing efficient protocols,'' in \emph{1\textsuperscript{st} ACM
  Conference on Computer and Communications Security}, Nov. 1993.

\bibitem{bloom1970:space}
B.~H. Bloom, ``Space/time trade-offs in hash coding with allowable errors,''
  \emph{Communications of the ACM}, vol.~13, no.~7, Jul. 1970.

\bibitem{bonneau2012:guessing}
J.~Bonneau, ``The science of guessing: analyzing an anonymized corpus of 70
  million passwords,'' in \emph{33\textsuperscript{th} IEEE Symposium on
  Security and Privacy}, May 2012.

\bibitem{certicom2000:sec2v1}
{Certicom Research}, ``{SEC} 2: Recommended elliptic curve domain parameters,''
  \url{http://www.secg.org/SEC2-Ver-1.0.pdf}, 2000, standards for Efficient
  Cryptography.

\bibitem{chakraborty2016:honeyword}
N.~Chakraborty and S.~Mondal, ``Toward improving storage cost and security
  features of honeyword based approaches,'' \emph{Procedia Computer Science},
  vol.~93, 2016.

\bibitem{chatterjee2016:typo}
R.~Chatterjee, A.~Athayle, D.~Akhawe, A.~Juels, and T.~Ristenpart, ``{pASSWORD
  tYPOS} and how to correct them securely,'' in \emph{37\textsuperscript{th}
  IEEE Symposium on Security and Privacy}, May 2016.

\bibitem{chaum1992:wallet}
D.~Chaum and T.~P. Pedersen, ``Wallet databases with observers,'' in
  \emph{Advances in Cryptology -- CRYPTO'92}, ser. LNCS, vol. 740, 1993.

\bibitem{cramer1994:proofs}
R.~Cramer, I.~Damg{\aa}rd, and B.~Schoenmakers, ``Proofs of partial knowledge
  and simplified design of witness hiding protocols,'' in \emph{Advances in
  Cryptology -- CRYPTO '94}, ser. LNCS, vol. 839, 1994.

\bibitem{culafi2021:mandiant}
A.~Culafi, ``{Mandiant}: Compromised {Colonial Pipeline} password was reused,''
  \url{https://www.techtarget.com/searchsecurity/news/252502216/Mandiant-Compromised-Colonial-Pipeline-password-was-reused},
  09 Jun. 2021.

\bibitem{das2014:tangled}
A.~Das, J.~Bonneau, M.~Caesar, N.~Borisov, and X.~Wang, ``The tangled web of
  password reuse,'' in \emph{21\textsuperscript{st} ISOC Network and
  Distributed System Security Symposium}, 2014.

\bibitem{deblasio2017:tripwire}
J.~{DeBlasio}, S.~Savage, G.~M. Voelker, and A.~C. Snoeren, ``Tripwire:
  Inferring internet site compromise,'' in \emph{17\textsuperscript{th}
  Internet Measurement Conference}, Nov. 2017.

\bibitem{dionysiou2022:lethe}
A.~Dionysiou and E.~Athanasopoulos, ``Lethe: Practical data breach detection
  with zero persistent secret state,'' in \emph{7\textsuperscript{th} IEEE
  European Symposium on Security and Privacy}, Jun. 2022.

\bibitem{elgamal1985:public-key}
T.~{ElGamal}, ``A public-key cryptosystem and a signature scheme based on
  discrete logarithms,'' \emph{IEEE Transactions on Information Theory},
  vol.~31, no.~4, 1985.

\bibitem{erguler2016:flatness}
I.~Erguler, ``Achieving flatness: Selecting the honeywords from existing user
  passwords,'' \emph{IEEE Transactions on Parallel and Distributed Systems},
  vol.~13, no.~2, 2016.

\bibitem{fan2014:cuckoo}
B.~Fan, D.~G. Andersen, M.~Kaminsky, and M.~D. Mitzenmacher, ``Cuckoo filter:
  Practically better than {Bloom},'' in \emph{10\textsuperscript{th} ACM
  Conference on Emerging Networking Experiments and Technologies}, 2014.

\bibitem{fiat1986:fiatshamir}
A.~Fiat and A.~Shamir, ``How to prove yourself: Practical solutions to
  identification and signature problems,'' in \emph{Advances in Cryptology --
  CRYPTO 1986}, ser. LNCS, vol. 263, Aug. 1986.

\bibitem{florencio2014:guide}
D.~Flor\^{e}ncio, C.~Herley, and P.~C. {van Oorschot}, ``An administrator's
  guide to internet password research,'' in \emph{28\textsuperscript{th} Large
  Installation System Administration Conference}, Nov. 2014.

\bibitem{freedman2005:oprf}
M.~J. Freedman, Y.~Ishai, B.~Pinkas, and O.~Reingold, ``Keyword search and
  oblivious pseudorandom functions,'' in \emph{2\textsuperscript{nd} Theory of
  Cryptography Conference}, ser. LNCS, vol. 3378, Feb. 2005.

\bibitem{golla2018:psm}
M.~Golla and M.~D\"{u}rmuth, ``On the accuracy of password strength meters,''
  in \emph{25\textsuperscript{th} ACM Conference on Computer and Communications
  Security}, Oct. 2018.

\bibitem{ibm2021:breach}
{IBM Security}, ``Cost of a data breach report 2021,''
  \url{https://www.ibm.com/security/data-breach}, 2021.

\bibitem{ikeda2021:colonial}
S.~Ikeda, ``{Colonial Pipeline} hack connected to password leak of 8.4 billion
  accounts; attackers got in via an old {VPN} account,''
  \url{https://www.cpomagazine.com/cyber-security/colonial-pipeline-hack-connected-to-password-leak-of-8-4-billion-accounts-attackers-got-in-via-an-old-vpn-account/},
  14 Jun. 2021.

\bibitem{juels2013:honeywords}
A.~Juels and R.~L. Rivest, ``Honeywords: Making password-cracking detectable,''
  in \emph{20\textsuperscript{th} ACM Conference on Computer and Communications
  Security}, Nov. 2013.

\bibitem{malone2012:investigating}
D.~Malone and K.~Maher, ``Investigating the distribution of password choices,''
  in \emph{21\textsuperscript{st} International World Wide Web Conference},
  Apr. 2012.

\bibitem{mayer2022:managers}
P.~Mayer, C.~W. Munyendo, M.~L. Mazurek, and A.~J. Aviv, ``Why users (don't)
  use password managers at a large educational institution,'' in
  \emph{31\textsuperscript{st} USENIX Security Symposium}, Aug. 2022.

\bibitem{mazurek2013:guessability}
M.~L. Mazurek, S.~Komanduri, T.~Vidas, L.~Bauer, N.~Christin, L.~F. Cranor,
  P.~G. Kelley, R.~Shay, and B.~Ur, ``Measuring password guessability for an
  entire university,'' in \emph{20\textsuperscript{th} ACM Conference on
  Computer and Communications Security}, Nov. 2013.

\bibitem{mitzenmacher2005:probability}
M.~Mitzenmacher and E.~Upfal, \emph{Probability and Computing: Randomization
  and Probabilistic Techniques in Algorithms and Data Analysis}.\hskip 1em plus
  0.5em minus 0.4em\relax Cambridge University Press, 2005.

\bibitem{pearman2017:habitat}
S.~Pearman, J.~Thomas, P.~E. Naeini, H.~Habib, L.~Bauer, N.~Christin, L.~F.
  Cranor, S.~Egelman, and A.~Forget, ``Let's go in for a closer look: Observing
  passwords in their natural habitat,'' in \emph{24\textsuperscript{th} ACM
  Conference on Computer and Communications Security}, Oct. 2017.

\bibitem{reuter2016:alibaba}
Reuters, ``Hackers attack 20 million accounts on {Alibaba}'s {Taobao} shopping
  site,'' \url{https://www.reuters.com/article/us-alibaba-cyber-idUKKCN0VD14X},
  04 Feb. 2016.

\bibitem{rohatgi2021:webplotdigitizer}
A.~Rohatgi, ``{WebPlotDigitizer}: Version 4.5,''
  \url{https://automeris.io/WebPlotDigitizer}, 2021.

\bibitem{shape2018:spill}
{Shape Security}, ``2018 credential spill report,''
  \url{https://info.shapesecurity.com/rs/935-ZAM-778/images/Shape_Credential_Spill_Report_2018.pdf},
  2018.

\bibitem{tan2020:blocklist}
J.~Tan, L.~Bauer, N.~Christin, and L.~F. Cranor, ``Practical recommendations
  for stronger, more usable passwords combining minimum-strength,
  minimum-length, and blocklist requirements,'' in \emph{27\textsuperscript{th}
  ACM Conference on Computer and Communications Security}, Nov. 2020.

\bibitem{thomas2017:credential}
K.~Thomas, F.~Li, A.~Zand, J.~Barrett, J.~Ranieri, L.~Invernizzi, Y.~Markov,
  O.~Comanescu, V.~Eranti, A.~Moscicki, D.~Margolis, V.~Paxson, and
  E.~Bursztein, ``Data breaches, phishing, or malware? {Understanding} the
  risks of stolen credentials,'' in \emph{24\textsuperscript{th} ACM Conference
  on Computer and Communications Security}, 2017.

\bibitem{tsiounis1998:elgamalcpa}
Y.~Tsiounis and M.~Yung, ``On the security of {ElGamal} based encryption,'' in
  \emph{Public Key Cryptography --- PKC 1998}, ser. LNCS, vol. 1431, 1998.

\bibitem{wang2018:domino}
C.~Wang, S.~T.~K. Jan, H.~Hu, D.~Bossart, and G.~Wang, ``The next domino to
  fall: Empirical analysis of user passwords across online services,'' in
  \emph{8\textsuperscript{th} ACM Conference on Data and Application Security
  and Privacy}, Mar. 2018.

\bibitem{wang2017:zipf}
D.~Wang, H.~Cheng, P.~Wang, X.~Huang, and G.~Jian, ``Zipf’s law in
  passwords,'' in \emph{IEEE Transactions on Information Forensics and
  Security}, vol.~12, no.~11, Nov. 2017.

\bibitem{wang2018:honeywords}
D.~Wang, H.~Cheng, P.~Wang, J.~Yan, and X.~Huang, ``A security analysis of
  honeywords,'' in \emph{25\textsuperscript{th} ISOC Network and Distributed
  System Security Symposium}, Feb. 2018.

\bibitem{wang2016:targeted}
D.~Wang, Z.~Zhang, P.~Wang, J.~Yan, and X.~Huang, ``Targeted online password
  guessing: An underestimated threat,'' in \emph{23\textsuperscript{rd} ACM
  Conference on Computer and Communications Security}, 2016.

\bibitem{wang2022:honeywords}
D.~Wang, Y.~Zou, Q.~Dong, Y.~Song, and X.~Huang, ``How to attack and generate
  honeywords,'' in \emph{43\textsuperscript{rd} IEEE Symposium on Security and
  Privacy}, May 2022.

\bibitem{wang2021:amnesia}
K.~C. Wang and M.~K. Reiter, ``Using {Amnesia} to detect credential database
  breaches,'' in \emph{30\textsuperscript{th} USENIX Security Symposium}, Aug.
  2021.

\bibitem{xu2021:chunk}
M.~Xu, C.~Wang, J.~Yu, J.~Zhang, K.~Zhang, and W.~Han, ``Chunk-level password
  guessing: Towards modeling refined password composition representations,'' in
  \emph{28\textsuperscript{th} ACM Conference on Computer and Communications
  Security}, Nov. 2021.

\end{thebibliography}
